\documentclass[12pt,notitlepage]{report}
\usepackage[cp1250]{inputenc}
\usepackage{a4wide}
\usepackage{latexsym}
\usepackage{graphicx}
\usepackage{amssymb}
\usepackage{amsmath}
\usepackage{mathrsfs}
\usepackage{slashed}
\usepackage{feynmp}
\usepackage{fancyhdr}
\begin{document}

\bibliographystyle{unsrt}
\newcommand{\beq}{\begin{equation}}
\newcommand{\eeq}{\end{equation}}
\newcommand{\bra}[1]{\langle #1|}
\newcommand{\ket}[1]{|#1\rangle}

\begin{titlepage}
\begin{center}
\ \\

\vspace{15mm}

\large
Charles University in Prague\\

\vspace{5mm}

{\Large\bf DIPLOMA THESIS}

\vspace{10mm}

\includegraphics[scale=0.3]{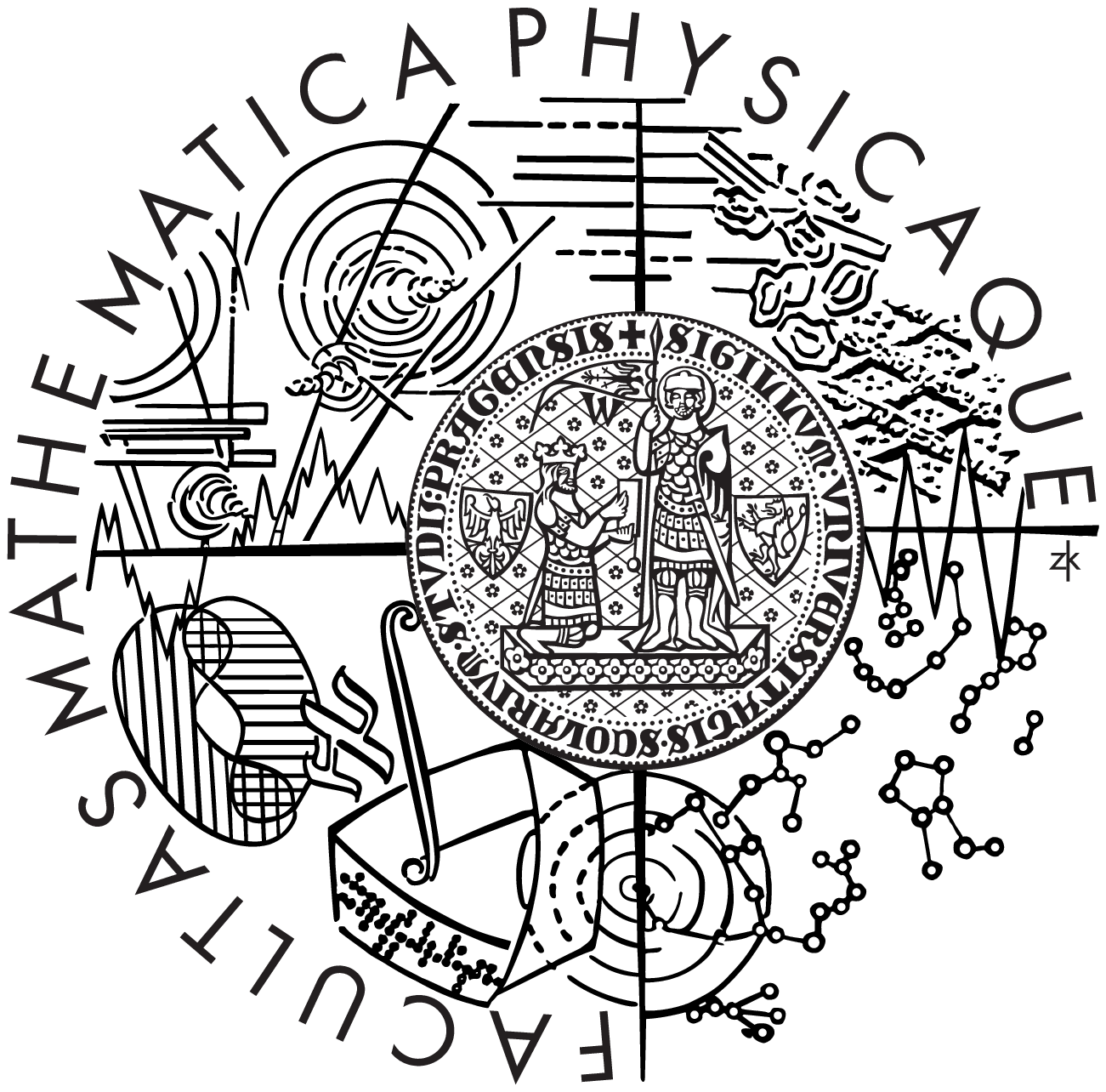} 

\vspace{15mm}

{\Large Sergej Moroz}\\ 
\vspace{5mm}
{\Large\bf Perturbative and non-perturbative aspects \\ of $QED_{3}$}\\ 
\end{center}
\vspace{20mm}

\large
\begin{center}
\noindent Diploma supervisor: Jiří Hošek \end{center} 
\begin{center} Department of Theoretical Physics \end{center} 
\begin{center}Nuclear Physics Institute \end{center} 
\begin{center}Academy of Sciences of the Czech Republic, Řež \end{center}  
\hskip20mm   
\vspace{1mm} 

\hskip20mm 

\vspace{10mm}

\begin{center}
2007 
\end{center}

\end{titlepage} 






\newpage

\pagenumbering{roman}
\tableofcontents

\newpage

\begin{abstract}
In this work we review various findings in the planar quantum physics with the special emphasis on the two-component quantum electrodynamics in three-dimensio\-nal spacetime ($QED_{3}$) with the Chern-Simons (CS) term. First the classical electrodynamics is briefly outlined. Then we construct and classify the unitary irreducible representations of the Poincar\'e group and connect them to various covariant fields. Subtleties with discrete symmetries arising in $QED_{3}$ are also discussed. Quantization of the free gauge field with and without the CS term is performed in Coulomb gauge. Subsequently we discuss the ultraviolet behavior of the theory in perturbative regime and perform some one-loop calculations. The induced CS term is examined using the effective action formalism. The non-relativistic $e^-$-$e^-$ interactions are considered in the MCS theory and the possibility of bound states is briefly discussed. We also examine the non-relativistic scattering of two anyons in the pure CS theory to one-loop order.
\end{abstract}
\chapter*{Acknowledgments}
I would like to express the deepest gratitude to my diploma thesis supervisor Jiří Hošek for countless hours of discussions with me. I am also obliged to him for the permanent support and deep insight in physics, which he shared with me during the last two years. 

I am also indebted to Jiří Hošek, Jiří Adam and Tomáš Brauner for careful reading of the manuscript and important remarks and suggestions.

This work was supported in part by the Institutional Research Plan AV0Z10480505, and by the GACR Grant No. 202/06/0734.

And last but not least I would like to express my personal thanks to my parents and Diana without whose love and patience this work would never have appeared.    

\newpage


\pagestyle{fancy}
\renewcommand{\chaptermark}[1]{\markboth{#1}{}}
\fancyhf{} 
\fancyhead[RE,RO]{\bfseries\thepage}
\fancyhead[LE,LO]{\bfseries\leftmark}
\renewcommand{\headrulewidth}{0.5pt}
\renewcommand{\footrulewidth}{0pt}
\addtolength{\headheight}{0.5pt} 
\fancypagestyle{plain}{%
\fancyhead{} 
\renewcommand{\headrulewidth}{0pt} 
\fancyfoot[C]{\thepage}
}

\pagenumbering{arabic}

\chapter{Introduction}
\label{ch:intro}

Studying physics in two spatial dimensions is a great intellectual challenge. There are numerous differences between the well-known three-dimensional world and the world in the plane. Some distinctions occur already on classical level, but the most drastic differences take place in quantum physics. Due to the odd dimensionality of spacetime novel striking effects are generated in this theory.

In this thesis we review some basic findings made in planar physics. We pay special attention to classical and quantum electrodynamics ($QED_{3}$). It is possible to introduce two types of quantum electrodynamics in the plane, which have different properties. Here we develop the most natural version of $QED_{3}$ from our point of view, which is called the Maxwell-Chern-Simons quantum field theory with two-component fermions \cite{DJT:1981}. There is, however, an alternative formulation of planar $QED_{3}$, where one uses the four-component fermion fields. In this formalism one doubles the number of fermions. Two fermions of two different flavors are described by one four-component object and the Dirac matrices become four dimensional. In this formulation chiral symmetry is possible and one may investigate an interesting question of dynamical chiral-symmetry breaking \cite{Cornwall:1980,Pisarski:1984,Appelquist:2004}.  

Electrodynamics in the plane has many unusual properties already on the classical level. There are some rather obvious ones such as that the magnetic field is a pseudoscalar quantity in this theory and that the static potential is logarithmic, i.e., confining. More subtle distinction is that the well-known Huygens principle is violated in the plane. This fact yields rather interesting consequences.

The basics of relativistic quantum theory in two spatial dimensions are rather unusual. Spin is a pseudoscalar operator and the spin Hilbert space of a particle is always one-dimensional. Free fields acquire different spin quantum numbers with respect to the three-dimensional world. For example, the electron has spin $1/2$, its antiparticle positron has spin $-1/2$ and massless photon has spin $0$. These findings have deep connection with the analysis of unitary irreducible representations of the Poincar\'e group \cite{Binegar:1981}. Another interesting fact is that particles with generalized statistics, called anyons, are theoretically possible in the plane \cite{Dunne:1998,Wilczek:1990}.

$QED_{3}$ is a superrenormalizable field theory, i.e., all couplings have positive dimension of mass and there are only a few Feynman diagrams, which have ultraviolet divergences. These divergences may be cured by the renormalization procedure of parameters of the Lagrangian. However, $QED_{3}$ suffers from severe infrared divergences \cite{Jackiw:1980}.

The most interesting thing about electrodynamics in two spatial dimensions, however, is that it is possible to introduce a new term into the Lagrangian, called  the Chern-Simons term (CS), which is Lorentz-invariant, bilinear in the gauge field and does not spoil the gauge invariance and renormalizability \cite{DJT:1981}. This extended version of electrodynamics is called the Maxwell-Chern-Simons theory (MCS). Classically, this new term generates a magnetic field in the form of a tube in the vicinity of a point charge \cite{Moura-Melo:2000}. It is not straightforward to quantize the Maxwell-Chern-Simons theory because it is a gauge theory and gauge fixing is needed. Working in Coulomb gauge and using Dirac quantization procedure yields the rewarding result- MCS is a theory of massive photons with spin one \cite{Haller:1994}. Massiveness of photon partially solves the infrared problems of $QED_{3}$, but some subtleties persist. One-loop quantum corrections lead to a gauge variance of the physical mass. In order to get rid of all infrared divergences we have to work in a special type of gauge (Landau gauge). It is possible to show that the Chern-Simons term is generated perturbatively by the fermion mass term, even if it is absent in the original Lagrangian \cite{DJT:1981}. One also may investigate the pure CS theory coupled to fermions and it was shown that this theory has anyon degrees of freedom \cite{Dunne:1998}.

Two-body bound states were also studied in the frame of the non-relativistic quantum-mechani\-cal reduction of $QED_{3}$ with the CS term. It was conjectured that the electron-electron bound states are possible for some values of the mass and the coupling parameters \cite{Dobroliubov:1993}.  

Although it is not apparent at the first sight, the planar physics has numerous applications. First, it may be considered as a toy model for investigating confinement \cite{Feynman:1981}. Second, three-dimensional superrenormalizable field theories may be considered as the effective theories of the 4d world at high temperature, i.e., some equilibrium properties of high-temperature 4d field theory may be calculated from the 3d effective theory    \cite{Appelquist:1981,Kajantie:1995}. Third, the pure CS theory, which is anyonic, possibly describes the high $T_{c}$ superconductivity \cite{Lykken:1991,Wilczek:1990}. And finally the most prominent physical application is the quantum Hall effect. Anyons play an important role in explaining the fractional quantum Hall effect \cite{Wilczek:1990}. It is really encouraging that the experimental evidence of the existence of the anyons has been recently published \cite{Goldman:2005, Goldman:2006, Goldman:2007}.

This thesis is organized as follows: In Chapter \ref{ch:classical} we discuss various properties of classical electrodynamics with and without the Chern-Simons term. In Chapter \ref{ch:Lorentz} Lorentz group and Dirac equation are briefly reviewed. In Chapter \ref{ch:Poincare} we closely follow \cite{Binegar:1981} analyzing the unitary irreducible representations of the Poincar\'e group M(2,1). Field quantization of free fields (with main emphasis on the MCS gauge field) is performed in Chapter \ref{ch:quantization}. There are subtleties with discrete symmetries in the planar world and they are considered in Chapter \ref{ch:discrete}. The following Chapter \ref{ch:perturbative} is devoted to some one-loop perturbative calculations in the MCS $QED_{3}$. We follow \cite{Redlich:1983} in elegant calculation of the one-loop effective action employing the "proper time" method due to J. Schwinger in Chapter \ref{ch:effect}. In Chapter \ref{ch:non-perturbative} we have a look at a possible formation of the bound states in the MCS theory \cite{Dobroliubov:1993}. In the subsequent Chapter \ref{ch:CS} spin transmutation (fermions become anyons due to one-loop vertex correction) is examined in the pure CS theory. We also calculate non-relativistic anyon-anyon scattering to the one-loop order and discuss the result. The last Chapter \ref{ch:QHE} is devoted to the quantum Hall effect, arguably the most successful physical application of the planar physics. In the Appendix A some interesting aspects of classical and quantum mechanics are briefly discussed. The Appendix B is devoted to the solution of the Landau problem in the symmetric gauge.
\chapter{Classical physics}
\label{ch:classical}
In order to investigate quantum electrodynamics in three-dimensional spacetime we should first understand the classical version of the theory. This Chapter is devoted to this task.\footnote{ Few words to notation: we introduce a 3-potential $A^{\mu}=(\phi, \mathbf{A})$ and a 3-current $j^{\mu}=(\rho,\mathbf{j})$. We also define the gradient symbol as $\nabla ^{j} \equiv \frac{\partial}{\partial x^{j}}$. The Levi-Civita tensor is defined as totally antisymmetric tensor with $\epsilon^{012}=1$. The permutation symbol in two dimensions is $\epsilon^{jk}\equiv \epsilon^{0jk}$. The metric tensor is $g_{\mu\nu}=diag(1,-1,-1)$.} First the basic properties of the Maxwell electrodynamics are briefly reviewed. We derive the retarded propagator of the theory and demonstrate that the Huygens' principle is violated. Subsequently, we introduce the the Chern-Simons extension of the Maxwell electrodynamics, which is specific for the three-dimensional spacetime. The retarded Green's function and electromagnetic field generated by the static point source are calculated.  
\section{Classical electrodynamics}
The electromagnetic tensor is:
$$ F^{\mu \nu}=\partial^{\mu}A^{\nu}-\partial^{\nu}A^{\mu}=\left( \begin{array}{ccc}  0 & -E^{1} & -E^{2} \\
																					E^{1} & 0 & B \\
																					E^{2} & -B & 0 \\ 

\end{array} \right). $$
Magnetic and electric fields can be expressed as:
\beq \label{eq:CLE0}
B=-\epsilon_{ij}\nabla^{i}A^{j} \qquad E^{i}=-\frac{\partial A^{i}}{\partial t}- \nabla^{i} \phi,
\eeq
thus the magnetic field $B$ is a pseudoscalar field in contrast to four-dimensional world. \\
The Lorentz force equation may be written in the following form:
\beq \label{eq:CLE0a} \dot{p^{i}}=eE^{i}+eB\epsilon^{i}_{\ j}v^{j}. \eeq
The dynamics of the electromagnetic field is described by the well-known relativistic Lagrangian:
\beq \label{eq:ClE0a}
\mathscr{L}=-\frac{1}{4}F^{\mu\nu}F_{\mu\nu}-j_{\mu}A^{\mu}.
\eeq
Using the Euler-Lagrange equations and the antisymmetry property of the electromagnetic tensor $F^{\mu\nu}$ we readily obtain the Maxwell equations in the covariant form:
\beq \label{eq:ClE1}
  \partial_{\mu}F^{\mu \nu}=j^{\nu} \qquad \epsilon^{\nu\rho\sigma}\partial_{\nu}F_{\rho \sigma}=0.
\eeq
In our notation they can be written as three 2-dim equations (there is no analogy of $\mathbf{\nabla}\cdot \mathbf{B}=0$ in two spatial dimensions)\footnote{It is now easy to obtain the well-known results: relativistic invariant is $\mathbf{E}^{2}-B^{2}$, the energy density of the electromagnetic field is $\omega=\frac{\mathbf{E}^{2}+B^{2}}{2}$ and Poynting vector is $S^{i}=B\epsilon^{i}_{\ j}E^{j}$.}:
\beq \label{eq:CLE2}
 \mathbf{\nabla}\cdot \mathbf{E}=\rho \qquad -\epsilon_{j}^{\ \thinspace i}\nabla^{j}B=j^{i}+\frac{\partial{E^{i}}}{\partial t} \qquad -\frac{\partial B}{\partial t}+\epsilon_{jk}\nabla^{j}E^{k}=0.
\eeq

As in the three-dimensional theory the electromagnetic waves can propagate in space. The wave equation is (we use the special Lorentz gauge $\mathbf\nabla\cdot \mathbf{A}=0$ and $\phi=0$):
$$ \Delta \mathbf{A}-\partial_{t}^{2}\mathbf{A}=0. $$
The plane wave is transverse:
$$ \mathbf{E}\cdot \mathbf{n}=0. $$
The magnetic field in the plane wave can be written as:
$$ B=-\epsilon_{ij}E^{j}n^{k}. $$

It is a well known fact that in odd-dimensional spacetimes the Huygens' principle is \emph{violated}. By that we mean that the retarded Green's function $G^{\beta\gamma}_{R}(x,x^{\prime})$ support is not restricted (as a function of $x^{\prime}$ for fixed $x$) to the light cone. As the result there is no analogue of the Lienard-Wiechert formula. Let us show it. In the Maxwell electrodynamics in (2+1) dimensions the retarded Green's function $G_{R}^{\beta\gamma}(x,x^{\prime})$ is a function of only $x-x^{\prime}$, which is due to translation invariance. It also must solve the following equation (with the boundary condition $G^{\beta\gamma}_{R}(x-x^{\prime})=0$ for $t^{\prime}>t$ ):
\beq \label{eq:CLE3}
\left(\Box_{x} g_{\alpha \beta}-\partial_{\alpha}\partial_{\beta}\right) G^{\beta\gamma}_{R}(x-x^{\prime})=-\delta_{\alpha}^{\gamma}\delta(x-x^{\prime}).
\eeq
The operator acting on $G_{R}^{\beta\gamma}(x-x^{\prime})$ is singular, so it is not possible to invert it. Thus we are dealing with a singular theory. This fact forces us to introduce the gauge fixing term (e.g., $-\frac{1}{2\xi}(\partial \cdot A)^{2}$) into the Lagrangian of the theory which makes the operator regular, i.e., invertible:
\beq \label{eq:CLE3a}
\left(\Box_{x}g_{\alpha \beta}-(1-\frac{1}{\xi})\partial_{\alpha}\partial_{\beta}\right) G^{\beta\gamma}_{R}(x-x^{\prime})=-\delta_{\alpha}^{\gamma}\delta(x-x^{\prime}).
\eeq
We must impose a subsidiary condition (in our case it is the Lorentz condition $\partial \cdot A=0$) on the solutions of the field equations to recover the original Maxwell theory.

To solve (\ref{eq:CLE3a}) it is convenient first to write the retarded propagator as a Fourier integral:
\beq \label{eq:CLE4a}
G^{\beta\gamma}_{R}(x-y)=\int\frac{d^{3}p}{(2\pi)^{3}}e^{-ip(x-y)}G^{\beta\gamma}_{R}(p).
\eeq
The differential equation (\ref{eq:CLE3a}) transforms into the algebraic equation for $G^{\beta\gamma}_{R}(p)$:
\beq \label{eq:CLE4b}
\left(p^{2}g_{\alpha\beta}-(1-\frac{1}{\xi})p_{\alpha}p_{\beta}\right)G_{R}^{\beta\gamma}(p)=\delta_{\gamma}^{\alpha}.
\eeq
After inverting (\ref{eq:CLE4b}) we finally obtain:
\beq \label{eq:CLE5}
G_{R}^{\beta\gamma}(p)=\frac{p^{2}g^{\beta\gamma}-p^{\beta}p^{\gamma}}{p^{4}}+\xi\frac{p^{\beta}p^{\gamma}}{p^{4}}.
\eeq
In denominators of (\ref{eq:CLE5}) we understand $p^{2}=(p_{0}+i\epsilon)^{2}-\mathbf{p}^{2}$, where $\epsilon$ is a tiny positive number. This pole prescription imposes the retarded boundary condition. Classically the electromagnetic field of a given current distribution $j^{\mu}(y)$ is (up to a homogeneous solution):
\beq \label{eq:CLE4}
A^{\mu}(x)=-\int d^{3}y G^{\mu\nu}_{R}(x,y) j_{\nu}(y).
\eeq
We can drop terms proportional to $p^{\beta}p^{\gamma}$ in (\ref{eq:CLE5}) in classical calculations because the current $j^{\mu}(x)$ is conserved.
Now we should make the inverse Fourier transformation of (\ref{eq:CLE5}) to get our final result: 
\begin{eqnarray} 
G^{\mu\nu}_{R}(x,x^{\prime})&=&g^{\mu\nu}\underbrace{\int \frac{d^{3}p}{(2\pi)^{3}}\frac{}{}e^{-ip\cdot(x-x^{\prime})}\frac{1}{p^{2}}}_{G^{2+1}(x-x^{\prime})} \nonumber \\
&=&-g^{\mu\nu}\frac{\Theta(x^{0}-(x^{\prime})^{0})}{2\pi}\int_{0}^{\infty}dp J_{0}(p|\mathbf{x}-\mathbf{x^{\prime}}|)\sin(p[x^{0}-(x^{\prime})^{0}]), \label{eq:CLE5a0}
\end{eqnarray}
where $J_{0}(x)$ is a zeroth order Bessel function of the first kind. It is possible to calculate the one-dimensional integral in (\ref{eq:CLE5a0}) and our result is:
\beq \label{eq:CLE5a}
G^{\mu\nu}_{R}(x,x^{\prime})=-\frac{g^{\mu\nu}}{2\pi}\frac{1}{\sqrt{(x-x^{\prime})^{2}}}\Theta(x^{0}-(x^{\prime})^{0}),
\eeq 
for $\sqrt{(x-x^{\prime})^{2}}\geq0$ and $G_{R}(x,x^{\prime})=0$ otherwise. We realize that the retarded Green's function is located not only on the light cone, but also inside of it. Hence there is no analogy of the Lienard-Wiechert potential of a point charge. The field $A^{\mu}(x)$ depends \emph{not only on retarded position} of a point source $x^{\prime}(\tau_{R})$ \footnote{ where $ (x-x^{\prime}(\tau_{R}))$ is a null 3-vector.}
, but on the \emph{whole history} of its motion for $\tau<\tau_{R}$!

Let us now calculate the static Coulomb potential of a point charge $q$ using the method of Green's functions. A point charge moving on the world line $z(t)$ generates the current distribution:
\beq \label{eq:CLE5b} j^{\mu}(y)=q\int^{+\infty}_{-\infty} dt \dot{z}^{\mu}(t)\delta^{(2+1)}(y-z(t)). \eeq
The static potential is created by a charge with 3-velocity $\dot{z}^{\mu}(t)=(1,\mathbf{0})$. Now we can substitute the current distribution (\ref{eq:CLE5b}) into (\ref{eq:CLE4}). The integral diverges so we must regulate it by cutting it off at some large $\tau$. In the limit $\tau\to \infty$ we obtain:
\beq \label{eq:CLE6}
A^{0}(r,t)=-\frac{q}{2\pi}\ln(r)+\frac{q}{2\pi}\lim_{\tau\to\infty}(\ln|\tau+\sqrt{\tau^2-r^{2}}|).
\eeq
We can drop the second divergent term because it does not depend on $r$ and merely shifts the scalar potential. The vector potential $\mathbf{A}$ is absent in the static case.
\section{Maxwell-Chern-Simons classical theory} \label{CMCS}
It is an amazing feature of the three-dimensional spacetime that it is possible to introduce a novel term (called the Chern-Simons term) to the Lagrangian of electrodynamics, which does not violate gauge symmetry \cite{DJT:1981}. Our new Lagrangian is:
\beq \label{eq:MCS1}
\mathscr{L}=-\frac{1}{4}F^{\mu\nu}F_{\mu\nu}+\frac{m}{2}\epsilon^{\mu\nu\rho}A_{\mu}\partial_{\nu}A_{\rho}-j_{\mu}A^{\mu},
\eeq
where $m$ is a real parameter, absolute value of which will be identified with the mass of photon in the quantum version of the theory. This Lagrangian describes the celebrated Maxwell-Chern-Simons (MCS) theory because it contains both the Maxwell and the Chern-Simons terms. The Lagrangian is not gauge invariant but changes up to a divergence under gauge transformations (if the current is conserved). This leads to the gauge-invariant action, assuming that the fields decrease sufficiently fast at infinity and the surface term may be neglected. 

The quantum version of the MCS theory will be of our main interest in this thesis. That it is why it is worth to spend some time with the classical version of this theory. We will follow closely \cite{Moura-Melo:2000}, where this analysis was made. The Euler-Lagrange equations for the Lagrangian (\ref{eq:MCS1}) are:
\beq \label{eq:MCS2}
\partial_{\rho}F^{\rho\nu}+m\epsilon^{\nu\rho\mu}\partial_{\rho}A_{\mu}=j^{\nu}.
\eeq
Assuming that the current is conserved and taking the divergence of (\ref{eq:MCS2}) we get the Bianchi identity for the dual field strength, which alternatively follows from the antisymmetry of $F_{\mu\nu}$:  \beq \label{eq:MCS2a} ^{*}F^{\mu}\equiv \frac{1}{2}\epsilon^{\mu\nu\rho}F_{\nu\rho} \eeq
\beq \label{eq:MCS3}
\partial_{\mu}\ ^{*}F^{\mu}=0.
\eeq 
Let us now construct the retarded Green's function of the MCS theory. We shall follow the same steps as in the pure electrodynamics: add the gauge fixing term $-\frac{1}{2\xi}(\partial \cdot A)^{2}$, perform the Fourier transformation of the field equations, invert the algebraic operator in p-representation. Going through this procedure we obtain the propagator $G_{R}^{\mu\nu}(p)$:
\beq \label{eq:MCS4}
G_{R}^{\mu\nu}(p)=\frac{p^{2}g^{\mu\nu}-p^{\mu}p^{\nu}-im\epsilon^{\mu\nu\rho}p_{\rho}}{p^{2}(p^{2}-m^{2})}+\xi \frac{p^{\mu}p^{\nu}}{p^{4}},
\eeq
where the retarded pole prescription is understood; i.e., $p^{2}=(p_{0}+i\epsilon)^{2}-\mathbf{p}^{2}$ in denominators of (\ref{eq:MCS4}). We can drop the terms proportional to $p_{\mu}p_{\nu}$ as before, but the inverse Fourier transform is more complicated now:
\begin{eqnarray}
G_{R}^{\mu\nu}(x-y)&=& g^{\mu\nu}\underbrace{\int \frac{d^{3}p}{(2\pi)^{3}}e^{-ip\cdot(x-y)}\frac{1}{p^{2}-m^{2}}}_{G^{mass}(x-y)}+ \nonumber \\ &&+\frac{m}{m^{2}}\int\frac{d^{3}p}{(2\pi)^{3}}e^{-ip\cdot(x-y)}\left[\frac{1}{p^{2}}-\frac{1}{p^{2}-m^{2}}\right]\epsilon^{\mu\nu\alpha}(ip_{\alpha}) \nonumber\\
&=&g^{\mu\nu}G^{mass}(x-y)+\frac{m}{m^{2}}\epsilon^{\mu\nu\alpha}(\partial_{y})_{\alpha}\left[G^{2+1}(x-y)-G^{mass}(x-y)\right],  \label{eq:MCS4a}
\end{eqnarray}
where $G^{2+1}(x-y)$ was defined in (\ref{eq:CLE5a0}). The retarded Green's function is used in integrals like (\ref{eq:CLE4}). The current $j_{\nu}(y)$ in (\ref{eq:CLE4}) is usually well localized, so it is possible to perform integration by parts in (\ref{eq:MCS4a}). Our final result is:
\beq \label{eq:MCS5}
G_{R}^{\mu\nu}(x-y)=G^{mass}(x-y)g^{\mu\nu}+\frac{m}{m^{2}}\left[G^{mass}(x-y)-G^{2+1}(x-y)\right]\epsilon^{\mu\nu\rho}(\partial_{y})_{\rho},
\eeq
where $G^{2+1}(x-y)$ was calculated in (\ref{eq:CLE5a}), thus it remains to calculate $G^{mass}(x-y)$ now. From this point let us focus our attention on the case $m>0$:
\beq \label{Gm1}
G^{mass}(t,\mathbf{r})=\int \frac{d^{2}p}{(2\pi)^{2}}e^{i\mathbf{p}\cdot \mathbf{r}}\int\frac{dp^{0}}{2\pi}e^{-ip^{0}t}\frac{1}{p^{2}-m^{2}},
\eeq
where $t=x^{0}-y^{0}$ and $\mathbf{r}=\mathbf{x}-\mathbf{y}$. First, we perform the $p^{0}$ integration. Using the retarded pole prescription we find that the integrand of the inner integral (\ref{Gm1}) has two poles $p^{0}=\pm\sqrt{\mathbf{p}^{2}+m^{2}}-i\epsilon$, which both lie under the real axis. We may enclose the contour of integration in the lower half of the complex $p^{0}$-plane and use the Cauchy theorem. Doing so we obtain the integral, written in polar coordinates as follows:
\beq \label{Gm2}
G^{mass}(t,\mathbf{r})=-\theta(t)\int^{\infty}_{0}\frac{dp}{(2\pi)^{2}}\int_{0}^{2\pi}d\theta e^{ipr\cos\theta}\frac{p}{\sqrt{p^{2}+m^{2}}}\sin(t\sqrt{p^{2}+m^{2}}),
\eeq
where $p\equiv |\mathbf{p}|$ and $r\equiv |\mathbf{r}|$. The angular integration can be easily performed using the following integral representation of the Bessel function:
\beq \label{Gm3}
J_{0}(x)=\frac{1}{2\pi}\int^{2\pi}_{0}d\theta e^{ix\cos{\theta}}.
\eeq
Thus it remains to perform the radial integration:
\beq \label{Gm4}
G^{mass}(t,\mathbf{r})=-\frac{\theta(t)}{2\pi}\int^{\infty}_{0}dp J_{0}(pr)\frac{p}{\sqrt{p^{2}+m^{2}}}\sin(t\sqrt{p^{2}+m^{2}}).
\eeq
To calculate the last integral we use the general result due to Sonine \cite{Watson:1944}:
\begin{eqnarray} 
\int^{\infty}_{0}J_{\mu}(rp)\frac{J_{\nu}(t\sqrt{p^{2}+m^{2}})}{(p^{2}+m^{2})^{\nu/2}}p^{\mu+1}dp&=&0 \qquad t<r \label{Gm5} \\
\int^{\infty}_{0}J_{\mu}(rp)\frac{J_{\nu}(t\sqrt{p^{2}+m^{2}})}{(p^{2}+m^{2})^{\nu/2}}p^{\mu+1}dp&=&\frac{r^{\mu}}{t^{\nu}}\left\{ \frac{\sqrt{t^{2}-r^{2}}}{m} \right\}^{\nu-\mu-1}J_{\nu-\mu-1}\left\{m\sqrt{t^{2}-r^{2}}\right \} \quad t>r, \nonumber
\end{eqnarray}
where $Re(\nu)>Re(\mu)>-1$ and $r$ and $t$ must be positive. Now let us substitute $\mu=0$ and $\nu=1/2$ into the last expression and recall that:
\beq \label{Gm6}
J_{1/2}(x)=\sqrt{\frac{2}{\pi x}}\sin(x) \qquad J_{-1/2}(x)=\sqrt{\frac{2}{\pi x}}\cos(x)
\eeq
this gives us:
\beq \label{Gm7}
\int^{\infty}_{0}dp J_{0}(pr)\frac{p}{\sqrt{p^{2}+m^{2}}}\sin(t\sqrt{p^{2}+m^{2}})=\frac{\cos(m\sqrt{t^{2}-r^{2}})}{\sqrt{t^{2}-r^{2}}}.
\eeq
Our final result for (\ref{Gm4}) is:
\beq \label{Gm8} G_{mass}(x-y)=-\frac{1}{2\pi}\frac{\theta((x-y)^2)\cos(m\sqrt{(x-y)^2})}{\sqrt{(x-y)^2}}\theta(x^{0}-y^{0})
\eeq
and is in accordance with \cite{Moura-Melo:2000}.

The Huygens principle is \emph{violated} in the MCS electrodynamics, but it is rather natural here because, as we will show in quantum version of the theory, the photons are massive. Retarded Green's functions of the massive theories in (3+1) dimensions violate the Huygens principle, too.
 
Now we can calculate the Coulomb potential of a point charge. We anticipate novel effects. Taking the zeroth component of (\ref{eq:MCS2}) and writing it in an integral form we get the Gauss law:
\beq \label{eq:MCS6a}
rE(r)+m\int_{0}^{r} dr^{\prime}r^{\prime}B(r^{\prime})=\frac{q}{2\pi}.
\eeq
Maybe a magnetic field is generated by a static object in this theory! Let us compute $E^{i}(\mathbf{x})$ and $B(\mathbf{x})$ created by a static charge using the Green's function (\ref{eq:MCS5}) and (\ref{eq:CLE4}). The computation is straightforward but tedious ( for more details see \cite{Moura-Melo:2000}). The result is:
\beq \label{eq:MCS6}
\begin{array} {c}
\Phi(\mathbf{x})=\frac{q}{2\pi}K_{0}(m|\mathbf{x}|) \qquad A^{i}(\mathbf{x})=\frac{q}{2\pi}\frac{1}{m}\frac{e^{ij}x^{j}}{|\mathbf{x}|}\left( \frac{1}{|\mathbf{x}|}-mK_{1}(m|\mathbf{x}|) \right) \\
E^{i}(x)=\frac{q}{2\pi}\frac{mx^{i}}{|\mathbf{x}|}K_{1}(m|\mathbf{x}|) \qquad B(x)=\frac{q}{2\pi}mK_{0}(m|\mathbf{x}|)=m\Phi(\mathbf{x}),
\end{array}
\eeq
where $K_{0}(x)$ and $K_{1}(x)$ are the modified Bessel functions of the second kind, which are positive, monotonically decreasing functions. Their asymptotic behavior is roughly $e^{-|m\mathbf{x}|}/\sqrt{|m\mathbf{x}|}$. We see that the \emph{magnetic field} is indeed generated by a point charge. It is easy to demonstrate that the Gauss law (\ref{eq:MCS6a}) is really satisfied. Loosely speaking static point charge generates a magnetic tube of width $m^{-1}$ and strength $q/m$. Although the static charge generates both magnetic and electric fields, it is easy to show that it does not radiate: the flux of the Poynting vector is zero.

Substituting the Green's function (\ref{eq:MCS5}) and the static current distribution (\ref{eq:CLE5b}) (with $\dot{z}(t)=(1,\mathbf{0})$) into (\ref{eq:CLE4}) we easily see that under the transformation $m\to -m$ the static potential $\Phi(\mathbf{x})$ remains unchanged and the vector potential $A^{i}(\mathbf{x})$ changes its sign. Thus using (\ref{eq:CLE0}) we conclude that  $m\to-m$ yields $B(\mathbf{x})\to -B(\mathbf{x})$ and  $E(\mathbf{x})\to E(\mathbf{x})$. 

We can rewrite the field equation in the dual form to illustrate that the quantum version of the theory will describe massive particles. Using (\ref{eq:MCS2a}) and (\ref{eq:MCS2}) we can write:
\beq \label{eq:MCS7}
(\Box+m^2)^{*}F_{\alpha}=(g_{\alpha \gamma}m+\epsilon_{\alpha \beta \gamma}\partial^{\beta})j^{\gamma}.
\eeq
The dual field strength solves the massive Klein-Gordon equation with the nontrivial right-hand side.

The energy-momentum tensor $\Theta_{\mu\nu}$ has the form of the pure Maxwell electrodynamics \cite{DJT:1981} because the new Chern-Simons term contains no explicit metric factors. Using Landau prescription $\Theta^{\mu\nu}(x)=\frac{2}{\sqrt{-g}}\frac{\delta}{\delta g_{\mu\nu}(x)}\int d^{3}x \sqrt{-g} \mathscr{L}$ we obtain no additional terms to the electrodynamic energy-momentum tensor:
\beq \label{eq:MCS8}
\Theta^{\mu \nu}=-F^{\mu\alpha}F_{\alpha}^{ \ \nu}+\frac{1}{4}g^{\mu\nu}F^{\alpha \beta}F_{\alpha \beta}.
\eeq

\chapter{Lorentz group and Dirac equation}
\label{ch:Lorentz} 
In this Chapter we shall discuss some properties of the Lorentz group SO(2,1) and its finite-dimensional representations. This will help us to construct the covariant fields, which are the building blocks of the local field theories. The Dirac equation will be also examined and the non-relativistic Pauli equation will be derived. 

The Lorentz group has two antihermitian boost generators $N_{1}$,\ $N_{2}$ and one hermitian rotation generator $M$. The $so(2,1)$ algebra is defined by its commutation relations:
\beq \label{eq:RQM1}
[N_{1}, N_{2}]=-iM \qquad
[M,N_{1}]=iN_{2} \qquad
[M,N_{2}]=-iN_{1}. \qquad
\eeq
There is only one Casimir operator $C$:
$$ C=-(N_{1}^{2}+N_{2}^{2})+M^{2}. $$
We obtain the group element $\Lambda$ by exponentiating its algebra:
$$ \Lambda(a,b_{1},b_{2})=\exp(i(aM+b_{1}N_{1}+b_{2}N_{2})),  $$
where $b_{i}$ are the boost parameters (rapidity) and $a$ is a rotation parameter (angle).

Same relations could be written in a somewhat different form. Let us define an antisymmetric matrix of the generators:
$$ I^{12}=-I^{21}=M \qquad I^{0k}=-I^{k0}=N_{k}. $$
Commutation relations (\ref{eq:RQM1}) take the form:
\beq \label{eq:RQM1a}
[I^{\alpha \beta},I^{\gamma \delta}]=i(g^{\alpha \delta}I^{\beta \gamma}-g^{\alpha \gamma}I^{\beta \delta}+g^{\beta \gamma}I^{\alpha \delta}-g^{\beta \delta}I^{\alpha \gamma}).
\eeq
Let us also define an antisymmetric matrix $\omega_{\mu \nu}$:
$$ \omega_{12}=-\omega_{21}=a \qquad \omega_{0k}=-\omega_{k0}=b_{k}. $$
General group element $\Lambda$ may be written as:
\beq \label{eq:RQM1b} \Lambda(\omega)=\exp(\frac{i}{2}\omega_{\mu \nu}I^{\mu\nu}). \eeq

The low-dimensional irreducible representations of $so(2,1)$ are of particular interest in planar physics. Hence let us first construct the fundamental irreducible representation of $so(2,1)$, which is a two-component spinorial representation.  First, we define the Dirac matrices $\gamma^{\mu}$ as a set of matrices, which satisfy
\beq \label{eq:RQM1ba}
\{\gamma^{\alpha},\gamma^{\beta}\}=2g^{\alpha\beta}.
\eeq 
In (3+1) dimensions this definition fixes $\gamma$ matrices up to a regular transformation; i.e., if two sets $\gamma_{\mu}$ and $\gamma^{\prime}_{\mu}$ satisfy (\ref{eq:RQM1ba}), than $\gamma^{\prime}_{\mu}=U\gamma_{\mu}U^{-1}$ for some regular $U$ and they are in the \emph{same} equivalence class. We also assume that:
\beq \label{eq:RQM1ba1}
\gamma_{0}^{\dag}=\gamma_{0} \qquad \gamma_{i}^{\dag}=-\gamma_{i}.
\eeq
Thus the matrix $U$ must be unitary to preserve (\ref{eq:RQM1ba1}). In (2+1) dimensions there are \emph{two equivalence classes} of $\gamma$-matrices (see Appendix of \cite{Sohnius:1985}). If $\{ \gamma \}$ is in the first equivalence class than $\{ -\gamma \}$ is in the second. We shall prove this assertion later.  

In two spatial dimensions it is possible to find  \emph{three $2\times 2$ matrices}, which satisfy (\ref{eq:RQM1ba}) and (\ref{eq:RQM1ba1}). Different equivalent realizations of $\gamma$ matrices are used in the literature:
\beq \label{eq:RQM1bb}
\begin{array}{cc}
 \gamma^{0}=\sigma^{3} \quad \gamma^{j}=\sigma^{3}\sigma^{j} &  Dirac \\
 \gamma^{0}=\sigma^{3} \quad \gamma^{j}=i\sigma^{j} & Jackiw \\
 \gamma^{0}=\sigma^{2} \quad \gamma^{1}=i\sigma^{3} \quad \gamma^{2}=i\sigma^{1} & Appelquist \\ 
\end{array}
\eeq
Two-dimensional $\gamma$-matrices satisfy the commutation relations:
\beq \label{eq:RQM1bc} [\gamma^{\mu}, \gamma^{\nu}]=-2i\epsilon^{\mu \nu \rho}\gamma_{\rho}, \eeq
hence from (\ref{eq:RQM1bc}) and (\ref{eq:RQM1ba}) it follows that it is possible to reduce the number of $\gamma$ matrices in different products, e.g.:
\beq \label{eq:RQM1bd}
\gamma^{\mu}\gamma^{\nu}=g^{\mu\nu}-i\epsilon^{\mu\nu\rho}\gamma_{\rho}.
\eeq
If we define $\Sigma^{\mu\nu}$:
\beq \label{eq:RQM2a0} \Sigma^{\mu\nu}\equiv\frac{i}{2}[\gamma^{\mu},\gamma^{\nu}]=\epsilon^{\mu\nu\rho}\gamma_{\rho}, \eeq
it is straightforward to show that $\Sigma^{\mu\nu}/2$ satisfy (\ref{eq:RQM1a}). Thus we have constructed the generators of the fundamental representation of $SO(2,1)$ group. A general group element $\Lambda(\omega)$ (\ref{eq:RQM1b}) may be written in the fundamental representation as:
\beq \label{eq:RQM2a}
\Lambda(\omega)=\exp(\frac{i}{4}\omega_{\mu \nu}\Sigma^{\mu\nu}).
\eeq
Boost generators $N_{i}$ and the rotation generator $M$ in the fundamental representation can be identified from (\ref{eq:RQM2a0}) :
\beq \label{eq:RQM2a1} M=\frac{\gamma_{0}}{2} \qquad N_{1}=\frac{\gamma_{2}}{2} \qquad N_{2}=-\frac{\gamma_{1}}{2}. \eeq

Let us go back to the proof of the existence of two equivalence classes of $\gamma$ matrices in (2+1) dimensions. First we introduce the matrix $\gamma_{4}$:
\beq \label{gamma1}
\gamma_{4}\equiv\gamma_{0}\gamma_{1}\gamma_{2},
\eeq 
which due to (\ref{eq:RQM1ba}) \emph{commutes} with all $\gamma$ matrices. Furthermore, it also commutes with the generators of the Lorentz group (\ref{eq:RQM2a0}), therefore due to the Schur's lemma  it must be a multiple of the unit matrix. Imagine now that there is a unitary matrix $U$ with the property:
\beq \label{gamma2}
U\gamma_{\mu}U^{-1}=-\gamma_{\mu}
\eeq
and hence
\beq \label{gamma3}
U\gamma_{4}U^{-1}=(-1)^{3}\gamma_{4}=-\gamma_{4}.
\eeq
However, this is in contradiction that $\gamma_{4}$ is a multiple of the unit matrix, hence there is \emph{no} such a matrix $U$ with the property (\ref{gamma2}). And at least two equivalence classes exist.

In two spatial dimensions the two-component spinorial fields are used as building blocks of the Dirac theory. The Dirac equation looks exactly the same as in the (3+1) theory:
\beq \label{eq:RQM2}
(i\slashed{\partial}-m)\psi(x)=0,
\eeq
where $\psi(x)$ is a two-component field, which transforms under the Lorentz transformations as a fundamental spinorial representation of $SO(2,1)$ group. For example for the rotation by an angle $\phi$:
\beq \label{eq:RQM3}
\psi^{\prime}(x^{\prime})=\exp(-i\phi M)\psi(\Lambda^{-1}x),
\eeq
where the spin matrix $M$ in Dirac and Jackiw realizations is:
$$ M=\frac{\gamma_{0}}{2}=\frac{1}{2} \left ( \begin{array}{cc}
1 & 0\\
0 & -1\end{array}\right). $$
Thus \emph{the Dirac particles have spin $\pm 1/2$}. 

Now we shall explicitly find the plane-wave solutions of the free Dirac equation (\ref{eq:RQM2}).  Let us work in the Dirac realization of $\gamma$-matrices. Positive-energy solutions are:
$$ \psi(x)=u(p)e^{-ipx} \qquad p^{2}=m^{2} \qquad p^{0}>0 $$
$$ (\gamma^{\mu}p_{\mu}-m)u(p)=0 \qquad u^{+}(p)u(p)=2E $$
\beq \label{eq:RQM4}
u_{1}=\frac{p^{1}-ip^{2}}{\sqrt{E-m}} \qquad u_{2}=\sqrt{E-m}.
\eeq
Negative-energy solution can be easily found:
$$ \psi(x)=v(p)e^{ipx} \qquad p^{2}=m^{2} \qquad p^{0}>0 $$
$$ (\gamma^{\mu}p_{\mu}+m)v(p)=0 \qquad v^{+}(p)v(p)=2E $$
\beq \label{eq:RQM5}
v_{1}=\frac{p^{1}-ip^{2}}{\sqrt{E+m}} \qquad v_{2}=\sqrt{E+m}.
\eeq
In the non-relativistic region $u_{1}>>u_{2}$ and $v_{1}<<v_{2}$, thus the Dirac realization is useful for non-relativistic calculations. In order to illustrate this fact we shall derive the \emph{Pauli equation}, which is a first non-relativistic approximation of the Dirac equation for electron in an external electromagnetic field. External electromagnetic field is introduced into the free Dirac equation (\ref{eq:RQM2}) by the minimal coupling:
\beq \label{Pe1}
\partial_{\mu}\to D_{\mu}=\partial_{\mu}+ieA_{\mu} \qquad A_{\mu}=(\Phi,-\mathbf{A}),
\eeq
where $e$ is the charge of the electron.
  
The Dirac equation is now (here we explicitly write $\hbar$ and $c$):
\beq \label{Pe2}
i\hbar \frac{\partial}{\partial t}\psi=e\Phi \psi+\alpha^{j} \underbrace{(-ic\hbar\frac{\partial}{\partial x^{j}}-e A^{j})}_{\pi^{j}}\psi+\beta mc^{2} \psi,
\eeq
where $\alpha^{j}=\gamma^{0}\gamma^{j}$ and $\beta=\gamma^{0}$. Now we write $\psi=\left(\begin{array}{c} \phi^{\prime} \\ \chi^{\prime}  \end{array}\right)$ and substitute it into (\ref{Pe2}):
\beq \label{Pe3}
\begin{array}{c}
i\hbar\frac{\partial}{\partial t}\phi^{\prime}=e\Phi \phi^{\prime}+(\pi^{1}-i\pi^{2})\chi^{\prime}+mc^{2}\phi^{\prime} \\
i\hbar\frac{\partial}{\partial t}\chi^{\prime}=e\Phi \chi^{\prime}+(\pi^{1}+i\pi^{2})\phi^{\prime}-mc^{2}\chi^{\prime}.
\end{array}
\eeq
To perform the non-relativistic approximation we write $\psi=\left(\begin{array}{c} \phi \\ \chi  \end{array}\right)e^{-imc^{2}t/\hbar}$ in order to subtract the relativistic rest energy. The coupled system of equations (\ref{Pe3}) now takes the form:
\beq \label{Pe4}
\begin{array}{c}
i\hbar\frac{\partial}{\partial t}\phi=e\Phi \phi+(\pi^{1}-i\pi^{2})\chi \\
i\hbar\frac{\partial}{\partial t}\chi=e\Phi \chi+(\pi^{1}+i\pi^{2})\phi-2mc^{2}\chi.
\end{array}
\eeq
In the weak external electromagnetic field ($mc^{2}>>e\Phi$ and $mc^{2}>>e|\mathbf{A}|$) we anticipate that $\chi$ is a small component; i.e., $\chi<<\phi$ so from the second equation (\ref{Pe4}):
\beq \label{Pe5}
\chi \approx \frac{1}{2mc^{2}}(\pi^{1}+i\pi^{2})\phi.
\eeq
Now we substitute the last expression into the first equation (\ref{Pe4}) and obtain:
\beq \label{Pe6}
i\hbar\frac{\partial}{\partial t}\phi=e\Phi \phi+\frac{1}{2mc^{2}}(\pi^{1}-i\pi^{2})(\pi^{1}+i\pi^{2})\phi.
\eeq
The second term on the RHS of the last equation may be rewritten as:
$$\frac{1}{2mc^{2}}((\pi^{1})^{2}+(\pi^{2})^{2}+i[\pi^{1},\pi^{2}])=\frac{1}{2mc^{2}}\left((c\mathbf{p}-e\mathbf{A})^{2}-e\hbar c(\underbrace{\frac{\partial A^{2}}{\partial x^{1}}-\frac{\partial A^{1}}{\partial x^{2}})}_{-B}\right). $$
Thus we obtained the well-known \emph{Pauli equation}:
\beq \label{RQM6}
i\hbar\frac{\partial}{\partial t}\phi=e\Phi\phi+\frac{(\mathbf{p}-\frac{e}{c}\mathbf{A})^{2}}{2m}\phi+\frac{e\hbar}{2mc}B\phi,
\eeq 
which is a \emph{one component} equation for electron in an external electromagnetic field.

There is \emph{no chiral $\gamma_5$ matrix} in two-component spinorial representation: It is not possible to find a $2\times 2$ matrix $\gamma_{5}$, which satisfies $\{\gamma^{\mu}, \gamma_{5}\}=0$ for $\mu=0,1,2$. Here are some useful relations for traces of $\gamma$ matrices, which will be useful for us later:
\beq \label{gamma}
\begin{array}{c}
 Tr\gamma^{\mu}=0 \qquad Tr(\gamma^{\mu}\gamma^{\nu})=2g^{\mu\nu} \qquad Tr(\gamma^{\mu}\gamma^{\nu}\gamma^{\rho})=-2i\epsilon^{\mu\nu\rho} \\
 Tr(\gamma^{\mu}\gamma^{\nu}\gamma^{\rho}\gamma^{\phi})=2(g^{\mu\nu}g^{\rho\phi}-g^{\mu\rho}g^{\nu\phi}+g^{\mu\phi}g^{\nu\rho}) \\
 \gamma^{\mu}\gamma_{\mu}=3 \qquad \gamma^{\mu}\gamma^{\nu}\gamma_{\mu}=-\gamma^{\nu} \\
 \gamma^{\mu}\gamma^{\nu}\gamma^{\phi}\gamma_{\mu}=2g^{\nu\phi}-\gamma^{\phi}\gamma^{\nu} \\
 \gamma^{\mu}\gamma^{\nu}\gamma^{\phi}\gamma^{\rho}\gamma_{\mu}=-2\gamma^{\nu}g^{\phi\rho}-\gamma^{\nu}\gamma^{\rho}\gamma^{\phi}+2\gamma^{\phi}\gamma^{\rho}\gamma^{\nu}. \\  
\end{array}
\eeq
Note that the trace of the odd number of $\gamma$ matrices does not vanish identically! It is not surprising for us because the proof of this statement in (3+1) dimensions uses the $\gamma^{5}$ matrix, which is absent here. Projectors on the states with definite momentum $p$ are:
$$ u(p)\bar{u}(p)=\slashed{p}+m \qquad v(p)\bar{v}(p)=\slashed{p}-m. $$ 
For the future references we note that the (2+1)-dimensional Gordon identity holds:
\beq \label{eq:RQM7}
\bar{u}(p^{\prime})\gamma_{\mu}u(p)=\bar{u}(p^{\prime})\left\{\frac{p_{\mu}+p^{\prime}_{\mu}}{2m}+i\epsilon_{\mu\nu\rho}\frac{q^{\nu}\gamma^{\rho}}{2m} \right\}u(p),
\eeq
where the fermions are on the mass shell ($p^{2}=(p^{\prime})^{2}=m^{2}$) and $q=p^{\prime}-p$.

It is useful to construct explicitly the irreducible 3-vector representation of the Lorentz group. Let us introduce a set of $3\times 3$ matrices:
\beq \label{eq:RQM2a2}
(I^{\alpha \beta})^{\mu}_{ \nu}=i(g^{\alpha \mu}g^{\beta}_{\ \nu}-g^{\alpha}_{\ \nu}g^{\beta \mu}).
\eeq
It is easy to show that these matrices satisfy the commutation relations (\ref{eq:RQM1a}) and therefore are the generators of $SO(2,1)$ in vector representation.

\chapter{Poincar\'e group M(2,1)}
\label{ch:Poincare}
In this Chapter we shall classify the unitary irreducible representations (UIRs) of the Poincar\'e group M(2,1). Also we shall connect the UIRs of the Poincar\'e group with the local covariant fields, which are used in quantum field theories. The detailed analysis was made in \cite{Binegar:1981} and we will entirely follow this analysis here.

\section{Classification of the unitary irreducible representations}
The Poincar\'e group is defined as a six-parameter group of real spacetime transformations:
\beq \label{eq:Poi0}
(\Lambda, a): x^{\mu}\to \Lambda^{\mu}_{ \ \nu}x^{\nu}+a^{\mu},
\eeq
where $a^{\mu}$ are the translation parameters and $\Lambda^{\mu}_{ \ \nu}$ is a matrix of the Lorentz transformation.
Due to translation and Lorentz invariance of the relativistic quantum theory the Poincar\'e group forms a symmetry group of the quantum system. That is why the Hilbert space of the quantum system is a representation space of the M(2,1). Elementary particles of the theory can be classified by the UIRs of the M(2,1). The group M(2,1) is a semidirect product of the (2+1)-dimensional translations and the Lorentz group $L \equiv SO(2,1)$. All Lorentz transformations form four disjoint sets, but only one of them will be of our interest here:
$$ L_{++}=\{ \Lambda \in L, \det \Lambda>0, \Lambda^{0}_{\ 0}>0 \}. $$
$L_{++}$ is the largest connected subgroup of $L$. In our further calculations we will restrict ourselves to the Poincar\'e subgroup $\pi _{++}$, which is a semi-direct product of the spacetime translations and of the $L_{++}$ group.

It is possible to relate $L_{++}$ to $SL(2,R)$ group. First, let us define the basis of the vector space $\mathscr{M}$ of real hermitian matrices:
$$ \tau_{0}=\left( \begin{array}{cc} 1 & 0 \\ 0 & 1 \end{array} \right) \qquad \tau_{1}=\left( \begin{array}{cc} 1 & 0 \\ 0 & -1 \end{array} \right) \qquad \tau_{2}=\left( \begin{array}{cc} 0 & 1 \\ 1 & 0 \end{array} \right). $$
To each $x^{\mu}$ from the (2+1) Minkowski spacetime $M$ there exist $\chi\in \mathscr{M}$:
$$ \chi=x \cdot \tau=\left( \begin{array}{cc} x^{0}+x^{1} & x^{2} \\ x^{2} & x^{0}-x^{1} \end{array} \right). $$
It is worth noting that the determinant of the $\chi$ matrix is simply the spacetime interval:
\beq \label{eq:Poi0a0}
\det{\chi}=(x^{0})^{2}-(x^{1})^{2}-(x^{2})^{2}.
\eeq
Now we define a transformation:
\beq \label{eq:Poi0a}
\chi\to \chi^{\prime}=\Omega \chi \Omega^{\top} \quad \Omega\in SL(2,R),
\eeq
which does not change the interval ($\det\chi^{\prime}=\det\chi$), hence it must correspond to some $\Lambda\in L_{++}$. It is possible to show that:
\beq \label{eq:Poi0b}
L_{++}=SL(2,R)/Z_{2}
\eeq
because for arbitrary $\Omega\in SL(2,R)$ the matrix $-\Omega$ yields the same transformation (\ref{eq:Poi0a}).

All $p^{\mu} \in M$ fall into six disjoint classes:
\beq \label{eq:Poi0c}
\begin{array}{c}
O^{+}_{m}=\{p\in M; \ p^{2}=m^{2}, \ p_{0}>0\}, \\
O^{-}_{m}=\{p\in M; \ p^{2}=m^{2}, \ p_{0}<0\}, \\
O^{+}_{0}=\{p\in M; \ p^{2}=0, \ p_{0}>0\}, \\
O^{-}_{0}=\{p\in M; \ p^{2}=0, \ p_{0}<0\}, \\
O_{im}=\{p\in M; \ p^{2}=-m^{2}\}, \\
O_{0}=(0,0,0). \\
\end{array}
\eeq
All elements of the class can be connected to each other by some $\Lambda\in L_{++}$.

The Poincar\'e algebra is defined by its commutation relations:
\beq \label{eq:Poi0c1}
[J^{\mu},J^{\nu}]=i\epsilon^{\mu\nu\rho}J_{\rho} \qquad [J^{\mu},P^{\nu}]=i\epsilon^{\mu\nu\rho}P_{\rho} \qquad [P^{\mu},P^{\nu}]=0,
\eeq
where we introduced $P^{\mu}\equiv(H,\mathbf{P})$ and $J^{\mu}\equiv(-M,\epsilon^{ij}N_{j})$ (see \ref{eq:RQM1}). It is possible to classify UIRs by the eigenvalues of the Casimir operators. It is straightforward to show that $P\cdot P$ and $P\cdot J$ commute with every generator of the Poincar\'e group as well as with each other. That is why the vectors from the representation space $\ket{p,j}$ ($p$ is the three-momentum and $j$ is a spin quantum number) satisfy the mass-shell condition:
\beq \label{eq:Poi0c2} (P\cdot P -m^{2})\ket{p,j}=0 \eeq
and the Pauli-Lubanski condition:
\beq \label{eq:Poi0c3} (P\cdot J+jm)\ket{p,j}=0. \eeq

Now we shall construct the UIRs of the Poincar\'e group using the Wigner method. First we should choose from every class (\ref{eq:Poi0c}) a standard reference vector $\hat{p}$. Each vector $p$ of the given class is connected to $\hat{p}$ by some Lorentz transformation, but this connection is not unique. We should choose some standard transformation $L(p)$ such that $p=L(p)\hat{p}$. Lorentz transformations, which leave $\hat{p}$ fixed form the group called the stability group. Finding each UIR $D_{\xi \xi^{\prime}}$ of the stability group we form the UIR of $\pi_{++}$ by:  
\beq \label{eq:Poi0d}
\hat{U}(\Lambda)\ket{p,\xi}=N\sum_{\xi^{\prime}}D_{\xi \xi^{\prime}}(W(\Lambda,p))\ket{\Lambda p, \xi^{\prime}},
\eeq
where $W(\Lambda,p)=L^{-1}(\Lambda p)\Lambda L(p)$ and is an element of the stability group. The factor $N=\sqrt{\frac{E_{\mathbf{\Lambda p}}}{E_{\mathbf{p}}}}$ is a normalization factor due to the normalization of the creation and annihilation operators used in the text. 

In order to find the UIRs of the $\pi_{++}$ we should construct UIRs $D_{\xi \xi^{\prime}}$ of the stability groups, which are different for different classes (\ref{eq:Poi0c}). Hence now we shall identify stability groups for each class:

\subsection*{$O^{+}_{m}$}
We choose the standard vector as $\hat{p}=(m,0,0)$. The stability subgroup can be found by imposing the stability condition:
$$ \Omega \left( \begin{array}{cc} m & 0 \\ 0 & m \end{array} \right) \Omega^{\top}=\left( \begin{array}{cc} m & 0 \\ 0 & m \end{array} \right). $$
This condition implies that the stability group is homomorphic to $SO(2)$. It is not simply connected. In fact, it is infinitely connected. We take its universal covering group $R$, which is an additive group of real numbers, in order to take all multivalued representations of $SO(2)$. The UIRs of $R$ are labeled by an arbitrary real number $j$:
$$ D^{j}:\theta \to e^{-ij\theta} \quad j\in R. $$
These representations of stability group induce UIRs of the $\pi_{++}$, which we shall label:
\beq \label{eq:Poi0e}
U^{m,+,j} \quad j\in R.
\eeq

\subsection*{$O^{-}_{m}$}
Our standard vector is $\hat{p}=(-m,0,0)$. The analysis is completely the same as in the previous case. We label the UIRs by $j\in R$:
\beq \label{eq:Poi0f}
U^{m,-,j} \quad j\in R.
\eeq

\subsection*{$O^{+}_{0}$}
Here we take $\hat{p}=(1/2, 1/2,0)$. Imposing the stability condition:
$$ \Omega \left( \begin{array}{cc}  1 & 0 \\ 0 & 0 \end{array} \right) \Omega^{\top}=\left( \begin{array}{cc}  1 & 0 \\ 0 & 0 \end{array} \right),  $$
we get a solution:
$$ \Omega=\pm \left( \begin{array}{cc}  1 & a \\ 0 & 1 \end{array} \right), $$
which can be parametrized by $(a,\pm1)$. We realize that the the stability group in this case is homomorphic to $Z_{2}\otimes R$. There are two inequivalent irreducible representations of $Z_{2}$:
$$ D^{0}_{\ Z}(\pm 1)= 1 \qquad D^{1}_{\ Z}(\pm 1)= \pm 1. $$
Finally we find the UIRs of the stability group:
$$ D^{\epsilon, j}=D^{ \ \epsilon}_{ Z} \otimes D^{\ j}_{ R} \qquad \epsilon=0,1 \quad j\in R. $$
These representations induce the UIR of the $\pi_{++}$, which we shall call:
\beq \label{eq:Poi0g}
U^{0,+,\epsilon,j} \quad \epsilon=0,1 \quad j\in R.
\eeq

\subsection*{$O^{-}_{0}$}
We choose $\hat{p}=(-1/2,-1/2,0)$. Repeating the analysis from the previous subsection we get the representations:
\beq \label{eq:Poi0h}
U^{0,-,\epsilon,j} \quad \epsilon=0,1 \quad j\in R.
\eeq

\subsection*{$O_{im}$}
The standard vector can be chosen as $\hat{p}=(0,m,0)$. This class is not of interest because it describes tachyons: the particles with space-like 3-velocity.

Not every UIR of $\pi_{++}$ is physically interesting. Representations used in the literature are:
\beq \label{eq:Poi2} U^{m,+,j} \quad m>0 \quad j\in R \quad \textrm{massive particles} \eeq
\beq \label{eq:Poi3} U^{0,+,\epsilon,0} \quad m=0 \quad \epsilon=0,1 \quad j=0 \quad \textrm{massless particles} \eeq

The stability group of a massive particle (\ref{eq:Poi2}) is $SO(2)$, thus \emph{every particle has only one spin state $j$}. Spin $j$ is an arbitrary real number. This fact rises an interesting theoretical possibility for \emph{anyons}; i.e., particles with \emph{fractional} statistics in planar physics \cite{Wilczek:1990}. In the massless case we shall limit ourselves to $j=0$ in (\ref{eq:Poi3}).

Now we shall explicitly construct the operators of infinitesimal rotations and boosts for the physical UIRs (\ref{eq:Poi2}) and (\ref{eq:Poi3}):

\subsection*{$U^{m,+,j}$}
In order to construct the operators of symmetry we have to specify the standard transformation $L(p)$. We can choose $L(p)$ as a pure boost, which takes $\hat{p}$ into $p$. It is not difficult to show that for a pure rotation $\Lambda=R(\theta)$:
\beq \label{eq:Poi3aa} L^{-1}(R(\theta)p)R(\theta)L(p)=R(\theta). \eeq
For a pure boost $\Lambda=B(u,\mathbf{n})=\exp(iu \mathbf{n} \cdot \mathbf{N})$ in the direction $\mathbf{n}$ with rapidity $u=\arctan v$ ($v$ is the magnitude of the velocity vector) we find:
\beq \label{eq:Poi3ab}  L^{-1}(B(u,\mathbf{n})p)B(u,\mathbf{n})L(p)=R(\theta_{b}) \qquad \theta_{b}=\frac{u(p^{1}n^{2}-p^{2}n^{1})}{E_{p}+m} \eeq
for details see \cite{Jackiw:1990}. From (\ref{eq:Poi3aa}) and (\ref{eq:Poi3ab}) we see that the Wigner transformation is always a rotation, which we denote $R_{(\Lambda ,p)}$.

 Following \cite{Jackiw:1990} we shall rewrite these results in a covariant form. In an arbitrary representation of the Lorentz group the transformation may be written as $\Lambda^{a}_{\ b}=\exp(i\omega\cdot j)^{a}_{\ b}$, where $j^{\mu}$ are the Lorentz generators in a given representation. These generators satisfy the commutation relations (\ref{eq:Poi0c1}). The Wigner rotation $R_{(\Lambda ,p)}$ for the infinitesimal transformation $\Lambda^{a}_{\ b}$ may be written as:
\beq \label{eq:Poi3ac}
R_{(\Lambda , p)}=\left( L^{-1}(\Lambda p)\Lambda L(p)\right) ^{a}_{\ b}=\delta^{a}_{\ b}+i(j\cdot \eta)^{a}_{\ b} \omega\cdot \Delta(p) \qquad \Delta^{\mu}(p)=\frac{p^{\mu}+\eta^{\mu}m}{p\cdot \eta+m},
\eeq
where $\eta=(1,0,0)$. It is easy to show that (\ref{eq:Poi3ac}) is a generalized compact form of (\ref{eq:Poi3aa}) and (\ref{eq:Poi3ab}).
 
For an infinitesimal transformation $\Lambda$ it is true that $(\Lambda p)^{\mu}=p^{\mu}-\epsilon^{\mu\nu\rho}\omega_{\nu}p_{\rho}$. Using this fact we can write:
\beq \label{Poi3ad}
L^{-1}(\Lambda p)=\exp(-\epsilon^{\mu\nu\rho}\omega_{\nu}p_{\rho}\frac{\partial}{\partial p^{\mu}})L^{-1}(p)\approx (1-\epsilon^{\mu\nu\rho}\omega_{\nu}p_{\rho}\frac{\partial}{\partial p^{\mu}})L^{-1}(p).
\eeq

With (\ref{eq:Poi3ac}) and (\ref{Poi3ad}) it is now straightforward to get a relation, which will be useful for us later:
\beq \label{eq:Poi3ae}
L^{-1}(p)(-i\epsilon^{\mu\nu\rho}p_{\nu}\frac{\partial}{\partial p^{\rho}}+j^{\mu})L(p)=j \cdot \eta \Delta^{\mu}.
\eeq

Due to (\ref{eq:Poi0d}) we can now write for the infinitesimal rotation operator:
\beq \label{eq:Poi3a}
\hat{R}^{m,+,j}(\theta)\ket{p,j}=ND^{j}(\theta)\ket{R(\theta)p,j}=N\left(1-ij\theta+\theta(p_{1}\frac{\partial}{\partial p^{2}}-p_{2}\frac{\partial}{\partial p^{1}})\right) \ket{p,j}.
\eeq

Infinitesimal boost operator $\hat{L}^{m,+,j}(u,\mathbf{n})$ can be also obtained:
\beq \label{eq:Poi3a1}
\begin{array}{c}
\hat{L}^{m,+,j}(u,\mathbf{n})\ket{p,j}=ND^{j}(\theta_{b})\ket{B(u,\mathbf{n})p,j}= \\
N\left(1-ij\theta_{b}-u[(\mathbf{p} \cdot \mathbf{n})\frac{\partial}{\partial p^{0}}+p^{0}n^{1}\frac{\partial}{\partial p^{1}}+p^{0}n^{2}\frac{\partial}{\partial p^{2}}] \right)\ket{p,j}, \end{array}
\eeq
where $\theta_{b}$ is from (\ref{eq:Poi3ab}).

It is possible to find an elegant and unified form of (\ref{eq:Poi3a}) and (\ref{eq:Poi3a1}) \cite{Jackiw:1990}. Let us define the generators $J^{\mu}=(-M,\epsilon^{ij}N_{j})$ of the Lorentz subgroup in $U^{m,+,j}$. From (\ref{eq:Poi3a}) and (\ref{eq:Poi3a1}) we obtain:
\beq \label{eq:Poi3a2}
J^{\mu}=-i\epsilon^{\mu\nu\rho}p_{\nu}\frac{\partial}{\partial p^{\rho}}-j\Delta^{\mu}(p),
\eeq  
where $j$ is the spin of the representation $U^{m,+,j}$. These generators $J^{\mu}$ satisfy the commutation relations (\ref{eq:Poi0c1}). $J^{\mu}$ is an operator in the one-particle Hilbert space. Its generalization to the many-particle Fock space is straightforward:
\beq \label{eq:Poi3a3}
J^{\mu}=\int d^{2}p \hat{a}^{\dag}_{\mathbf{p}}\left(-i\epsilon^{\mu\nu\rho}p_{\nu}\frac{\partial}{\partial p^{\rho}}-j\Delta^{\mu}(p)\right)\hat{a}_{\mathbf{p}}.
\eeq

\subsection*{$U^{0,+,0,0}$}
This representation has spin zero, thus we can immediately write the infinitesimal rotation operator:
\beq \label{eq:Poi3b}
\hat{R}^{0,+,0,0}(\theta)\ket{p}=N\left(1+\theta(p_{1}\frac{\partial}{\partial p^{2}}-p_{2}\frac{\partial}{\partial p^{1}})\right) \ket{p}
\eeq
and the boost operator:
\beq \label{eq:Poi3b1}
\hat{L}^{0,+,0,0}(u,\mathbf{n})\ket{p,j}=
N\left(1-u[(\mathbf{p} \cdot \mathbf{n})\frac{\partial}{\partial p^{0}}+p^{0}n^{1}\frac{\partial}{\partial p^{1}}+p^{0}n^{2}\frac{\partial}{\partial p^{2}}] \right)\ket{p,j}.
\eeq

\subsection*{$U^{0,+,1,0}$}
This representation transforms almost as the previous one. Its infinitesimal rotation operator and boost operators are the same as (\ref{eq:Poi3b}) and (\ref{eq:Poi3b1}), but finite rotations are different. We know that $\Omega(\theta+2\pi)=-\Omega(\theta),$
that is why we have:
\beq \label{eq:Poi3c}
\hat{R}^{0,+,1,0}(\theta)\ket{p}=\left\{ \begin{array}{c} N\ket{R(\theta)p} \quad 0 \le  \theta \le 2\pi \\
                                                 -N\ket{R(\theta)p} \quad 2\pi <  \theta < 4\pi. \\
                                    \end{array}  \right.            
\eeq

Let us summarize the results obtained so far. Following Wigner we have constructed and classified the unitary irreducible representations of the Poincar\'e group by constructing the representations of the stability subgroups. We have also identified the standard transformation with a pure boost. This fact helped us to construct the infinitesimal rotation and boost operators for various physically interesting representations. In field theory, however, we work with various field operators, which are some special linear combinations of creation and annihilation operators. In the next section we shall concentrate on the connection of the field operators with the UIRs.

\section{Covariant fields}
It is common to introduce the \emph{covariant} field operators $\phi(x)$, which transform under the Poincar\'e transformation as follows:
\beq \label{eq:Poi4-} (a,\Lambda): U(a,\Lambda)\phi(x)U^{-1}(a,\Lambda) =D(\Lambda^{-1})\phi(\Lambda x+a), \eeq
where 
$U(a,\Lambda)$ is a unitary operator in the Hilbert space of the particles, which accomplishes the Poincar\'e transformations of the state vectors between two Lorentz frames. $D(\Lambda)$ is a finite-dimensional representation of the Lorentz group (instead of representation of a stability group). Now we can Fourier-transform the field operators $\phi(x)$:
\beq \label{eq:Poi4a} \phi(p)=\int d^{3}x \exp(-ip\cdot x) \phi(x). \eeq 
The Poincar\'e transformation can now be written as:
\beq \label{eq:Poi4} (a,\Lambda): U(a,\Lambda)\phi(p)U^{-1}(a,\Lambda)= e^{-ip\cdot a}D(\Lambda^{-1})\phi(\Lambda p). \eeq
The covariant fields are not generally irreducible. The field equation (and a subsidiary condition perhaps) is needed to remove unphysical degrees of freedom. The solution of the field equation provides a unitary irreducible representation of $\pi _{++}$ or is decomposable into these representations. We shall show this on examples of scalar, Dirac and vector fields.     

\subsection*{Scalar fields} 
This field is the simplest. Its transformation properties and wave equation in p-representa\-tion are:
\beq \label{eq:Poi5}
U(\Lambda)\phi(p)U^{-1}(\Lambda)=\phi(\Lambda p) \qquad (p^{2}-m^{2})\phi(p)=0.
\eeq
From these properties we conclude that a massive scalar field belongs to $U^{m,+,0}$, while a massless scalar field belongs to $U^{0,+,0,0}$. Thus the scalar fields describe \emph{massive and massless particles with zero spin}.

\subsection*{Dirac fields}
$\psi(p)$ transforms according to the fundamental representation of $SO(2,1)$:
\beq \label{eq:Poi6}
U(\Lambda)\psi(p)U^{-1}(\Lambda)= \exp(-\frac{1}{2}i\omega \cdot \gamma)\psi(\Lambda p),
\eeq
where $\omega^{\rho}=\frac{1}{2}\omega_{\mu\nu}\epsilon^{\mu \nu \rho}$ (see (\ref{eq:RQM2a0}) and (\ref{eq:RQM2a})), and satisfies the Dirac field equation:
\beq \label{eq:Poi7}
(p\cdot \gamma-m)\psi(p)=0. 
\eeq

We shall deal with massive and massless case separately:

\subsubsection*{\textit{Massive case}}
The Hamiltonian in p-representation is:
$$ H=\pmb{\alpha} \cdot \mathbf{p}+\gamma^{0}m, $$
where $\pmb{\alpha}=\gamma^{0}\pmb{\gamma}$. This Hamiltonian can be diagonalized by means of the well-known Foldy-Wouthuysen unitary transformation $U$:
$$ U=\exp\left(\frac{\lambda \mathbf{p}\cdot \pmb{\gamma}}{2|\mathbf{p}|} \right) \qquad \lambda=\arctan(\frac{|\mathbf{p}|}{m}). $$
The transformation acts as:
$$ \begin{array}{c} H^{\prime}=UHU^{-1}=\epsilon_{\mathbf{p}}\gamma^{0} \\
										U\psi(p)= \left( \begin{array}{c} \phi^{+}(p) \\ \phi^{-}(p) \end{array} \right) \\
										H^{\prime}\phi^{\pm}(p)=\pm \epsilon_{\mathbf{p}} \phi^{\pm}(p),
		\end{array} $$
where $\epsilon_{\mathbf{p}}=\sqrt{\mathbf{p}^{2}+m^{2}}$. 

Let us now find out how positive-energy $\phi^{+}(p)$ and negative-energy $\phi^{-}(p)$ solutions transform under the infinitesimal Lorentz transformations. First we shall transform the infinitesimal rotation generator:
\beq \label{eq:Poi7a} \begin{array}{c}
R^{\prime}=URU^{-1}=\exp(\frac{\lambda \mathbf{p}\cdot \pmb{\gamma}}{2|\mathbf{p}|})\left(1-\frac{1}{2}i\theta\gamma^{0}+\theta(p_{1}\frac{\partial}{\partial p^{2}}-p_{2}\frac{\partial}{\partial p^{1}})\right)\exp(-\frac{\lambda \mathbf{p}\cdot \pmb{\gamma}}{2|\mathbf{p}|})=\\
=\left(1-\frac{1}{2}i\theta\gamma^{0}+\theta(p_{1}\frac{\partial}{\partial p^{2}}-p_{2}\frac{\partial}{\partial p^{1}})\right), \end{array}
\eeq
where the relation:
$$ [\frac{\mathbf{p}\cdot \pmb{\gamma}}{|\mathbf{p}|},R]=0 $$
was used. We see that the energy eigenstates transform as:
\beq \label{eq:Poi7b}
\phi^{\pm}(p)\to\left\{1 \mp \frac{1}{2}i\theta+\theta(p_{1}\frac{\partial}{\partial p^{2}}-p_{2}\frac{\partial}{\partial p^{1}})\right\}\phi^{\pm}(p).
\eeq
From (\ref{eq:Poi3a}) we see that $\phi^{+}(p)\in U^{m,+,1/2}$ and $\phi^{-}(p)\in U^{m,+,-1/2}$. Our final result is that \emph{the massive Dirac fields describe particles with spin $\pm 1/2$}.  
\subsubsection*{\textit{Massless case}}
The analysis is very similar. The massless Hamiltonian is:
$$ H=\pmb{\alpha} \cdot \mathbf{p} $$
and the Foldy-Wouthuysen transformation is now:
$$ U=\exp\left(\frac{\pi \mathbf{p}\cdot \pmb{\gamma}}{4|\mathbf{p}|} \right). $$
Transforming the infinitesimal rotation operator we get as before:
$$ \phi^{\pm}(p)\to\left\{1 \mp \frac{1}{2}i\theta+\theta(p_{1}\frac{\partial}{\partial p^{2}}-p_{2}\frac{\partial}{\partial p^{1}})\right\}\phi^{\pm}(p). $$
So far it seems that we are dealing with the particles with spin $\pm 1/2$. It is possible and necessary, however, to make the redefinition of the fields:
\beq \label{eq:Poi7c}
\Phi^{\pm}(p)\equiv(p_{1}\pm ip_{2})^{1/2}\phi^{\pm}(p),
\eeq
which was not possible in the massive case because this redefinition would be singular. By saying this we mean that for the massive field every momentum vector $p^{\mu}$ satisfies the mass-shell condition $p^{2}=m^{2}$. This implies that it is always possible to find a reference frame, where $p^{\mu}=(m,0,0)$. In this reference frame the redefinition (\ref{eq:Poi7c}) is singular for the selected $p^{\mu}$. The transformation properties of the new fields $\Phi^{\pm}(p)$ under the infinitesimal rotation are:
\beq \label{Poi7d}
\Phi^{\pm}(p)\to \Phi^{\pm}(Rp),
\eeq
while under finite rotations the new field transforms as:
\beq \label{Poi7e}
\Phi^{\pm}(p)\to \{ \begin{array}{c} \Phi^{\pm}(Rp) \quad 0\le \theta \le 2\pi \\
																		-\Phi^{\pm}(Rp) \quad 2\pi < \theta < 4\pi, \end{array}
\eeq 
due to double-valuedness of $(p_{1}\pm ip_{2})^{1/2}$. We see that massless Dirac field is connected with $\Phi^{+}(p) \in U^{0,+,1,0}$ and $\Phi^{-}(p) \in U^{0,-,1,0}$ representations which are \emph{almost scalars}. One may rightfully ask why we chose to make the redefinition (\ref{eq:Poi7c}) for the massless Dirac particles. As for rotation it seemed that $\phi^{\pm}(p)$ are the particles with spin $\pm 1/2$. The answer lies in infinitesimal boost generators. After the redefinition (\ref{eq:Poi7c}) the boosts take the correct form (\ref{eq:Poi3a1}) with $j=0$ \cite{Binegar:1981}. 

\subsection*{Vector fields} A vector field $A^{\nu}(p)$ transforms as follows:
\beq \label{eq:Poi8}
U(\Lambda) A_{\nu}(p) U^{-1}(\Lambda) = (\Lambda^{-1})_{\nu}^{\ \mu}A_{\mu}(\Lambda p).
\eeq
\subsubsection*{\textit{Real massive case}}
Massive vector field $A^{\mu}$ satisfies the field equation:
\beq \label{eq:Poi9}
(p^{2}-m^{2})A^{\nu}(p)=0
\eeq
and the subsidiary condition:
\beq \label{eq:Poi10}
p_{\nu}A^{\nu}(p)=0.
\eeq
Let us now try to extract the UIRs following \cite{Binegar:1981}. First we shall explicitly solve the subsidiary condition. We redefine our field as follows:
\beq \label{eq:Poi11}
B_{\mu}=[L^{-1}(p)]_{\mu}^{\ \nu} A_{\nu}(p),
\eeq
where as always $L(p)\hat{p}=p$. It is easy to show, that (\ref{eq:Poi10}) implies:
$$ B_{0}(p)=0 $$
and the transformation properties from (\ref{eq:Poi8}) are:
\beq \label{eq:Poi12}
U(\Lambda) B_{\nu}(p) U^{-1}(\Lambda) = [W(\Lambda^{-1},\Lambda p)]_{\nu}^{\ \mu}B_{\mu}(\Lambda p),
\eeq
where $W(\Lambda^{-1},\Lambda p)=L^{-1}(p)\Lambda^{-1}L(\Lambda p)$. Now we shall construct an infinitesimal rotation by an angle $\theta$. From (\ref{eq:Poi3aa}) we conclude that:
\beq \label{eq:Poi13}
U(\theta)B_{\rho}U^{-1}(\theta)=[R^{-1}(\theta)]_{\rho}^{\ \mu}B_{\mu}(Rp)
\eeq
and:
\beq \label{eq:Poi13a} U(\theta)B_{0}(p)U^{-1}(\theta)=0. \eeq
Writing (\ref{eq:Poi13}) infinitesimally for the spatial components we get:
$$ U(\theta)B_{i}(p)U^{-1}(\theta)=\left\{\delta_{ij}-\theta \epsilon_{ij}+\delta_{ij}\theta(p_{1}\frac{\partial}{\partial p^{2}}-p_{2}\frac{\partial}{\partial p^{1}})\right\}B_{j}(p).$$
The transformation properties of the hermitian fields $B_{i}(p)$ do not coincide with (\ref{eq:Poi3a}). Now we shall define a complex combination:
\beq \label{eq:Poi14}
\Phi(p)\equiv B_{1}(p)-iB_{2}(p).
\eeq
The transformation properties of this field under infinitesimal rotations are:
$$ U(\theta)\Phi(p)U^{-1}(\theta)=\left\{1-i\theta+\theta(p_{1}\frac{\partial}{\partial p^{2}}-p_{2}\frac{\partial}{\partial p^{1}})\right\}\Phi(p),$$
so $\Phi(p)\in U^{m,+,1}$. Thus the real vector field describes \emph{massive particles with spin 1}.

\subsubsection*{\textit{Massless case}}
Massless vector field satisfy the Maxwell equation:
\beq \label{eq:Poi15}
p^{2}A^{\mu}(p)-p^{\mu}p_{\lambda}A^{\lambda}(p)=0.
\eeq
Our aim is to identify the UIR, so we have to remove the gauge ambiguity. To do it let us fix the gauge covariantly taking the Lorentz condition:
\beq \label{eq:Poi16}
p_{\mu}A^{\mu}=0.
\eeq
From (\ref{eq:Poi15}) and (\ref{eq:Poi16}) it is easy to see that $A(p)\ne 0$ only if $p^{2}=0$.
Due to (\ref{eq:Poi15}) and (\ref{eq:Poi16}) the general form of $A(p)$ is:
\beq \label{eq:Poi17}
A^{\mu}=p^{\mu}\alpha(p)+p_{\perp}^{\mu}\beta(p),
\eeq
where $p_{\perp}^{\mu}=(0,p^{2},-p^{1})$ is orthogonal to $p$; $\alpha(p)$ and $\beta(p)$ are the scalar functions. Obviously the Lorentz condition does not remove the gauge freedom completely and $p^{\mu}\alpha(p)$ is a residual gauge freedom left after imposing the Lorentz gauge condition (\ref{eq:Poi16}). Consequently, the field has only \emph{one degree of freedom}. Let us now connect it with the UIR. To do it we define a scalar field $\phi(p)$ as:
\beq \label{eq:Poi18}
p_{\mu}\phi(p)=\epsilon_{\mu\nu\rho}p^{\nu}A^{\rho}(p).
\eeq
This new field does not change under any gauge transformation and solves the Klein-Gordon equation:
\beq
p^{\mu}p_{\mu}\phi=p^{\mu}(p)\epsilon_{\mu\nu\rho}p^{\nu}A^{\rho}(p)=0.
\eeq
As a result we have found the transformation, which connects $A^{\mu}(p)$ field to the \emph{scalar field} $\phi(p)$ that belongs to $U^{0,+,0,0}$ and has \emph{zero spin}.
\subsection*{MCS theory}
We began our discussion of the classical MCS theory in Chapter \ref{ch:classical}. It would be interesting to find out what type of particles this theory describes in its quantum version. For this reason we shall follow \cite{Jackiw:1990} closely. Let us remind that the covariant field equations for the free MCS theory is (\ref{eq:MCS2}):
\beq \label{eq:Poi19}
 \partial_{\rho}F^{\rho\nu}+\frac{m}{2}\epsilon^{\nu\rho\mu}F_{\rho\mu}=0.
\eeq
In p-representation the field equations in the dual form are:
\beq \label{eq:Poi20}
ip^{\mu}\epsilon^{\mu\nu\rho}\ ^{*}F_{\rho}+m^{*}F^{\nu}=0,
\eeq
where $^{*}F^{\nu}=\frac{1}{2}\epsilon^{\nu\alpha\beta}F_{\alpha\beta}$.
This implies the transversality condition:
\beq \label{eq:Poi21}
^{*}F^{\mu}p_{\mu }=0.
\eeq
In oder to get the UIR we shall solve the field equations (\ref{eq:Poi20}). The solution is:
\beq \label{eq:Poi22}
^{*}F^{\mu}=\frac{1}{\sqrt{2}}\left[ \left( \begin{array}{c} 0\\ 1\\ i\\ \end{array} \right)+\frac{p^{1}+ip^{2}}{m(E+m)} \left( \begin{array}{c} E+m\\ p^{1} \\p^{2} \end{array} \right) \right]\Psi(p),
\eeq
where $\Psi(p)$ is an unknown function. We shall now show that $\Psi(p)$ transforms as a massive spin-1 particle under the Lorentz transformation.

First, we know how the pseudovector field $^{*}F^{\alpha}(p)$ transforms under the Lorentz transformation. Lorentz generators for a pseudovector field are well-known:
\beq \label{eq:Poi23}
(j^{\mu})^{\alpha}_{\ \beta}=-i\epsilon^{\mu\nu\rho}p_{\nu}\frac{\partial}{\partial p_{\rho}}\delta^{\alpha}_{\ \beta}+i\epsilon^{\alpha\mu}_{\ \ \beta}.
\eeq
The question is now how the function $\Psi(p)$ transforms under Lorentz transformation. To answer this question it is useful first to represent $^{*}F^{\mu}(p)$ in the following form:
\beq \label{Poi24}
^{*}F^{\mu}(p)=B(p)^{\mu}_{\ \nu}N^{\nu}_{\ \rho} (^{*}F^{\rho}_{0}(p)),
\eeq
where $B(p)^{\mu}_{\ \nu}$ is a boost transformation in the 3-vector representation, which takes $\hat{p}\to p$ and $N$ is the numerical unitary matrix:
\beq \label{eq:Poi25}
N^{\nu}_{\ \rho}=\left( \begin{array}{ccc} \sqrt{2} & 0 & 0 \\
																						0 & 1 & 1 \\
																						0 & -i & i \\ \end{array}  \right)
\eeq
and $^{*}F^{\rho}_{0}(p)$ is:
\beq \label{eq:Poi26}
^{*}F^{\rho}_{0}(p)=\left( \begin{array}{c} 0\\ 0\\ \Psi(p) \\ \end{array} \right).
\eeq

The representation (\ref{Poi24}) is convenient because the transversality condition (\ref{eq:Poi21}) is satisfied manifestly.

The Lorentz generator acts on $^{*}F^{\rho}_{0}(p)$ as:
\beq \label{eq:Poi27}
(J^{\mu})^{\alpha}_{\ \beta}=\left(N^{-1}B^{-1}(p)j^{\mu}B(p)N(p)\right)^{\alpha}_{ \ \beta}.
\eeq
Now we can substitute (\ref{eq:Poi23}) into (\ref{eq:Poi27}) and use the relation (\ref{eq:Poi3ae}). Straightforward calculation lead us to:
\beq \label{eq:Poi28}
(J^{\mu})^{\alpha}_{\ \beta}=-i\epsilon^{\mu\nu\rho}p_{\nu}\frac{\partial}{\partial p^{\rho}}\left( \begin{array}{ccc} 1 & 0 & 0 \\ 0 & 1 & 0 \\ 0 & 0 & 1 \\ \end{array} \right)+ \Delta^{\mu} \left( \begin{array}{ccc} 0 & 0 & 0 \\ 0 & 1 &  0 \\ 0 & 0 & -1 \end{array} \right),
\eeq
where $\Delta^{\mu}$ is defined in (\ref{eq:Poi3ac}). The action on $\Psi(p)$ is:
\beq \label{eq:Poi29}
J^{\mu}=-i\epsilon^{\mu\nu\rho}p_{\nu}\frac{\partial}{\partial p^{\rho}}-\Delta^{\mu}.
\eeq
We notice readily that it is exactly the transformation of spin-1 particle with nonzero mass (\ref{eq:Poi3a2}). Thus we conclude that \emph{photons are massive spin-1 particles} in the MCS theory! 
\chapter{Field quantization}
\label{ch:quantization}
In this Chapter we shall quantize various \emph{free} fields. We shall pay special attention to quantization of the gauge fields (with and without the Chern-Simons term). The Coulomb gauge will be used to fix the gauge freedom.

\textbf{Scalar field:} We shall not carry out the canonical quantization procedure for the real scalar field in three dimensions because it is completely analogous to the conventional four-dimensional case. Here we shall only summarize the results. The field Lagrangian is:
\beq \label{eq:QF00}
\mathscr{L}=\frac{1}{2}\partial_{\mu} \phi \partial^{\mu} \phi-\frac{1}{2}m^{2}\phi^{2}.
\eeq
The free Heisenberg scalar field $\phi(\mathbf{x},t)$ can be written as the following linear combination of the creation operators $a^{\dag}_{\mathbf{p}}$ and annihilation operators $a_{\mathbf{p}}$:
\beq \label{eq:QF0A} \phi(\mathbf{x},t)=\int \frac{d^{2}p}{(2\pi)}\frac{1}{\sqrt{2E_{\mathbf{p}}}}(a_{\mathbf{p}}e^{-ip\cdot x}+a^{\dag}_{\mathbf{p}}e^{ip\cdot x}) \eeq
and it solves the Klein-Gordon equation:
$$ (\Box +m^{2})\phi=0. $$
For the future references the Feynman propagator $D_{F}=\bra{0}T\phi(x)\phi(y)\ket{0}$ is
\beq \label{eq:QF1}
D_{F}(x,y)=\int\frac{d^{3}p}{(2\pi)^{3}}\frac{i}{p^{2}-m^{2}+i\epsilon}e^{-ip\cdot(x-y)}.
\eeq

\textbf{Massive Dirac fields:} Quantization of the free massive Dirac field $\psi(\mathbf{x},t)$ in three dimensions follows completely the same lines as in four dimensions. For the future references we shall only write the results of this procedure here. The Dirac Lagrangian is
\beq \label{eq:QF1A0}
\mathscr{L}=\bar{\psi}(i\slashed{\partial}-m)\psi.
\eeq
The field $\psi$ may be written as a linear combination of the electron annihilation operators $a_{\mathbf{p}}$ and positron creation operators $b^{\dag}_{\mathbf{p}}$:
\beq \label{eq:QF1A} \psi(\mathbf{x},t)=\int \frac{d^{2}p}{(2\pi)}\frac{1}{\sqrt{2E_{\mathbf{p}}}}(a_{\mathbf{p}}u(\mathbf{p})e^{-ip\cdot x}+b^{+\dag}_{\mathbf{\mathbf{p}}}v(p)e^{ip\cdot x}). \eeq
The Feynman propagator $S_{F}=\bra{0}T\psi(x)\bar{\psi}(y)\ket{0}$ is
\beq \label{eq:QF2}
S_{F}(x,y)=\int\frac{d^{3}p}{(2\pi)^{3}}\frac{i(\slashed{p}+m)}{p^{2}-m^{2}+i\epsilon}e^{-ip\cdot(x-y)}.
\eeq

\textbf{Real massive vector field:} We shall only sketch quantization of a free massive vector field $B^{\mu}$ following \cite{Horejsi:2002}. The Lagrangian we start with is:
\beq \label{eq:QF2aa} \mathscr{L}=-\frac{1}{4}F_{\mu\nu}F^{\mu\nu}+\frac{1}{2}m^{2}B_{\mu}B^{\mu}. \eeq
Using the Euler-Lagrange equations it is easy to obtain the field equations:
\beq \label{eq:QF2a}
\partial_{\mu}F^{\mu\nu}+m^{2}B^{\nu}=0,
\eeq
which are equivalent to the following two equations:
\beq \label{eq:QF2b}
(\Box+m^{2})B^{\mu}=0 \qquad \partial_{\mu}B^{\mu}=0.
\eeq
The first equation tells us that the field $B^{\mu}$ describes some massive particles. The second equation in (\ref{eq:QF2b}) is a constraint, so a massive vector boson has only \emph{two degrees of freedom} in (2+1) physics. Let us find the plane wave solutions of (\ref{eq:QF2b}) now:
$$ B_{\mu}(x)=\epsilon_{\mu}(k)e^{-ik\cdot x}. $$
Substituting this plane wave into (\ref{eq:QF2b}) we obtain:
\beq \label{eq:QF2c} k^{2}=m^{2} \qquad k_{\mu}\epsilon^{\mu}=0. \eeq
There are two independent solutions of (\ref{eq:QF2c}), which satisfy the normalization condition:
\beq \label{eq:QF2d} \epsilon(k)\cdot \epsilon^{*}(k)=-1. \eeq
They are:
\beq \label{eq:QF2e} \epsilon^{\mu}(k,\lambda=1)=\epsilon^{\mu} _{T}(k)=(0,\mathbf{\epsilon}\thinspace(\mathbf{k})) \qquad \epsilon^{\mu}(k,\lambda=2)=\epsilon^{\mu} _{L}(k)=(\frac{|\mathbf{k}|}{m},\frac{k_{0}}{m}\frac{\mathbf{k}}{|\mathbf{k}|}),
\eeq
where $\mathbf{\epsilon}\thinspace(\mathbf{k})\cdot \mathbf{k}=0$.
We can also write the completeness relation:
\beq \label{eq:QF2f}
\sum_{\lambda=1}^{2}\epsilon_{\mu}(k,\lambda)\epsilon^{*}_{\nu}(k,\lambda)=-g_{\mu\nu}+\frac{1}{m^{2}}k_{\mu}k_{\nu}.
\eeq
Now the field $B_{\mu}(x)$ can be written in terms of a plane-wave expansion:
\beq \label{eq:QF2g}
B_{\mu}(x)=\sum^{2}_{\lambda=1}\int\frac{d^{2}k}{(2\pi)\sqrt{2E_{\mathbf{k}}}}\left[a(\mathbf{k},\lambda)\epsilon_{\mu}(\mathbf{k},\lambda)e^{-ik\cdot x}+a^{\dag}(\mathbf{k},\lambda)\epsilon^{*}_{\mu}(\mathbf{k},\lambda)e^{ik\cdot x}\right]
\eeq
and the commutation relations for the creation and the annihilation operators follow from the canonical quantization procedure:
\beq \label{eq:QF2h}
[a(\mathbf{k},\lambda),a^{+\dag}(\mathbf{k}^{\prime},\lambda^{\prime})]=\delta_{\lambda, \lambda^{\prime}}\delta^{(2)}(\mathbf{k}-\mathbf{k^{\prime}}).
\eeq
Thus we are dealing with bosons, which is in accordance with our previous result from Chapter \ref{ch:Poincare}.

For the future references we write here the covariant version of the Feynman propagator (for more details \cite{Horejsi:2002}):
\beq \label{eq:QF2j}
(D_{F})_{\mu\nu}(q)=i\frac{-g_{\mu\nu}+m^{-2}q_{\mu}q_{\nu}}{q^{2}-m^{2}+i\epsilon}.
\eeq

\textbf{Massless gauge vector field in Coulomb (radiation) gauge:} In this subsection our main aim is to quantize the free electromagnetic gauge field without the Chern-Simons term. We closely follow the treatment of \cite{Bjorken:RQF} here. It is nontrivial to quantize the electromagnetic field because there are more variables than there are degrees of freedom. The Maxwell Lagrangian is:
\beq \label{QFF0} \mathscr{L}=-\frac{1}{4}F_{\mu\nu}F^{\mu\nu}. \eeq
In order to perform the canonical quantization we first construct the conjugate momenta:
\beq \label{eq:QFF1}
\pi^{0}=\frac{\partial\mathscr{L}}{\partial \dot{A_{0}}}=0 \qquad \pi^{k}=\frac{\partial\mathscr{L}}{\partial \dot{A_{k}}}=-\dot{A^{k}}-\frac{\partial A_{0}}{\partial x^{k}}=E^{k}.
\eeq
We have obtained the primary constraint $\pi^{0}=0$, due to the fact that the Lagrangian (\ref{QFF0}) does not depend on $\dot{A}^{0}$. We can find the secondary constraint in the theory if we calculate the Euler-Lagrange field equation for the time component of the electromagnetic field:   
\beq \label{eq:QFF1a}
\partial_{k}\underbrace{\frac{\partial \mathscr{L}}{\partial(\partial_{k}A_{0})}}_{E^{k}}=\frac{\partial \mathscr{L}}{\partial A_{0}}=0,
\eeq
which does not contain the second order time derivative. We easily recognize the Gauss law.

Let us now construct the Hamiltonian:
\beq \label{eq:QFF2}
H=\int d^{2}x (\pi^{k}\dot{A_{k}}-\mathscr{L})=\int d^{2}x \frac{1}{2}(\mathbf{E}^{2}+B^{2}).
\eeq
We attempt to follow the canonical quantization procedure imposing the following equal-time commutators:
\beq \label{eq:QFF3}
[A^{\mu}(\mathbf{x},t),A^{\nu}(\mathbf{x^{\prime}},t)]=0
\eeq
\beq \label{eq:QFF4} 
[\pi^{k}(\mathbf{x},t),\pi^{j}(\mathbf{x^{\prime}},t)]=0
\eeq
\beq \label{eq:QFF5}
[\pi^{k}(\mathbf{x},t),A^{0}(\mathbf{x^{\prime}},t)]=0.  
\eeq
It is evident that $A^{0}$ is a $c$-number because it commutes with both $\pi^{k}$ (\ref{eq:QFF5}) and $A^{\mu}$ (\ref{eq:QFF3}). Thus it is not a physical degree of freedom. According to the canonical quantization procedure the commutator of $\pi^{i}$ and $A_{j}$ should take the following form:
\beq \label{eq:QFF5a}
[\pi^{i}(\mathbf{x},t),A_{j}(\mathbf{x^{\prime}},t)]=[E^{i}(\mathbf{x},t),A_{j}(\mathbf{x^{\prime}},t)]=-i\delta_{ij}\delta(\mathbf{x}-\mathbf{x^{\prime}}).
\eeq
There is a difficulty, however, that the last relation does not satisfy the Gauss constraint (\ref{eq:QFF1a}). Taking the divergence $\nabla^{i}_{x}$ from the RHS of (\ref{eq:QFF5a}) we do not get zero:
\beq \label{eq:QFF5b}
-i\delta_{ij}\frac{\partial}{\partial x^{i}}\int\frac{d^{2}k}{(2\pi)^{2}}\exp\left(i\mathbf{k}\cdot(\mathbf{x}-\mathbf{x^{\prime}})\right)=\delta_{ij}\int\frac{d^{2}k}{(2\pi)^{2}}\exp(i\mathbf{k}\cdot(\mathbf{x}-\mathbf{x^{\prime}}))k^{i}.
\eeq
Thus we should modify the RHS of the last commutator to be consistent with the Gauss law. In order to achieve this aim we introduce the transverse delta function:
\beq \label{eq:QFF5c} \delta^{tr}_{ij}(\mathbf{x}-\mathbf{x}^{\prime})=\int \frac{d^{2}k}{(2\pi)^{2}}e^{i\mathbf{k}\cdot(\mathbf{x}-\mathbf{x^{\prime}})}\left(\delta_{ij}-\frac{k_{i}k_{j}}{\mathbf{k}^{2}}  \right). \eeq
It is straightforward to show that this modification is consistent with the Gauss constraint. Thus our final result for the last commutator (\ref{eq:QFF5a}) is:
\beq \label{eq:QFF6}
[\pi^{i}(\mathbf{x},t),A^{j}(\mathbf{x^{\prime}},t)]=i\delta^{tr}_{ij}(\mathbf{x}-\mathbf{x^{\prime}}).
\eeq 
Acting with $\nabla_{\mathbf{x}^{\prime}}$ on (\ref{eq:QFF6}) we see that the operator $\nabla\cdot\mathbf{A}$ is a $c$-number because it commutes with the canonical fields (\ref{eq:QFF3}) and the conjugate momenta (\ref{eq:QFF6}). It is possible to find a gauge, where $\nabla\cdot\mathbf{A}=0$ and $A_{0}=0$. This gauge is called the radiation or Coulomb gauge. It has an advantage that only one real transverse degree of freedom appears in the formalism. The problem is, however, that it is not manifestly covariant, so we must check the covariance of the theory. To do so we should find the transformation properties of the field under the Poincar\'e transformations and show that it transforms as a 3-vector up to gauge terms.

To accomplish this aim we shall construct the generators of the Lorentz and translation transformations with the help of Noether's theorem working in the radiation gauge. For temporal translations we obtain:  
\beq \label{eq:QFF7}
P^{0}=H=\int d^{2}x \mathscr{H}(x)=\frac{1}{2}\int d^{2}x ((\dot A)^{2}+B^{2}),
\eeq
which is our Hamiltonian. For spatial translations Noether's theorem yields:
\beq \label{eq:QFF8}
P^{i}=-\int d^{2}x\sum^{2}_{j=1}(\dot A_{j}\partial^{i}A_{j}),
\eeq
which is the momentum of the field. It can be explicitly verified that the operators $A^{\mu}(x)$ transform as fields under the infinitesimal translations:
\beq \label{eq:QFF8a}
i[P^{\mu},A_{\nu}(x)]=\frac{\partial A_{\nu}(x)}{\partial x_{\mu}}.
\eeq

Let us find the generator of spatial rotations in the radiation gauge:
\beq \label{eq:QFF9}
M=\underbrace{\int d^{2}x \left( \sum^{2}_{j=1}\dot{A}^{j}(x^{1}\frac{\partial}{\partial x^{2}}-x^{2}\frac{\partial}{\partial x^{1}})A^{j} \right)}_{M_{I}}+ \underbrace{\int d^{2}x(\dot{A}^{1}A^{2}-\dot{A}^{2}A^{1})}_{M_{II}}.
\eeq
It can be shown that under an infinitesimal spatial rotation characterized by the infinitesimal parameter $\epsilon_{\mu\nu}$ ($\epsilon_{ij}=-\epsilon_{ji}, \ \epsilon_{\mu 0}=0 \ \mu=0,1,2$ ) the electromagnetic potential $\mathbf{A}$ transforms as a 2-vector field:
$$ U(\epsilon)A^{j}U^{-1}(\epsilon)=A^{j}(x^{\prime})-\epsilon^{j}_{\ k}A^{k}(x^{\prime})=[S^{-1}]^{j}_{\ i}(\epsilon)A^{i}(x^{\prime}), $$
where $x^{\prime \thinspace i}=(\delta^{i}_{j}+ \epsilon^{i}_{\ j})x^{j}$ and $S^{-1}(\epsilon)^{j}_{\ i}=(\delta^{j}_{i}- \epsilon^{j}_{\ i})$.

There are two boost generators, which can be found by Noether's theorem:
\beq \label{eq:QFF10}
B^{k}=\int d^{2}x \left( t\sum^{2}_{r=1}\dot{A}^{r}\frac{\partial A^{r}}{\partial x_{k}}-x^{k} \mathscr{H}(x) \right).
\eeq
Infinitesimal boosts with the parameters $\epsilon^{\mu\nu}$ ($e^{0\mu}=-e^{\mu 0}, e^{ij}=0, \ i,j=1,2$) are:
\beq \label{eq:QFF11}
U(\epsilon)A^{\mu}(x)U^{-1}(\epsilon)=A^{\mu}(x^{\prime})-\epsilon^{\mu\nu}A_{\nu}(x^{\prime})+\frac{\partial \Lambda (x^{\prime},\epsilon)}{\partial x^{\prime}_{\mu}},
\eeq
where the gauge term is:
\beq \label{eq:QFF12}
\frac{\partial \Lambda (x^{\prime},\epsilon)}{\partial x^{\prime}_{\mu}}=\epsilon_{0k}\frac{\partial}{\partial x^{\prime}_{\mu}}\int \frac{d^{2}x}{2\pi}\ln(|\mathbf{x}-\mathbf{x^{\prime}}|)\frac{\partial A^{k}(\mathbf{x},t)}{\partial t}.
\eeq 
Thus the electromagnetic field transforms under the Lorentz transformation as a 3-vector up to the gauge term (\ref{eq:QFF12}). It is also possible to show \cite{Bjorken:RQF} that the radiation gauge is invariant under the Lorentz transformations and that the commutators (\ref{eq:QFF3})-(\ref{eq:QFF6}) are also invariant. Hence the quantization procedure is covariant.

Maxwell equations for $\mathbf{A}(x)$ in the radiation gauge take the form $\Box \mathbf{A}=0$, so we can make the expansion:
\beq \label{eq:QFF13}
A^{i}(\mathbf{x},t)=\int \frac{d^{2}k}{2\pi\sqrt{2\omega}}\epsilon^{i}(\mathbf{k})[a(\mathbf{k})e^{-ik\cdot x}+a^{+\dag}(\mathbf{k})e^{ik\cdot x}],
\eeq
where $\epsilon^{i}(\mathbf{k})$ is a real transverse polarization vector, which satisfies $\mathbf{\epsilon}(k)\cdot \mathbf{k}=0$ and $\omega=|\mathbf{k}|$. With the relation (\ref{eq:QFF13}) and an analogous decomposition for the conjugate fields $\pi^{i}$, the operators $a(\mathbf{k})$ and $a^{\dag}(\mathbf{k})$ can be expressed in terms of $\mathbf{A}(x)$ and $\pmb{\pi}(x)$. Using the relations (\ref{eq:QFF3})-(\ref{eq:QFF6}) we easily get:
\beq \label{eq:QFF14}
[a(\mathbf{k}),a(\mathbf{k}^{\prime})]=[a^{\dag}(\mathbf{k}),a^{\dag}(\mathbf{k}^{\prime})]=0 \qquad [a(\mathbf{k}),a^{\dag}(\mathbf{k}^{\prime})]=\delta(\mathbf{k}-\mathbf{k^{\prime}}),
\eeq
which are the well-known commutation relations for the creation and annihilation operators of bosons. 

Now we shall demonstrate that photons have zero spin in the free Maxwell theory in three dimensions. To do so, let us calculate:
$$ M a^{\dag}(\mathbf{k})\ket{0}=[M, a^{\dag}(\mathbf{k})]\ket{0}, $$
where $M$ is given by (\ref{eq:QFF9}). The operator $M$ consists of two terms, the first of them $M_{I}$, which contains partial derivatives of the electromagnetic field, can be identified with the orbital part of the angular momentum operator. The second term $M_{II}=\int d^{2}x (\dot{A}^{1}A^{2}-\dot{A}^{2}A^{1})$ can be identified with the spin part of the angular momentum. Now writing $M_{II}$ in terms of $a(\mathbf{k})$ and $a^{\dag}(\mathbf{k})$ and using (\ref{eq:QFF14}) we get:
$$ [M_{II}, a^{\dag}(\mathbf{k})]=0. $$ 
This means that the spin of photon is \emph{zero}. This result agrees with our findings in Chapter \ref{ch:Poincare} that the real gauge field describes a particle with zero spin.  

Now we will construct the Feynman propagator in the radiation gauge. Using (\ref{eq:QF1}) we obtain:
\beq \label{eq:QFF15}
D_{F}(x,y)_{\mu \nu}=\bra{0}T(A_{\mu}(x)A_{\nu}(y))\ket{0}=\int\frac{d^{3}k}{(2\pi)^{3}}\frac{i}{k^{2}+i\epsilon}e^{-ik\cdot(x-y)}\epsilon_{\mu}(k)\epsilon_{\nu}(k).
\eeq
In the reference frame, where we performed the quantization let us define two vectors $\eta^{\mu}$ and ${\hat{k}^{\mu}}$:
$$ \eta^{\mu}=(1,0,0) \qquad \hat{k}^{\mu}=\frac{k^{\mu}-(k\cdot \eta)\eta^{\mu}}{\sqrt{(k\cdot \eta)^{2}-k^{2}}}. $$
These two vectors and the polarization vector $\epsilon^{\mu}(\mathbf{k})=(0, \pmb{\epsilon}(\mathbf{k}))$ form a complete set of unit orthogonal vectors in the Minkowski space. We can write:
$$ \epsilon_{\mu}(k)\epsilon_{\nu}(k)=-g_{\mu \nu}+\eta_{\mu}\eta_{\nu}-\hat{k}_{\mu}\hat{k}_{\nu} $$
Using the last relation in (\ref{eq:QFF15}) the Feynman propagator in the Coulomb gauge is:
\beq \label{eq:QFF16}
D_{F}(x,y)_{\mu\nu}=g_{\mu \nu}D_{F}(x,y)+i\int \frac{d^{3}k}{(2\pi)^{3}} \frac{e^{-ik\cdot (x-y)}}{(k^{2}+i\epsilon)} \frac{k^{2}\eta_{\mu}\eta_{\nu}-(k\cdot \eta)(k_{\mu}\eta_{\nu}+\eta_{\mu}k_{\nu})+k_{\mu}k_{\nu}}{(k\cdot \eta)^{2}-k^{2}}.
\eeq

\textbf{Free Maxwell-Chern-Simons theory:}
Here we shall quantize the free Maxwell-Chern-Simons theory in the Coulomb gauge following closely \cite{DJT:1981} and \cite{Haller:1994}. First we shall implement the Coulomb gauge condition explicitly into the Lagrangian, which we write here in the noncovariant form:
\beq \label{QMCS1}
\mathscr{L}=-\frac{1}{4}F_{ln}F^{ln}-\frac{1}{2}F_{0l}F^{0l}+\frac{1}{4}m\epsilon_{ln}(F_{ln}A_{0}+2F_{0l}A_{n})-G\partial_{l}A^{l},
\eeq 
where $l,n=1,2$, the Einstein summation is assumed and $G$ is a new auxiliary canonical variable. Using the Euler-Lagrange equation we easily obtain the field equation (\ref{eq:MCS2}) and the Coulomb gauge condition $\nabla\cdot \mathbf{A}=0$. To construct the Hamiltonian the canonical momenta are first calculated:
\beq \label{QMCS2}
\Pi_{i}=-F_{0i}-\frac{1}{2}m\epsilon_{in}A_{n} \quad \Pi_{0}=0 \quad \Pi_{G}=0.
\eeq 
It turns out that only $A^{i}$ are the canonical variables with nontrivial conjugate momenta. The remaining fields $A^{0}$ and $G$ must be the solutions of some constraints. We also recognize two primary constraints $C_{1}=\Pi_{0}$ and $C_{2}=\Pi_{G}$. So the total Hamiltonian in the Coulomb gauge takes the form:
\begin{eqnarray} \label{QMCS3}
H^{T}_{c}&=&\int d^{2}x (\Pi_{i}\dot{A^{i}}-\mathscr{L}-U_{1}C_{1}-U_{2}C_{2})= \nonumber \\
&=& \int d^{2}x[\frac{1}{2}\Pi_{l}\Pi_{l}+\frac{1}{4}F_{ln}F_{ln}+A_{0}\partial_{i}\Pi_{i}+
\frac{1}{8}m^{2}A_{l}A_{l}+G\partial_{l}A_{l}- \nonumber \\ 
&& \ \ \ \ \ \ \ \ \ \frac{1}{4}m\epsilon_{ln}F_{ln}A_{0}+\frac{m}{2}\epsilon_{ln}\Pi_{l}A_{n}-U_{1}C_{1}-U_{2}C_{2}],
\end{eqnarray}
where $U_{1}$ and $U_{2}$ are yet unknown functions of spacetime. Primary constraints must be satisfied at all times, which implies that
\beq \label{QMCS4} \partial_{0}C_{i}=[H^{T}_{c},C_{i}]_{P}=0, \eeq
where $[,]_{P}$ denotes the Poisson bracket. This condition might imply secondary constraints and it is indeed the case. Using (\ref{QMCS4}) for $C_{1}$ yields a Gauss constraint $C_{3}$:
\beq \label{QMCS5}
C_{3}=\partial_{l}\Pi_{l}-\frac{1}{4}m\epsilon_{ln}F_{ln}
\eeq
and for the evolution of $C_{2}$ we obtain the gauge constraint $C_{4}$
\beq \label{QMCS6}
C_{4}=\partial_{l}A_{l}.
\eeq
Now we must apply (\ref{QMCS4}) to secondary constraints $C_{3}$ and $C_{4}$. For $C_{3}$ a new constraint is obtained:
\beq \label{QMCS7}
C_{6}=\partial_{l}\partial_{l}G.
\eeq
For the gauge constraint $C_{4}$ we readily get:
\beq \label{QMCS8}
C_{5}=\partial_{l}\Pi_{l}-\partial_{l}\partial_{l}A_{0}+\frac{1}{4}m\epsilon_{ln}F_{ln},
\eeq
hence the scalar part of the vector potential $A_{0}$ can be expressed as follows:
\beq \label{QMCS9}
A_{0}=\frac{1}{\nabla^{2}}(\partial_{l}A_{l}+\frac{1}{4}m\epsilon_{ln}F_{ln}).
\eeq
Now we should repeat the procedure for the new secondary constraints $C_{5}$ and $C_{6}$, but it is straightforward to show that they do not generate any new constraints but fix two unknown functions $U_{1}$ and $U_{2}$.

Now we shall construct the matrix $\mathscr{M}_{ij}(\mathbf{x},\mathbf{y})=[C_{i}(\mathbf{x}),C_{j}(\mathbf{y})]_{P}$ that will be useful for us later. This matrix is:
\beq \label{QMCS10}
\mathscr{M}(\mathbf{x}-\mathbf{y})= \left(
\begin{array}{cccccc}
0&0&0&0&\nabla^{2}&0 \\
0&0&0&0&0&-\nabla^{2} \\
0&0&0&-\nabla^{2}&0&0 \\
0&0&\nabla^{2}&0&\nabla^{2}&0 \\
-\nabla^{2}&0&0&-\nabla^{2}&0&0 \\
0&\nabla^{2}&0&0&0&0 \\
\end{array} \right) \delta^{(2)}(\mathbf{x}-\mathbf{y}).
\eeq
It has the following inverse $\mathscr{Y}(\mathbf{x}-\mathbf{y})$:
\beq \label{QMCS10a}
\mathscr{Y}(\mathbf{x}-\mathbf{y})= \left(
\begin{array}{cccccc}
0&0&\frac{1}{\nabla^{2}}&0&-\frac{1}{\nabla^{2}}&0 \\
0&0&0&0&0&\frac{1}{\nabla^{2}} \\
-\frac{1}{\nabla^{2}}&0&0&\frac{1}{\nabla^{2}}&0&0 \\
0&0&-\frac{1}{\nabla^{2}}&0&0&0 \\
\frac{1}{\nabla^{2}}&0&0&0&0&0 \\
0&-\frac{1}{\nabla^{2}}&0&0&0&0 \\
\end{array} \right) \delta^{(2)}(\mathbf{x}-\mathbf{y}).
\eeq 

The existence of the nonsingular matrix $\mathscr{M}$ implies that the theory has only the second-type constraints and we can employ the Dirac-Bargmann procedure to work in the Hamiltonian formalism. Following Dirac the equation of motion for the canonical variables $z(x)=\{{A^{i}(x),\Pi_{i}(x)}\}$ is:
\beq \label{QMCS11}
\dot{z}(x)=[z(x),H_{c}]_{D},
\eeq
where $H_{c}$ is the constrained Coulomb Hamiltonian and $[,]_{D}$ denotes the Dirac bracket defined by the following relation:
\beq \label{QMCS12}
[F(\mathbf{x}),G(\mathbf{y})]_{D}=[F(\mathbf{x}),G(\mathbf{y})]_{P}-\int d\mathbf{z_{1}}d\mathbf{z_{2}}\sum_{i,j}[F(\mathbf{x}),C_{i}(\mathbf{z_{1}})]_{P}\mathscr{Y}_{ij}(\mathbf{z_{1}},\mathbf{z_{2}})[C_{j}(\mathbf{z_{2}}),G(\mathbf{y})]_{P}.
\eeq
We construct the constrained Coulomb Hamiltonian applying all the found constraints $C_{1}-C_{6}$ and the relation (\ref{QMCS9}) to the unconstrained Hamiltonian (\ref{QMCS3}). The result is:
\beq \label{QMCS13}
H_{c}=\frac{1}{2}\int d^{2}x \left[(\Pi^{T})^{2}+A^{T}_{i}(-\nabla^{2}+m^{2})A^{T}_{i}\right],
\eeq
where $\Pi^{T}_{i}$ and $A^{T}_{i}$ are the transverse parts of the canonical momentum and the field:
\beq \label{QMCS14}
\Pi^{T}_{i}=(\delta_{ij}-\frac{\partial_{i}\partial_{j}}{\nabla^{2}})\Pi_{j} \qquad
A^{T}_{i}=(\delta_{ij}-\frac{\partial_{i}\partial_{j}}{\nabla^{2}})A_{j}.
\eeq
Thus the theory possesses only one dynamical degree of freedom, which is the transverse part of the field.
To quantize the theory we simply make the substitution:
\beq \label{QMCS15}
i[A,B]_{D}\to [\hat{A},\hat{B}].
\eeq
Using (\ref{QMCS12}) it is not difficult to  obtain the following equal-time canonical commutators:
\beq \label{QMCS16}
[A^{T}_{i}(\mathbf{x}),A^{T}_{j}(\mathbf{y})]=0,
\eeq
\beq \label{QMCS17}
[\Pi^{T}_{i}(\mathbf{x}),\Pi^{T}_{j}(\mathbf{y})]=0,
\eeq
\beq \label{QMCS18}
[(A^{T})^{i}(\mathbf{x}),\Pi^{T}_{j}(\mathbf{y})]=i(\delta^{i}_{j}-\frac{\partial_{i}\partial_{j}}{\nabla^{2}})\delta(\mathbf{x}-\mathbf{y}).
\eeq
The last commutator is identical to (\ref{eq:QFF6}) that we obtained for the free  Maxwell Lagrangian without  the Chern-Simons term. 
Having the canonical commutators at hand, we are now able to construct the Heisenberg equations of motion for the canonical variables:
\beq \label{QMCS19}
(\dot{A}^{T})^{i}=-i[(A^{T})^{i},H_{c}]=\Pi^{T}_{i} \qquad \dot{\Pi}^{T}_{i}=-i[\Pi^{T}_{i},H_{c}]=(\nabla^{2}-m^{2})(A^{T})^{i}.
\eeq
Now we can combine these two equations to obtain a field equation for the massive physical transverse field $\mathbf{A}^{T}$:
\beq \label{QMCS20}
(\Box+m^{2})\mathbf{A}^{T}=0.
\eeq

In Chapter \ref{ch:Poincare} it has already been shown that the Maxwell-Chern-Simons field carries spin 1. It is instructive, however, to derive the same result in a somewhat different fashion. We shall closely follow \cite{DJT:1981} here. The transverse field $\mathbf{A}^{T}$ satisfies (\ref{QMCS20}), which can be explicitly solved:
\beq \label{QMCS21}
\mathbf{A}^{T}=\int\frac{d^{2}p}{(2\pi)\sqrt{2E_{\mathbf{p}}}}(\pmb{\epsilon}_{\mathbf{p}}a_{\mathbf{p}}e^{-ip\cdot x}+\pmb{\epsilon}^{*}_{\mathbf{p}}a^{\dag}_{\mathbf{p}}e^{ip\cdot x}),
\eeq
where the dispersion relation for a massive photon is $E_{\mathbf{p}}=\sqrt{\mathbf{p}^{2}+m^{2}}$. The Coulomb gauge constraint and the normalization condition $\pmb{\epsilon}_{\mathbf{p}}\cdot\pmb{\epsilon}_{\mathbf{p}}=1$ almost determine the polarization vector $\pmb{\epsilon}_{\mathbf{p}}$. Thus, our first guess for it is:
\beq \label{QMCS21a}
\epsilon^{i}_{\mathbf{p}}=\frac{\epsilon^{ij}p^{j}}{\mathbf{p}^{2}}.
\eeq

 Similar to the pure Maxwell theory the operators $a^{\dag}_{\mathbf{p}}$ and $a_{\mathbf{p}}$ satisfy the canonical commutation relations for Bose particles. It is tempting to identify these operators with the creation and annihilation operators of a massive photon. We have to be cautious, however. For the identification to be correct, \emph{the particles must transform correctly under translations and Lorentz transformations}.

We begin with the translation symmetry. Using Noether's theorem the canonical energy-momentum tensor is obtained:
\beq \label{QMCS22}
T^{\mu}_{\ \nu}=-g^{\mu}_{\ \nu}\mathscr{L}+(\frac{m}{2}\epsilon^{\mu\rho\alpha}A_{\alpha}-F^{\mu\rho})\partial_{\nu}A_{\rho}.
\eeq
The total momentum operator is $P^{\mu}=\int_{x^{0}=t} d^{2}x T^{0\mu}(\mathbf{x},t)$ and we shall write it in terms of $a^{\dag}_{\mathbf{p}}$ and $a_{\mathbf{p}}$. According to (\ref{QMCS22}) the Hamiltonian is:
\beq \label{QMCS23}
H=P^{0}=\frac{1}{2}\int d^{2}x(\mathbf{E}^{2}+B^{2})=\int d^{2}x (-\frac{1}{2}F^{0j}F_{0j}+\frac{1}{4}F^{ik}F_{ik}).
\eeq
Our next aim is to express the Hamiltonian $H$ in terms of the creation and annihilation operators. For this purpose it is convenient to introduce the spatial Fourier transformation:
\beq \label{QMCS24}
\widetilde{F}^{\alpha \beta}(\mathbf{p},t)=\int d^{2}x F^{\alpha \beta}(\mathbf{x},t)e^{-i \mathbf{p}\cdot \mathbf{x}}.
\eeq
The Hamiltonian is now:
\beq \label{QMCS25}
H=\int d^{2}p \left[-\frac{1}{2}\widetilde{F}^{0j}(\mathbf{p})\widetilde{F}_{0j}(\mathbf{-p})+\frac{1}{4}\widetilde{F}^{ik}(\mathbf{p})\widetilde{F}_{ik}(\mathbf{-p})\right].
\eeq
The following relations will be useful in the subsequent calculation:
\beq \label{QMCS26}
\begin{array}{c}
\widetilde{F}^{ik}=i(p^{i}\widetilde{A}^{k}-p^{k}\widetilde{A}^{i}) \\
\widetilde{F}^{0k}=\partial^{0}\widetilde{A}^{k}-ip^{k}\widetilde{A}^{0}) \\
\widetilde{A}^{i}(\mathbf{p})=\frac{1}{\sqrt{2E_{\mathbf{p}}}}(a_{\mathbf{p}}\epsilon^{i}_{\mathbf{p}}e^{-ip^{0}t}+a^{+\dag}_{\mathbf{-p}}(\epsilon^{*})^{i}_{\mathbf{-p}}e^{ip^{0}t})\\
\widetilde{A}^{0}=\frac{m\widetilde{B}}{\mathbf{p}^{2}}, \\ \end{array}
\eeq
where the last relation is simply the Fourier transformation of (\ref{QMCS9}). Substituting (\ref{QMCS26}) into (\ref{QMCS25}) we obtain:
\beq \label{QMCS27}
H=\int d^{2}p E_{\mathbf{p}}a^{\dag}_{\mathbf{p}}a_{\mathbf{p}}.
\eeq
Very similar calculations yield the spatial momentum operator $P^{i}$:
\beq \label{QMCS28}
P^{i}=\int d^{2}p \ p^{i}a^{\dag}_{\mathbf{p}}a_{\mathbf{p}}.
\eeq
So far we have obtained the anticipated results, so let us look at Lorentz symmetry.

The Lagrangian (\ref{QMCS1}), which we used for the quantization is not Lorentz-invariant. It is so because we introduced the auxiliary variable $G$ to implement the Coulomb gauge. The real theory is covariant, so we shall use (\ref{eq:MCS1}) with $j^{\mu}=0$  as our covariant Lagrangian. Noether's theorem for Lorentz symmetries leads to the existence of the conserved currents $\mathscr{M}^{\mu(\alpha \beta)}$:
\beq \label{QMCS29}
\mathscr{M}^{\mu(\alpha \beta)}=x^{\alpha}T^{\mu\beta}-x^{\beta}T^{\mu\alpha}-{i}\frac{\partial \mathscr{L}}{\partial(\partial_{\mu}A_{\phi})}(\Sigma^{(\alpha \beta)})^{\phi}_{\ \rho}A^{\rho},
\eeq
where $(\Sigma^{(\alpha \beta)})^{\mu}_{\ \nu}$ are the generators of the Lorentz group in the vector representation (\ref{eq:RQM4}). The rotation generator $M$ may be written as:
\beq \label{QMCS30}
M=\int_{x^{0}=t} d^{2}x \mathscr{M}^{0(12)}(x)=\int d^{2}x \left( x^{1}T^{02}-x^{2}T^{01}-F^{01}A^{2}+F^{02}A^{1}+\frac{m}{2}A^{i}A_{i}  \right).
\eeq
Now we substitute $A^{0}$ from (\ref{QMCS9}) into the last expression. We also assume that the fields are so small at infinity that the surface term can be neglected in integrations by parts. Doing so we readily obtain:
\beq \label{QMCS31}
M=\int_{x^{0}=t} d^{2}x \dot{A}^{j}(x^{2}\partial^{1}-x^{1}\partial^{2})A_{j}.
\eeq
Repeating techniques we used in the calculation of the total momentum $P^{\mu}$, the rotation generator can be written as follows:
\beq \label{QMCS32}
M=\int d^{2}p a^{+\dag}_{\mathbf{p}}\frac{1}{i}\frac{\partial}{\partial \theta} a_{\mathbf{p}},  \qquad \theta=\arctan(\frac{p^{2}}{p^{1}}).
\eeq
Looking at (\ref{eq:Poi3a3}) it might seem that photon has zero spin in the MCS theory. We should, however, construct the boost generators first. From (\ref{QMCS29}) they are:
\beq \label{QMCS33}
N_{j}=\int_{x^{0}=t} d^{2}x \mathscr{M}^{0(j0)}=\int_{x^{0}=t} d^{2}x \left( x^{j}\mathscr{H}-x^{0}\mathscr{P}^{j}-\partial^{0}(A^{j})A^{0}\right),
\eeq
where $\mathscr{H}=T^{00}$ and $\mathscr{P}^{i}=T^{0i}$. Now we would like to express the boost generators in terms of $a^{\dag}_{\mathbf{p}}$ and $a_{\mathbf{p}}$. First we take $t=0$, so the second term in the previous formula can be dropped. It is convenient to separate the boost generator $N_{j}$ into two distinct parts $N^{I}_{j}=\int d^{2}x (-\partial^{0}(A^{j})A^{0})$ and $N^{II}_{j}=\int d^{2}x (x^{j}\mathscr{H})$. We shall calculate them separately. Let us start with $N^{I}_{j}$. Fourier-transforming and using (\ref{QMCS26}) we get:
\beq \label{QMCS34}
N^{I}_{j}=-\int d^{2}p \partial^{0}(\widetilde{A}^{j}(\mathbf{p}))\frac{m\widetilde{B}(-\mathbf{p})}{\mathbf{p}^{2}}=-m\int d^{2}p \frac{\epsilon^{jk}p^{k}}{\mathbf{p}^{2}}a^{\dag}_{\mathbf{p}}a_{\mathbf{p}}.
\eeq  
Now we can proceed with $N^{II}_{j}$. Using (\ref{QMCS13}) and (\ref{QMCS19}), Fourier-transforming and then using (\ref{QMCS26}):
\beq \label{QMCS35} \begin{array}{c}
N^{II}_{j}=\int d^{2}x x^{j} \mathscr{H}= \\ \frac{1}{2}\int d^{2}x x^{j} A^{i}(m^{2}-\partial_{0}^{2}-\nabla^{2})A^{i}= \\ 
\int d^{2}p (-i\partial^{j}\widetilde{A}^{i}(\mathbf{p}))E^{2}_{\mathbf{p}}\widetilde{A}^{i}(-\mathbf{p})= \\ \frac{i}{2}\int d^{2}p E_{\mathbf{p}}(a^{\dag}_{\mathbf{p}}\partial_{j}a_{\mathbf{p}}-[\partial_{j}a^{\dag}_{\mathbf{p}}]a_{\mathbf{p}}).
\end{array}
\eeq
Our final expression for the boost generator is:
\beq \label{QMCS36}
N_{j}=\frac{i}{2}\int d^{2}p E_{\mathbf{p}}\left(a^{\dag}_{\mathbf{p}}\partial_{j}a_{\mathbf{p}}-\partial_{j}[a^{\dag}_{\mathbf{p}}]a_{\mathbf{p}}\right)-m\int d^{2}p \frac{\epsilon^{jk}p^{k}}{\mathbf{p}^{2}}a^{\dag}_{\mathbf{p}}a_{\mathbf{p}},
\eeq
which poses a problem: The last formula for the boost generator does not coincide with our previous result (\ref{eq:Poi3a3})! Let us remind that for a general spin $j$ particle we obtained in (\ref{eq:Poi3a3}):
\beq \label{QMCS36a}
N_{j}=\frac{i}{2}\int d^{2}p E_{\mathbf{p}}\left(a^{\dag}_{\mathbf{p}}\partial_{j}a_{\mathbf{p}}-\partial_{j}[a^{\dag}_{\mathbf{p}}]a_{\mathbf{p}}\right)-j\int d^{2}p \frac{\epsilon^{jk}p^{k}}{E_{\mathbf{p}}+m}a^{\dag}_{\mathbf{p}}a_{\mathbf{p}}
\eeq
that does not coincide with (\ref{QMCS36}) for any value of $j$.

Something must be wrong. The problem can be solved by a redefinition of the creation and annihilation operators \cite{DJT:1981}. Let us define new operators $\widetilde{a}^{\dag}_{\mathbf{p}}$ and $\widetilde{a}_{\mathbf{p}}$ by:
\beq \label{QMCS37}
\widetilde{a}^{\dag}_{\mathbf{p}}\equiv e^{-i(m/|m|)\theta} a^{\dag}_{\mathbf{p}} \qquad \widetilde{a}_{\mathbf{p}}\equiv e^{i(m/|m|)\theta} a_{\mathbf{p}}, 
\eeq
where $\theta$ is defined in (\ref{QMCS32}). The redefinition does not change the canonical commutation relations. Let us substitute the new operators into (\ref{QMCS27}), (\ref{QMCS28}), (\ref{QMCS32}) and (\ref{QMCS36}):
\beq \label{QMCS38}
\begin{array}{c}
P^{\mu}=\int d^{2}p \ p^{\mu}\widetilde{a}^{\dag}_{\mathbf{p}}\widetilde{a}_{\mathbf{p}} \\
M=\int d^{2}p \widetilde{a}^{\dag}_{\mathbf{p}}\frac{1}{i}\frac{\partial}{\partial \theta} \widetilde{a}_{\mathbf{p}}+
\frac{m}{|m|}\int d^{2}p \widetilde{a}^{\dag}_{\mathbf{p}}\widetilde{a}_{\mathbf{p}} \\
N_{j}=\frac{i}{2}\int d^{2}p E_{\mathbf{p}}\left(\widetilde{a}^{\dag}_{\mathbf{p}}\partial_{j}\widetilde{a}_{\mathbf{p}}-\partial_{j}[\widetilde{a}^{\dag}_{\mathbf{p}}]\widetilde{a}_{\mathbf{p}}\right)+\frac{m}{|m|}\int d^{2}p \frac{\epsilon^{jk}p^{k}}{E_{\mathbf{p}}+|m|}\widetilde{a}^{\dag}_{\mathbf{p}}\widetilde{a}_{\mathbf{p}}. \\
\end{array}
\eeq
The boost generators are now the same as those in (\ref{eq:Poi3a3}) with $j=\frac{m}{|m|}$. So we conclude that $\widetilde{a}^{\dag}_{\mathbf{p}}$ and $\widetilde{a}_{\mathbf{p}}$ are the proper creation and annihilation operators. Photon is indeed massive and has non-vanishing spin 1 or -1 depending on the sign of the mass parameter $m$. One may ask what is a weak point of our initial derivation (\ref{QMCS21})-(\ref{QMCS36})? The answer resides in the definition of the polarization vector $\epsilon^{i}_{\mathbf{p}}$ (\ref{QMCS21a})\cite{Devecchi:1994}. Obviously our original guess:
\beq \label{QMCS38a}
\epsilon^{i}_{\mathbf{p}}=\frac{\epsilon^{ij}p^{j}}{\mathbf{p}^{2}}
\eeq
is wrong. The Coulomb gauge constraint and the normalization condition do not specify the complex phase of $\mathbf{\epsilon}^{i}_{\mathbf{p}}$. The correct formula for the polarization vector is:
\beq \label{QMCS39}
\epsilon^{i}_{\mathbf{p}}=e^{-i\frac{m}{|m|}\theta}\frac{\epsilon^{ij}p^{j}}{\mathbf{p}^{2}}
\eeq
and $\theta$-dependence of this additional phase leads to the non-vanishing spin.

\chapter{Discrete symmetries of $QED_{3}$}
\label{ch:discrete}
In this Chapter we try to construct the operators of parity $P$ and time inversion $T$ for Dirac fermions with some reasonable properties. While for the massless fermions $P$ and $T$ can be constructed, it turns out that for the massive fermions there is no good parity and time reversal transformation in three-dimensional spacetime. We define the combined $PT$ and charge-conjugate $C$ transformations, which work well for both massive and massless fermions.  The question of the invariance of $QED_{3}$ with respect to these symmetries is also addressed in this Chapter.
\section{Naive approach to discrete symmetries}
Parity in two spatial dimensions is defined as an operation, which inverts only one spatial direction. It is a reasonable definition because inversion of both spatial directions is a rotation by the angle $\pi$. Jackiw and Templeton \cite{Jackiw:1980}, \cite{DJT:1981} introduced the following realization of the parity operator $P$ in the Hilbert space of the vector and the fermion fields:
$$ P A^{0}(t,\mathbf{r})P^{-1}=A^{\prime0}(t,\mathbf{r})=A^{0}(t,\mathbf{r^{\prime}}), $$
\beq \label{eq:DS1}
P A^{1}(t,\mathbf{r})P^{-1}=A^{\prime1}(t,\mathbf{r})=-A^{1}(t,\mathbf{r^{\prime}}),
\eeq 
$$ P A^{2}(t,\mathbf{r})P^{-1}=A^{\prime2}(t,\mathbf{r})=A^{2}(t,\mathbf{r^{\prime}}), $$
\beq \label{eq:DS2}
P \psi(t,\mathbf{r})P^{-1}=\psi^{\prime}(t,\mathbf{r})=\sigma^{1} \psi(t,\mathbf{r^{\prime}}),
\eeq
$$ \mathbf{r}=(x,y) \qquad \mathbf{r^{\prime}}=(-x,y). $$
By stating that some equation (e.g., the Dirac equation) is invariant under some symmetry operation we mean that if the fields $\psi(x)$ and $A(x)$ satisfy the given equation in some coordinate system $S$ then the transformed fields $\psi^{\prime}(x^{\prime})$ and $A^{\prime}(x^{\prime})$ also satisfy the equation in a transformed system $S^{\prime}$. For example, for the Dirac equation the invariance means:
$$ (i\slashed{\partial}+e \slashed{A}(x) -m)\psi(x)=0 \Rightarrow (i\slashed{\partial}^{\prime}+e \slashed{ A}^{\prime}(x^{\prime}) -m)\psi^{\prime}(x^{\prime})=0.  $$
It is easy to show that the massless Dirac and Maxwell equations are invariant under parity transformation (\ref{eq:DS1}), (\ref{eq:DS2}). It is important to note, however, that the mass term in the Dirac equation is odd under the parity transformation. It is then tempting to work with massless fermions in order to have a parity-conserving theory.

Jackiw and Templeton also introduced their realization of the time-inversion operator $T$ in the Hilbert space of the vector and the fermion fields \cite{Jackiw:1980}, \cite{DJT:1981}:
$$ T A^{0}(t,\mathbf{r})T^{-1}=A^{0}(-t,\mathbf{r}), $$
\beq \label{eq:DS3}
T \mathbf{A}(t,\mathbf{r})T^{-1}=-\mathbf{A}(-t,\mathbf{r}),
\eeq 
\beq\label{eq:DS4}
T \psi(t,\mathbf{r})T^{-1}=\sigma^{2}\psi(-t,\mathbf{r}),
\eeq
where $\sigma^{2}$ is understood as an antilinear operator in the last equation. It is also not difficult to show that the massless Dirac and Maxwell equations are invariant under T-inversion.

Let us investigate parity more closely and consider only fermions. Where did the transformation (\ref{eq:DS2}) come from? We may take (\ref{eq:QF1A}) and ask ourselves how does parity act on the creation operators $b^{\dag}_{\mathbf{p}}$ and the annihilation operators $a_{\mathbf{p}}$? According to Chapter \ref{ch:Poincare} the massless fermion is "almost" scalar, so it would be natural if:
\beq \label{eq:DS5}
Pa_{\mathbf{p}}P^{-1}=\eta_{a} a_{\mathbf{p^\prime}} \qquad Pb_{\mathbf{p}}P^{-1}=\eta_{b} b_{\mathbf{p^\prime}},
\eeq  
where $\mathbf{p^{\prime}}=(-p^{1},p^{2})$ and $\eta_{a}$ and $\eta_{b}$ are some complex phases. We assume also that the parity operator is linear and unitary. Implementing our intuitive definition we get for the field $\psi(x)$:
\begin{eqnarray} P\psi(x)P^{-1}=\int \frac{d^{2}p}{(2\pi)^2}\frac{1}{\sqrt{2E}}\left(\eta_{a}a_{\mathbf{p^{\prime}}}u(\mathbf{p})e^{-ip\cdot x}+(\eta_{b})^{*}b^{\dag}_{\mathbf{p^{\prime}}}v(\mathbf{p})e^{ip\cdot x}  \right)= \nonumber \\ 
=\int \frac{d^{2}p}{(2\pi)^2}\frac{1}{\sqrt{2E}}\left(\eta_{a}a_{\mathbf{p}}u(\mathbf{p^{\prime}})e^{-ip\cdot x^{\prime}}+(\eta_{b})^{*}b^{\dag}_{\mathbf{p}}v(\mathbf{p^{\prime}})e^{ip\cdot x^{\prime}}  \right), \nonumber
\end{eqnarray}
where $x^{\prime}=(t,-x^{1},x^{2})$. It would be nice, if a momentum-independent matrix $O$ existed such that:
\beq \label{eq:DS5A}\eta_{a}u(\mathbf{p^{\prime}})=Ou(\mathbf{p}) \qquad (\eta_{b})^{*}v(\mathbf{p^{\prime}})=Ov(\mathbf{p}). \eeq
In that case we could write $P\psi(x)P^{-1}=O\psi(x^{\prime})$. Positive and negative solutions $u(\mathbf{p})$ and $v(\mathbf{p})$ can be easily found for the Dirac equation in the $Jackiw$ realization (\ref{eq:RQM1bb}) of $\gamma$ matrices (it is valid for both the massive and the massless cases) \cite{Hand:1993} :
\beq \label{eq:DS6}
u(\mathbf{p})e^{-ip\cdot x}=\frac{1}{\sqrt{\epsilon_{\mathbf{p}}+m}}\left(\begin{array}{c} \epsilon_{\mathbf{p}}+m \\ ip_{1}-p_{2} \end{array} \right) e^{-ip\cdot x} \quad \bar{u}(\mathbf{p})u(\mathbf{p})=2m \quad u^{+}(\mathbf{p})u(\mathbf{p})=2E ,
\eeq
\beq \label{eq:DS7}
v(\mathbf{p})e^{ip\cdot x}=\frac{1}{i\sqrt{\epsilon_{\mathbf{p}}+m}}\left(\begin{array}{c} ip_{1}+p_{2}  \\ -\epsilon_{\mathbf{p}}-m  \end{array} \right) e^{ip\cdot x} \quad \bar{v}(\mathbf{p})v(\mathbf{p})=-2m \quad v^{+}(\mathbf{p})v(\mathbf{p})=2E,
\eeq 
where $\epsilon_{\mathbf{p}}=\sqrt{p_{1}^{2}+p_{2}^{2}+m^{2}}$. Having this at hand we can easily show that there is \emph{no $p$-independent matrix $O$}, which would satisfy (\ref{eq:DS5A}). The argumentation may be reversed. If we take (\ref{eq:DS2}) as a definition of parity action on the fields, we get \emph{nonlinear} action on the annihilation and the creation operators:
\begin{eqnarray} P \psi(t,\mathbf{r})P^{-1}=\sigma^{1} \psi(t,\mathbf{r^{\prime}})=
\int \frac{d^{2}p}{(2\pi)^2}\frac{1}{\sqrt{2E}}\left(a_{\mathbf{p}}\sigma^{1}u(\mathbf{p})e^{-ip\cdot x^{\prime}}+b^{\dag}_{\mathbf{p}}\sigma^{1}v(\mathbf{p})e^{ip\cdot x^{\prime}}  \right)= \nonumber \\
=\int \frac{d^{2}p}{(2\pi)^2}\frac{1}{\sqrt{2E}}\left(a_{\mathbf{p^{\prime}}}\sigma^{1}u(\mathbf{p^{\prime}})e^{-ip\cdot x}+b^{\dag}_{\mathbf{p^{\prime}}}\sigma^{1}v(\mathbf{p^{\prime}})e^{ip\cdot x}  \right)= \nonumber \\
=\int \frac{d^{2}p}{(2\pi)^2}\frac{1}{\sqrt{2E}}\left(-a_{\mathbf{p^{\prime}}}v(\mathbf{p})e^{-ip\cdot x}-b^{\dag}_{\mathbf{p^{\prime}}}u(\mathbf{p})e^{ip\cdot x}  \right), \nonumber
\end{eqnarray}
where we used $ \sigma^{1}u(\mathbf{p^{\prime}})=-v(\mathbf{p})$ and $\sigma^{1}v(\mathbf{p^{\prime}})=-u(\mathbf{p})$. Thus we see that our natural definition (\ref{eq:DS5}) of parity is in contradiction with the definition of Jackiw and Templeton. The same contradiction exists for T-inversion. Something must be wrong with our analysis of the massless Dirac theory.
\section{Systematic approach to discrete symmetries}
Here we shall take more systematic approach to the discrete symmetries.
\subsection*{P-inversion}
Let us now try to introduce the operator $P$ on the UIRs of the Poincar\'e group $\pi_{++}$, which satisfies general requirements for the reasonable parity operator. The UIRs were constructed by us in Chapter \ref{ch:Poincare}. The operator $P$ is defined by its commutation relations:
\beq \label{eq:DS8} 
\{ P, J \}=\{ P, N_{1} \}=[ P, N_{2} ]=0
\eeq 
\beq \label{eq:DS9}
[P, H]=\{ P, P^{1} \}=[ P, P^{2} ]=0. 
\eeq
The operator $P$ must be linear because it satisfies $P\exp(-iHt)P^{-1}=\exp(-iHt)$.
From now on we shall distinguish two different cases: $m\ne 0$ and $m=0$.
\subsubsection*{\textit{Massive case}}
Let us denote the standard vector of $U^{m,+,j}$ by $\ket{k,j}$, where the three-vector $k=(1,0,0)$. Using the relations (\ref{eq:DS8}) and (\ref{eq:DS9}) we get:
\beq \label{eq:DS10}
P\ket{k,j}=\eta_{P}\ket{k,-j},
\eeq
where $\eta_{P}$ is a complex phase. Our next question is how does the $P$ act on a general state $\ket{p,j}$? Let us define the standard transformation $L(p)=R(\theta)B_{2}(|\mathbf{p}|)$ (we remember that $p=L(p)k$ in the Wigner construction). By this we mean that first we perform a boost in the $y$-direction in order the particle to have right magnitude of the momentum $|\mathbf{p}|$. Next we make an appropriate rotation in order the particle to have right direction of the momentum $\mathbf{p}$. From the relations (\ref{eq:DS8}) and (\ref{eq:DS9}) we see that:
\beq \label{eq:DS11}
P\ket{p,j}=\eta_{P}\ket{p^{\prime},-j},
\eeq
where $p^{\prime}=(p^{0},-p^{1},p^{2})$. Thus we see that \emph{it is not possible} to introduce $P$ acting within the given $U^{m,+,j}$. We need \emph{two representations} $U^{m,+,j}$ and $U^{m,+,-j}$ in order to define the $P$. Parity would then transform a state vector from the $U^{m,+,j}$ to the $U^{m,+,-j}$. Applying it to the massive Dirac fields we obtain:
\beq \label{eq:DS12}
Pa^{\dag}_{\mathbf{p}}P^{-1}=\eta^{P}_{\ a}b^{\dag}_{\mathbf{p^{\prime}}} \quad Pb^{\dag}_{\mathbf{p}}P^{-1}=\eta^{P}_{\ b}a^{\dag}_{\mathbf{p^{\prime}}}.
\eeq
So parity transforms electrons into positrons and vice versa! Assuming that $P$ is a linear operator of symmetry leads us to the fact that the field $\psi(x)$ (\ref{eq:QF1A}) does not transform into itself. We do not obtain (\ref{eq:DS2})! Thus loosely speaking \emph{in the massive QED in three dimensions there is no good parity transformation}. In literature this fact is formulated as a statement that the mass term in the Dirac equation is parity violating \cite{DJT:1981}. We prefer to say that parity is not well defined for massive particles. Let us turn our attention to the massless case. 
\subsubsection*{\textit{Massless case}}
Let us take now the representation $U^{0,+,1,0}$. A standard vector is now denoted as $\ket{k}$, where $k=(1,1,0)$. From the commutation relations (\ref{eq:DS8}) and (\ref{eq:DS9}) we obtain:
\beq \label{eq:DS14}
P\ket{k}=\eta_{P}\ket{k^{\prime}}.
\eeq
Taking the same standard transformation $L(p)$ as in the massive case we get for an arbitrary vector $\ket{p}$:
\beq \label{eq:DS15}
\hat{P}\ket{p}=\eta_{P}\ket{p^{\prime}}.
\eeq
We observe that $P$ can be defined in $U^{0,+,1,0}$. Let us now apply our results to the massless Dirac theory. The decomposition (\ref{eq:QF1A}) is also valid in the massless Dirac theory, but from (\ref{eq:Poi7c}) we know that $a_{\mathbf{p}}$ and $b^{\dag}_{\mathbf{p}}$ do not transform by the UIR $U^{0,+,1,0}$. Their transformation rules under P-inversion are more complicated (see (\ref{eq:Poi7c})). We start with the decomposition of massless fermion field: 
\beq \label{eq:DS15a}
\psi(\mathbf{x},t)=\int \frac{d^{2}p}{(2\pi)^{2}}\frac{1}{\sqrt{2E_{\mathbf{p}}}}(a_{\mathbf{p}}u(\mathbf{p})e^{-ip\cdot x}+b^{\dag}_{\mathbf{p}}v(\mathbf{p})e^{ip\cdot x})
\eeq
and perform the Fourier transformation due to (\ref{eq:Poi4a}): 
\begin{eqnarray} 
\psi(q)&=&\int d^{3}x \exp(-iq\cdot x)\psi(x)\nonumber \\&=&\frac{2\pi}{\sqrt{2E_{\mathbf{q}}}}\left(a_{-\mathbf{q}}u(-\mathbf{q})\delta(\sqrt{\mathbf{q}^{2}+m^{2}}+q^{0})+b^{\dag}_{\mathbf{q}}v(\mathbf{q})\delta(\sqrt{\mathbf{q}^{2}+m^{2}}-q^{0})\right). \label{eq:DS15b}
\end{eqnarray}
Now in accordance with (\ref{eq:Poi7c}) we make the redefinition of the massless fermion fields:
\begin{eqnarray} \label{eq:DS15c}
\psi^{+}(q)\to \Psi^{+}(q)=\frac{2\pi}{\sqrt{2E_{\mathbf{q}}}}\underbrace{(q_{1}+iq_{2})^{1/2}a_{-\mathbf{q}}}_{\widetilde{a}_{-\mathbf{q}}}u(-\mathbf{q})\delta(\sqrt{\mathbf{q}^{2}+m^{2}}+q^{0}) \\
\psi^{-}(q)\to\Psi^{-}(q)=\frac{2\pi}{\sqrt{2E_{\mathbf{q}}}}\underbrace{(q_{1}-iq_{2})^{1/2}b^{\dag}_{\mathbf{q}}}_{\widetilde{b}^{\dag}_{\mathbf{q}}}v(\mathbf{q})\delta(\sqrt{\mathbf{q}^{2}+m^{2}}-q^{0}).
\end{eqnarray}
Due to (\ref{eq:Poi7c}) the operators $\widetilde{a}^{\dag}_{\mathbf{q}}$ and $\widetilde{b}^{\dag}_{\mathbf{q}}$ are identified as  the correct creation operators; i.e., $\ket{q}=\widetilde{a}^{\dag}_{\mathbf{q}}\ket{0}$ in the Hilbert space of the massless electron and $\ket{q}=\widetilde{b}^{\dag}_{\mathbf{q}}\ket{0}$ in the Hilbert space of the massless positron. According to (\ref{eq:DS15}) the action of the parity operator on these creation operators is:
\beq \label{eq:DS15d}
P\widetilde{a}^{\dag}_{\mathbf{q}}P^{-1}=\eta^{P}_{a}\widetilde{a}^{\dag}_{\mathbf{q}^{\prime}} \qquad P\widetilde{b}^{\dag}_{\mathbf{q}}P^{-1}=\eta^{P}_{b}\widetilde{b}^{\dag}_{\mathbf{q}^{\prime}}
\eeq
and using (\ref{eq:DS15c}):
\beq \label{eq:DS15e}
\widetilde{a}_{\mathbf{q}}=i(q_{1}+iq_{2})^{1/2}a_{\mathbf{q}} \qquad \widetilde{b}^{\dag}_{\mathbf{q}}=(q_{1}-iq_{2})^{1/2}b^{\dag}_{\mathbf{q}}
\eeq
we easily get the action of the parity operator on $a_{\mathbf{q}}$ and $b^{\dag}_{\mathbf{q}}$: 
\beq \label{eq:DS16}
Pa_{\mathbf{q}}P^{-1}=\eta^{*P}_{\ a}\frac{iq_{1}+q_{2}}{\epsilon_{\mathbf{q}}}a_{\mathbf{q^{\prime}}} \qquad
Pb^{\dag}_{\mathbf{q}}P^{-1}=\eta^{P}_{\ b}\frac{iq_{1}-q_{2}}{\epsilon_{\mathbf{q}}}b^{\dag}_{\mathbf{q^{\prime}}}.
\eeq
So our "natural" definition (\ref{eq:DS5}) \emph{was wrong}.
Straightforward calculation leads us to the transformation law of the field operator (we take $\eta^{*P}_{a}=-1$ and $\eta^{P}_{b}=1$):
\beq \label{eq:DS17}
P\psi(x)P^{-1}=\sigma^{1}\psi(x^{\prime})
\eeq
that coincide with the definition (\ref{eq:DS2}) of the P-inversion for the massless fermion field. It is easy to show that $P\mathscr{L}(x)P^{-1}=\mathscr{L}(x^{\prime})$, so \emph{parity is a symmetry of the massless Dirac theory}.
\subsection*{T-inversion}
Now we shall try to introduce an operator $T$ of time-reversal on the UIRs of $\pi_{++}$. The $T$ operator must be antilinear because time inversion reverses time evolution, i.e., we have $T\exp(-iHt)T^{-1}=\exp(iHt)$. The commutation relations are:
\beq \label{eq:DS18} 
\{ T, J \}=[T, N_{1}]=[ T, N_{2} ]=0
\eeq 
\beq \label{eq:DS19}
[T, H]=\{ T, P^{1} \}=\{ T, P^{2} \}=0. 
\eeq
\subsubsection*{\textit{Massive case}}
We shall construct the action of $T$ on the massive UIRs $U^{m,+,j}$. Using our commutation relations it is not difficult to show that:
\beq \label{eq:DS20}
T\ket{k,j}=\eta^{T}\ket{k,-j},
\eeq
where $\eta^{T}$ is a complex phase. For a general state $\ket{p,j}$ taking the same standard transformation as in the previous subsection yields:
\beq \label{eq:DS21}
T\ket{p,j}=\eta^{T}\ket{\tilde{p},-j},
\eeq
where $\tilde{p}=(p^{0},-\mathbf{p})$. As parity the operator $T$ mixes two massive representations. Specially for the massive Dirac field we get:
\beq \label{eq:DS22}
Ta^{\dag}_{\mathbf{p}}T^{-1}=\eta^{T}_{\ a}b^{\dag}_{\mathbf{-p}} \quad Tb^{\dag}_{\mathbf{p}}T^{-1}=\eta^{T}_{\ b}a^{\dag}_{\mathbf{-p}}.
\eeq
For the transformation of the field operator $\psi(x)$ we have the same problem as before. The field \emph{does not transform into itself}; i.e., the relation (\ref{eq:DS4}) is not satisfied.
\subsubsection*{\textit{Massless case}}
The action of the $T$ in $U^{0,+,1,0}$ on a standard state vector $\ket{k}$ and a general vector $\ket{p}$ is (we are using relations (\ref{eq:DS18}) and (\ref{eq:DS19})):
\beq \label{eq:DS23}
T\ket{k}=\eta^{T}\ket{\tilde{k}} \qquad T\ket{p}=\eta^{T}\ket{\tilde{p}},
\eeq
where $\tilde{p}=(p^{0},-\mathbf{p})$. Thus the $T$ is well-defined in $U^{0,+,1,0}$. Now we shall apply our results to the massless Dirac theory. From (\ref{eq:Poi7c}) we know that $a_{\mathbf{p}}$ and $b^{\dag}_{\mathbf{p}}$ do not transform under the irreducible representations. Following the same steps as for parity we get:
\beq \label{eq:DS24}
Ta_{\mathbf{p}}T^{-1}=\eta^{*T}_{\ a}\frac{p_{1}+ip_{2}}{\epsilon_{\mathbf{p}}}a_{-\mathbf{p}} \qquad
Tb^{\dag}_{\mathbf{p}}T^{-1}=\eta^{T}_{\ b}\frac{p_{1}-ip_{2}}{\epsilon_{\mathbf{p}}}b^{\dag}_{-\mathbf{p}},
\eeq
where $\eta^{T}_{a}$ and $\eta^{T}_{b}$ are complex phases. Applying the $T$ operator to the field operator $\psi(x)$ we obtain:
\beq \label{eq:DS25}
T\psi(t,\mathbf{x})T^{-1}=\sigma^{2}\psi(-t,\mathbf{x}),
\eeq
so we recovered the definition of Jackiw (\ref{eq:DS4}) of time-reversal for massless fermions. Applying $T$ to $\mathscr{L}$ we obtain that \emph{time-reversal is a symmetry of the massless Dirac theory}.
\subsection*{Combined symmetry $O$}
In preceding subsections we showed that there are no reasonable parity and time-reversal operators for the massive fermion field. However, it is possible to define an operator $O=TP$. We shall call it a combined symmetry. For the massive fermions we have:
\beq \label{eq:DS26}
Oa^{\dag}_{\mathbf{p}}O^{-1}=\eta^{O}_{\ a}b^{\dag}_{\mathbf{-p^{\prime}}} \quad Ob^{\dag}_{\mathbf{p}}O^{-1}=\eta^{O}_{\ b}a^{\dag}_{\mathbf{-p^{\prime}}}.
\eeq
It is straightforward to show that the massive fermion field $\psi(\mathbf{x},t)$ transforms as:
\beq \label{eq:DS27}
O\psi(\mathbf{x},t)O^{-1}=\gamma^{0}\psi(\mathbf{x}^\prime,-t),
\eeq
if we choose $\eta^{O}_{\ a}=1$ and $\eta^{O}_{\ b}=-1$. \emph{The combined symmetry $O$ is a good symmetry of massive fermions}.
\subsection*{Charge conjugation}
We should note that there are no difficulties in defining the charge conjugation symmetry $C$ for both the massive and the massless fermions. The operator $C$ is defined by its action on the creation operators:
\beq \label{eq:DS28}
C a^{\dag}_{\mathbf{p}}C^{-1}=\eta^{C}_{a}b^{\dag}_{\mathbf{p}} \qquad C b^{\dag}_{\mathbf{p}}C^{-1}=\eta^{C}_{b}a^{\dag}_{\mathbf{p}},
\eeq
where $\eta^{C}_{a}$ and $\eta^{C}_{b}$ are complex phases. This definition \emph{works well} for both the massive and the massless fermions because the charge-conjugation operator $C$ is the internal symmetry operator and \emph{does not transform spacetime coordinates}. From (\ref{eq:DS28}) and the linearity of $C$ we get the transformation of the fields $\psi(x)$:
\beq \label{eq:DS29}
C \psi(\mathbf{x},t)C^{-1}=\gamma^{1}\psi^{*}(\mathbf{x},t),
\eeq
where we used $(\eta_{a}^{C})^{*}=\eta_{b}^{C}=1$.

\chapter{Perturbative Maxwell-Chern-Simons $QED_{3}$}
\label{ch:perturbative}
In this Chapter few one-loop perturbative results in the MCS theory is presented. We start with the Feynman rules and the UV structure of the theory. It turns out that the theory is superrenormalizable. Subsequently, the one-loop vacuum polarization diagram and electron self-energy are calculated in the bare perturbation theory. We finish this Chapter calculating the transverse running coupling to the one-loop order.   

\section{Feynman rules and power counting}
In this Chapter we shall study $QED_{3}$, which is the Maxwell-Chern-Simons theory coupled minimally to the two-component Dirac fermions, in its perturbative regime. The Lagrangian of the theory is\footnote{\textrm{Note that we have changed the notation here. The photon mass parameter is denoted by $\mu$ instead of $m$.}}:
\beq \label{$QED_{3}$}
\mathscr{L}=\underbrace{-\frac{1}{4}F^{\mu\nu}F_{\mu\nu}+\frac{\mu}{2}\epsilon^{\mu\nu\rho}A_{\mu}\partial_{\nu}A_{\rho}-\frac{1}{2\alpha}(\partial_{\mu}A^{\mu})^{2}}_{\mathscr{L}_{G}}+\underbrace{\bar{\psi}(i\slashed{\partial}-m)\psi}_{\mathscr{L}_{\psi}}\underbrace{-e\bar{\psi}\gamma^{\mu}\psi A_{\mu}}_{\mathscr{L}_{I}},
\eeq
where $-\frac{1}{2\alpha}(\partial_{\mu}A^{\mu})^{2}$ is a gauge fixing term, $m$ is the fermion bare mass and $\mu$ is an axial photon parameter, the absolute value of which is the bare photon mass. First we write the Feynman rules for this Lagrangian; i.e., we calculate the photon and the fermion free propagators and the bare interaction vertex.
  
 Although it was illuminating to quantize the MCS theory in the Coulomb gauge, for calculation it is more convenient to use the covariant version of the free photon propagator $\mathscr{D}_{\mu\nu}(x,y)$. In order to find it, we start with the Lagrangian $\mathscr{L}_{G}$ (\ref{$QED_{3}$}).
 
The Euler-Lagrange equations for the Lagrangian $\mathscr{L}_{G}$ are:
\beq \label{VP0a}
\underbrace{(-\Box g_{\nu\rho}+(1-\frac{1}{\alpha})\partial_{\nu}\partial_{\rho}+\mu\epsilon_{\nu\rho\phi}\partial^{\phi})}_{O_{\nu\rho}}A^{\rho}=0.
\eeq
The propagator is a Green's function $G^{\beta\rho}(x-y)$ of this equation:
\beq \label{VP0b}
O_{x}^{\alpha\beta}G_{\beta\rho}(x-y)=-ig^{\alpha}_{\rho}\delta(x-y).
\eeq
It is convenient to perform the Fourier transformation of $G^{\beta\rho}(x-y)$:
\beq \label{VP0c}
G^{\beta\rho}(x-y)=\int\frac{d^{3}p}{(2\pi)^{3}}e^{ip(x-y)}G^{\beta\rho}(p).
\eeq
The differential equations (\ref{VP0b}) transform into the algebraic equations for $G^{\beta\rho}(p)$:
\beq \label{VP0d}
[p^{2}g^{\nu\rho}-(1-\frac{1}{\alpha})p^{\nu}p^{\rho}+i\mu\epsilon^{\nu\rho\alpha}p_{\alpha}](iG_{\rho\mu}(p))=g^{\nu}_{\mu}.
\eeq
Solving it and introducing the Feynman pole prescription we end up with the Feynman covariant propagator for the electromagnetic field $\mathscr{D}_{\mu\nu}(p)$:
\beq \label{VP1}
\mathscr{D}_{\mu\nu}(p)=\frac{-i}{p^{2}-\mu^{2}+i\epsilon}[P_{\mu\nu}(p)-\frac{i\mu\epsilon_{\mu\nu\rho}p^{\rho}}{p^{2}+i\epsilon}]-\frac{i\alpha p_{\mu}p_{\nu}}{(p^{2}+i\epsilon)^{2}},
\eeq
where $\mu$ is the bare photon mass, $P^{\mu\nu}(p)=g^{\mu\nu}-p^{\mu}p^{\nu}/p^{2}$ and $\alpha$ is the gauge parameter.
\begin{figure}[!htb]
\begin{center}
\begin{tabular}{c}

\begin{fmffile}{graph1}
\begin{fmfgraph*}(60,50)
\fmfleft{i1}
\fmfright{o1}
\fmflabel{$\nu$}{i1}
\fmflabel{$\mu$}{o1}
\fmf{photon,label=${ \leftarrow p}$}{i1,o1}

\end{fmfgraph*}
\end{fmffile}

\end{tabular}
\caption{Free photon propagator $\mathscr{D}_{\mu\nu}(p)$}\label{graph1}
\end{center}
\end{figure}
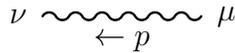

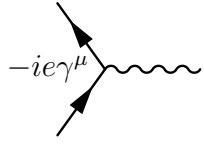
\begin{figure}[!htb]
\begin{center}
\begin{tabular}{c}

\begin{fmffile}{graph2}
\begin{fmfgraph*}(60,50)
\fmfleft{i1,i2}
\fmfright{i3}
\fmf{fermion}{i1,v1,i2}
\fmf{photon}{v1,i3}
\fmflabel{$-ie\gamma^{\mu}$}{v1}

\end{fmfgraph*}
\end{fmffile}

\end{tabular}
\caption{Bare vertex}\label{graph2}
\end{center}
\end{figure}

The Feynman graph for the photon propagator is shown in Fig. \ref{graph1}. The direction of the momentum arrow is important here because the axial part of the propagator (\ref{VP1}) is odd in the momentum variable $p^{\mu}$. The free fermion propagator is conventional:
\beq \label{VP2}
\mathscr{S}(p)=\frac{i}{\slashed{p}-m+i\epsilon}=\frac{i(\slashed{p}+m)}{p^{2}-m^{2}+i\epsilon},
\eeq
where $m$ is the free fermion mass. Fig. \ref{graph2} illustrates the bare vertex and its Feynman prescription.

Next we examine the ultraviolet (UV) properties of various Feynman diagrams. It is sufficient to study the amputated 1PI Feynman diagrams because the external lines do not affect the UV properties and the reducible diagrams are the simple products of the 1PI diagrams. For a large momentum $k$ the photon propagator behaves as $1/k^{2}$ and the fermion propagator as $1/k$. Every loop yields a three dimensional integration over the loop momentum and effectively brings three powers of momentum $p$ into the numerator of the Feynman integral. We define the superficial degree of divergence $D$ of a Feynman diagram as:
\beq \label{$QED_{3}$a}    
D=3L-P_{e}-2P_{\gamma},
\eeq
where $L$ is a number of loops, $P_{e}$ is a number of the electron propagators and $P_{\gamma}$ is a number of the photon propagators. For a momentum cutoff $\Lambda$ we anticipate that the diagram diverges as  $\Lambda^{D}$ for $D>0$, diverges as $\log\Lambda$ for $D=0$ and is convergent for $D<0$. These naive expectations might be often wrong due to the divergent subdiagrams and symmetries.

For further analysis it is worth to write $D$ as a function of number of the vertices $V$, external electron lines $N_{e}$ and external photon lines $N_{\gamma}$. To do so we use the formulas:
\beq \label{$QED_{3}$b}
L=P_{e}+P_{\gamma}-V+1 \qquad P_{\gamma}=\frac{1}{2}(V-N_{\gamma}) \qquad P_{e}=V-\frac{1}{2}N_{e}
\eeq
and obtain our final result:
\beq \label{$QED_{3}$c}
D=-\frac{1}{2}V-N_{e}-\frac{1}{2}N_{\gamma}+3.
\eeq
The theory is superrenormalizable because the presence of vertices decreases the superficial degree $D$. We can also show that the MCS is superrenormalizable by calculating the mass dimension of the couplings. From the free part of the Lagrangian (\ref{$QED_{3}$}) we get:
\beq \label{SR1}
[A]=\frac{1}{2}  \qquad [\psi]=1,
\eeq
where $[x]$ denotes the mass dimension of the quantity $x$. Now it is easy to calculate the mass dimension of the couplings:
\beq \label{SR2}
[m]=1 \qquad [\mu]=1 \qquad [e]=\frac{1}{2}.
\eeq
All couplings have positive dimension of mass, thus the theory is superrenormalizable.

At this point we should make a remark about the mass dimension of the interaction coupling in the MCS theory. As we have already shown the theory is superrenormalizable. To do reasonable perturbative expansion we need to find some dimensionless small parameter. What is this parameter in the superrenormalizable MCS theory? First, let us analyze the process of scattering of two-particle initial state into two-particle final state in the pure massless $QED_{3}$, i.e., $m=0$ and $\mu=0$. It is easy to show that the invariant amplitude $M_{fi}$ of scattering 2-particle initial state into $n$-particle final state in $d$ dimensions has the mass dimension:
\beq \label{ptt1}
[M_{fi}]=4-n(d-2),
\eeq 
In our special case we get $M_{fi}=2$. The naive perturbative expansion can be now written for $M_{fi}$:
\beq \label{ptt2}
M_{fi}=\sum^{\infty}_{k=1}a_{k}(e^{2})^{k},
\eeq
where $a_{k}$ has mass dimension $2-k$. We shall rewrite last expansion using the dimensionless quantities. In the massless $QED_{3}$ there is no parameter with dimension of mass in the Lagrangian, so the only dimensionful parameter, which can compensate the dimension of $e^{2}$ in the expansion is the energy $E_{CMS}=\sqrt{s}$. Thus (\ref{ptt2}) takes the form:
\beq \label{ptt3}
M_{fi}=s\sum^{\infty}_{k=1}(\frac{e^{2}}{\sqrt{s}})^{k}\widetilde{a}_{k},
\eeq
where $\widetilde{a}_{k}$ are dimensionless. We factored out $s$ in the last expression to guarantee the correct dimension of $M_{fi}$. In this case the dimensionless perturbative parameter can be identified with $\frac{e^{2}}{\sqrt{s}}$. Thus at sufficiently low energy it seems that the theory becomes \emph{strongly coupled}. It is believed that the mass is generated \emph{non-perturbatively} in this case \cite{Cornwall:1996}. It was shown \cite{Jackiw:1980} that the naive perturbative expansion (\ref{ptt2}) leads to infrared divergences even for arbitrary $s$. These divergences can be cured solving the truncated Schwinger-Dyson equations. This procedure leads to non-analytic (logarithmic) terms in the coupling constant in the expansion (\ref{ptt2}). In the MCS theory we have two additional mass scales in the Lagrangian: the photon mass $\mu$ and the fermion mass $m$. Due to kinematics $\sqrt{s}>min(m,\mu)$. It is \emph{reasonable} to assume that the series (\ref{ptt2}) organizes itself in such a way that $\frac{e^{2}}{m}$ and $\frac{e^{2}}{\mu}$ are dimensionless expansion parameters, which are found in (\ref{ptt2}). We shall see that it actually happens in perturbative results obtained in this Chapter.

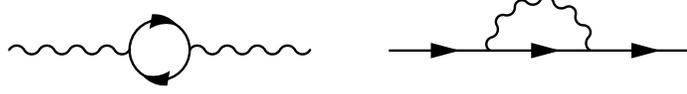
\begin{figure}[!htb]
\begin{center}
\begin{tabular}{cc}    
\parbox{35mm}{\begin{fmffile}{vp} 
  \fmfframe(1,4)(1,4){ 
   \begin{fmfgraph*}(40,22)
   \fmfleft{i1}
   \fmfright{i2}
   \fmf{photon}{i1,v1}
   \fmf{photon}{v2,i2}
   \fmf{fermion,left}{v1,v2,v1} 
   \end{fmfgraph*}
  }
\end{fmffile}}
& \hspace{10mm}
\parbox{35mm}{\begin{fmffile}{ferm1} 	
   \fmfframe(1,4)(1,4){
   \begin{fmfgraph}(40,22) 
   \fmfleft{i1}
   \fmfright{i2}
   \fmf{fermion}{i1,v1}
   \fmf{fermion}{v1,v2}
   \fmf{fermion}{v2,i2}
   \fmf{photon, left, tension=0}{v1,v2}
\end{fmfgraph}
  }
\end{fmffile}} 
\\
\end{tabular}
\caption{Vacuum polarization (a), fermion self-energy (b) }\label{graph3}
\end{center}
\end{figure}

There are only a few Feynman diagrams, which have $D\ge 0$. For the vacuum bubble diagrams with $N_{e}=0$ and $N_{\gamma}=0$ we get $D=4-L$. Thus the vacuum bubble diagrams with no more than four independent loops have nonnegative $D$. For the vacuum polarization diagrams with $N_{e}=0$ and $N_{\gamma}=1$ the relations (\ref{$QED_{3}$b}) and (\ref{$QED_{3}$c}) give $D=2-L$, i.e., the diagrams with no more than two independent loops superficially diverge. Finally, for the corrections to the electron propagator with $N_{e}=2$ and $N_{\gamma}=0$ we get $D=1-L$, which gives rise to a single one-loop superficially divergent diagram. All other types of Feynman diagrams have $D<0$. The only one-loop diagrams with $D\ge 0$ are depicted in Fig. \ref{graph3}- they are the vacuum polarization ($D=1$) and the electron self-energy diagram ($D=0$).

\unitlength = 1mm

The vacuum bubble does not contribute to the S-matrix, so we shall now explicitly calculate the only two remaining divergent one-loop 1PI diagrams in the MCS theory.
\section{Vacuum polarization}
Let us start with the vacuum polarization diagram. The inverse of the full photon propagator $D_{\mu\nu}(p)$ may be expressed as:
\beq \label{VP3}
(D^{-1})_{\mu\nu}(p)=(\mathscr{D}^{-1})_{\mu\nu}(p)-i\Pi_{\mu\nu}(p),
\eeq
where $\Pi_{\mu\nu}$ is the polarization tensor, which includes all 1PI corrections to the free photon propagator. The Ward identity $p^{\mu}\Pi_{\mu\nu}=0$ helps us to specify the tensor structure of the polarization tensor:
\beq \label{VP3a}
\Pi_{\mu\nu}(p)=P_{\mu\nu}\Pi^{(1)}(p^{2})+im\epsilon_{\mu\nu\rho}p^{\rho}\Pi^{(2)}(p^{2}),
\eeq
where we factored out $m$ in the second term of the last expression for the later convenience. The second axial term in (\ref{VP3a}) is obviously absent in the case of (3+1) dimensions. Let us now calculate the scalar functions $\Pi^{(1)}(p^{2})$ and $\Pi^{(2)}(p^{2})$ to the order $O(e^{2})$.
In the one-loop approximation $\Pi^{\mu\nu}(p)$, denoted here by $\Pi_{2}^{\mu\nu}(p)$, is:
\beq \label{VP4}
\Pi^{\mu\nu}_{2}(q)=-ie^{2}\int\frac{d^{3}k}{(2\pi)^{3}}tr\left(\gamma^{\mu}\frac{i}{\slashed{k}-m}\gamma^{\nu}\frac{i}{\slashed{k}+\slashed{q}-m}\right).
\eeq
It can be readily simplified by computing the traces with the help of (\ref{gamma}), introducing the Feynman parameter and performing a shift in the integration variable. The result is:
\begin{eqnarray} 
\Pi^{\mu\nu}_{2}(p) & = & 2ie^{2}\int \frac{d^{3}l}{(2\pi)^{3}}\int^{1}_{0}dy \frac{-g^{\mu\nu}l^{2}+2l^{\mu}l^{\nu}}{(l^{2}-\Delta)^{2}}+{} \label{VP5}   \\
& & 2ie^{2}\int \frac{d^{3}l}{(2\pi)^{3}}\int^{1}_{0}dy \frac{2y(y-1)p^{\mu}p^{\nu}+g^{\mu\nu}(m^{2}-y(y-1)p^{2})-im\epsilon^{\mu\nu\rho}p_{\rho}}{(l^{2}-\Delta)^{2}}, \nonumber
\end{eqnarray}
where $\Delta=m^{2}-y(1-y)p^{2}-i\epsilon$. By power-counting the first integral $(\Pi_{I})^{\mu\nu}_{2}(p)$  in (\ref{VP5}) is linearly divergent and the second integral $(\Pi_{II})^{\mu\nu}_{2}(p)$ is convergent. Some regularization procedure is necessary to regularize the first integral. We prefer to use the dimensional regularization. It is not a priori clear how to generalize the Levi-Civita $\epsilon$-tensor to $d$ dimensions \cite{Delbourgo:1992}. Fortunately, we do not need to do it because the axial part of (\ref{VP5}) is convergent. After the Wick rotation the integral $(\Pi_{I})^{\mu\nu}_{2}(p)$ in $d$ dimensions is:
\beq \label{VP6}
(\Pi_{I})^{\mu\nu}_{2}(p)=-2e^{2}\int \frac{d^{d}l_{E}}{(2\pi)^{d}}\int^{1}_{0} dy \frac{g^{\mu\nu}l_{E}^{2}(1-\frac{2}{d})}{(l_{E}^{2}+\Delta)^{2}}.
\eeq
To calculate it we shall use the general integration formulae: 
\begin{eqnarray}
\int \frac{d^{d}l_{E}}{(2\pi)^{d}}\frac{1}{(l_{E}^{2}+\Delta)^{n}} &= &\frac{1}{(2\pi)^{d/2}}\frac{\Gamma(n-\frac{d}{2})}{\Gamma(n)} (\frac{1}{\Delta})^{n-\frac{d}{2}} \nonumber \\
\int \frac{d^{d}l_{E}}{(2\pi)^{d}}\frac{l_{E}^{2}}{(l_{E}^{2}+\Delta)^{n}} & = & \frac{1}{(2\pi)^{d/2}}\frac{d}{2}\frac{\Gamma(n-\frac{d}{2}-1)}{\Gamma(n)} (\frac{1}{\Delta})^{n-\frac{d}{2}-1}. \label{VP7}
\end{eqnarray}
From the second formula in (\ref{VP7}) it appears that the integral (\ref{VP6}) in the limit $d\to 3$ is perfectly convergent! The result is:
\beq \label{VP8}
(\Pi_{I})^{\mu\nu}_{2}(p)=\frac{2e^{2}}{8\pi}\int^{1}_{0}dy \Delta^{1/2}g^{\mu\nu}.
\eeq
It is a known fact \cite{Greiner:1992} that the generalized integrals may have another type of divergences than the original integrals. In this case the dimensional regularization changed the linearly divergent integral into the convergent integral. Using the first formula in (\ref{VP7}) it is straightforward to calculate also $(\Pi_{II})^{\mu\nu}_{2}(p)$. The total result is in accordance with (\ref{VP3a}) and two scalar functions of our interest are:
\begin{eqnarray}
\Pi^{(1)}(p^{2})& = & \frac{-e^{2}p^{2}}{4\pi}\int^{1}_{0}dy \frac{2y(1-y)}{\sqrt{\Delta}}+O(e^{4}) \nonumber \\
\Pi^{(2)}(p^{2})& = & \frac{e^{2}}{4\pi}\int^{1}_{0}dy \frac{1}{\sqrt{\Delta}}+O(e^{4}). \label{VP9}
\end{eqnarray}
These integrals are elementary and for $p^{2}>0$ and $p^{2}<4m^{2}$ we obtain \cite{Delbourgo:1992}: 
\begin{eqnarray}
\Pi^{(1)}(p^{2})& = & \frac{e^{2}}{16\pi} \left[ \ln(\frac{2m+\sqrt{p^{2}}}{2m-\sqrt{p^{2}}})(\sqrt{p^{2}}+\frac{4m^{2}}{\sqrt{p^{2}}})-4m  \right]+O(e^{4}) \nonumber \\
\Pi^{(2)}(p^{2})& =& \frac{e^{2}}{4\pi \sqrt{p^{2}}}\ln\left(\frac{2m+\sqrt{p^{2}}}{2m-\sqrt{p^{2}}}\right)+O(e^{4}). \label{VP10}
\end{eqnarray}
For $p^{2}<0$ the result (\ref{VP10}) is also correct if we analytically continue the logarithm and the square root functions to the complex plane. For future needs we compute the limits of $\Pi^{(1)}(p^{2})$ and $\Pi^{(2)}(p^{2})$ as $p^{2}\to 0$:
\beq \label{VP10a}
\Pi^{(1)}(p^{2}\to 0)=\frac{e^{2}p^{2}}{12\pi m}  \qquad
\Pi^{(2)}(p^{2}\to 0)=\frac{e^{2}}{4\pi m}.
\eeq

First we would like to examine the case when $m=0$ and $\mu=0$. In this special case we drop the Chern-Simons and fermion mass terms from the Lagrangian (\ref{$QED_{3}$}); i.e., we are working with massless bare fermions and photons. The relation (\ref{VP3a}) reduces to $\Pi_{\mu\nu}(p)=P_{\mu\nu}\Pi(p^{2})$, where $\Pi(p^{2})$ is given by (\ref{VP10}): 
\beq \label{VP11}
\Pi(p^{2})=\Pi^{(1)}(p^{2})=\frac{e^{2}\sqrt{-p^{2}}}{16}.
\eeq
This result is in accordance with \cite{Jackiw:1980}. Inverting (\ref{VP3}) we easily obtain the corrected propagator:
\beq \label{VP12}
D^{\mu\nu}(p)=\frac{-iP^{\mu\nu}}{p^{2}-\Pi(p^{2})}-\frac{i\alpha p^{\mu}p^{\nu}}{p^{4}}.
\eeq
The propagator has a pole at $p^{2}=0$, so the photon remains massless and quantum corrections do not generate the photon mass. Instead, the quantum correction renormalizes the field. We expand the gauge-independent part of the propagator near the pole at $p^{2}=0$, i.e., write it in the form:
\beq \label{VP13a}
\frac{-iP^{\mu\nu}Z_{G}}{p^{2}}+c.c.,
\eeq
where $Z_{G}$ denotes the gauge field renormalization and $c.c.$ denotes continuum contributions.
The gauge-independent part of the propagator can be written as: 
\beq \label{VP13}
\frac{-iP^{\mu\nu}}{p^{2}-\Pi(p^{2})}=\frac{-iP^{\mu\nu}}{p^{2}-\Pi(0)-p^{2}\Pi^{\prime}(0)+O(p^{4})}=\frac{-iP^{\mu\nu}(1+\Pi^{\prime}(0))}{p^{2}}+c.c.,
\eeq
where we used $\Pi(0)=0$. The gauge field renormalization factor $Z_{G}=1+\Pi^{\prime}(0)$ is divergent, so it must be regularized.
   
Now we shall calculate the corrected propagator in the general case $m\ne 0$ and $\mu \ne 0$. According to (\ref{VP3}) we should first calculate the inverse of the bare propagator (\ref{VP1}). The propagator has a general tensor form:
\beq \label{VP14}
\mathscr{D}_{\mu\nu}(p)=AP_{\mu\nu}(p)+B\frac{\mu\epsilon_{\mu\nu\rho}p^{\rho}}{p^{2}}+C\frac{ p_{\mu}p_{\nu}}{p^{2}},
\eeq
where $A$, $B$ and $C$ are the scalar functions of $p^{2}$ with the mass dimension -2. We assume that the inverse of the propagator has the same tensor form:
\beq \label{VP15}
\mathscr{D}^{-1}_{\mu\nu}(p)=XP_{\mu\nu}(p)+Y\frac{\mu\epsilon_{\mu\nu\rho}p^{\rho}}{p^{2}}+Z\frac{ p_{\mu}p_{\nu}}{p^{2}},
\eeq
where $X$, $Y$ and $Z$ are the yet unknown scalar functions of $p^{2}$ with the mass dimension 2. To determine these functions we substitute (\ref{VP14}) and our ansatz (\ref{VP15}) into the following relation:
\beq \label{VP15a} \mathscr{D}_{\mu\nu}(p)(\mathscr{D}^{-1})^{\nu\rho}(p)=\delta^{\rho}_{\mu}. \eeq
This yields:
\beq \label{VP16}
X=\frac{Ap^{2}}{A^{2}p^{2}+B^{2}\mu^{2}} \qquad Y=-\frac{BX}{A}=-\frac{Bp^{2}}{A^{2}p^{2}+B^{2}\mu^{2}} \qquad Z=\frac{1}{C}.
\eeq
Using these results we get the inverse bare propagator:
\beq \label{VP17}
\mathscr{D}^{-1}_{\mu\nu}(p)=ip^{2}P_{\mu\nu}(p)-p^{2}\frac{\mu\epsilon_{\mu\nu\rho}p^{\rho}}{p^{2}}+\frac{ip^{2}}{\alpha}\frac{ p_{\mu}p_{\nu}}{p^{2}}.
\eeq
Now according to (\ref{VP3}) and (\ref{VP3a}) the inverse of the corrected propagator is:
\beq \label{VP18}
D^{-1}_{\mu\nu}(p)=[\underbrace{ip^{2}-i\Pi^{(1)}(p^{2})}_{A^{\prime}}]P_{\mu\nu}(p)+\underbrace{\left[\frac{m}{\mu}\Pi^{(2)}(p^{2})p^{2}-p^{2}\right]}_{B^{\prime}}\frac{\mu\epsilon_{\mu\nu\rho}p^{\rho}}{p^{2}}+\underbrace{\frac{ip^{2}}{\alpha}}_{C^{\prime}}\frac{ p_{\mu}p_{\nu}}{p^{2}}.
\eeq
Now we must solve the same problem as we were facing above. The corrected propagator has the general tensor form:
\beq \label{VP19}
D_{\mu\nu}(p)=X^{\prime}P_{\mu\nu}(p)+Y^{\prime}\frac{\mu\epsilon_{\mu\nu\rho}p^{\rho}}{p^{2}}+Z^{\prime}\frac{ p_{\mu}p_{\nu}}{p^{2}}.
\eeq
Using the primed version of (\ref{VP16}) we obtain the final result:
\begin{eqnarray}
X^{\prime}& = & \frac{i(p^{2}-\Pi^{(1)})}{p^{2}(m\Pi^{(2)}-\mu)^{2}-(p^{2}-\Pi^{(1)})^{2}} \nonumber \\
Y^{\prime}& = & -\frac{B^{\prime}X^{\prime}}{A^{\prime}}=\frac{1}{\mu}\frac{\mu-m\Pi^{(2)}}{(m\Pi^{(2)}-\mu)^{2}-p^{2}(1-\Pi^{(1)}/p^{2})^{2}} \nonumber \\
Z^{\prime} & = & -\frac{i\alpha}{p^{2}}. \label{VP20}
\end{eqnarray}
These functions are the complicated functions of $p^{2}$, so we shall extract the pole structure of the corrected propagator. We anticipate that the vacuum polarization renormalizes the mass and the wave function. We are interested in the pole structure of $X^{\prime}$, so let us write it in the following form:
\beq \label{VP21}
X^{\prime}=\frac{-iZ_{G}}{p^{2}-\mu_{ph}^{2}}+c.c.,
\eeq
where $Z_{G}$ is a residue of the pole, $\mu_{ph}$ is renormalized mass of photon and c.c. denotes continuum contributions. In order to calculate these quantities we write $X^{\prime}$ as:
\beq \label{VP22}
X^{\prime}=\frac{-i}{p^{2}-\underbrace{(\Pi^{(1)}+\frac{p^{2}(m\Pi^{(2)}-\mu)^{2}}{p^{2}-\Pi^{(1)}})}_{g(p^{2})}}.
\eeq
The results obtained by one-loop calculations are valid only to $O(e^{2})$, that is why we expand $g(p^{2})$ in powers of $e^{2}$ (note that $\Pi^{(1)}(p^{2})$ and $\Pi^{(2)}(p^{2})$ are already of the first order in $e^{2}$):
\beq \label{VP22a}
g(p^{2})=\mu^{2}+\Pi^{(1)}-2\mu m\Pi^{(2)}+\frac{\mu^{2}}{p^{2}}\Pi^{(1)}+O(e^{4}).
\eeq
So the square of the physical mass to the first order in $e^{2}$ is:
\beq \label{VP23}
\mu_{ph}^{2}=g(\mu_{ph}^{2})\approx g(\mu^{2}) =\mu^{2}+2\Pi^{(1)}(\mu^{2})-2\mu m\Pi^{(2)}(\mu^{2})+O(e^{4}).
\eeq
For the physical mass we finally obtain:
\beq \label{VP23a}
\mu_{ph}=\mu\left(1-\frac{m}{\mu}\Pi^{(2)}(\mu^{2})+\frac{\Pi^{(1)}(\mu^{2})}{\mu^{2}}\right)+O(e^{4}).
\eeq
To get $Z_{G}$ we expand $X^{\prime}$  near the pole:
\beq \label{VP24}
X^{\prime}=\frac{-i}{p^{2}-g(p^{2})}\approx \frac{-i}{p^{2}-\mu_{ph}^{2}-g^{\prime}\mid _{p^{2}=\mu_{ph}^{2}}(p^{2}-\mu_{ph}^{2})} \approx \frac{-i(1+g^{\prime}\mid_{p^{2}=\mu_{ph}^{2}})}{p^{2}-\mu_{ph}^{2}}. 
\eeq
That is why the result for $Z_{G}$ is:
\beq \label{VP25}
Z_{G}=1+g^{\prime}\mid_{p^{2}=\mu^{2}}+O(e^{4}).
\eeq
We must also extract the pole structure of $Y^{\prime}$. Using the same expansion in $e^{2}$ we readily get:
\beq \label{VP25a}
Y^{\prime}=-\frac{B^{\prime}X^{\prime}}{A^{\prime}}=\frac{-\mu_{ph}Z_{G}}{p^{2}-\mu_{ph}^{2}}+c.c.
\eeq
Our result is an expansion near the pole for the corrected propagator:
\beq \label{VP26}
D_{\mu\nu}(p)=\frac{-iZ_{G}}{p^{2}-\mu_{ph}^{2}}\left[P_{\mu\nu}(p)-\frac{i\mu_{ph}\epsilon_{\mu\nu\rho}p^{\rho}}{p^{2}}\right]-\frac{i\alpha p_{\mu}p_{\nu}}{p^{4}}+c.c.,
\eeq
so our guess was actually correct. The only effect of the one-loop corrections is the mass and the wave function renormalization. It is important that \emph{the generic structure of the corrected propagator is the same as for the free propagator}. We should mention here that it was shown in \cite{DJT:1981} that the Pauli-Villars regularization procedure yields \emph{the different result} for the physical mass $\mu_{ph}$ (\ref{VP23a}) in terms of bare mass $\mu$. 

The last aspect we shall address here is a special case of $m\ne0$ and $\mu=0$, i.e., interacting theory with massive bare fermions and massless bare photons. The bare propagator in this case is:
\beq \label{VP26a}
\mathscr{D}_{\mu\nu}(p)=\frac{-i}{p^{2}+i\epsilon}P_{\mu\nu}(p)-\frac{i\alpha p_{\mu}p_{\nu}}{(p^{2}+i\epsilon)^{2}}.
\eeq

According to (\ref{VP20}) in this case we obtain:
\begin{eqnarray}
X^{\prime}& = & \frac{i(p^{2}-\Pi^{(1)})}{p^{2}m^{2}(\Pi^{(2)})^{2}-(p^{2}-\Pi^{(1)})^{2}} \nonumber \\
Y^{\prime} & = &-\frac{B^{\prime}X^{\prime}}{A^{\prime}}=\frac{m}{\mu}\frac{ip^{2}\Pi^{(2)}}{p^{2}-\Pi^{(1)}}X^{\prime}. \label{VP27}
\end{eqnarray}
Using these results the corrected propagator may be rewritten in more elegant form \cite{DJT:1981}:
\beq \label{VP28}
D_{\mu\nu}(p)=\frac{-i}{p^{2}-\Pi(p^{2})}\left[ P_{\mu\nu}(p)-\frac{i\epsilon_{\mu\nu\rho}p^{\rho}}{p^{2}}\mathscr{M}(p^{2}) \right]
\eeq
\beq \label{VP29}
\Pi(p^{2})=\Pi^{(1)}(p^{2})+\frac{[m\Pi^{(2)}(p^{2})]^{2}}{1-\Pi^{(1)}(p^{2})/p^{2}}
\eeq
\beq \label{VP30}
\mathscr{M}(p^{2})=-\frac{m\Pi^{(2)}(p^{2})}{1-\Pi^{(1)}(p^{2})/p^{2}}.
\eeq
The crucial observation is that t\emph{he axial part of the propagator was generated by the fermion loop}. The question arises: does the correction generate a non zero photon mass? To answer this question we shall use our results from (\ref{VP10a}).  To the first order in $\alpha$:
\beq \label{VP31}
\mu_{ph}^{2}=\Pi(0)=(\frac{e^{2}}{4\pi})^{2}+O(e^{4}) \qquad \mathscr{M}=-\frac{e^{2}}{4\pi}+O(e^{4}).
\eeq
Thus the corrected propagator has the following pole structure:
\beq \label{VP32}
D_{\mu\nu}(p)=\frac{-iZ_{G}}{p^{2}-\mu_{ph}^{2}}\left[P_{\mu\nu}(p)+\frac{i\mu_{ph}\epsilon_{\mu\nu\rho}p^{\rho}}{p^{2}}\right]-\frac{i\alpha p_{\mu}p_{\nu}}{p^{4}}+c.c.,
\eeq
where $\mu_{ph}=\frac{e^{2}}{4\pi}$.
\emph{The fermion loop indeed generates the photon mass} in this case, although initially photon was massless. The calculation reveals a deep connection between the Chern-Simons term in the gauge part of the Lagrangian and the fermion mass term in the matter part of the Lagrangian.  If one of them is present in the bare Lagrangian, the second will be generated by quantum corrections even in the case it was absent in the original bare Lagrangian\footnote{We shall elucidate this finding in Chapter \ref{ch:effect} using the effective action formalism.}. The fact that the mass parameter of photon is negative in (\ref{VP31}) is not a problem. According to (\ref{QMCS38}) it means only that the particle has mass $|\mu_{ph}|$ and spin $-1$. The corrected propagator has the form (\ref{VP26}) with a real mass $|\mu_{ph}|=\frac{e^{2}}{4\pi}$.
\section{Fermion propagator}
Now we shall calculate the first quantum corrections to the fermion propagator. In general the inverse of the corrected fermion propagator $S(p)$ may be written as:
\beq \label{FP1}
S^{-1}(p)=\mathscr{S}^{-1}(p)+i\Sigma(p),
\eeq
where $\mathscr{S}(p)$ is given in (\ref{VP2}) and $\Sigma(p)$ is a $2\times2$ matrix called the electron mass operator, which includes all 1PI corrections to the free fermion propagator. Inverting (\ref{FP1}) yields:
\beq \label{FP2}
S(p)=\frac{i}{\slashed{p}-m-\Sigma(p)+i\epsilon}.
\eeq
The one-loop approximation for $\Sigma(p)$, denoted here as $\Sigma_{2}(p)$, is given by the last Feynman graph given in Fig. \ref{graph3}
and may be analytically written as:
\beq \label{FP3}
\Sigma_{2}(p)=-ie^{2}\int\frac{d^{3}k}{(2\pi)^{3}}\gamma_{\mu}\mathscr{S}(p+k)\gamma_{\nu}\mathscr{D}^{\mu\nu}(k).
\eeq
The free gauge propagator $\mathscr{D}^{\mu\nu}$ (\ref{VP1}) has three distinct parts, so we can write $\Sigma_{2}(p)$ naturally as:
\beq \label{FP4}
\Sigma_{2}(p)=\underbrace{\Sigma^{I}_{2}(p)}_{transverse}+\underbrace{\Sigma^{II}_{2}(p)}_{axial}+\underbrace{\Sigma^{III}_{2}(p)}_{gauge}.
\eeq

The transverse part $\Sigma^{I}_{2}(p)$ is:
\beq \label{FP5}
\Sigma^{I}_{2}(p)=-ie^{2}\int\frac{d^{3}k}{(2\pi)^{3}}\gamma_{\mu}\frac{i}{\slashed{p}+\slashed{k}-m+i\epsilon}\gamma_{\nu}\frac{-i}{k^{2}-\mu^{2}+i\epsilon}P^{\mu\nu}(k),
\eeq
where $P^{\mu\nu}(k)=g^{\mu\nu}-\frac{k^{\mu}k^{\nu}}{k^{2}}$. Power-counting tells us that the transverse integral is UV logarithmically divergent. We shall show, however, that the integral is in fact convergent due to the symmetric integration. We shall use the dimensional regularization:
\beq \label{FP6}
\Sigma^{I}_{2}(p)=-ie^{2}\int\frac{d^{d}k}{(2\pi)^{d}}\frac{\left(2p_{\mu}-(\slashed{p}+\slashed{k}-m)\gamma_{\mu}\right)\gamma_{\nu}}{[(p+k)^{2}-m^{2}+i\epsilon][k^{2}-\mu^{2}+i\epsilon]}\left[g^{\mu\nu}-\frac{k^{\mu}k^{\nu}}{k^{2}+i\epsilon}\right].
\eeq
In the numerator of the last expression we used the anticommutation relations for $\gamma$ matrices. After introducing the Feynman parameters and shifting the integration variable we find that the only potentially UV divergent terms in the integrand are linear or trilinear in the integration variable. Obviously, due to the symmetric integration we can drop them. The final integral is convergent and after standard calculations we obtain a rather complicated result written as a sum of three parametric integrals:
\begin{eqnarray}
\Sigma^{I}_{2}(p) & = & \frac{e^{2}}{8\pi}\left[ \int^{1}_{0}dx \frac{2x\slashed{p}+2m}{\sqrt{(x-1)xp^{2}+xm^{2}+(1-x)\mu^{2}-i\epsilon}}- \right. \nonumber \\
& & \left. -\slashed{p} \int^{1}_{0}dx\int^{1-x}_{0} dy \frac{1}{\sqrt{x(x-1)p^{2}+xm^{2}+y\mu^{2}-i\epsilon}} + \right.  \nonumber \\ 
& & \left.+\slashed{p}p^{2}\int^{1}_{0}dx \int^{1-x}_{0} dy \frac{x^{2}}{\left[x(x-1)p^{2}+xm^{2}+y\mu^{2}-i\epsilon\right]^{3/2}}\right]. \label{FP7}
\end{eqnarray}

Next we calculate the axial part $\Sigma^{II}_{2}(p)$:
\beq \label{FP8}
\Sigma^{II}_{2}(p)=-ie^{2}\int\frac{d^{3}k}{(2\pi)^{3}}\gamma_{\mu}\frac{i}{\slashed{p}+\slashed{k}-m+i\epsilon}\gamma_{\nu}\frac{-\mu \epsilon_{\mu\nu\alpha}k^{\alpha}/k^{2}}{k^{2}-\mu^{2}+i\epsilon}.
\eeq
By power-counting this integral is UV convergent, so the standard techniques including the Feynman parameters, shift of the variable and  Wick rotation yield the final result:
\begin{eqnarray}
\Sigma^{II}_{2}(p) & = & \frac{e^{2}}{8\pi}\left[ \int^{1}_{0}dx\int^{1-x}_{0}dy\frac{3\mu}{\sqrt{(x-1)xp^{2}+xm^{2}+y\mu^{2}-i\epsilon}}- \right. \nonumber \\
& & \left.   -\mu\int^{1}_{0} dx  \int^{1-x}_{0}dy\frac{mx\slashed{p}+x(x-1)p^{2}}{\left[(x-1)xp^{2}+xm^{2}+y\mu^{2}-i\epsilon\right]^{3/2}}\right].  \label{FP9}
\end{eqnarray}
For future needs it is important to know how $\Sigma^{I}_{2}(p)+\Sigma^{II}_{2}(p)$ behave as the functions of the momentum $p$. For $p^{2}<m^{2}$ the integrals in (\ref{FP7}) and (\ref{FP9}) are obviously convergent because the expressions in the denominators under the square roots are positive. On the mass shell $p^{2}=m^{2}$ two suspect integrals $I_{1}$ and $I_{2}$ arise, which may spoil the convergence:
\beq \label{FP9a}
I_{1}=\int^{1}_{0}dx \int^{1-x}_{0} dy \frac{1}{\sqrt{x^{2}+y\frac{\mu^{2}}{m^{2}}-i\epsilon}} \qquad I_{2}=\int^{1}_{0}dx \int^{1-x}_{0} dy \frac{x^{2}}{(x^{2}+y\frac{\mu^{2}}{m^{2}}-i\epsilon)^{3/2}}. 
\eeq
These integrals are elementary and convergent, so $\Sigma^{I}_{2}(m)+\Sigma^{II}_{2}(m)$ is finite. For $p^{2}>m^{2}$ the analysis is not so straightforward, but the integrals are also convergent because at most the integrable singularities are encountered. Thus the sum $\Sigma^{I}_{2}(p)+\Sigma^{II}_{2}(p)$ is \emph{finite} for arbitrary value of $p^{2}$.
 
Finally we calculate the gauge part $\Sigma^{III}(p)$:
\beq \label{FP10}
\Sigma^{III}_{2}(p)=-ie^{2}\int\frac{d^{3}k}{(2\pi)^{3}}\gamma_{\mu}\frac{i}{\slashed{p}+\slashed{k}-m+i\epsilon}\gamma_{\nu}\left( -\frac{i\alpha k^{\mu}k^{\nu}}{(k^{2}+i\epsilon)^{2}} \right).
\eeq
At the first sight it seems that the integral is logarithmically UV divergent, but due to the symmetric integration we encounter no UV divergences in this integral and the final result written as a parametric integral is:
\beq \label{FP11}
\Sigma^{III}_{2}(p)=\frac{\alpha e^{2}}{16\pi}\left[ \int^{1}_{0} dx \frac{(1-x)[3m-(1+5x)\slashed{p}]}{\sqrt{(x-1)xp^{2}+xm^{2}-i\epsilon}}-p^{2}\int^{1}_{0} dx \frac{x^{2}(1-x)[m+(1-x)\slashed{p}]}{\left[(x-1)xp^{2}+xm^{2}-i\epsilon\right]^{3/2}} \right].
\eeq
It is interesting to examine the last expression for different values of the momentum $p$. For $p^{2}<m^{2}$ both integrals in (\ref{FP11}) are convergent because the denominators in (\ref{FP11}) are positive for $x\in[0,1]$. On the mass shell $p^{2}=m^{2}$ we can easily calculate $\Sigma^{III}(m)$ and obtain:
\beq \label{FP11a}
\Sigma_{2}^{III}(m)=-\frac{\alpha e^{2}}{8\pi}.
\eeq
For $p^{2}>m^{2}$, however, there are problems with the second integral in (\ref{FP11}). The integrand has nonintegrable pole at $x=1-\frac{m^{2}}{p^{2}}$ and \emph{the integral is infrared divergent}!

The formulae (\ref{FP7}), (\ref{FP9}) and (\ref{FP11}) may be simplified in the special case of the massless $QED_{3}$, i.e., $m=0$ and $\mu=0$. The integrations in (\ref{FP7}) and (\ref{FP11}) are now easy and we obtain:
\beq \label{FP11b}
\Sigma_{2}(p)=-\frac{e^{2}\alpha}{16}\frac{\slashed{p}}{\sqrt{-p^{2}-i\epsilon}}.
\eeq
The result is gauge-dependent and is in accordance with \cite{Jackiw:1980}. 
 
At this point we explore the pole structure of the corrected fermion propagator in order to find the mass and the wave function renormalizations of the fermion. The matrix structure of $\Sigma(p)$ may be written as:
\beq \label{FP12}
\Sigma(p)=A(p^{2})+B(p^{2})\slashed{p},
\eeq
where $A(p^{2})$ and $B(p^{2})$ are the scalar functions of $p^{2}$ and are are at least $O(e^{2})$. The corrected propagator (\ref{FP2}) is now (we do not write explicitly the $i\epsilon$ term here): 
\beq \label{FP13}
S(p)=\frac{i}{\slashed{p}-m-A(p^{2})-B(p^{2})\slashed{p}}=\frac{i(\slashed{p}(1-B(p^{2}))+(m+A(p^{2})))}{p^{2}[1-B(p)]^{2}-[m+A(p^{2})]^{2}}.
\eeq
To find the location of the propagator's pole we set the denominator of (\ref{FP13}) equal zero:
\beq \label{FP14}
0=\left[ p^{2}(1-B(p))^{2}-(m+A(p^{2}))^{2}\right]|_{p^{2}=m_{ph}}= p^{2}-\underbrace{\left[m^{2}+2A(m^{2})+2m^{2}B(m^{2})\right]}_{m^{2}_{ph}}+O(e^{4}).
\eeq
The last relation tells us that to the first order in $e^{2}$ the physical mass may be written as:
\beq \label{FP15}
m_{ph}=m+A(m^{2})+mB(m^{2})+O(e^{4})=m+\Sigma_{2}(m)+O(e^{4}),
\eeq
so we showed here that the one-loop corrections shift the mass by:
\beq \label{FP15a}
\delta m=\Sigma_{2}(m).
\eeq
We could have calculated $\delta m$ performing the parametric integrals in (\ref{FP7}), (\ref{FP9}) and using (\ref{FP11a}), but the result is rather lengthy and is not illuminating. For us the important observation is that $\delta m$ is a \emph{gauge-dependent} quantity\footnote{See (\ref{FP11a})}.

Now we are ready to expand the corrected propagator (\ref{FP2}) near its pole:
\beq \label{FP16}
S(p)=\frac{i}{\slashed{p}-m-\Sigma(p)}\approx \frac{i}{\slashed{p}-m-\Sigma(m_{ph})-(\slashed{p}-m_{ph})\Sigma^{\prime}(p)|_{p=m_{ph}}}.
\eeq
Our final result reads:
\beq \label{FP17}
S(p)=\frac{iZ_{\psi}}{\slashed{p}-m_{ph}}+c.c. \qquad Z_{\psi}=1+\Sigma^{\prime}(p)|_{p=m}+O(e^{4}).
\eeq
We should emphasize at this point that there were two types of difficulties in the calculations of the corrected fermion propagator. First, according to (\ref{FP11a}) and (\ref{FP15}) the \emph{renormalized fermion mass is gauge-dependent}. This result seems alarming because the physical mass should not depend on the choice of gauge. Second, we encountered \emph{the infrared divergences} for $p^{2}>m^{2}$, which might be a problem in some higher order calculations.

In their original paper \cite{DJT:1981} Deser, Jackiw and Templeton proposed to work in the Landau transverse gauge ($\alpha=0$). This special choice of gauge solves our infrared problems. It was shown in \cite{Tyutin:1997} and \cite{Tyutin:1998} that all Feynman diagrams are infrared-finite in the transverse gauge. It seems unnatural, however, that we should prefer some special type of gauge in our calculations. For the general choice of the gauge parameter $\alpha$ we must deal with infrared divergences, so some regularization procedure is needed. In \cite{Tyutin:1997} and \cite{Tyutin:1998} Tyutin and Zeitlin introduced this regularization as a Proca term $\frac{\theta^{2}}{2}A^{\mu}A_{\mu}$ directly into the Lagrangian. The virtue of the new term is that it cures the infrared divergences in any gauge $\alpha$, makes the renormalized mass a gauge-invariant quantity and the Greens functions are gauge-invariant for the external fermions on the mass shell. The regularization may be taken off by $\theta\to0$. By doing so we come back to the Landau gauge of original theory. This lengthy procedure justifies the exceptional status of the Landau gauge.
\section{Running coupling constant}
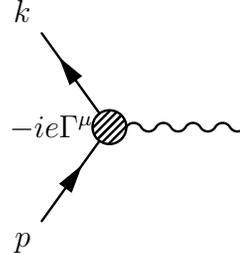
\begin{figure}[!htb]
\begin{center}
\begin{tabular}{c}

\begin{fmffile}{graph5}
\begin{fmfgraph*}(30,25)
\fmfleft{i1,i2}
\fmfright{i3}
\fmf{fermion}{i1,v1}
\fmf{fermion}{v1,i2}
\fmf{photon}{v1,i3}
\fmfblob{.15w}{v1}
\fmflabel{$-ie\Gamma^{\mu}$}{v1}
\fmflabel{$p$}{i1}
\fmflabel{$k$}{i2}
\end{fmfgraph*}
\end{fmffile}

\end{tabular}
\caption{Amputated vertex function $\Gamma^{\mu}(p,k)$}\label{graph5}
\end{center}
\end{figure}

Now we calculate the running coupling $e(p)$ in the MCS theory with the Lagrangian (\ref{$QED_{3}$}): One-loop corrections modify the tree level-interaction. Some of these modifications may be absorbed into the coupling, which makes the coupling momentum-dependent. We have seen already (\ref{VP26}), (\ref{FP17}) that the one-loop corrections generate the field strength-renormalizations and shift the masses of the particles. Here we shall work in the renormalized perturbation theory, however, so we introduce the renormalized fields $\psi_{r}=Z_{\psi}^{-1/2}\psi$ and $A^{\mu}_{r}=Z_{G}^{-1/2}A^{\mu}$  into the Lagrangian:
\beq \label{RC1}
\mathscr{L}=-\frac{1}{4}Z_{G}F_{r}^{\mu\nu}(F_{r})_{\mu\nu}+\frac{\mu}{2}Z_{G}\epsilon^{\mu\nu\rho}(A_{r})_{\mu}\partial_{\nu}(A_{r})_{\rho}+Z_{\psi}\bar{\psi_{r}}(i\slashed{\partial}-m)\psi_{r}-Z_{\psi}\sqrt{Z_{G}}e\bar{\psi_{r}}\gamma^{\mu}\psi_{r} (A_{r})_{\mu}.
\eeq
We also write the Lagrangian in terms of the physical parameters $m_{ph}$, $\mu_{ph}$ and the physical electric charge $e_{ph}$, measured at zero momentum transfer ($q^{2}=0$):
\beq \label{RC2}
\begin{array}{c}
\mathscr{L}=-\frac{1}{4}F_{r}^{\mu\nu}(F_{r})_{\mu\nu}+\frac{\mu_{ph}}{2}\epsilon^{\mu\nu\rho}(A_{r})_{\mu}\partial_{\nu}(A_{r})_{\rho}+\bar{\psi_{r}}(i\slashed{\partial}-m_{ph})\psi_{r}-e_{ph}\bar{\psi_{r}}\gamma^{\mu}\psi_{r} (A_{r})_{\mu}- \\
-\frac{1}{2}\delta_{A}F_{r}^{\mu\nu}(F_{r})_{\mu\nu}+\delta_{\mu}\epsilon^{\mu\nu\rho}(A_{r})_{\mu}\partial_{\nu}(A_{r})_{\rho}+\bar{\psi_{r}}(i\delta_{\psi}\slashed{\partial}-\delta_{m})\psi_{r}-\delta_{e}e_{ph}\bar{\psi_{r}}\gamma^{\mu}\psi_{r} (A_{r})_{\mu},
\end{array}
\eeq
where $\delta_{A}$, $\delta_{\mu}$, $\delta_{\psi}$, $\delta_{m}$ and $\delta_{e}$ are the position-independent counterterms, which are at least $O(e^{2})$. In order to simplify the notation we write $A^{\mu}$ instead of $(A_{r})^{\mu}$ and $\psi$ instead of $\psi_{r}$. We shall also drop the subscript $_{ph}$ and write simply $m$, $\mu$ and $e$ for the physical parameters. The Feynman rules for these counterterms are depicted in Fig. \ref{graph4}.
\begin{figure}[!htb]
\begin{center}

\begin{eqnarray} \label{RC2a}
\parbox{25mm}{
    \begin{fmffile}{g1}
    \begin{fmfgraph}(25,12) 
    \fmfleft{i1,i2}
    \fmfright{i3}
    \fmf{fermion}{i1,v1}
    \fmf{fermion}{v1,i2}
    \fmf{photon}{v1,i3}
    \fmfv{decor.shape=square,decor.filled=full,decor.size=5}{v1}
    \end{fmfgraph}
    \end{fmffile}} & \qquad -ie\gamma^{\mu}\delta_{e} \\
\parbox{25mm}{
   \begin{fmffile}{g2}
   \begin{fmfgraph}(25,12) 
   \fmfleft{i1}
   \fmfright{i2}
   \fmf{fermion}{i1,v1}
   \fmf{fermion}{v1,i2}
   \fmfv{decor.shape=square,decor.filled=full,decor.size=5}{v1}
   \end{fmfgraph}
   \end{fmffile}} & \qquad  i(\slashed{p} \delta_{\psi}-\delta_{m})  \\
\parbox{25mm}{
   \begin{fmffile}{g3}
   \begin{fmfgraph*}(25,12) 
   \fmfleft{i1}
   \fmfright{i2}
   \fmflabel{$\mu$}{i1}
   \fmflabel{$\nu$}{i2}
   \fmf{photon, label=$\to p$}{i1,v1}
   \fmf{photon}{v1,i2}
   \fmfv{decor.shape=square,decor.filled=full,decor.size=5}{v1}
   \end{fmfgraph*}
   \end{fmffile}} & \qquad -i\delta_{A}(g^{\mu\nu}p^{2}-p^{\mu}p^{\nu})+\delta_{\mu}\epsilon^{\mu\rho\nu}p_{\rho}
\end{eqnarray}

\caption{Feynman rules for the counterterms}\label{graph4}
\end{center}
\end{figure}
Each of these five counterterms must be now fixed by the renormalization conditions, which demand that $m$, $\mu$ and $e$ are the real physical parameters and that $Z_{G}=1$ and $Z_{\psi}=1$ holds for the renormalized fields. The following renormalization conditions fix three counterterms $\delta_{\psi}$, $\delta_{m}$ and $\delta_{e}$:
\begin{eqnarray}
\Sigma(\slashed{p}=m) & = & 0 \nonumber \\
\frac{d}{dp}\Sigma(\slashed{p})|_{\slashed{p}=m} & = & 0 \nonumber \\
\Gamma^{\mu}(p,p) & = & \gamma^{\mu}, \label{RC3}
\end{eqnarray}
where $\Gamma^{\mu}(p,k)$ is a corrected vertex function. It is depicted in Fig. \ref{graph5}, where the gray blob denotes a bare vertex plus all 1PI corrections to it.

It remains to fix the counterterms $\delta_{A}$ and $\delta_{\mu}$. As it follows from (\ref{VP3a}) the polarization tensor $\Pi_{\mu\nu}(p)$ can be expressed in terms of two scalar functions $\Pi^{(1)}(p^{2})$ and $\Pi^{(2)}(p^{2})$ as:
\beq \label{RC3a}
\Pi_{\mu\nu}(p)=P_{\mu\nu}\Pi^{(1)}(p^{2})+im\epsilon_{\mu\nu\rho}p^{\rho}\Pi^{(2)}(p^{2}).
\eeq
These functions were calculated explicitly in the bare perturbation theory to one loop in (\ref{VP10}). Photon's counterterm vertices from Fig. \ref{graph4} modify the scalar functions $\Pi^{(1)}(p^{2})$ and $\Pi^{(2)}(p^{2})$ as follows:
\beq \label{RC4}
\Pi^{(1)}(p^{2}) \to \Pi^{(1)}(p^{2})-\delta_{A}p^{2} \qquad \Pi^{(2)}(p^{2})\to \Pi^{(2)}(p^{2})+\frac{\delta_{\mu}}{m}.
\eeq
The corrected photon propagator has been already calculated (\ref{VP19}, \ref{VP20}) in the bare perturbation theory. We demand now that $Z_{G}=1$ and $\mu_{ph}=\mu$ in (\ref{VP21}) for the renormalized field; i.e., the transverse part of the corrected photon propagator $X^{\prime}$ (\ref{VP26}), (\ref{VP27}) takes the form:
\beq \label{RC4a}
X^{\prime}=\frac{i(p^{2}-\Pi^{(1)})}{p^{2}m^{2}(\Pi^{(2)})^{2}-(p^{2}-\Pi^{(1)})^{2}}=\frac{-i}{p^{2}-\mu^{2}}+c.c.,
\eeq
where the last equality holds near the pole.
From (\ref{VP22}), (\ref{VP24}) and (\ref{VP25}) these requirements lead to the following renormalization conditions, which are valid to the first order in $e^{2}$:
\begin{eqnarray}
m\mu\Pi^{(2)}(\mu^{2}) & = & \Pi^{(1)}(\mu^{2}) \nonumber \\
\Pi^{(1)}(\mu^{2}) & = & -\mu^{2}\left(2m\mu\frac{d\Pi^{(2)}}{dp^{2}}|_{\mu^{2}}-2\frac{d\Pi^{(1)}}{dp^{2}}|_{\mu^{2}}\right). \label{RC5}
\end{eqnarray}
These conditions fix the remaining counterterms $\delta_{A}$ and $\delta_{\mu}$. It is straightforward, but rather tedious to extract the counterterms from (\ref{RC5}), (\ref{RC4}) and (\ref{VP10}):
\beq	 \label{RC6}
\delta_{A}=\frac{e^{2}}{4\pi}\left[ \frac{4m^{2}}{\mu^{2}(2m+\mu)}-\frac{m(2m-\mu)}{\mu^{3}}\ln\frac{2m+\mu}{2m-\mu}  \right]
\eeq
\beq \label{RC7}  
\delta_{\mu}=\frac{e^{2}}{4\pi}\left[ \frac{24m^{3}-4m^{2}\mu-6m\mu^{2}+\mu^{3}}{4(2m+\mu)\mu^{2}}\ln\frac{2m+\mu}{2m-\mu}-\frac{6m^{2}+m\mu}{(2m+\mu)\mu}  \right].
\eeq
The result looks quite cumbersome mainly because of the existence of two mass scales $m$ and $\mu$ in the theory.

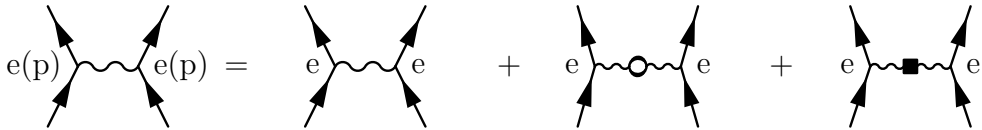
\begin{figure}[!htb]
\begin{center}
\begin{tabular}{ccccccccc}    

\parbox{20mm}{
\begin{fmffile}{RC1} 	%
   \begin{fmfgraph*}(20,16) 
   \fmfleft{l1,l2}
   \fmfright{r1,r2}
   \fmf{fermion}{l1,v1,l2}
   \fmf{fermion}{r1,v2,r2}
   \fmf{photon}{v1,v2}
   \fmflabel{e(p)}{v1}
   \fmflabel{e(p)}{v2}
\end{fmfgraph*}
\end{fmffile}}
\quad = \quad
\parbox{22mm}{
\begin{fmffile}{RC2} 	%
   \begin{fmfgraph*}(20,16) 
   \fmfleft{l1,l2}
   \fmfright{r1,r2}
   \fmf{fermion}{l1,v1,l2}
   \fmf{fermion}{r1,v2,r2}
   \fmf{photon}{v1,v2}
   \fmflabel{e}{v1}
   \fmflabel{e}{v2}
\end{fmfgraph*}
\end{fmffile}
}
\quad + \quad
\parbox{22mm}{
\begin{fmffile}{RC3} 	%
   \begin{fmfgraph*}(20,16) 
   \fmfleft{l1,l2}
   \fmfright{r1,r2}
   \fmf{fermion}{l1,v1,l2}
   \fmf{fermion}{r1,v2,r2}
   \fmf{photon}{v1,v3}
   \fmf{photon}{v4,v2}
   \fmf{fermion,left}{v3,v4,v3}
   \fmflabel{e}{v1}
   \fmflabel{e}{v2}
\end{fmfgraph*}
\end{fmffile}
}
\quad + \quad
\parbox{22mm}{
\begin{fmffile}{RC4} 	%
   \begin{fmfgraph*}(20,16) 
   \fmfleft{l1,l2}
   \fmfright{r1,r2}
   \fmf{fermion}{l1,v1,l2}
   \fmf{fermion}{r1,v2,r2}
   \fmf{photon}{v1,v3}
   \fmfv{decor.shape=square,decor.filled=full,decor.size=5}{v3}
   \fmf{photon}{v2,v3}
   \fmflabel{e}{v1}
   \fmflabel{e}{v2}
\end{fmfgraph*}
\end{fmffile}
}
\end{tabular}
\caption{Running coupling for the transverse part of the photon propagator}\label{graph6}
\end{center}
\end{figure}

Finally, let us calculate the running coupling $e(p)$ to the first order in $e^{2}$. In Fig. \ref{graph6} we illustrate the notion of the running coupling to the first order in $e^{2}$ for the fermion-fermion scattering.
\unitlength = 1mm

As mentioned before the quantum corrections to the photon propagator are  absorbed into the coupling. For the transverse part of the propagator we get:
\beq \label{RC8}
e^{2}(p)\frac{-i}{p^{2}-\mu^{2}}=e^{2}X^{\prime}=e^{2}\frac{i(p^{2}-\Pi^{(1)})}{p^{2}m^{2}(\Pi^{(2)})^{2}-(p^{2}-\Pi^{(1)})^{2}}.
\eeq
We should remark here that in the MCS theory we should have introduced, in fact, two different running couplings one for the transverse part (\ref{RC8}) and another for the axial part. These two running couplings are distinct, because they flow differently. We shall not do it here and we shall examine only the transverse running coupling. 

\begin{figure}[!htb]
\begin{center}
\includegraphics[angle=0, width=0.5\textwidth, height=0.5\textwidth]{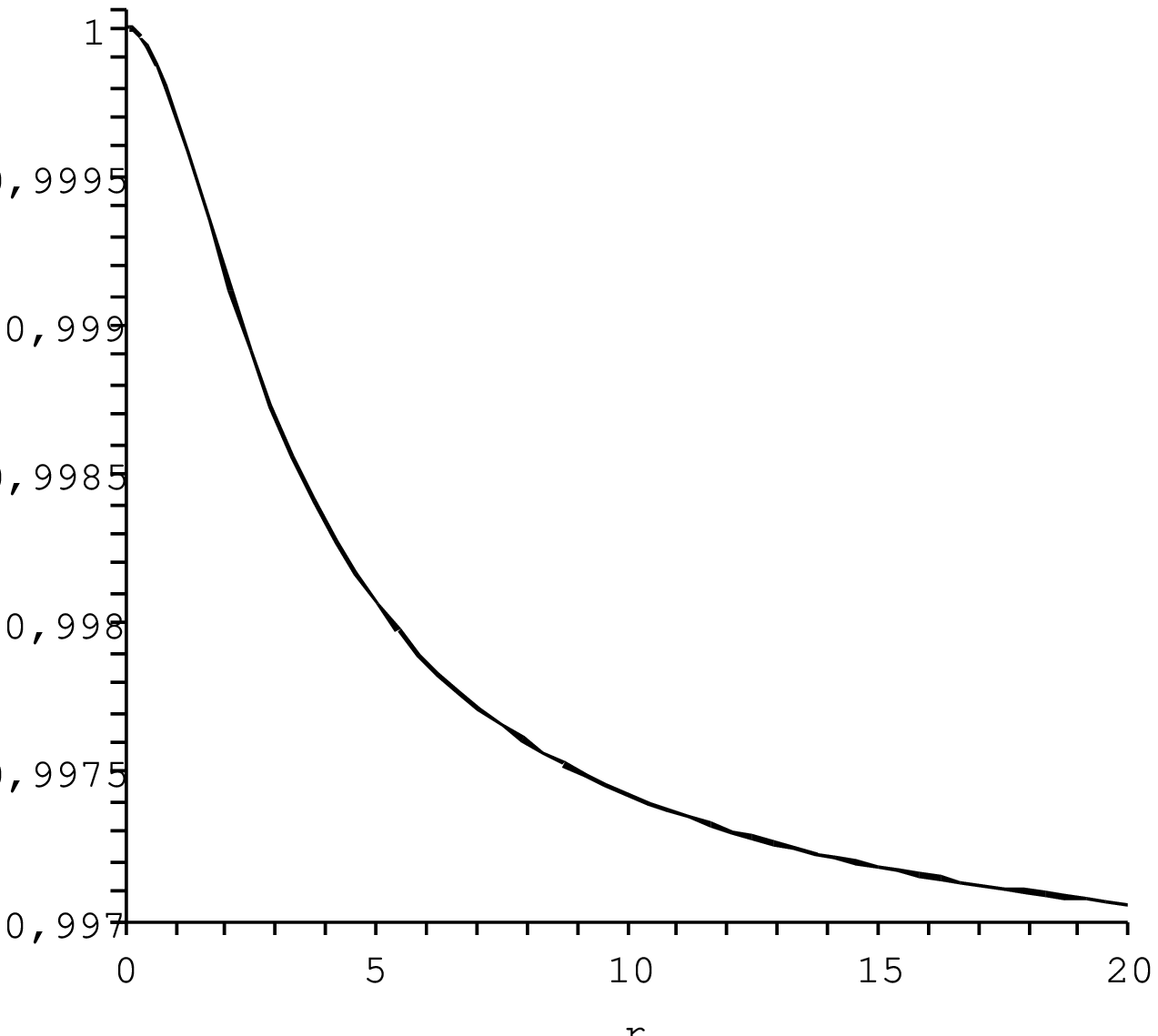}
\caption{Running coupling $e^{2}(r)$}\label{graph7}
\end{center}
\end{figure}

With the help of the mathematical program Maple we obtained $e^{2}(p)$ for general parameters $m$ and $\mu$ to $O(e^{2})$, but the result is cumbersome. Instead we look closer at the special case of massless photons, i.e., $\mu\to 0$. Assuming this we get rid of the mass scale $\mu$. For $p^{2}<0$, which is of our interest for fermions on the mass shell in Fig. \ref{graph6} we obtain (with the help of Maple) a rather compact formula:
\beq \label{RC9}
\frac{e^{2}(r)}{e^{2}}=1+\frac{e^{2}}{4\pi}\left( \frac{m}{r^{2}}-\frac{1}{3m}-\frac{4m^{2}-r^{2}}{2r^{3}}\arctan(\frac{r}{2m}) \right),
\eeq     
where $r$ is a real positive parameter such that $r^{2}=-p^{2}$. In Fig. \ref{graph7} we plotted $e^{2}(r)$ in units of $e^{2}$ for $m=1$ and $\alpha=\frac{e^{2}}{4\pi m}=\frac{1}{100}$.
We see that for zero momentum transfer $r=0$ the running coupling $e(r)$ equals the physical charge $e$ from the Lagrangian. Then for increasing $r$ the running coupling decreases and asymptotically tends to:
\beq \label{RC10}
\frac{e^{2}(\infty)}{e^{2}}=1-\frac{e^{2}}{12\pi m},
\eeq
which makes sense only for sufficiently large fermion mass, where the perturbation theory works well.

It seems that \emph{the MCS theory with the light photon $\mu\to0$ remains in the perturbative regime for an arbitrary energy of collision if the theory was set into the perturbative regime (i.e., $\alpha=\frac{e^{2}}{4\pi m} <<\infty$) for zero momentum transfer}. The statement holds at least in the first order of the perturbation theory.  
\chapter{Effective action}
\label{ch:effect}
It was demonstrated in Chapter \ref{ch:perturbative} using the one-loop calculation that the massive fermions generate the CS term even if it is absent in the original Lagrangian. In this Chapter we elaborate on these matters more thoroughly. To that end the concept of the \emph{one-loop effective action} will be introduced. We also develop the \emph{perturbative} expansion of the effective action \footnote{in the theory of quantized fermions interacting with the background classical electromagnetic field} in the coupling constant $e$. Using the result of Chapter \ref{ch:perturbative} we convince ourselves that the CS term is really present in the effective action. Subsequently, the powerful \emph{non-perturbative} "proper time" method due to J. Schwinger will be introduced and employed in the case of the electrodynamics with quantized massive fermions.  

\section{One-loop effective action and perturbative expansion} 
Let us consider the massive $QED_{3}$ without the CS term. The generating functional for the electromagnetic Greens' functions $Z[j]$ is given by the following functional integral:
\beq \label{pea1}
Z[j]=\int \mathfrak{D}A\mathfrak{D}\overline{\psi}\mathfrak{D}\psi \exp\left[i(-\frac{1}{4}F^{\mu\nu}F_{\mu\nu}+\overline{\psi}(i\slashed{\partial}-m-e\slashed{A})\psi+j_{\mu}A^{\mu}) \right].
\eeq
In order to take into account the quantum fluctuations, which come from the virtual fermions, we integrate out the fermion fields in $Z[j]$ and obtain:
\beq \label{pea2}
Z[j]=\int \mathfrak{D}A \exp\left[i(-\frac{1}{4}F^{\mu\nu}F_{\mu\nu}+\Gamma^{1loop}[A]+j_{\mu}A^{\mu}) \right],
\eeq
where the one-loop effective action $\Gamma^{1loop}[A]$ is given by:
\beq \label{pea3}
\Gamma^{1loop}[A]=-i\ln \mathsf{Det}(i\slashed{\partial}-e\slashed{A}-m),
\eeq
where $\mathsf{Det}$ denotes the functional determinant of the Dirac operator. Relation (\ref{pea2}) suggests that we can treat $\Gamma^{1loop}[A]$ as a modification of the classical electromagnetic action due to the quantum fermion fluctuations. Whence the name \emph{effective action} makes a good sense.

The superscript $^{1loop}$ originates from the \emph{perturbative diagrammatic} expansion of (\ref{pea3}), which we are now going to construct. Let us start with the determinant in (\ref{pea3}) and first factor out its "free" part:
\beq \label{pea4}
\mathsf{Det}(i\slashed{\partial}-m-e\slashed{A})=\mathsf{Det}(i\slashed{\partial}-m)\mathsf{Det}(1-\frac{e}{i\slashed{\partial}-m}\slashed{A}).
\eeq
Assuming that the well-known matrix identity (by $\mathsf{Tr}$ we denote here the trace operation for the discreet indices and integration for the continues indices):
\beq \label{pea5}
\mathsf{Det}A=\exp(\mathsf{Tr}\ln(A))
\eeq
applies also to infinite-dimensional matrices we end up with the following expression:
\beq \label{pea6}
\mathsf{Det}(1-\frac{e}{i\slashed{\partial}-m}\slashed{A})=\exp\left( \mathsf{Tr}\left[ -\sum_{n=1}^{\infty}\frac{1}{n}\left\{i\frac{-ie}{i\partial-m}\slashed{A} \right\}^{n} \right] \right),
\eeq
where we used the power expansion of $\ln(1+x)$ near $x=1$ and assumed that the coupling constant $e$ is sufficiently small to make the expansion valid.

Putting all peaces together the one-loop effective action is:
\beq \label{pea7}
i\Gamma^{1loop}[A]=\ln \mathsf{Det}(i\slashed{\partial}-m)-\sum_{n=1}^{\infty}\frac{1}{n}\mathsf{Tr}\left\{\frac{i}{i\slashed{\partial}-m}(-ie\slashed{A}) \right\}^{n}.
\eeq
The sum in the last expression has a beautiful diagramatic interpretation. Indeed, recall that the free Feynman propagator solves the following equation with appropriate boundary condition:
\beq \label{pea8}
(i\slashed{\partial}-m)S_{F}(x-y)=i\delta(x-y)
\eeq
and $\mathsf{Tr}$ denotes in (\ref{pea7}) the spacetime integration and spinor tracing. The effective action thus can be written as:
\beq \label{pea9}
i\Gamma^{1loop}[A]=\ln \mathsf{Det}(i\slashed{\partial}-m)+\sum_{n=1}^{\infty}\mbox{}
\parbox{20mm}{
\begin{fmffile}{ep}
\begin{fmfgraph*}(30,30)
\fmfsurroundn{v}{6}
\fmf{photon}{v1,i1} 
\fmf{photon}{v2,i2} 
\fmf{photon}{v3,i3}
\fmf{phantom}{v4,i4} 
\fmf{photon}{v5,i5}
\fmf{photon}{v6,i6} 
\fmfcyclen{fermion,right=0.25, tension=6/8}{i}{6}
\fmfv{l=.,l.a=180,l.d=6thick}{i3}
\fmfv{l=.,l.a=180,l.d=6thick}{i5}
\fmfv{l=.,l.a=180,l.d=3thick}{i4}
\fmfdot{v1}
\fmfdot{v2}
\fmfdot{v3}
\fmfdot{v5}
\fmfdot{v6} 
\end{fmfgraph*}
\end{fmffile}}
\eeq
The electromagnetic field is treated as a classical c-field in the last diagram. The elctromagnetic vertex used in the previous expression is illustrated in Fig \ref{graph16}. The minus sign in front of the sum in (\ref{pea7}) is compensated by the minus sign from the loop in (\ref{pea9}). The factor $\frac{1}{n}$ in the sum in (\ref{pea7}) is exactly the symmetry factor of the digram with $n$ external photon lines in (\ref{pea9}). Hence, albeit the trivial $A$-independent term in (\ref{pea9}),  the effective action is given by the sum of all fermion one-loop diagrams with arbitrary number of photon legs. The superscript $^{1loop}$ is well-justified. 
\begin{figure}[!htb]
\begin{center}
\beq
\parbox{25mm}{
    \begin{fmffile}{vertex}
    \begin{fmfgraph}(25,12) 
    \fmfleft{i1,i2}
    \fmfright{i3}
    \fmf{fermion}{i1,v}
    \fmf{fermion}{v,i2}
    \fmf{photon}{v,i3}
    \fmfdot{i3}
    \end{fmfgraph}
    \end{fmffile}} \qquad -ei\gamma^{\mu}\int d^{3}x A_{\mu}(x)
\eeq
\caption{Interaction vertex}\label{graph16}
\end{center}
\end{figure}

Having developed the perturbative expansion for the one-loop effective action, we are now in position to calculate some few lowest terms. The tadpole diagram gives the linear contribution to the $\Gamma^{1loop}[A]$:
\beq \label{pea10}
\parbox{25mm}{
\begin{fmffile}{tadpole}
\begin{fmfgraph}(20,12) 
    \fmfleft{i1} \fmfright{i2}
    \fmf{photon}{v,i1}
    \fmf{fermion,left}{v,i2,v}
    \fmfdot{i1}
    \end{fmfgraph}
\end{fmffile}}=ie\int d^{3}xA^{\mu}(x)\int \frac{d^{3}k}{(2\pi)^{3}}\mathsf{tr}\left[\gamma_{\mu}\frac{i(\slashed{k}+m)}{k^{2}-m^{2}+i\epsilon} \right],
\eeq
which vanishes due to the symmetric integration in $k$. The next step is the quadratic contribution to $\Gamma^{1loop}[A]$:
\begin{eqnarray}
\parbox{25mm}{
\begin{fmffile}{pea11}
\begin{fmfgraph}(20,12) 
    \fmfleft{i1} \fmfright{i2}
    \fmf{photon}{v,i1}
    \fmf{photon}{v1,i2}
    \fmf{fermion,left}{v,v1,v}
    \fmfdot{i1}
    \fmfdot{i2}
    \end{fmfgraph}
\end{fmffile}}&=&\frac{1}{2}\int\int d^{3}x d^{3}y A_{\mu}(x)\Pi_{2}^{\mu\nu}(x-y)A_{\nu}(y)= \nonumber \\
&=&\frac{1}{2}\int d^{3}p A_{\mu}(-p)\Pi_{2}^{\mu\nu}(p)A_{\nu}(p). \label{pea11}
\end{eqnarray}
The polarization tensor $\Pi_{\mu\nu}(p)$ was defined in (\ref{VP3}) and has the tensor structure:
\beq \label{pea12}
\Pi_{\mu\nu}(p)=P_{\mu\nu}\Pi^{(1)}(p^{2})+im\epsilon_{\mu\nu\rho}p^{\rho}\Pi^{(2)}(p^{2}).
\eeq
At this stage we shall take only the axial part of the polarization tensor and work in the region of the $p$-space, where $m^{2}>>p^{2}$. In this region according to (\ref{VP9}) we have:
\beq \label{pea13}
\Pi_{\mu\nu}(p)_{A}=im\epsilon_{\mu\nu\rho}p^{\rho}\Pi^{(2)}(p^{2})\approx i\frac{e^{2}m}{4\pi |m|}\epsilon^{\mu\nu\rho}p^{\rho}.
\eeq
Substituting this result into (\ref{pea11}) we finally obtain the axial quadratic contribution to the effective action:
\begin{eqnarray}
\left(i\Gamma^{1loop}\right)_{2,axial}&=&\frac{ie^{2}}{8\pi}\frac{m}{|m|}\int d^{3}p A_{\mu}(-p)\epsilon^{\mu\nu\rho}A_{\nu}(p)p_{\rho} \nonumber \\
&=&\frac{e^{2}}{8\pi}\frac{m}{|m|}\int d^{3}x\epsilon^{\mu\nu\rho}A_{\mu}(x)\partial_{\nu}A_{\rho}(x). \label{pea14}
\end{eqnarray}
This is nothing else but the  \emph{CS term}. We have just demonstrated that the massive fermions generate at one-loop order the CS term in the effective action. It must be noted here that there is an inconsistency in our calculations: the expression (\ref{pea13}) works only for $p^{2}<<m^{2}$, but we substituted it into (\ref{pea11}) and integrated over the \emph{whole} $p$ spacetime. To our defense it seems that the procedure works well for the \emph{slowly} varying external fields $A^{\mu}(x)$. Using (\ref{VP9}), the correct integral takes the form:
\beq \label{pea15}
\left(i\Gamma^{1loop}\right)_{2,axial}=\frac{ie^{2}}{8\pi}\int_{0}^{\infty} dy\int d^{3}p \frac{m}{\sqrt{m^{2}+y(1-y)p^{2}-i\epsilon}}A_{\mu}(-p)\epsilon^{\mu\nu\rho}A_{\nu}(p)p_{\rho}.
\eeq
For $p^{2}<<m^{2}$ we get essentially the integrand of (\ref{pea14}), while for large $p^{\mu}$ we expect the Fourier components $A^{\mu}(p)$ to be sufficiently small and we can drop this part of the integration domain. In the next section we use completely different technique to obtain the same result (\ref{pea14}) for the slowly varying external electromagnetic field. 
\section{Non-perturbative "proper time" method}
So far we have performed only the perturbative calculations of the one-loop effective action in the planar electrodynamics with massive quantized fermions. We demonstrated that the CS term appears in the effective action even if it is absent in the original Lagrangian. However, more than fifty years ago J. Schwinger developed a famous "proper time" method for doing \emph{exact} calculations of the one-loop effective action in the \emph{constant} background electromagnetic field \cite{Schwinger:1951}. It is plausible to assume that the effective action, obtained in this way, is a good approximation of the effective action for the slowly varying electromagnetic field. In this section we shall demonstrate how the CS term appears in the "proper-time" approach \cite{Redlich:1983}. The result will coincide with our finding from the last section. In addition the non-perturbative Maxwell part of the effective action will be also deduced \cite{Redlich:1983}. Calculation of the effective action for $QED_{3}$ in the external inhomogeneous electromagnetic field can be found in \cite{Gies1,Gies2}.

As has already been shown in the last section the one-loop effective action of the electrodynamics in two spatial dimension with massive quantized electrons is given by\footnote{The notion of the covariant derivative is introduced for the convenience in the subsequent computations}:
\beq \label{pt1}
\Gamma^{1loop}[A]=i\mathsf{Tr}\ln(\slashed{D}+m),
\eeq 
where $\mathsf{Tr}$ denotes the trace over spinor indices and the spacetime integration, the covariant derivative is defined by:
\beq \label{pt2}
\slashed{D}\equiv-i\slashed{\partial}-e\slashed{A}.
\eeq
Let us introduce the background gauge c-field $A_{\mu}$ with the \emph{constant} field strength tensor $F_{\mu\nu}=const$. The gauge can be found, in which the electromagnetic potential is expressed as:
\beq \label{pt3}
A_{\mu}=\frac{1}{2}x^{\alpha}F_{\alpha\mu}=-\frac{1}{2}x^{\alpha}\epsilon_{\mu\alpha\beta}\mbox{}^{\ *}F^{\beta},
\eeq
where the field strength dual is given by $^{*}F^{\mu}=\frac{1}{2}\epsilon^{\mu\nu\rho}F_{\nu\rho}$.

The gauge invariance dictates the possible appearance of both the Maxwell (some function of $F^{\alpha\beta}F_{\alpha\beta}$) and the CS (some function of $\epsilon^{\alpha\beta\gamma}A_{\alpha}F_{\beta\gamma}$) parts in the effective action for the \emph{general} field $A_{\mu}$. For the \emph{constant} electromagnetic field, however, the possible CS term vanishes identically:
\beq \label{pt4}
\epsilon^{\alpha\beta\gamma}A_{\alpha}F_{\beta\gamma}=^{*}F^{\alpha}\epsilon_{\alpha\beta\gamma}\ ^{*}F^{\beta}x^{\gamma}\equiv0,
\eeq
where (\ref{pt3}) was used. Thus, it is not sufficient to calculate only (\ref{pt1}) for the \emph{constant} field in order to obtain the CS-dependent part of the effective action. On the other hand, by calculating (\ref{pt1}) we shall obtain the Maxwell part of the effective action.  In order to get the CS part, we shall also calculate the vacuum expectation value of the modified Dirac current $j^{\mu}(x)$, defined by the relation:
\beq \label{pt5}
j^{\mu}(x)\equiv\frac{1}{2}e(\gamma^{\mu})_{ab}[\overline{\psi}_{a}(x),\psi_{b}(y)],
\eeq
where $a,b$ are spinor indices. The modified current differs from the standard Dirac current by x-independent, however, divergent term. The virtue of the modified current is that it can be conveniently expressed in terms of the T-product:
\beq \label{pt6}
j^{\mu}(x)=e\gamma_{ab}\overline{\lim}_{y\to x}T[\overline{\psi}_{a}(x)\psi_{b}(y)],
\eeq
where $T$ is a conventional time-ordering operation and $\overline{\lim}_{y\to x}$ denotes the average of two limits: one with $y$ approaching $x$ from the future and the other with $y$ approaching $x$ from the past. The vacuum expectation value can be neatly expressed as follows:
\beq \label{pt7}
<j^{\mu}(x)>=-e\overline{\lim}_{y\to x}\mathsf{tr}[\gamma^{\mu}\underbrace{<T\psi(x)\bar{\psi}(y)>}_{-iG(x,y)}],
\eeq
where $\mathsf{tr}$ denotes here the trace over the spinor indices and $G(x,y)$ is the Feynman Green's function, which solves the equation:
\beq \label{pt8}
(\slashed{D}+m)G(x,y)=\delta(x-y).
\eeq
From the last section it is clear that the vacuum expectation value (\ref{pt7}) can be expressed as a functional derivative of the one-loop effective action:
\beq \label{pt9}
<j^{\mu}>=\frac{\delta \Gamma^{1loop}}{\delta A_{\mu}}.
\eeq
It has already been  demonstrated in the last section that the CS term \emph{is} present in the effective action of the massive $QED_{3}$. At the same time it \emph{cannot} be obtained from (\ref{pt1}) for the constant background field. To deduce the presence of the CS term in the effective action we note that the part of $<j^{\mu}>$, corresponding to the CS term, is non-vanishing. Thus, functionally integrating (\ref{pt9}) we shall get the CS part of the effective action. On the other hand, the Maxwell part of  $\Gamma^{1loop}[A]$ cannot be obtained from (\ref{pt9}), because the functional derivative (\ref{pt9}) of the Maxwell part vanishes for the constant field. To summarize, in order to get both parts of the $\Gamma^{1loop}[A]$ we must calculate \emph{both} (\ref{pt1}) and (\ref{pt9}) for the constant electromagnetic field. To that end it is convenient to introduce the operator notation at this stage. The Green's operator is defined by its matrix elements in x-representation:
\beq \label{pt10}
G(x,y)=(x|G|y)=(x|\frac{1}{\slashed{D}+m}|y) \qquad (x|y)=\delta^{(3)}(x-y),
\eeq
where $m$ has infinitesimal negative imaginary part. We can exploit this fact by writing $G$ in the elegant integral form:
\beq \label{pt11}
G=i\int^{\infty}_{0}ds \exp\left[-is(-\slashed{D}\slashed{D}+m^{2})\right](-\slashed{D}+m).
\eeq
Thus, substituting the last relation into (\ref{pt7}) we obtain a useful expression for $<j^{\mu}>$:
\beq \label{pt11a}
<j^{\mu}(x)>=-e\overline{\lim}_{y\to x}\int^{\infty}_{0}ds(x|\mathsf{tr}\{\gamma^{\mu}\exp\left[-is(-\slashed{D}\slashed{D}+m^{2})\right](-\slashed{D}+m)\}|y).
\eeq
As far as $\Gamma^{1loop}[A]$ is concerned it is advantageous to express it as follows:
\begin{eqnarray} 
\Gamma^{1loop}[A]&=&i\int d^{3}x \mathsf{tr} \ln(\slashed{D}(x)+m)=i\int d^{3}x \mathsf{tr} \ln(-\slashed{D}(x)+m) \nonumber \\
&=&\frac{i}{2}\int d^{3}x \mathsf{tr} \ln(-\slashed{D}(x)\slashed{D}(x)+m^{2})=\int d^{3}x \mathscr{L}_{eff}(x), \label{pt11b}
\end{eqnarray}
where the second equality holds because  $\slashed{D}(x)=-\slashed{D}(-x)$ for the field (\ref{pt3}). It is useful now to write $\mathscr{L}_{eff}$ in the integral form:
\beq \label{pt12}
\mathscr{L}_{eff}(x)=\frac{i}{2}\mathsf{tr}(x|\int^{\infty}_{0}\frac{ds}{s}\left[\exp(-is[-\slashed{D}\slashed{D}+m^2])-\exp(-is[-\Box+m^2])\right]|x),
\eeq
which can be easily verified. We introduced the second term in the integrand of the last relation to make $\mathscr{L}_{eff}$ vanish for zero electromagnetic background field.

We are approaching the main point of the Schwinger's method. Note that both (\ref{pt11a}) and (\ref{pt12}) contain in the integrands a nontrivial factor of the form $(x|\exp(i\slashed{D}\slashed{D}s)|y)$. If we now identify $\mathscr{H}=-\slashed{D}\slashed{D}$ with a one-particle Hamiltonian operator and $s$ with the "time", the factor is expressed as:
\beq \label{pt13}
(x|\exp(i\slashed{D}\slashed{D}s)|y)=(x|U(s)|y)\equiv(x,s|y,0) \qquad U(s)=\exp(-i\mathscr{H}s).
\eeq
In our new language (\ref{pt11a}) and (\ref{pt12}) can be expressed as:
\begin{eqnarray}
<j^{\mu}(x)>&=&-e\mathsf{tr} \gamma^{\mu}\int^{\infty}_{0}ds \exp(-im^{2}s)\overline{\lim}_{x\to y}(x,s|(-\slashed{D}(0)+m)|y,0) \label{pt14} \\
\mathscr{L}_{eff}(x)&=&\frac{i}{2}\int^{\infty}_{0}\frac{ds}{s}\exp(-im^2s)\left[\lim_{x\to y} \mathsf{tr}(x,s|y,0)-\frac{1}{4\pi^{3/2}}\frac{e^{3i\pi/s}}{s^{3/2}} \right], \label{pt15}
\end{eqnarray}
where we calculated the second term in the integrand of (\ref{pt12}) using $p$-representation and and analytical continuation. It is evident from the last two equations that it is sufficient to find both $(x,s|y,0)$ and $(x,s|D_{\mu}(0)|y,0)$ and then to substitute them into (\ref{pt14}) and (\ref{pt15}).
Exploiting the properties of $\gamma$-matrices we are able to simplify the Hamiltonian $\mathscr{H}$:
\beq \label{pt16}
\mathscr{H}=-\slashed{D}\slashed{D}=-D_{\mu}D^{\mu}-\frac{e}{2}F_{\mu\nu}\sigma^{\mu\nu} \qquad \sigma_{\mu\nu}=\frac{i}{2}[\gamma_{\mu},\gamma_{\nu}].
\eeq
The Heisenberg equations of motion for the operators $x^{\mu}$ and $D^{\mu}$ are:
\begin{eqnarray}
\frac{dx^{\mu}}{ds}&=&i[\mathscr{H},x^{\mu}]=-2D^{\mu} \label{pt17} \\
\frac{dD_{\mu}}{ds}&=&i[\mathscr{H},D_{\mu}]=-2eF_{\mu\nu}D^{\nu}-ie\frac{\partial F_{\nu\mu}}{\partial x_{\nu}}+\frac{e}{2}\sigma_{\rho\nu}\frac{\partial F^{\rho\nu}}{\partial x^{\mu}} \label{pt18}
\end{eqnarray}
for the general background field. For the constant field only the first term on the RHS of (\ref{pt18}) is present.

The matrix element $(x,s|y,0)$ solves the following partial differential equation:
\begin{eqnarray}
i\partial_{s}(x,s|y,0)&=&(x|\mathscr{H}U(s)|y)=(x|U(s)U^{\dagger}(s)\mathscr{H}U(s)|y)= \nonumber \\
&=&(x,s|\mathscr{H}(x(s),D(s))|y,0) \label{pt19}
\end{eqnarray}
with the boundary condition:
\beq \label{pt20}
(x,s|y,0)\to_{s\to 0}\delta(x-y).
\eeq
If we are lucky enough we can solve the Heisenberg equations (\ref{pt17}) and (\ref{pt18}) and express the Hamiltonian as a function of appropriately ordered $x(s)$ and $y(0)$. The equation (\ref{pt19}) then takes the form:
\beq \label{pt21}
i\partial_{s}(x,s|y,0)=f(x,y,s)(x,s|y,0),
\eeq
where $f(x,y,s)$ is a c-function. Thus the equation can be readily integrated:
\beq \label{pt22}
(x,s|y,0)=C(x,y)\exp(-i\int_{0}^{s}ds^{\prime} f(x,y,s^{\prime})).
\eeq
The s-independent function $C(x,y)$ can be found by solving the following partial differential equations:       
\begin{eqnarray}
(-i\partial_{\mu}^{x}-eA_{\mu}(x))(x,s|y,0)&=&(x,s|D_{\mu}(s)|y,0) \label{pt23} \\
(i\partial_{\mu}^{y}-eA_{\mu}(y))(x,s|y,0)&=&(x,s|D_{\mu}(0)|y,0) \label{pt24}
\end{eqnarray}
the second equation (\ref{pt24}) follows form the first (\ref{pt23}) by complex conjugation.

Let us apply the aforementioned general procedure to the special case of the  \emph{constant} background electromagnetic field. The operator equations of motion (\ref{pt17}) and (\ref{pt18}) simplify considerably in this case:
\begin{eqnarray}
\frac{dx^{\mu}}{ds}&=&i[\mathscr{H},x^{\mu}]=-2D^{\mu} \label{pt25} \\
\frac{dD_{\mu}}{ds}&=&i[\mathscr{H},D_{\mu}]=-2eF_{\mu\nu}D^{\nu} \label{pt26}
\end{eqnarray}
The solutions of these equations are:
\begin{eqnarray}
D(s)&=&\exp(-2esF)D(0) \label{pt27} \\
x(s)-y(0)&=&\frac{\exp(-2esF)-1}{eF}D(0), \label{pt28}
\end{eqnarray}
where the matrix notation is used and we do not write explicitly the spacetime indices. Solving the last equation for $D(0)$ and substituting it in (\ref{pt27}) we get:
\begin{eqnarray}
D(s)&=&-\frac{e}{2}F\exp(-esF)\sinh^{-1}(esF)(x(s)-y(0)) \nonumber \\
&=&-\frac{e}{2}F(x(s)-y(0))\exp(esF)\sinh^{-1}(esF), \label{pt29}
\end{eqnarray}
where the second equality holds because of antisymmetry property $F_{\mu\nu}x^{\nu}=-x^{\nu}F_{\nu\mu}$. We are now ready to express the Hamiltonian $\mathscr{H}(s)$ in terms of $x(s)$ and $y(0)$:
\beq \label{pt30}
\mathscr{H}(s)+\frac{e}{2}\sigma F=-D(s)D(s)=-(x(s)-y(0))K(x(s)-y(0)), 
\eeq
with $K$ given by
\beq \label{pt31}
K=\frac{e^{2}}{4}F^{2}\sinh^{-2}(esF).
\eeq
The Hamiltonian in (\ref{pt30}) is not ordered correctly. We need to move all $x(s)$ to the left and $y(0)$ to the right. The correct form is achieved exploiting the commutation relation:
\beq \label{pt32}
[y_{\mu}(0),x_{\nu}(s)]=i\left(\frac{\exp(-2esF)-1}{eF} \right)_{\nu\mu},
\eeq
whence the Hamiltonian looks:
\beq \label{pt33}
\mathscr{H}(s)+\frac{e}{2}\sigma F=-x(s)Kx(s)-y(0)Ky(0)+2x(s)Ky(0)-\frac{i}{2}\mathsf{tr}\left[eF \coth(esF) \right].
\eeq
The Schr\"odinger equation (\ref{pt21}) now takes the form:
\beq \label{pt34}
i\partial_{s}(x,s|y,0)=\left[-\frac{e}{2}\sigma F-xKx-yKy+2xKy-\frac{i}{2}\mathsf{tr}\left[eF \coth(esF)\right]\right](x,s|y,0),
\eeq
which can be easily integrated with the outcome:
\beq \label{pt35}
(x,s|y,0)=C(x,y)s^{-3/2}\exp\left(-\frac{1}{2}L(s) \right)\exp\left[ -\frac{ie}{4}(x-y)F\coth(esF)(x-y) \right] \exp\left\{ \frac{ie}{2}s \sigma F \right\},
\eeq
where $L(s)$ is defined by
\beq \label{pt36}
L(s)=\mathsf{tr}\left[\ln\left\{(esF)^{-1}\sinh(esF) \right\} \right].
\eeq
It remains to fix the s-independent function $C(x,y)$, which we shall achieve using (\ref{pt23}) and (\ref{pt24}). First we substitute our solution (\ref{pt35}) into (\ref{pt23}) and (\ref{pt24}) to get the differential equations for $C(x,y)$:
\begin{eqnarray}
(-i\partial_{\mu}^{x}-eA_{\mu}(x)-\frac{e}{2}F_{\mu\nu}[x-y]^{\nu})C(x,y)&=&0 \label{pt37} \\
(i\partial_{\mu}^{y}-eA_{\mu}(y)-\frac{e}{2}F_{\mu\nu}[x-y]^{\nu})C(x,y)&=&0 \label{pt38}
\end{eqnarray}
The solution of the first equation (\ref{pt37}) is expressed as a contour integral:
\beq \label{pt39}
C(x,y)=\widetilde{C}(y)\exp\left(i\int^{x}_{y}dz^{\mu}e\{A_{\mu}(z)+\frac{1}{2}F_{\mu\nu}(z-y)^{\nu} \} \right)
\eeq 
with $\widetilde{C}(y)$ yet unspecified function. The integral under the exponential is path-independent because the integral has a vanishing rotation. The path of integration in the contour integral, thus, can be taken arbitrary and we use it choosing the path as a straight line between $y$ and $x$. Due to the antisymmetry of the constant field strength $F_{\mu\nu}$ we end up with:
\beq \label{pt40}
C(x,y)=\widetilde{C}(y)\exp\left(i\int^{x}_{y}dz^{\mu}eA_{\mu}(z) \right).
\eeq
Following exactly the same lines we can solve the second equation (\ref{pt38}) with the result:
\beq \label{pt41}
C(x,y)=\widetilde{C}(x)\exp\left(i\int^{x}_{y}dz^{\mu}eA_{\mu}(z) \right).
\eeq
Two solutions must be multiplicatively equivalent, so we deduce that $\widetilde{C}(x)=\widetilde{C}(y)=C$:
\beq \label{pt42}
C(x,y)=C\exp\left(i\int^{x}_{y}dz^{\mu}eA_{\mu}(z) \right).
\eeq
To fix the constant $C$ we exploit the boundary condition (\ref{pt20}). In the limit $s\to0$ we integrate (\ref{pt35}) over the variable $x$ and obtain:
\beq \label{pt43}
C=\frac{e^{3i\pi/4}}{(4\pi)^{3/2}}.
\eeq
It is now easy to calculate $(x,s|D_{\mu}(0)|y,0)$ using $(\ref{pt28})$:
\beq \label{pt44}
(x,s|D_{\mu}(0)|y,0)=-\left[\frac{eF}{2}\left(\coth(esF)+1\right) \right]_{\mu\nu}(x-y)^{\nu}(x,s|y,0).
\eeq
Finally, we have calculated both $(x,s|y,0)$ and $(x,s|D_{\mu}(0)|y,0)$, which are given by formulae (\ref{pt35}), 	(\ref{pt36}), (\ref{pt42}), (\ref{pt43}) and (\ref{pt44}). We are ready to substitute our results into (\ref{pt14}) and (\ref{pt15}) to calculate $<j^{\mu}(x)>$ and $\mathscr{L}_{eff}(x)$:
\begin{eqnarray}
<j^{\mu}(x)>&=&-emC\int\frac{ds}{s^{5/2}}e^{-im^{2}s}e^{-L(s)}\mathsf{tr}\left[ \gamma^{\mu}\exp(\frac{ies}{2}\sigma F) \right] \label{pt45} \\
\mathscr{L}_{eff}(x)&=&iC\int^{\infty}_{0}\frac{ds}{s^{5/2}}e^{-im^{2}s}\left[\frac{1}{2}e^{-L(s)}\mathsf{tr}\left\{\exp(\frac{ies}{2}\sigma F) \right\}-1 \right]. \label{pt46}
\end{eqnarray}
The traces in (\ref{pt45}) and (\ref{pt46}) can be calculated using the properties of the $\gamma$-matrices:
\begin{eqnarray}
\mathsf{tr}\exp(\frac{ies}{2}\sigma_{\mu\nu}F^{\mu\nu})=2\cos(es|^{*}F|) \label{pt47} \\
\mathsf{tr}\exp(\gamma^{\rho}\frac{ies}{2}\sigma_{\mu\nu}F^{\mu\nu})=2i\sin(es|^{*}F|)\frac{^{*}F^{\rho}}{|^{*}F|}, \label{pt48}
\end{eqnarray}
where $|^{*}F|^{2}=^{*}F_{\mu}^{\ *}F^{\mu}$. Last but not least, it remains to compute  $e^{-L(s)}$:
\beq \label{pt49}
e^{-L(s)}=\exp\left( -\frac{1}{2}\mathsf{tr}\left[\ln\left\{\frac{\sinh(esF)}{esF} \right\} \right]\right)=\det\left(\left\{\frac{\sinh(esF)}{esF} \right\}^{-1/2} \right).
\eeq
Equivalently the determinant can be written as:
\beq \label{pt50}
e^{-L(s)}=\left[\frac{\sinh(es\lambda_{1})}{es\lambda_{1}}\frac{\sinh(es\lambda_{2})}{es\lambda_{2}}\frac{\sinh(es\lambda_{3})}{es\lambda_{3}}\right]^{-1/2},
\eeq
where $\lambda_{i}$ are the eigenvalues of the constant field strength matrix, i.e., solutions of:
\beq \label{pt51}
F_{\mu}^{\ \nu}\psi^{(i)}_{\nu}=\lambda_{i}\psi^{(i)}_{\mu},
\eeq
where $\psi^{(i)}_{\mu}$ are corresponding eigenvectors. The eigenvalues $\lambda_{i}$ are found with the help of the following trick, which works in three dimensions. Consider the third power of the eigenvalue equation:
\beq \label{pt52}
\underbrace{F_{\mu}^{\ \rho}F_{\rho}^{\ \sigma}}_{-|^{*}F|^{2}\delta^{\mu}_{\nu}}\underbrace{F_{\sigma}^{\ \nu}\psi^{(i)}_{\nu}}_{\lambda_{i}\psi^{(i)}_{\sigma}}=\lambda_{i}^{\ 3}\psi^{(i)}_{\mu},
\eeq
whence
\beq \label{pt53}
\lambda_{i}^{\ 3}+|^{*}F|^{2}\lambda_{i}=0,
\eeq
with the solutions $\lambda_{i}=\{0,\pm i|^{*}F| \}$. Substitution of $\lambda_{i}$ into (\ref{pt50}) gives us:
\beq \label{pt52a}
e^{-L(s)}=\frac{es|^{*}F|}{\sin(es|^{*}F|)}.
\eeq
The last step in our quite lengthy calculations is to perform $s$-integration in (\ref{pt45}) and (\ref{pt46}). The current $<j^{\mu}(x)>$ is given by:
\beq \label{pt53a}
<j^{\mu}(x)>=-2ie^{2}mC\int^{\infty}_{0}\frac{ds}{s^{1/2}}e^{-im^{2}s}\mbox{}^{\ *}F^{\mu}=\frac{e^{2}}{4\pi}\frac{m}{|m|}\mbox{}^{*}F^{\mu}.
\eeq
The functional integration of the equation (\ref{pt9}) yields:
\beq \label{pt54}
\Gamma^{1loop}_{CS}=\frac{e^{2}}{8\pi}\frac{m}{|m|}\int d^{3}x\epsilon_{\mu\nu\rho} A^{\mu}\partial^{\nu}A^{\rho}.
\eeq
We readily recognize the \emph{CS term} with the photon parameter $\mu=\frac{e^{2}}{4\pi}\frac{m}{|m|}$. The result \emph{coincides} with the perturbative derivation (\ref{pea14}) from the last section.

Substituting (\ref{pt47}) and (\ref{pt52a}) into (\ref{pt46}) we get the integral expression for the $\mathscr{L}_{eff}$:
\begin{eqnarray} \label{pt55}
\mathscr{L}_{eff}&=&iC\int^{\infty}_{0}\frac{ds}{s^{5/2}}e^{-im^{2}s}\left[es|^{*}F|\cot(es|^{*}F|)-1 \right]= \nonumber \\
&=&\frac{1}{8\pi^{3/2}}\int^{\infty}_{0}\frac{ds}{s^{5/2}}e^{-m^{2}s}\left[es|^{*}F|\coth(es|^{*}F|)-1 \right],
\end{eqnarray}
where we made a substitution $s\to-is$ and used $\cot(-ix)=i\coth(x)$ to get the second equality. The computation of the above integral is a bit more involved than the previous one. First we introduce a substitution $l=es|^{*}F|$ and use the following series expansion:
\beq \label{pt56}
l\coth(l)-1=2\sum_{n=1}^{\infty}\frac{l^{2}}{l^{2}+(\pi n)^{2}}.
\eeq
The Maxwell part of the effective Lagrangian $\mathscr{L}_{eff}$ takes the form:
\beq \label{pt56a}
\mathscr{L}_{eff}=\left( \frac{e|^{*}F|}{4\pi} \right)^{3/2}\int^{\infty}_{0} dl \sum_{n=1}^{\infty}\frac{e^{-\alpha l}}{\sqrt{l}[l^{2}+(\pi n)^{2}]},
\eeq
where $\alpha=\frac{m^{2}}{e|^{*}F|}$. Assuming we can interchange the sum and the integral, we first calculate the integrals $\mathscr{I}_{n}$:
\beq \label{pt57}
\mathscr{I}_{n}=\int^{\infty}_{0}dl\frac{e^{-\alpha l}}{\sqrt{l}(l^{2}+(\pi n)^{2})}=\frac{1+i}{2\sqrt{\pi i}n^{3/2}}\left\{\cos(\alpha \pi n)-\sin(\alpha\pi n)\right\},
\eeq
where initially a substitution $r=\sqrt{l}$ was performed and then the residue theorem was employed. We are able to sum the series in the special case $\alpha=0$, i.e., in the limit of zero fermion mass $m=0$:
\beq \label{pt58}
\mathscr{L}_{eff}=\frac{1}{2\pi^{2}}\left(\frac{e|^{*}F|}{2} \right)^{3/2}\underbrace{\sum^{\infty}_{n=1}\frac{1}{n^{3/2}}}_{\zeta(3/2)},
\eeq
where $\zeta(x)$ is the Riemann zeta function. Recall that in three dimensions $3|^{*}F|^{2}=F_{\mu\nu}F^{\mu\nu}$, whence we have just calculated the Maxwell part of the effective Lagrangian. For the slowly varying background field the Maxwell part of the effective action $\Gamma^{1loop}_{M}[A]$ for the massless fermions is given by:
\beq \label{pt59}
\Gamma^{1loop}_{M}[A]=\frac{1}{2\pi^{2}}\zeta(3/2)\int d^{3}x \left( \frac{e|^{*}F|}{2}\right)^{3/2}.
\eeq
Let us stress that the dependence on the coupling constant $e$ is non-polynomial, hence the result is non-perturbative.
The complete effective action in the limit of massless fermions, obtained by the Schwinger "proper time" method is finally:
\beq \label{pt60}
\Gamma^{1loop}[A]=\frac{1}{2\pi^{2}}\zeta(3/2)\int d^{3}x \left( \frac{e|^{*}F|}{2}\right)^{3/2}\pm\frac{e^{2}}{8\pi}\int d^{3}x\epsilon_{\mu\nu\rho} A^{\mu}\partial^{\nu}A^{\rho}.
\eeq

\chapter{Non-perturbative Maxwell-Chern-Simons $QED_{3}$}
\label{ch:non-perturbative}
In this Chapter we consider some non-perturbative phenomena in the Maxwell-Chern-Simons $QED_{3}$. In particular we are interested in non-relativistic electron-electron interactions. It was shown in Chapter \ref{ch:classical} that classically a static point charge generates in its vicinity not only the electric field, but also a magnetic tube of width $\mu$ and strength $e/\mu$, where $\mu$ is a photon mass parameter. Let us now consider a system of two interacting electrons. In accordance with the Lorentz force law (\ref{eq:CLE0a}) the two electrons feel not only the Coulomb repulsion, but also the magnetic interaction, which might be attractive. So classically an interesting theoretical possibility of the \emph{electron-electron bound states} arises. Inspired by this classical motivation we shall investigate the problem of the electron-electron effective interaction first by using the quantum-mechanical Pauli equation (\ref{RQM6}). Then we shall examine the problem field-theoretically using the Breit approach. Having the effective potential at hand we shall try to find the electron-electron bound states.   
\section{Pauli equation} \label{sec:Pauli}
The non-relativistic electron-electron interaction in the MCS theory was first addressed by Kogan in \cite{Kogan:1989}. The two-body problem may be transformed into the problem of a particle with the reduced mass $\frac{m}{2}$ in an external electromagnetic field. Pauli Hamiltonian (\ref{RQM6}) takes the form (here we use $\hbar=c=1$ ):
\beq \label{NP1}
H=e\Phi+\frac{1}{m}(i\partial_{i}-e A_{i})^{2}+\frac{e}{m}B.
\eeq
We derived the electromagnetic field of a static point charge in the MCS theory in Chapter \ref{ch:classical} and for the positive photon parameter it reads:
\begin{eqnarray}
\Phi(\mathbf{x})=\frac{e}{2\pi}K_{0}(\mu|\mathbf{x}|) & & A^{i}(\mathbf{x})=\frac{e}{2\pi}\frac{1}{\mu}\frac{e^{ij}x^{j}}{|\mathbf{x}|}\left( \frac{1}{|\mathbf{x}|}-\mu K_{1}(\mu|\mathbf{x}|) \right) \nonumber \\
E^{i}(x)=\frac{e}{2\pi}\frac{\mu x^{i}}{|\mathbf{x}|}K_{1}(\mu |\mathbf{x}|) & & B(x)=\frac{e}{2\pi}\mu K_{0}(\mu |\mathbf{x}|)=\mu \Phi(\mathbf{x}). \label{NP2}
\end{eqnarray}
We also demonstrated in section \ref{CMCS} that under the transformation $\mu\to-\mu$ both $B(\mathbf{x})$ and $A^{i}(\mathbf{x})$ change their signs while $\Phi(\mathbf{x})$ and $E^{i}(\mathbf{x})$ remain unchanged. As will be obvious further in order to get a possible attraction we must take the negative photon parameter. In this Chapter we investigate the MCS theory with the negative photon parameter and by $\mu$ we denote the absolute value of the photon parameter, thus $\mu>0$. Using the expression (\ref{NP2}) and the last remark we can rewrite the Hamiltonian (\ref{NP1}) as:
\beq \label{NP3}
H=e(1-\frac{\mu}{m})\Phi+\frac{1}{m}(i\partial_{i}+e A_{i})^{2},
\eeq
where $\Phi$ and $A_{i}$ are given in (\ref{NP2}).
Let us rearrange the second term of (\ref{NP3}):
\beq \label{NP4} (i\partial_{i}-eA_{i})^{2}=-\nabla^{2}+e^{2}A_{i}A_{i}+2ieA_{i}\partial_{i}+i[\partial_{i},eA_{i}].
\eeq
The axial structure of $A^{i}(\mathbf{x})$ in (\ref{NP2}) implies that the commutator in the last expression vanishes. The Laplacian in polar coordinates $(r,\theta)$ is:
\beq \label{NP5}
\nabla^{2}=\frac{\partial^{2}}{\partial r^{2}}+\frac{1}{r}\frac{\partial}{\partial r}+\frac{1}{r^{2}}\frac{\partial^{2}}{\partial \theta^{2}}
\eeq
and the second term in (\ref{NP4}) is:
\beq \label{NP6}
e^{2}A_{i}A_{i}=e^{2}r^{2}f^{2}(r) \qquad f(r)=-\frac{e}{2\pi}\frac{1}{r^{2}\mu}\left(1-\mu r K_{1}(\mu r)\right).
\eeq
Finally the third term in (\ref{NP4}) may be rewritten in polar coordinates as:
\beq \label{NP7}
2ieA_{i}\partial_{i}=-2ief(r)\partial_{\theta}.
\eeq
Combining all these terms together we get the Hamiltonian in polar coordinates:
\beq \label{NP8}
H=e(1-\frac{\mu}{m})\Phi+\frac{1}{m}\left[-\frac{\partial^{2}}{\partial r^{2}}-\frac{1}{r}\frac{\partial}{\partial r}-\frac{1}{r^{2}}\frac{\partial^{2}}{\partial \theta^{2}}+e^{2}r^{2}f^{2}(r)-2ief(r)\partial_{\theta}\right]
\eeq
or written in the general X-representation:
\beq \label{NP8a}
\hat{H}=-\frac{\nabla^2{}}{m}+e(1-\frac{\mu}{m})\Phi+\frac{1}{m}\left[e^{2}r^{2}f^{2}(r)+2ef(r)\hat{l}\right],
\eeq
where $\hat{l}$ is the angular momentum operator corresponding to the relative motion.

It is straightforward to show that $[\hat{H}, \hat{l}]=0$. That is why we try to get the solutions of the stationary Schr\"odinger equation $H\psi_{E}=E\psi_{E}$ as the eigenfunctions of the orbital angular momentum with eigenvalues $l$:
\beq \label{NP9}
\psi_{E,l}(r,\theta)=\psi_{l,E}(r)\exp(il\theta).
\eeq
After inserting this form into the Schr\"odinger equation we obtain the $l$-dependent effective Hamiltonian for $\psi_{E,l}(r)$:
\begin{eqnarray}
H_{l}=\underbrace{-\frac{1}{mr}\frac{\partial}{\partial r}\left(r\frac{\partial}{\partial r}\right)}_{kinetic}+  \underbrace{\left(1-\frac{\mu}{m}\right)e\Phi}_{electromagnetic}+\underbrace{\frac{1}{mr^{2}}l^{2}(r)}_{centrifugal} \nonumber \\
l(r)=l-\frac{e^{2}}{2\pi \mu}(1-\mu r K_{1}(\mu r)) \qquad
\Phi(r)=\frac{e}{2\pi}K_{0}(\mu r).  \label{NP10}
\end{eqnarray}
Every term in (\ref{NP10}) has a clear physical interpretation: the first one is simply the kinetic term. The second term comes from the electrostatic and Pauli interactions. For $\mu<m$ the Coulomb repulsion dominates over the magnetic attraction. For $\mu>m$, however, the second term \emph{changes its sign} and this \emph{opens a possibility for the existence of bound states}. We must be careful, however, because the Pauli equation was derived under the assumption that $eB<<m^{2}$ ((\ref{RQM6})). It is not clear if the given assumption is satisfied in our case. The third term is a well-known centrifugal barrier, which is, however, $r$-dependent in our case. For $r\to \infty$ we get the effective angular momentum quantum number $l_{eff}=l-\frac{e^{2}}{2\pi \mu}$. The same result may be obtained for arbitrary $r$ in the so called \emph{anyon limit} \cite{Dobroliubov:1993}, that is $e^{2}\to \infty$, $\mu\to \infty$ with $k=\frac{2\pi\mu}{e^{2}}$ fixed. In this special limit the first term in (\ref{NP3}) becomes an attractive delta function $-\frac{2\pi}{km}\delta^{2}(\mathbf{r})$ and the effective angular momentum quantum number shifts by a real constant $l\to l_{eff}=l-\frac{1}{k}$. Thus the effective Hamiltonian (\ref{NP10}) takes the form:
\beq \label{NP11}
H_{l}=-\frac{1}{mr}\frac{\partial}{\partial r}\left(r\frac{\partial}{\partial r}\right)-\frac{1}{km}\frac{1}{r}\delta(r)+\frac{1}{mr^{2}}\left(l-\frac{1}{k}\right)^{2}.
\eeq 
As was shown in Chapter \ref{ch:Poincare} the existence of massive particles with an arbitrary spin is acceptable in the plane. While the canonical angular momentum $l$ in (\ref{NP11}) must be integer (in the theory of bosons) or half-integer (in the theory of fermions), the kinetic angular momentum $l-\frac{1}{k}$ may take non-integral values. It was shown in \cite{Goldhaber:1988} that exactly the kinetic angular momentum may be identified as the angular moment of the composite particle. Here we have just encountered the quantum mechanical Chern-Simons realization of anyons \cite{Wilczek:1990}.   

\section{Breit equation} \label{sec:Breit}
\begin{figure}[!htb]
\begin{center}
\begin{tabular}{ccccccccc}    

\begin{fmffile}{t-channel} 	
   \fmfframe(1,12)(1,12){
   \begin{fmfgraph*}(30,25) 
   \fmfleft{l1,l2}
   \fmfright{r1,r2}
   \fmf{fermion}{l1,v1,r1}
   \fmf{fermion}{l2,v2,r2}
   \fmf{photon}{v1,v2}
   \fmflabel{$p^{\prime}$}{r1}
   \fmflabel{$k^{\prime}$}{r2}
   \fmflabel{$p$}{l1}
   \fmflabel{$k$}{l2}   
\end{fmfgraph*}
  }
\end{fmffile}

\begin{fmffile}{u-channel} 	
   \fmfframe(1,12)(1,12){
   \begin{fmfgraph*}(30,25) 
   \fmfleft{l1,l2}
   \fmfright{r1,r2}
   \fmf{fermion}{l1,v1,r1}
   \fmf{fermion}{l2,v2,r2}
   \fmf{photon}{v1,v2}
   \fmflabel{$k^{\prime}$}{r1}
   \fmflabel{$p^{\prime}$}{r2}
   \fmflabel{$p$}{l1}
   \fmflabel{$k$}{l2}
\end{fmfgraph*}
  }
\end{fmffile}

\end{tabular}
\caption{Tree-level ee $\to$ ee interaction: t-channel and u-channel diagrams}\label{graph8}
\end{center}
\end{figure}
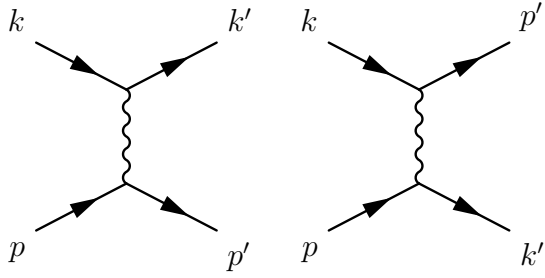

In this section we rederive the effective Hamiltonian (\ref{NP8a}) by means of the quantum field theory. We shall start with the relativistic amplitude $M_{fi}$ of the process $e+e\to e+e$ (M\o ller scattering). Tree-level contributions are depicted in Fig. \ref{graph8} and $M_{fi}$ is given by:
\beq \label{BE1}
M_{fi}=M_{fi}^{t}-M_{fi}^{u},
\eeq   
where the minus sign is due to the Fermi statistics.
\unitlength = 1mm

In the non-relativistic approximation the amplitude $M_{fi}$ is simply the first Born amplitude, which is proportional to the Fourier transform of the effective interaction. So let us first calculate $M_{fi}^{t}$:
\beq \label{BE2}
iM_{fi}^{t}=-e^{2}\bar{u}(p^{\prime})\gamma^{\mu}u(p)\bar{u}(k^{\prime})\gamma^{\nu}u(k)\mathscr{D}_{\mu\nu}(q) \qquad q=p-p^{\prime},
\eeq
where $\mathscr{D}_{\mu\nu}(q)$ is a free photon propagator in covariant gauge \footnote{Note that the axial part of the propagator has opposite sign to (\ref{VP1}). If we limit ourselves to $\mu>0$ it describes the massive photon with mass $\mu$ and spin $s=-1$.}:
\beq \label{BE3}
\mathscr{D}_{\mu\nu}(q)=\frac{-i}{q^{2}-\mu^{2}}[P_{\mu\nu}(q)+i\mu\epsilon_{\mu\nu\rho}q^{\rho}/q^{2}]-\frac{i\alpha q_{\mu}q_{\nu}}{q^{4}}.
\eeq
The terms in the propagator which are proportional to $q^{\mu} q^{\nu}$ will not contribute to the amplitude because they are contracted with the conserved currents. The Dirac spinors $u(p)$ in $Jackiw$ realization were calculated in (\ref{eq:DS6}) and we shall need the non-relativistic $O((\frac{\mathbf{p}}{m})^{2})$ approximation here:
\beq \label{BE4}
u(\mathbf{p})=\frac{1}{\sqrt{\epsilon_{\mathbf{p}}+m}}\left(\begin{array}{c} \epsilon_{\mathbf{p}}+m \\ ip_{1}-p_{2} \end{array} \right)\to\sqrt{2m}\left(\begin{array}{c} 1+\frac{\mathbf{p}^{2}}{8m^{2}} \\ \frac{ip_{1}-p_{2}}{2m} \end{array} \right).
\eeq
Let us also calculate the non-relativistic $O((\frac{\mathbf{p}}{m})^{2})$ approximation for the currents:
\begin{eqnarray}
\bar{u}(p^{\prime})\gamma^{0}u(p) & \to &  2m \left[ 1+\frac{\mathbf{p}^{2}}{2m^{2}}+\frac{\mathbf{q}^{2}}{8m^{2}}-\frac{\mathbf{p}\cdot \mathbf{q}}{2m^{2}}+\frac{i\epsilon_{ij}p_{i}q_{j}}{4m^{2}}  \right] \nonumber \\
\bar{u}(p^{\prime})\gamma^{i}u(p) & \to &  -2p_{i}+q_{i}-i\epsilon_{ij}q_{j} \nonumber \\
\bar{u}(p^{\prime})u(p) & \to &  2m \left[ 1+\frac{\mathbf{q}^{2}}{8m^{2}}-\frac{i\epsilon_{ij}p_{i}q_{j}}{4m^{2}}  \right], \label{BE5}
\end{eqnarray}
where $\mathbf{q}=\mathbf{p}-\mathbf{p^{\prime}}$. Now we can substitute (\ref{BE5}) into (\ref{BE2}) in order to calculate the non-relativistic approximation of the t-channel amplitude. The symmetric part of the propagator (\ref{BE3}) contributes by:
\beq \label{BE6}
iM_{fi}^{tI}=-\frac{4im^{2}e^{2}}{\mathbf{q}^{2}+\mu^{2}}\left( 1-\frac{\mathbf{p}\cdot\mathbf{k}}{m^{2}}+\frac{\mathbf{q}\cdot(\mathbf{k}-\mathbf{p})}{2m^{2}}-i\frac{\epsilon_{ij}k_{i}q_{j}}{2m^{2}}+i\frac{\epsilon_{ij}p_{i}q_{j}}{2m^{2}} \right),
\eeq
where we used $q^{2}-\mu^{2}=(p-p^{\prime})^{2}-\mu^{2}\approx -\mathbf{q}^{2}-\mu^{2}$.
The axial part of the photon propagator contributes to the amplitude by:
\beq \label{BE7}
iM_{fi}^{tII}=-\frac{4m^{2}e^{2}\mu}{\mathbf{q}^{2}(\mathbf{q}^{2}+\mu^{2})}\left(-i\frac{\mathbf{q}^{2}}{m}+\epsilon_{ij}\frac{(k-p)_{i}q_{j}}{m} \right).
\eeq
Having these results at hand we finally obtain the t-channel amplitude $M_{fi}^{t}$:
\beq \label{BE8}
M_{fi}^{t}=M_{fi}^{tI}+M_{fi}^{tII}=-\frac{4m^{2}e^{2}}{\mathbf{q}^{2}+\mu^{2}}\left( 1-\frac{\mu}{m} \right)-\frac{4ie^{2}\mu m}{\mathbf{q}^{2}(\mathbf{q}^{2}+\mu^{2})}\epsilon_{ij}(p_{i}-k_{i})q_{j},
\eeq
where only terms $O(\frac{\mathbf{p}}{m})$ were taken into account. Now we are able to get the effective interaction potential using the well-known Born formula:
\beq \label{BE9}
\widetilde{V}(q)=-\frac{M_{fi}^{t}}{4m^{2}},
\eeq
where $\widetilde{V}(q)$ is a Fourier transform of the non-relativistic potential. All we need to do now is to perform the inverse Fourier transformation. The first term on the RHS of (\ref{BE8}) transforms into:
\beq \label{BE10}
V^{I}(r)=\int \frac{d^{2}q}{(2\pi)^{2}}\widetilde{V}(q)e^{i\mathbf{q}\cdot\mathbf{r}}=e^{2}\left(1-\frac{\mu}{m}\right)\underbrace{\int \frac{d^{2}q}{(2\pi)^{2}}\frac{1}{\mathbf{q}^{2}+\mu^{2}}e^{i\mathbf{q}\cdot\mathbf{r}}}_{\frac{1}{2\pi}K_{0}(|\mu|r)}.
\eeq 
The second term on the RHS of (\ref{BE8}) is more tricky:
\beq \label{BE11}
V^{II}(r)=\frac{ie^{2}\mu}{m}\epsilon_{ij}(p_{i}-k_{i})\underbrace{\int\frac{d^{2}q}{(2\pi)^{2}}\frac{q_{j}}{\mathbf{q}^{2}(\mathbf{q}^{2}+\mu^{2})}e^{i\mathbf{q}\cdot \mathbf{r}}}_{I_{j}},
\eeq 
where we introduced $I_{j}$, which will be calculated as follows:
\begin{eqnarray} 
I_{j} &=& i\frac{\partial}{\partial r^{j}}\int\frac{d^{2}q}{(2\pi)^{2}}\frac{1}{\mathbf{q}^{2}(\mathbf{q}^{2}+\mu^{2})}e^{i\mathbf{q}\cdot \mathbf{r}}\nonumber \\
&=&i\frac{\partial}{\partial r^{j}}\left( -\frac{1}{2\pi \mu^{2}}{\ln(r)+K_{0}(\mu r)}  \right)=\frac{i}{2\pi \mu^{2}r^{2}}(1-r\mu K_{1}(\mu r))r_{j}, \label{BE12}
\end{eqnarray}
where we used the following relations:
\beq \label{BE13}
\int\frac{d^{2}q}{(2\pi)^{2}}\frac{1}{\mathbf{q}^{2}(\mathbf{q}^{2}+\mu^{2})}e^{i\mathbf{q}\cdot \mathbf{r}}=-\frac{1}{2\pi \mu^{2}}{\ln(r)+K_{0}(\mu r)} \qquad \frac{d}{dr}K_{0}(\mu r)=-\mu K_{1}(\mu r).
\eeq
Now we substitute (\ref{BE12}) into (\ref{BE11}) to obtain the final result:
\beq \label{BE14}
V^{II}(r)=-\frac{1}{mr^{2}}\frac{e^{2}}{2\pi\mu}(1-r\mu K_{1}(\mu r))\underbrace{\epsilon_{ij}(p_{i}-k_{i})r_{j}}_{2l},
\eeq
where $l$ is a relative angular momentum of two particles in their CMS. Thus the effective interaction potential is (\ref{BE10}), (\ref{BE14}):
\begin{eqnarray} 
V(r)&=&e^{2}(1-\frac{\mu}{m})\frac{1}{2\pi}K_{0}(|\mu|r)-\frac{1}{mr^{2}}\frac{e^{2}}{\pi\mu}\left[1-r\mu K_{1}(\mu r)\right]l= \nonumber \\
&=&\left(1-\frac{\mu}{m}\right)e\Phi+\frac{2ef(r)l}{m}. \label{BE15}
\end{eqnarray}
We may now compare the last formula with the Hamiltonian obtained via the Pauli equation (\ref{NP8a}). The electrostatic term and the $l$-dependent term are reproduced by the field-theoretic methods, but the first term in brackets in (\ref{NP8a}) \emph{is lacking}. It is no surprise for us because the given term is proportional to $e^{4}$ and \emph{can not be reproduced by tree level calculations}. It may be obtained only from higher order diagrams. The non-relativistic Hamiltonian (\ref{NP1}) is quadratic in momenta and this term is desperately needed to guarantee \emph{the gauge invariance} of the Hamiltonian \cite{Dobroliubov:1993}.

A remark is in order here. So far we have calculated only the t-channel diagram and thus we obtained the effective non-relativistic Hamiltonian (\ref{NP8a}) for two  distinguishable particles. To take into account that the fermions in the process are indistinguishable we should have also calculated the u-channel diagram. There is no need to do it, however. In non-relativistic quantum mechanics the interaction of the indistinguishable particles is given by the same operator as for distinguishable ones, but we must only antisymmetrize the wave function.

\section{Bound states}  
We have already shown that for $\frac{\mu}{m}>1$ the electromagnetic interaction in (\ref{NP10}) becomes attractive and this opens a possibility for the  existence of electron-electron bound states. This question was first examined by Kogan \cite{Kogan:1989} for the effective Hamiltonian (\ref{NP10}). It was shown that there are bound states for different values of $l$ in the strong coupling domain $\frac{e^{2}}{\mu}>>1$.

It is easy to show that for $l=1$ and $\frac{e^{2}}{\mu}=2\pi$ the centrifugal barrier disappears at large distances and the effective potential takes the form:
\beq \label{BS1}
V(r)=\frac{\mu^{2}}{m}\left[ K^{2}_{1}(\mu r)-\frac{\mu-m}{\mu}K_{0}(\mu r) \right].
\eeq
It was shown in \cite{Kogan:1989} that there is a bound state for this potential with the exponential binding energy:
\beq\label{BS2}
E\propto \frac{\mu^{2}}{m}\exp\left[-c(1-\frac{m}{\mu})^{-2}\right]
\eeq
for $\mu-m<<\mu$ and $c\propto 1$.

We should be careful at this point. The effective potential (\ref{NP10}) is only approximate. It was obtained using only tree and one-loop Feynman diagrams for electron-electron scattering. As we have already mentioned in Chapter 7 we can trust perturbation theory only if the dimensionless parameter $\frac{e^{2}}{\mu}<<1$. The bound states found in \cite{Kogan:1989} do not satisfy this condition. The question of the existence of the electron-electron bound states was studied numerically in \cite{Dobroliubov:1993} (slightly different effective potential was used there, however) using the WKB approximation. It was shown that no electron-electron bound states can be formed in the perturbative regime $\frac{e^{2}}{\mu}<<1$ in the non-relativistic limit.   
\chapter{One-loop calculations in the Chern-Simons theory}
\label{ch:CS}
As was mentioned at the end of the section \ref{sec:Pauli} the effective quantum mechanical Hamiltonian in the anyon limit looks simple and is given by (\ref{NP11}). In this section we show that the field theory, which describes the anyon limit is the pure Chern-Simons (CS) theory. We also calculate the scattering amplitude of two distinguishable fermions up to one loop in the non-relativistic (Pauli) version of this theory. Subsequently, one-loop calculation of the non-relativistic scattering is also performed in the fully relativistic Chern-Simons theory. The amplitudes, obtained by these two different approaches, differ by a term, which comes from the modification of the spin of the fermion \footnote{This is a celebrated spin transmutation, which occurs in the pure Chern-Simons theory.}.

\section{Basics of the Chern-Simons theory}
Let us take the MCS Lagrangian (\ref{$QED_{3}$}) (without the gauge-fixing term) and perform the anyon limit, i.e., take $e^{2}\to\infty$, $\mu\to \infty$ and hold $\frac{e^{2}}{2\pi \mu}$ fixed. To do so we first rescale the electromagnetic field:
\beq \label{CS1} \widetilde{A^{\mu}}\equiv\frac{e}{e_{CS}}A^{\mu},  \eeq
where a finite Chern-Simons charge $e_{CS}$ was defined. The Lagrangian (\ref{$QED_{3}$}) now looks:
\beq \label{CS2}
\mathscr{L}=-\frac{1}{4}\frac{e^{2}_{CS}}{e^{2}}\widetilde{F}^{\mu\nu}\widetilde{F}_{\mu\nu}+\frac{\mu e^{2}_{CS}}{2e^{2}}\epsilon^{\mu\nu\rho}\widetilde{A}_{\mu}\partial_{\nu}\widetilde{A}_{\rho}+\mathscr{L}_{\psi}-e_{CS} \bar{\psi}\gamma^{\mu}\psi\widetilde{A}_{\mu},
 \eeq
where $\mathscr{L}_{\psi}$ denotes the Dirac Lagrangian, which is unaffected by the rescaling. In the anyon limit the first Maxwell term in (\ref{CS2}) vanishes, while $\frac{\mu e^{2}_{CS}}{e^{2}}$ is finite and we shall call it simply $\mu_{CS}$. Thus we end up with the \emph{pure Chern-Simons Lagrangian} with finite parameters $e_{CS}$ and $\mu_{CS}$. From now on we shall drop subscript $_{CS}$ and write the rescaled electromagnetic field simply as $A^{\mu}$. Thus the Lagrangian is:
\beq \label{CS3}
\mathscr{L}_{CS}=\frac{\mu}{2}\epsilon^{\mu\nu\rho}A_{\mu}\partial_{\nu}A_{\rho}+\mathscr{L}_{\psi}-e\bar{\psi}\gamma^{\mu}\psi A_{\mu}.
\eeq
The Chern-Simons theory is interesting and intriguing in its own right \cite{Dunne:1998, Wilczek:1990}. To see this we first derive the classical field equations:
\beq \label{CS4}
\frac{\mu}{2}\epsilon^{\mu\nu\rho}F_{\nu\rho}=j^{\mu},
\eeq  
where $j^{\mu}$ is the electromagnetic matter current. The source-free solutions $F^{\mu\nu}=0$ are pure gauges $A^{\mu}(x)=\partial^{\mu}\omega(x)$, where $\omega(x)$ is an arbitrary scalar field. These solutions are \emph{not physical} and it follows that there are \emph{no} electromagnetic waves in the vacuum in this theory. Thus the CS field has no its own, independent dynamics. Its dynamics is governed by the matter field. To understand this statement better we rewrite the field equations (\ref{CS4}) in components:
\begin{eqnarray} 
\rho & = & \mu B \label{CS4a} \\
j^{i} & = & \mu \epsilon^{ij} E_{j}. \label{CS5}
\end{eqnarray}
For a point charge with $\rho(\mathbf{y},t)=e \delta(\mathbf{y}-\mathbf{x}(t))$ and $j^{i}(\mathbf{y},t)=eu^{i}(t)\delta(\mathbf{y}-\mathbf{x}(t))$, where $\mathbf{x}(t)$ is a trajectory of the point charge and $u^{i}(t)$ are spatial components of the 3-velocity vector, we obtain from the last expression:
\begin{eqnarray}
B(\mathbf{y},t) & = & \frac{e}{\mu} \delta(\mathbf{y}-\mathbf{x}(t)) \label{CS6} \\
E^{i}(\mathbf{y},t) & = & \frac{e}{\mu} \epsilon^{ij} u^{j}(t)\delta(\mathbf{y}-\mathbf{x}(t)).  \label{CS7}
\end{eqnarray}
Thus, the infinitely thin magnetic tube, which carries the finite magnetic flux $\frac{e}{\mu}$ , is attached to the point charge and the electric field follows exactly the motion of the charge. 

Let us find the electromagnetic potential $A^{\mu}(\mathbf{y},t)$ of the point source. The pure Chern-Simons theory is a gauge theory, thus the vector potential $A^{i}(\mathbf{y},t)$ is not a gauge-invariant quantity and we should fix the gauge first in order to get $A^{i}(\mathbf{y},t)$. Here we shall work in the Coulomb gauge ($\partial_{i}A^{i}=0$). From (\ref{CS4a}) and (\ref{CS6}) we get:
\beq \label{CS8}
e\delta(\mathbf{y}-\mathbf{x}(t))=\mu\left[\partial_{2}A^{1}(\mathbf{y},t)-\partial_{1}A^{2}(\mathbf{y},t)\right].
\eeq
Now we differentiate the last expression ($\frac{\partial}{\partial y^{i}}$) and use the Coulomb condition $\frac{\partial}{\partial y^{1}}A^{1}(\mathbf{y},t)+\frac{\partial}{\partial y^{2}}A^{2}(\mathbf{y},t)=0$ to get:
\beq \label{CS9}
e\epsilon_{ij}\frac{\partial}{\partial y^{j}}\delta(\mathbf{y}-\mathbf{x}(t))=\mu \nabla^{2}_{\mathbf{y}}A^{i}(\mathbf{y},t).
\eeq
Recalling that the Green's function of the two-dimensional Laplace equation is:
\beq \label{CS10}
\nabla^{2}_{\mathbf{y}}\left( \frac{1}{2\pi}\ln|\mathbf{y}-\mathbf{z}| \right)=\delta^{(2)}(\mathbf{y}-\mathbf{z})
\eeq
we are able to express the vector potential from (\ref{CS9}):
\beq \label{CS11}
A^{i}(\mathbf{y},t)=\frac{e}{2\pi \mu}\epsilon^{ij}\frac{y^{j}-x^{j}(t)}{|\mathbf{y}-\mathbf{x}(t)|^{2}}.
\eeq
It is important to note that this potential can be expressed as a \emph{gradient}. To see this we recall that $\partial_{i} arg(\mathbf{y})=-\epsilon_{ij}\frac{y^{j}}{|\mathbf{y}^{2}|}$, where the argument function is defined as $\arg(\mathbf{y})\equiv \arctan (\frac{y^{2}}{y^{1}})$. Hence (\ref{CS11}) can be written as:
\beq \label{CS12}
A_{i}(\mathbf{y},t)=\frac{e}{2\pi \mu}\frac{\partial}{\partial y^{i}} arg(\mathbf{y}-\mathbf{x}(t)).
\eeq
At first sight $A^{\mu}(\mathbf{y},t)$ represents a pure gauge, which can be transformed away using the following gauge transformation:
\begin{eqnarray} 
A_{i}(\mathbf{y},t) & \to & A_{i}(\mathbf{y},t)-\frac{\partial}{\partial y^{i}}\omega(\mathbf{y},t) \nonumber \\ \psi(\mathbf{y},t) &\to & \exp[-ie\omega(\mathbf{y},t)]\psi(\mathbf{y},t) \nonumber \\ \omega(\mathbf{y},t)& = & \frac{e}{2\pi \mu} arg(\mathbf{y}-\mathbf{x}(t)) \label{CS13},
\end{eqnarray}
where $\psi(\mathbf{y},t)$ is a non-relativistic matter field. For the general value of $\frac{e^{2}}{\mu}$ the transformed matter field becomes the multivalued function of the angle. To summarize, the gauge transformation (\ref{CS13}) helps us to get rid of the gauge field, but in this case the presence  of the Chern-Simons term in the Lagrangian manifests itself by \emph{multi-valuedness} of the matter field $\psi$.

Now we shall show how the fractional statistics comes into play. Let us imagine two identical point particles (I and II), which interact through the Chern-Simons field. Consider the adiabatic circular motion of the particle I around the static particle II. Due to Aharonov and B\"ohm \cite{Aharonov:1959} the wave function acquires the magnetic phase $\phi_{AB}$ after one rotation:
\beq \label{CS14}
\exp(i\phi_{AB})=\exp(ie\oint \mathbf{A}\cdot d\mathbf{y}).
\eeq
Using (\ref{CS12}) we can easily calculate $\phi_{AB}$ to get:
\beq \label{CS15}
\phi_{AB}=-\frac{e^{2}}{\mu}.
\eeq
On the other hand we can interpret one rotation as double interchange of two identical particles. Thus under one interchange of two particles we get the change of phase:
\beq \label{CS16}
\Delta\phi=-\frac{e^{2}}{2\mu}.
\eeq
For general choice of $e^{2}$ and $\mu$ we can get an arbitrary real exchange phase. In the subsequent sections we show that the particles, interacting through the Chern-Simons field, have fractional spin and that the spin-statistics theorem, which connects spin of the particle with the phase exchange, is satisfied. Hence, the Chern-Simons theory coupled to the non-relativistic matter describes \emph{anyons}, i.e., particles with arbitrary spin.

\section{Non-relativistic Pauli field theory}
After this brief excursion into the Chern-Simons physics let us go back to our initial aim to describe the interaction of two anyons in the non-relativistic domain. The effective potential is given by (\ref{NP11}):
\beq \label{CS17}
V_{l}=-\frac{1}{km}\frac{1}{r}\delta(r)+\frac{1}{mr^{2}}\left(l-\frac{1}{k}\right)^{2},
\eeq
where $k=\frac{2\pi\mu}{e^{2}}$.
It is important to remark here that the scattering of a particle in this potential in the non-relativistic quantum mechanics (the Aharonov-B\"ohm effect of particles with spin 1/2) was studied by Hagen \cite{Hagen:1989,Hagen:1990} and the \emph{exact} solution for the scattering amplitude $f(\phi)$ was obtained. The Hagen's result for the CS photon with the negative $\mu$-parameter and spin $s=-1$ (this choice is in accord with our footnotes in sections \ref{sec:Pauli}, \ref{sec:Breit}) is:
\beq \label{CS18}
f(\phi)=-i\left( \frac{i}{2\pi p} \right)^{1/2}\frac{\sin\left(\frac{e^{2}}{2\mu}\right) }{\sin\left(\frac{\phi}{2}\right)} e^{-i\frac{\phi}{2}},
\eeq
where $p$ is the magnitude of the non-relativistic initial momentum and we assume that the incident wave is coming from the left.

It is interesting and instructive to derive the scattering amplitude of two distinguishable anyons using the perturbative methods of quantum field theory and to compare the result with (\ref{CS18}). This aim was accomplished in \cite{Bergman:1993,Girotti:1997} and we generally follow the same line here. First, we shall construct a non-relativistic field theory (second-quantized version of the Pauli Lagrangian) and calculate the tree-level and the one-loop contributions to the scattering process. In the next section, the same process will be examined in the fully relativistic CS theory defined by (\ref{CS3}) in its non-relativistic sector.

The non-relativistic field theory is defined by the Pauli Langrangian\footnote{Note that in accordance with our previous calculations the photon parameter is negative and $\mu$ denotes its absolute value.}:
\begin{eqnarray} 
\mathscr{L}_{P}&=&\phi^{\dag}\left( i\frac{\partial}{\partial t}-eA^{0} \right) \phi-\frac{1}{2m}(\nabla \phi-ie\mathbf{A}\phi)^{\dag}\cdot(\nabla\phi-ie\mathbf{A}\phi) \nonumber \\ &&-\frac{e}{2m}B\phi^{\dag}\phi-\frac{\mu}{2}\epsilon^{\alpha\beta\gamma}A_{\alpha}\partial_{\beta}A_{\gamma}, \label{CS19}
\end{eqnarray}
where $\phi$ is the anticommuting one-component fermionic field and $B=F^{12}$ is the magnetic CS field. The Lagrangian is invariant under the Galilean transformations (assuming that $A^{0}$, $\phi$ and $\phi^{\dag}$ are scalar fields and $A^{i}$ is a vector field). The Euler-Lagrange field equations for the fermionic field $\phi^{\dag}$ lead to the Pauli equation (\ref{RQM6}).

To make a perturbative analysis of the scattering of two anyons in the non-relativistic field theory (\ref{CS19}), we should first find the Feynman rules for the propagators and the vertices. The fermionic quantum field $\phi$ satisfies the canonical equal-time anticommutation relations:
\beq \label{CS20}
\{\phi(\mathbf{x},t),\phi^{\dag}(\mathbf{y},t)\}=\delta^{(2)}(\mathbf{x}-\mathbf{y})
\eeq 
and can be decomposed into the plane waves:
\beq \label{CS21}
\phi(\mathbf{x},t)=\int \frac{d^{2}p}{2\pi}a_{\mathbf{p}}\exp\left[ -i(\frac{\mathbf{p}^{2}}{2m}t-\mathbf{p}\cdot \mathbf{x})  \right]
\eeq
with:
\beq \label{CS22}
\{a_{\mathbf{p}}, a^{\dag}_{\mathbf{k}}\}=\delta^{(2)}(\mathbf{p}-\mathbf{k}).
\eeq
The Feynman propagator can be easily found:
\begin{eqnarray} 
G_{F}(x-y)&=&\bra{0}T \phi(x)\phi^{\dag}(y)\ket{0} \nonumber \\
&=&\int \frac{d^{3}p}{(2\pi)^{3}}\frac{i}{p^{0}-\mathbf{p}^{2}/2+i\epsilon}\exp[-ip^{0}(x^{0}-y^{0}-\eta)+i\mathbf{p}\cdot(\mathbf{x}-\mathbf{y})], \label{CS23}
\end{eqnarray}
where we introduced the infinitesimal positive $\eta$ to regulate the ambiguity of time-ordering operation for $x^{0}=y^{0}$. With our choice $\eta>0$ (which is motivated by the fact that in the Lagrangian (\ref{CS19}) the fields always appear in the order $\phi^{\dag}\phi$ and of course $\bra{0}\phi^{\dag}(\mathbf{x},t)\phi(\mathbf{y},t)\ket{0}=0$) the propagator (\ref{CS23}) \emph{vanishes} at $x^{0}=y^{0}$. The fermionic Feynman propagator is depicted in the Fig. \ref{graph9}.

As far as the photon propagator is concerned, the anyonic limit of its covariant gauge version is given by the relation:
\beq \label{CS24}
\mathscr{D}_{\mu\nu}(p)=-\frac{1}{\mu}\frac{\epsilon_{\mu\nu\rho}p^{\rho}}{p^{2}+i\epsilon}.
\eeq
In the non-relativistic calculations it is more convenient, however, to use the Coulomb version of the propagator \cite{Girotti:1997}:
\beq \label{CS25}
\mathscr{D}_{\mu\nu}(p)=\frac{1}{\mu}\epsilon_{\mu\nu\rho}\frac{\bar{p}^{\rho}}{\mathbf{p}^{2}-i\epsilon},
\eeq 
where $\bar{p}^{\rho}=(0,\mathbf{p})$ (see Fig. \ref{graph9}). 

Finally, it is convenient to define the following two-body Green's function:
\beq \label{CS26}
\Delta_{B}(x)=\langle B(x)A^{0}(0)\rangle=-\frac{i}{\mu}\delta^{(3)}(x),
\eeq
The Green's functions of the magnetic field with the spatial components of the electromagnetic potential vanish due to antisymmetry of the photon propagator.  
\begin{figure}[!htb]
\begin{center}

\begin{eqnarray}
\parbox{25mm}{
    \begin{fmffile}{prop1}
    \begin{fmfgraph*}(25,12) 
    \fmfleft{i1}
    \fmfright{i2}
    \fmf{fermion, label=$p$}{i1,i2}
    \end{fmfgraph*}
    \end{fmffile}} & \qquad G_{F}(p)=\frac{i}{p^{0}-\mathbf{p}^{2}/2m+i\epsilon} \\
\parbox{25mm}{
   \begin{fmffile}{prop2}
   \begin{fmfgraph*}(25,12) 
   \fmfleft{i1}
   \fmfright{i2}
   \fmflabel{$\mu$}{i1}
   \fmflabel{$\nu$}{i2}
   \fmfcmd{%
     style_def wiggly_arrow expr p =
     cdraw (wiggly p);
     shrink (1);
     cfill (arrow p);
     endshrink;
     enddef;}
   \fmf{wiggly_arrow, label=$\mathbf{p}$}{i1,i2}
   \end{fmfgraph*}
   \end{fmffile}} & \qquad  \mathscr{D}_{\mu\nu}(p)=\frac{1}{\mu}\epsilon_{\mu\nu\rho}\frac{\bar{p}^{\rho}}{\mathbf{p}^{2}-i\epsilon}  \\
\parbox{25mm}{
   \begin{fmffile}{prop3}
   \begin{fmfgraph}(25,12) 
   \fmfleft{i1}
   \fmfright{i2}
   \fmf{dashes}{i1,i2}
   \end{fmfgraph}
   \end{fmffile}} & \qquad \Delta_{B}(p)=-\frac{i}{\mu}
\end{eqnarray}

\caption{Feynman rules for the propagators in the non-relativistic Pauli theory}\label{graph9}
\end{center}
\end{figure}
The Feynman rules for the interaction vertices can be read off from the Lagrangian. They can be found in the Fig. \ref{graph10}. 
\begin{figure}[!htb]
\begin{center}

\begin{tabular}{cccc}
\parbox{35mm}{
    \begin{fmffile}{ver1}
    \begin{fmfgraph*}(35,15) 
    \fmfleft{i1,i3}
    \fmfright{i2}
    \fmf{fermion}{i1,v1}
    \fmf{fermion}{v1,i3}
    \fmf{dashes}{v1,i2}
    \fmfv{l=B,l.a=60,l.d=4thick}{v1}
    \fmfv{decor.shape=circle,decor.filled=full,decor.size=2}{v1}
    \end{fmfgraph*}
    \end{fmffile}} &  $\frac{ie}{2m}$ &
\parbox{35mm}{
   \begin{fmffile}{ver2}
   \begin{fmfgraph*}(35,15) 
   \fmfleft{i1,i3}
    \fmfright{i2}
    \fmf{fermion}{i1,v1}
    \fmf{fermion}{v1,i3}
    \fmf{photon}{v1,i2}
    \fmfv{l=$0$,l.a=60,l.d=4thick}{v1} 
    \fmfv{decor.shape=circle,decor.filled=full,decor.size=2}{v1}
   \end{fmfgraph*}
   \end{fmffile}} &  $-ie$ \\ 
\mbox{} &&& \\   
\mbox{} &&& \\ 
\parbox{35mm}{
   \begin{fmffile}{ver3}
   \begin{fmfgraph*}(35,15) 
   \fmfleft{i1,i3}
    \fmfright{i2}
    \fmf{fermion}{i1,v1}
    \fmf{fermion}{v1,i3}
    \fmf{photon}{v1,i2}
    \fmfv{l=k,l.a=60,l.d=4thick}{v1} 
    \fmflabel{$p$}{i1}
    \fmflabel{$p^{\prime}$}{i3}
    \fmfv{decor.shape=circle,decor.filled=full,decor.size=2}{v1}
   \end{fmfgraph*}
   \end{fmffile}}  & $\frac{ie}{2m}(p+p^{\prime})^{k}$ &
\parbox{35mm}{
   \begin{fmffile}{ver4}
   \begin{fmfgraph*}(35,15) 
   \fmfleft{i1,i3}
    \fmfright{i2,i4}
    \fmf{fermion}{i1,v1}
    \fmf{fermion}{v1,i3}
    \fmf{photon}{v1,i2}
    \fmf{photon}{v1,i4} 
    \fmflabel{$p$}{i1} 
    \fmflabel{$p^{\prime}$}{i3} 
    \fmflabel{$i$}{i2} 
    \fmflabel{$j$}{i4}
    \fmfv{decor.shape=circle,decor.filled=full,decor.size=2}{v1}
   \end{fmfgraph*}
   \end{fmffile}} & $-\frac{ie^{2}}{m}\delta^{ij}$ \\
\mbox{} &&& \\          
\end{tabular}

\caption{Feynman rules for the interaction vertices in the non-relativistic Pauli theory}\label{graph10}
\end{center}
\end{figure}

Now, having Feynman rules of the Pauli theory at hand, we first calculate the scattering of two destinguishable anyons in their center of mass frame in tree level, which is described by four Feynman diagrams of Fig. \ref{graph11} (with our momenta definitions illustrated in the Figure).  

It is straightforward to calculate the invariant amplitude:
\beq \label{CS27}
M_{tree}=-\frac{e^{2}}{m\mu}\left(1-\frac{2i\mathbf{p}\times \mathbf{q}}{\mathbf{q}^{2}}\right),
\eeq
where $\mathbf{q}=\mathbf{p}-\mathbf{p^{\prime}}$. The real part of the last expression originates from the Pauli interaction term in the Lagrangian (\ref{CS19}), while the imaginary part describes the Aharonov-B\"ohm interaction. Expressing (\ref{CS27}) in terms of the scattering angle $\phi$, we obtain:
\beq \label{CS28}
M_{tree}=-\frac{ie^{2}}{m\mu}\frac{e^{-i\frac{\phi}{2}}}{\sin(\frac{\phi}{2})}.
\eeq
At this point it is worth noting that the scattering amplitude is in fact (up to some angle-independent factors) the first term in the expansion in powers of $e^{2}$ of the exact quantum mechanical result (\ref{CS18}).
\begin{figure}[!htb]
\begin{center}
\beq
\parbox{30mm}{
\begin{fmffile}{tree1} 	
   \begin{fmfgraph*}(30,25) 
   \fmfleft{l1,l2}
   \fmfright{r1,r2}
   \fmf{fermion}{l1,v1,r1}
   \fmf{fermion}{l2,v2,r2}
   \fmfcmd{%
     style_def wiggly_arrow expr p =
     cdraw (wiggly p);
     shrink (1);
     cfill (arrow p);
     endshrink;
     enddef;}
   \fmf{wiggly_arrow, label=$\mathbf{q}$}{v1,v2}
   \fmflabel{$p^{\prime}$}{r1}
   \fmflabel{$k^{\prime}$}{r2}
   \fmflabel{$p$}{l1}
   \fmflabel{$k$}{l2}
   \fmflabel{$i$}{v1}
   \fmflabel{$0$}{v2}   
\end{fmfgraph*}
\end{fmffile}} \,
+
\parbox{30mm}{
\begin{fmffile}{tree2} 	
   \begin{fmfgraph*}(30,25) 
   \fmfleft{l1,l2}
   \fmfright{r1,r2}
   \fmf{fermion}{l1,v1,r1}
   \fmf{fermion}{l2,v2,r2}
   \fmf{photon}{v1,v2}
   \fmflabel{$0$}{v1}
   \fmflabel{$i$}{v2}   
\end{fmfgraph*}
\end{fmffile}} \,
+
\parbox{30mm}{
\begin{fmffile}{tree3} 	
   \begin{fmfgraph*}(30,25) 
   \fmfleft{l1,l2}
   \fmfright{r1,r2}
   \fmf{fermion}{l1,v1,r1}
   \fmf{fermion}{l2,v2,r2}
   \fmf{dashes}{v1,v2}
   \fmflabel{$0$}{v1}
   \fmflabel{$B$}{v2}   
\end{fmfgraph*}
\end{fmffile}} \,
+
\parbox{30mm}{
\begin{fmffile}{tree4} 	
   \begin{fmfgraph*}(30,25) 
   \fmfleft{l1,l2}
   \fmfright{r1,r2}
   \fmf{fermion}{l1,v1,r1}
   \fmf{fermion}{l2,v2,r2}
   \fmf{dashes}{v1,v2}
   \fmflabel{$B$}{v1}
   \fmflabel{$0$}{v2}   
\end{fmfgraph*}
\end{fmffile}} \,
\eeq
\caption{Tree-level scattering of two distinguishable anyons, $\mathbf{q}=\mathbf{p}-\mathbf{p^{\prime}}$}\label{graph11}
\end{center}
\end{figure}

Now we shall discuss the one-loop corrections to the tree-level scattering. First, we make a very important observation that the one-loop corrections to the propagators and the vertex vanish in the Pauli theory. To illustrate this fact the corrected photon propagator $D_{0i}$ will be calculated. The one-loop vacuum-polarization Feynman diagrams can be found in Fig. \ref{graph12}. It is important to note that the fermion loop is given by the expression:
\beq \label{CS29}
(-1)\int \frac{d^{3}l}{(2\pi)^{3}}\frac{i}{l^{0}-\mathbf{l}^{2}/2m+i\epsilon} \, \frac{i}{(p+l)^{0}-(\mathbf{p+l})^{2}/2m+i\epsilon},
\eeq
where $l$ is the momentum, which circulates in the loop. We perform the $l^{0}$-integration using the residue theorem and easily realize that the integral \emph{vanishes}. This is due to the fact that both poles of the integrand lie in the lower part of the complex plane and we can enclose the integration contour in the upper part of the complex plane. This particular result can be generalized to the statement that \emph{every closed fermion loop is zero} in the Pauli theory. This useful finding originates from the fact that the denominator of the fermionic propagator is linear in the energy, which reflects the \emph{classical} (non-relativistic) nature of the Pauli theory. By very similar calculations we find that the general components of the vacuum polarization vanish, i.e., $D_{\alpha \beta}(\mathbf{p})=\mathscr{D}_{\alpha \beta}(\mathbf{p})$ for $\alpha, \beta=0,1,2$ \footnote{where by $D_{\alpha\beta}$ we denote the corrected photon propagator and by $\mathscr{D}_{\alpha\beta}$ free photon propagator}. One may convince himself that the same arguments (and the antisymmetry of the photon propagator) lead to the vanishing of the one-loop correction to the fermion propagator and the vertex diagram.    
\begin{figure}[!htb]
\begin{center}
\beq
\parbox{40mm}{
\begin{fmffile}{vacpol1} 	
   \begin{fmfgraph*}(40,25) 
   \fmfleft{l1}
   \fmfright{r1}
   \fmfcmd{%
     style_def wiggly_arrow expr p =
     cdraw (wiggly p);
     shrink (1);
     cfill (arrow p);
     endshrink;
     enddef;}
   \fmf{wiggly_arrow, label=$\mathbf{p}$}{l1,v1}
   \fmf{wiggly_arrow, label=$\mathbf{p}$}{v2,r1}
   \fmf{fermion, left}{v1,v2,v1}
   \fmflabel{$i$}{r1}
   \fmflabel{$0$}{l1}
   \fmfv{label=$j$, label.angle=-120}{v1}
   \fmfv{label=$0$, label.angle=-60}{v2}   
\end{fmfgraph*}
\end{fmffile}} \qquad \qquad
\parbox{40mm}{
\begin{fmffile}{vacpol2} 	
   \begin{fmfgraph*}(40,25) 
   \fmfleft{l1}
   \fmfright{r1}
   \fmfcmd{%
     style_def wiggly_arrow expr p =
     cdraw (wiggly p);
     shrink (1);
     cfill (arrow p);
     endshrink;
     enddef;}
   \fmf{wiggly_arrow, label=$\mathbf{p}$}{v2,r1}
   \fmf{dashes}{l1,v1}
   \fmf{fermion, left}{v1,v2,v1}
   \fmflabel{$i$}{r1}
   \fmflabel{$0$}{l1}
   \fmfv{label=$B$, label.angle=-120}{v1}
   \fmfv{label=$0$, label.angle=-60}{v2}   
\end{fmfgraph*}
\end{fmffile}}
\eeq
\caption{Vacuum polarization diagrams for $D_{0,i}(\mathbf{p})$}\label{graph12}
\end{center}
\end{figure}

The absence of the above-mentioned corrections substantially reduces the number of Feynman diagrams, which contribute to the one-loop anyon scattering. The remaining diagrams can be found in the Fig. \ref{graph13}. Not all of them, however, have non-trivial contribution. All the crossed diagrams (E-H, Fig. \ref{graph13}) give zero, exactly for the same reason as in the case of the propagator corrections, i.e., both loop fermionic propagators have their $k^{0}$-poles in the lower complex half-plane. As for the mixed diagrams (C and D, Fig. \ref{graph13}), it can be shown, summing all the contributions for various possible vertex assignments ($^{00}_{iB}$, $^{0B}_{i\,0}$, $^{i \, 0}_{0B}$, $^{iB}_{00}$ for the diagram C), that they vanish either. Thus, the only non-vanishing one-loop Feynman diagrams, which contribute to the scattering of two non-identical anyons, are \emph{the photon box diagram} (A, Fig. \ref{graph13}), \emph{the magnetic box diagram} (B, Fig. \ref{graph13}) and \emph{the two triangle diagrams} (I and J, Fig. \ref{graph13}). We shall now calculate these diagrams in the center of mass frame of the anyons.

Let us start with the box photon diagram (A, Fig. \ref{graph13}). We assign the momenta $p_{1}=(p^{0},\mathbf{p})$, $p_{2}=(p^{0},-\mathbf{p})$ to the initial anyons and $p^{\prime}_{1}=(p^{0},\mathbf{p^{\prime}})$, $p^{\prime}_{2}=(p^{0},-\mathbf{p^{\prime}})$ to the final anyons. The scattering amplitude $M_{\gamma \Box}$ is:
\begin{eqnarray}
iM_{\gamma \Box}&=&4\frac{e^{4}}{4m^{2}}\int \frac{d^{3}k}{(2\pi)^{3}}G_{F}(k)G_{F}(p_{1}+p_{2}-k)(p+k)^{i}\mathscr{D}_{i0}(p_{1}-k)(p^{\prime}+k)^{j}\mathscr{D}_{j0}(k-p^{\prime}_{1}) \nonumber \\
&=&4i \frac{e^{4}}{m\mu^{2}}\int\frac{d^{2}k}{(2\pi)^{2}}\frac{1}{\mathbf{k}^{2}-\mathbf{p}^{2}-i\epsilon} \frac{\mathbf{p}\times\mathbf{k}}{(\mathbf{p}-\mathbf{k})^{2}}\frac{\mathbf{p^{\prime}}\times\mathbf{k}}{(\mathbf{p^{\prime}}-\mathbf{k})^{2}} \label{CS30},
\end{eqnarray}      
where $\mathbf{a}\times\mathbf{b}=\epsilon_{ij}a^{i}b^{j}$ and we performed $k^{0}$ integration in the second line using the residue theorem. Obviously, we have summed four photon box diagrams with various vertex assignments ($^{ij}_{00}$, $^{0j}_{i0}$, $^{i0}_{0j}$, $^{00}_{ij}$ for the diagram A), which happened to contribute equally. The factor $4$ in (\ref{CS30}) reflects this fact. Now we go to the polar coordinates and perform first the angular integration:
\beq \label{CS31}
M_{\gamma \Box}=\frac{e^{4}}{m\mu^{2}}\int^{\infty}_{0} \frac{d(k^{2})}{4\pi}\frac{1}{k^{2}-p^{2}-i\epsilon}\left(-1+\frac{(k^{2}+p^{2})|k^{2}-p^{2}|}{(k^{2}-p^{2}e^{i\phi})(k^{2}-p^{2}e^{-i\phi})} \right),
\eeq
where $\phi$ is a scattering angle, i.e., $\cos(\phi)=\frac{\mathbf{p}\cdot\mathbf{p^{\prime}}}{|\mathbf{p}||\mathbf{p^{\prime}}|}$. To get (\ref{CS31}) we made use of the following two-dimensional vector identity:
\beq \label{CS32}
(\mathbf{a}\times\mathbf{b})(\mathbf{c}\times\mathbf{d})=(\mathbf{a}\cdot\mathbf{c})(\mathbf{b}\cdot\mathbf{d})-(\mathbf{b}\cdot\mathbf{c})(\mathbf{a}\cdot\mathbf{d})
\eeq
and the two angular integrals:
\begin{eqnarray} 
\int\frac{d\theta_{k}}{(\mathbf{k}+\mathbf{p})^{2}}&=&\frac{2\pi}{|\mathbf{k}^{2}-\mathbf{p}^{2}|}\nonumber\\
 \int\frac{d\theta_{k}}{(\mathbf{k}+\mathbf{p})^{2}(\mathbf{k}+\mathbf{p^{\prime}})^{2}}&=&\frac{2\pi}{|\mathbf{k}^{2}-\mathbf{p}^{2}|}\frac{\mathbf{p}^{2}+\mathbf{k}^{2}}{(\mathbf{p}^{2}-\mathbf{k}^{2}e^{-i\phi})(\mathbf{p}^{2}-\mathbf{k}^{2}e^{i\phi})}. \label{CS33}
\end{eqnarray}
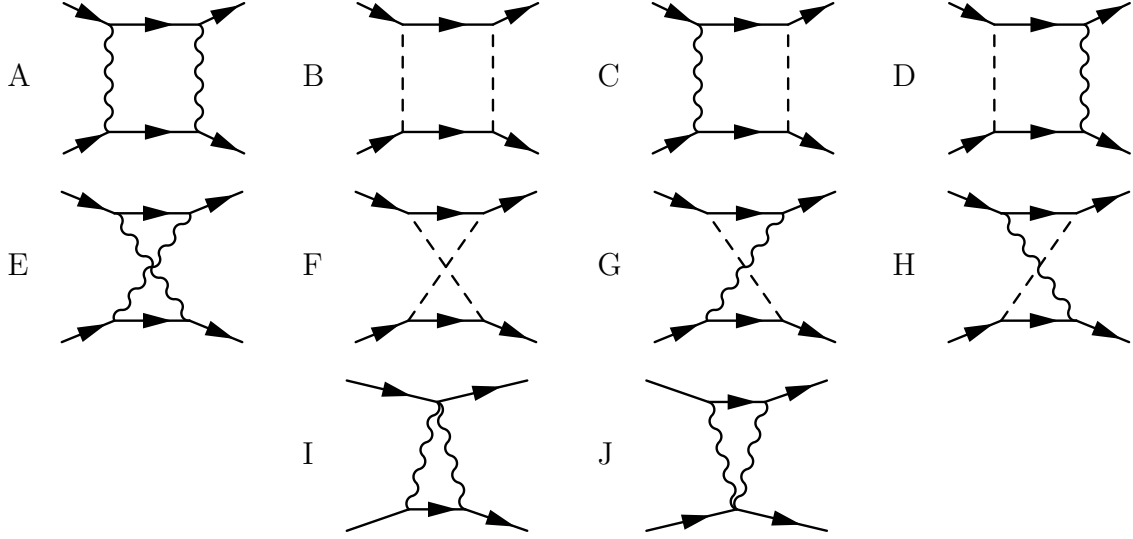
\begin{figure}[!htb]
\begin{center}
\begin{tabular}{cccc}
\parbox{35mm}{A \parbox{30mm}{
\begin{fmffile}{1la} 	
   \begin{fmfgraph*}(30,20) 
   \fmfleft{l1,l2}
   \fmfright{r1,r2}
   \fmf{fermion}{l1,v1}
   \fmf{fermion, tension=.5}{v1,v2}
   \fmf{fermion}{v2,r1}
   \fmf{fermion}{l2,v3}
   \fmf{fermion, tension=.5}{v3,v4}
   \fmf{fermion}{v4,r2}   
   \fmf{photon, tension=.2}{v1,v3}
   \fmf{photon, tension=.2}{v2,v4}
\end{fmfgraph*}
\end{fmffile}}} &
\parbox{35mm}{B \parbox{30mm}{
\begin{fmffile}{1lb} 	
   \begin{fmfgraph*}(30,20) 
   \fmfleft{l1,l2}
   \fmfright{r1,r2}
   \fmf{fermion}{l1,v1}
   \fmf{fermion, tension=.5}{v1,v2}
   \fmf{fermion}{v2,r1}
   \fmf{fermion}{l2,v3}
   \fmf{fermion, tension=.5}{v3,v4}
   \fmf{fermion}{v4,r2}      
   \fmf{dashes, tension=.2}{v1,v3}
   \fmf{dashes, tension=.2}{v2,v4}
\end{fmfgraph*}
\end{fmffile}}} &
\parbox{35mm}{C \parbox{30mm}{
\begin{fmffile}{1lc} 	
   \begin{fmfgraph*}(30,20) 
   \fmfleft{l1,l2}
   \fmfright{r1,r2}
   \fmf{fermion}{l1,v1}
   \fmf{fermion, tension=.5}{v1,v2}
   \fmf{fermion}{v2,r1}
   \fmf{fermion}{l2,v3}
   \fmf{fermion, tension=.5}{v3,v4}
   \fmf{fermion}{v4,r2}      
   \fmf{photon, tension=.2}{v1,v3}
   \fmf{dashes, tension=.2}{v2,v4}
\end{fmfgraph*}
\end{fmffile}}} &
\parbox{35mm}{D \parbox{30mm}{
\begin{fmffile}{1ld} 	
   \begin{fmfgraph*}(30,20) 
   \fmfleft{l1,l2}
   \fmfright{r1,r2}
   \fmf{fermion}{l1,v1}
   \fmf{fermion, tension=.5}{v1,v2}
   \fmf{fermion}{v2,r1}
   \fmf{fermion}{l2,v3}
   \fmf{fermion, tension=.5}{v3,v4}
   \fmf{fermion}{v4,r2}      
   \fmf{dashes, tension=.2}{v1,v3}
   \fmf{photon, tension=.2}{v2,v4}
\end{fmfgraph*}
\end{fmffile}}} \\
\mbox{}&&& \\
\parbox{35mm}{E \parbox{30mm}{
\begin{fmffile}{1le} 	
   \begin{fmfgraph*}(30,20) 
   \fmfleft{l1,l2}
   \fmfright{r1,r2}
   \fmf{fermion}{l1,v1}
   \fmf{fermion, tension=.5}{v1,v2}
   \fmf{fermion}{v2,r1}
   \fmf{fermion}{l2,v3}
   \fmf{fermion, tension=.5}{v3,v4}
   \fmf{fermion}{v4,r2}      
   \fmf{photon, tension=.2}{v1,v4}
   \fmf{photon, tension=.2}{v2,v3}
\end{fmfgraph*}
\end{fmffile}}} &
\parbox{35mm}{F \parbox{30mm}{
\begin{fmffile}{1lf} 	
   \begin{fmfgraph*}(30,20) 
   \fmfleft{l1,l2}
   \fmfright{r1,r2}
   \fmf{fermion}{l1,v1}
   \fmf{fermion, tension=.5}{v1,v2}
   \fmf{fermion}{v2,r1}
   \fmf{fermion}{l2,v3}
   \fmf{fermion, tension=.5}{v3,v4}
   \fmf{fermion}{v4,r2}      
   \fmf{dashes, tension=.2}{v1,v4}
   \fmf{dashes, tension=.2}{v2,v3}
\end{fmfgraph*}
\end{fmffile}}} &
\parbox{35mm}{G \parbox{30mm}{
\begin{fmffile}{1lg} 	
   \begin{fmfgraph*}(30,20) 
   \fmfleft{l1,l2}
   \fmfright{r1,r2}
   \fmf{fermion}{l1,v1}
   \fmf{fermion, tension=.5}{v1,v2}
   \fmf{fermion}{v2,r1}
   \fmf{fermion}{l2,v3}
   \fmf{fermion, tension=.5}{v3,v4}
   \fmf{fermion}{v4,r2}      
   \fmf{photon, tension=.2}{v1,v4}
   \fmf{dashes, tension=.2}{v2,v3}
\end{fmfgraph*}
\end{fmffile}}} &
\parbox{35mm}{H \parbox{30mm}{
\begin{fmffile}{1lh} 	
   \begin{fmfgraph*}(30,20) 
   \fmfleft{l1,l2}
   \fmfright{r1,r2}
   \fmf{fermion}{l1,v1}
   \fmf{fermion, tension=.5}{v1,v2}
   \fmf{fermion}{v2,r1}
   \fmf{fermion}{l2,v3}
   \fmf{fermion, tension=.5}{v3,v4}
   \fmf{fermion}{v4,r2}      
   \fmf{dashes, tension=.2}{v1,v4}
   \fmf{photon, tension=.2}{v2,v3}
\end{fmfgraph*}
\end{fmffile}}} \\
\mbox{}&&& \\
&
\parbox{35mm}{I \parbox{30mm}{
\begin{fmffile}{1li} 	
   \begin{fmfgraph*}(30,20) 
   \fmfleft{l1,l2}
   \fmfright{r1,r2}
   \fmf{fermon}{l1,v1}
   \fmf{fermion}{v1,v2}
   \fmf{fermion}{v2,r1}
   \fmf{fermion}{l2,v3}
   \fmf{fermion}{v3,r2}   
   \fmf{photon, tension=.2}{v1,v3}
   \fmf{photon, tension=.2}{v2,v3}
\end{fmfgraph*}
\end{fmffile}}} &
\parbox{35mm}{J \parbox{30mm}{
\begin{fmffile}{1lj} 	
 \begin{fmfgraph*}(30,20) 
   \fmfleft{l1,l2}
   \fmfright{r1,r2}
   \fmf{fermon}{l2,v1}
   \fmf{fermion}{v1,v2}
   \fmf{fermion}{v2,r2}
   \fmf{fermion}{l1,v3}
   \fmf{fermion}{v3,r1}   
   \fmf{photon, tension=.2}{v1,v3}
   \fmf{photon, tension=.2}{v2,v3} 
\end{fmfgraph*}
\end{fmffile}}} & \\
\end{tabular}
\caption{One-loop anyon scattering diagrams}\label{graph13}
\end{center}
\end{figure}
Finally, it remains to perform the radial part of the integration in (\ref{CS31}). We use the formula:
\beq \label{CS34}
\frac{1}{y-i\epsilon}=P\frac{1}{y}+i\pi\delta(y)
\eeq
and after straightforward, but tedious calculations we come to the result:
\beq \label{CS35}
M_{\gamma \Box}=-\frac{e^{4}}{4\pi m\mu^{2}}\left(\ln\frac{\mathbf{q}^{2}}{\mathbf{p}^{2}}+i\pi\right).
\eeq

We proceed with the magnetic box diagram (B, Fig. \ref{graph13}). The scattering amplitude $M_{B\Box}$ is given by:
\beq \label{CS36}
iM_{B\Box}=-4\frac{e^{4}}{4m^{2}\mu^{2}}\int \frac{d^{3}k}{(2\pi)^{3}}G_{F}(k)G_{F}(p_{1}+p_{2}-k),
\eeq 
which is UV-divergent and the UV cutoff $\Lambda$ is used to regulate the expression. The factor 4 has the similar origin as in the case of the photon box diagram. The integration is straightforward with the final result:
\beq \label{CS37}
M_{B\Box}=\frac{e^{4}}{4\pi m\mu^{2}}\left( \ln\frac{\Lambda^{2}}{\mathbf{p}^{2}}+i\pi \right).
\eeq

Finally, the two triangle diagrams (I and J, Fig. \ref{graph13}) are calculated. Both diagrams have an equal contribution $M_{I}=M_{J}=\frac{1}{2}M_{\vartriangle}$. The total amplitude is given by:
\beq \label{CS38}
iM_{\vartriangle}=2\frac{ie^{4}}{m}\int\frac{d^{3}k}{(2\pi)^{3}}G_{F}(k)\mathscr{D}_{0i}(p_{1}-k)\mathscr{D}_{i0}(p^{\prime}_{1}-k).
\eeq
It is straightforward to perform $k^{0}$-integration and we end up with the spatial integral:
\beq \label{CS39}
M_{\vartriangle}=-\frac{e^{4}}{m\mu^{2}}\int\frac{d^{2}k}{(2\pi)^{2}}\frac{\mathbf{k}\cdot(\mathbf{k}-\mathbf{p}+\mathbf{p}^{\prime})}{\mathbf{k}^{2}(\mathbf{k}-\mathbf{p}+\mathbf{p^{\prime}})^{2}}.
\eeq
Using (\ref{CS33}) and the integral:
\beq \label{CS40}
\int d\theta_{\mathbf{k}}\frac{\mathbf{k}\cdot\mathbf{q}}{(\mathbf{k}+\mathbf{q})^{2}}=\pi-\frac{\pi(\mathbf{k}^{2}+\mathbf{q}^{2})}{|\mathbf{k}^{2}-\mathbf{q}^{2}|}
\eeq
it is easy to calculate the angular part of the integration in (\ref{CS39}):
\beq \label{CS41}
M_{\vartriangle}=-\frac{e^{4}}{m\mu^{2}}\left[ \int\frac{dk}{2\pi}\frac{k}{|k^{2}-q^{2}|}+\frac{1}{2}\int\frac{dk}{2\pi}\frac{1}{k}\left\{1-\frac{k^{2}+q^{2}}{|k^{2}-q^{2}|} \right\} \right].
\eeq  
It seems that the final integral diverges near $k=0$ and $k=\infty$, thus we introduce the UV cutoff $\Lambda$ and the IR cutoff $\delta$. The outcome of the radial integration is:
\beq \label{CS42}
M_{\vartriangle}=\frac{e^{4}}{4\pi m\mu^{2}}\ln\frac{\mathbf{q}^{2}}{\Lambda^{2}},
\eeq
thus the IR divergences in (\ref{CS41}) were in fact artificial.  

We are now ready to state the final result. The total one-loop corrections $M_{1l}$ to the non-relativistic anyon-anyon scattering \emph{vanish}:
\beq \label{CS43}
M_{1l}=M_{\gamma \Box}+M_{B\Box}+M_{\vartriangle}=\frac{e^{4}}{4\pi m \mu^{2}}\left[ \ln\frac{\mathbf{p}^{2}}{\mathbf{q}^{2}}-i\pi+\ln\frac{\Lambda^{2}}{\mathbf{p}^{2}}+i\pi+\ln\frac{\mathbf{q}^{2}}{\Lambda^{2}} \right]=0.
\eeq
The obtained result is not a surprise for us. The expansion in $e^{2}$ of the exact quantum-mechanical amplitude (\ref{CS18}) has no $e^{4}$ term. Hence we obtain zero one-loop correction in the non-relativistic Pauli field theory in accordance with quantum mechanics. 

\section{Non-relativistic limit of the Chern-Simons theory}
\begin{figure}[!htb]
\begin{center}
\begin{tabular}{ccc}
\parbox{35mm}{A \parbox{30mm}{
\begin{fmffile}{CS0} 	
   \begin{fmfgraph*}(30,20) 
   \fmfleft{l1,l2}
   \fmfright{r1,r2}
   \fmf{fermion}{l1,v1,l2}
   \fmf{fermion}{r1,v2,r2}
   \fmf{photon}{v1,v2}
\end{fmfgraph*}
\end{fmffile}}} &
\parbox{35mm}{B \parbox{30mm}{
\begin{fmffile}{CS2} 	
   \begin{fmfgraph*}(30,20) 
   \fmfleft{l1,l2}
   \fmfright{r1,r2}
   \fmf{fermion}{l1,v3,v1,v4,l2}
   \fmf{fermion}{r1,v2,r2}
   \fmf{photon}{v1,v2}
   \fmf{photon, tension=0}{v3,v4}
\end{fmfgraph*}
\end{fmffile}}} &
\parbox{35mm}{C \parbox{30mm}{
\begin{fmffile}{CS3} 	
   \begin{fmfgraph*}(30,20) 
   \fmfleft{l1,l2}
   \fmfright{r1,r2}
   \fmf{fermion}{r1,v3,v1,v4,r2}
   \fmf{fermion}{l1,v2,l2}
   \fmf{photon}{v1,v2}
   \fmf{photon, tension=0}{v3,v4}
\end{fmfgraph*}
\end{fmffile}}} \\
\mbox{}&& \\
\parbox{35mm}{D \parbox{30mm}{
\begin{fmffile}{CS4} 	
   \begin{fmfgraph*}(30,20) 
   \fmfleft{l1,l2}
   \fmfright{r1,r2}
   \fmf{fermion}{l1,v1}
   \fmf{fermion, tension=.5}{v1,v2}
   \fmf{fermion}{v2,r1}
   \fmf{fermion}{l2,v3}
   \fmf{fermion, tension=.5}{v3,v4}
   \fmf{fermion}{v4,r2}      
   \fmf{photon, tension=.2}{v1,v4}
   \fmf{photon, tension=.2}{v2,v3}
\end{fmfgraph*}
\end{fmffile}}} &
\parbox{35mm}{E \parbox{30mm}{
\begin{fmffile}{CS1} 	
   \begin{fmfgraph*}(30,20) 
   \fmfleft{l1,l2}
   \fmfright{r1,r2}
   \fmf{fermion}{l1,v1}
   \fmf{fermion, tension=.5}{v1,v2}
   \fmf{fermion}{v2,r1}
   \fmf{fermion}{l2,v3}
   \fmf{fermion, tension=.5}{v3,v4}
   \fmf{fermion}{v4,r2}   
   \fmf{photon, tension=.2}{v1,v3}
   \fmf{photon, tension=.2}{v2,v4}
\end{fmfgraph*}
\end{fmffile}}}&
\parbox{35mm}{F \parbox{30mm}{
\begin{fmffile}{CS6} 	
   \begin{fmfgraph*}(30,20) 
   \fmfleft{l1,l2}
   \fmfright{r1,r2}
   \fmf{fermion}{l1,v1,l2}
   \fmf{fermion}{r1,v2,r2}
   \fmf{photon}{v1,v3}
   \fmf{photon}{v4,v2}
   \fmf{fermion, left}{v3,v4,v3}
\end{fmfgraph*}
\end{fmffile}}} \\
\mbox{}&& \\
\end{tabular}
\caption{Chern-Simons scattering diagrams}\label{graph14}
\end{center}
\end{figure}
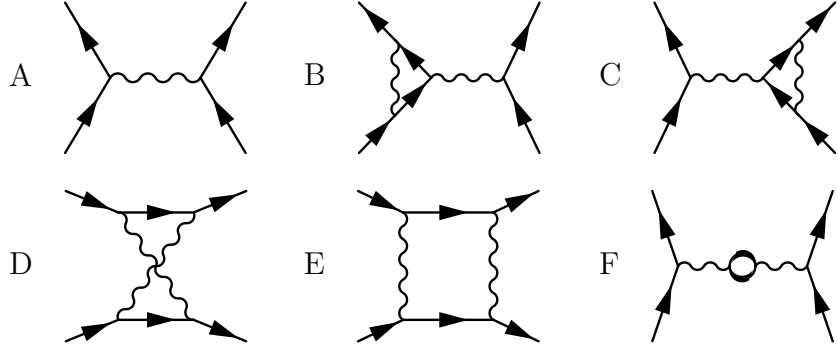

In previous section we examined the problem of the non-relativistic anyon-anyon scattering using the Pauli field theory, i.e., we first performed the classical limit going from the relativistic Chern-Simons theory to the Pauli theory and subsequently we calculated the scattering amplitude up to one-loop order. We demonstrated that the loop diagrams \emph{do not contribute} in the Pauli theory. In this section we tackle the same problem in a slightly different fashion: we shall first consider one-loop diagrams, which contribute to the amplitude in the fully relativistic Chern-Simons theory, which is defined by the Lagrangian:
\beq \label{CSS1}
\mathscr{L}=-\frac{\mu}{2}\epsilon^{\mu\nu\rho}A_{\mu}\partial_{\nu}A_{\rho}+\mathscr{L}_{\psi}-e\bar{\psi}\gamma^{\mu}\psi A_{\mu}
\eeq 
and calculate the non-relativistic (neglecting $O(\frac{\mathbf{p}^{2}}{m^{2}},\frac{\mathbf{q}^{2}}{m^{2}})$ terms) limit of the scattering amplitude of two anyons in their center of mass frame. The Feynman rules of the pure Chern-Simons theory coincide with the rules of the MCS theory, except for the photon propagator, which is given by:
\beq \label{CSS2}
\mathscr{D}_{\mu\nu}(q)=-\frac{1}{\mu}\frac{\epsilon_{\mu\nu\rho}p^{\rho}}{p^{2}+i\epsilon}
\eeq
in the covariant Landau gauge. The diagrams, which contribute non-trivially to the scattering amplitude of the two distinguishable anyons can be found in Fig. \ref{graph14}.

Contrary to the classical Pauli theory in the relativistic Chern-Simons theory there is no reason for absence of the quantum corrections to the propagators and the vertex. Let us calculate these corrections. First, we calculate the one-loop correction to the Chern-Simons vertex $\delta \Gamma^{\prime}_{\mu}(p,p^{\prime})$ (see Fig. \ref{graph15}):
\begin{eqnarray} 
\delta \Gamma^{\prime}_{\phi}(p,p^{\prime})&=&-e^{2}\int \frac{d^{3}l}{(2\pi)^{3}}\mathscr{D}_{\mu\nu}(l)\gamma^{\nu}\mathscr{S}(p^{\prime}-l)\gamma_{\phi}\mathscr{S}(p-l)\gamma^{\mu} \nonumber \\
&=&-\frac{e^{2}}{\mu}\int\frac{d^{3}l}{(2\pi)^{3}}\frac{\epsilon_{\mu\nu\rho}l^{\rho}\gamma^{\nu}(\slashed{p}^{\prime}-\slashed{l}+m)\gamma_{\phi}(\slashed{p}-\slashed{l}+m)\gamma^{\mu}}{[l^{2}+i\epsilon][(p^{\prime}-l)^{2}-m^{2}+i\epsilon][(p-l)^{2}-m^{2}+i\epsilon]} \label{CSS3}.
\end{eqnarray}
The vertex correction has a logarithmic UV divergence, hence we employ the dimensional regularization. First, using the identity (\ref{eq:RQM1bd}) for the product of the $\gamma$-matrices and the fact that the external fermions are on the mass shell ($p^{2}=(p^{\prime})^{2}=m^{2}$) we perform the reduction of the numerator of the integrand of (\ref{CSS3}):
\begin{eqnarray} 
Num_{\phi}&=&\underbrace{2il^{2}(p+p^{\prime})_{\phi}}_{I}-\underbrace{2il^{2}l_{\phi}}_{II}-\underbrace{2il\cdot (p+p^{\prime})l_{\phi}}_{III} \nonumber\\
&&+\underbrace{4\epsilon^{\mu\nu\rho}p_{\mu}p^{\prime}_{\nu}l_{\rho}\gamma_{\phi}}_{IV}+\underbrace{2l^{2}\epsilon_{\phi\lambda\beta}\gamma^{\lambda}(p-p^{\prime})^{\beta}}_{V}-\underbrace{2(p-p^{\prime})\cdot l \epsilon_{\phi\lambda\beta}\gamma^{\lambda}l^{\beta}}_{VI}. \label{CSS4}
\end{eqnarray}
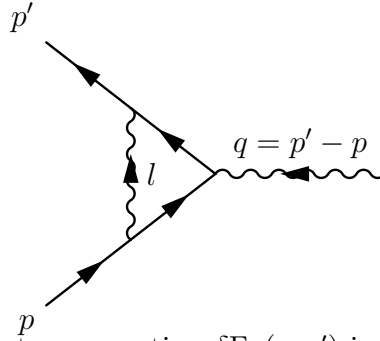
\begin{figure}
\begin{center}
\parbox{50mm}{
\begin{fmffile}{graph15} 	
   \begin{fmfgraph*}(50,35) 
   \fmfleft{l1,l2}
   \fmfright{r1}
   \fmfcmd{%
     style_def wiggly_arrow expr p =
     cdraw (wiggly p);
     shrink (1);
     cfill (arrow p);
     endshrink;
     enddef;}
   \fmf{fermion}{l1,v3,v1,v4,l2}
   \fmf{wiggly_arrow, label=$q=p^{\prime}-p$}{r1,v1}
   \fmf{wiggly_arrow, label=$l$, tension=0}{v3,v4}
   \fmflabel{$p^{\prime}$}{l2}
   \fmflabel{$p$}{l1}
\end{fmfgraph*}
\end{fmffile}}
\caption{One-loop vertex correction $\delta \Gamma_{\mu}(p,p^{\prime})$ in the Chern-Simons theory}\label{graph15}
\end{center}
\end{figure}
It is now straightforward to calculate all the integrals (I-VI) in the non-relativistic limit employing the standard techniques. For the illustrative reason we calculate here explicitly the integral III.  After introducing the Feynman parametrization we get:
\beq \label{CSS5}
\mathscr{I}^{III}_{\phi}=\frac{4ie^{2}}{\mu}(p+p^{\prime})^{\mu}\left\{\int \frac{d^{3}r}{(2\pi)^{3}}\frac{r^{2}g_{\mu\phi}}{3[r^{2}-\Delta]^{3}}+\int \frac{d^{3}r}{(2\pi)^{3}}\frac{(p^{\prime}x+py)_{\mu}(p^{\prime}x+py)_{\phi}}{[r^{2}-\Delta]^{3}} \right\},
\eeq  
where $\Delta=(xp^{\prime}+yp)^{2}-i\epsilon$. The first integral $\mathscr{I}^{III}_{\phi}(A)$ in (\ref{CSS5}) after the loop integration takes the form:
\beq \label{CSS6}
\mathscr{I}^{III}_{\phi}(A)=-\frac{e^{2}}{8\pi\mu m}(p+p^{\prime})_{\phi}\int^{1}_{0}dx\int^{1-x}_{0}dy\frac{1}{\sqrt{(x+y)^{2}-\delta xy}},
\eeq
where $\delta=\frac{q^{2}}{m^{2}}$. It is possible to calculate this integral exactly, but we are interested only in the non-relativistic limit $\delta<<1$, hence the power expansion in $\delta$ suffices:
\beq \label{CSS7}
\mathscr{I}^{III}_{\phi}(A)=-\frac{e^{2}}{8\pi\mu m}(p+p^{\prime})_{\phi}\left(1+\frac{1}{12}\delta+O(\delta^{2})\right).
\eeq
Similarly, for the second integral $\mathscr{I}^{III}_{\phi}(B)$ in (\ref{CSS5}) after the loop integration we obtain:
\beq \label{CSS8}
\mathscr{I}^{III}_{\phi}(B)=\frac{e^{2}}{4\pi\mu m}(p+p^{\prime})_{\phi}\left(1-\frac{1}{4}\delta \right)\int^{1}_{0}dx\int^{1-x}_{0}dy \frac{(x+y)x}{[(x+y)^{2}-\delta xy]^{3/2}},
\eeq
expanding in $\delta$ we get
\beq \label{CSS9}
\mathscr{I}^{III}_{\phi}(B)=\frac{e^{2}}{8pi\mu m}(p+p^{\prime})_{\phi}[1+O(\delta^{3})].
\eeq
The remaining integrals can be handled in the similar fashion. The final result for the unrenormalized vertex function $\Gamma^{\prime}_{\phi}(p,p^{\prime})=\gamma_{\phi}+\delta \Gamma_{\phi}(p,p^{\prime})$ can be obtained by calculating the remaining integrals and exploiting the (2+1)dimensional Gordon identity (\ref{eq:RQM7}). Dropping the relativistic corrections $O(\delta)$ we obtain:
\beq \label{CSS10}
\Gamma^{\prime}_{\phi}(p,p^{\prime})=\left(1+\frac{e^{2}}{4\pi\mu} \right)\gamma_{\phi}-\frac{ie^{2}}{4\pi m\mu}\epsilon_{\phi\mu\nu}q^{\mu}\gamma^{\nu},
\eeq
where $q=p^{\prime}-p$. We can use the Gordon identity (\ref{eq:RQM7}) once again to get rid of the $\gamma$ matrix in the axial term in (\ref{CSS10}):
\beq \label{CSS11}
\Gamma^{\prime}_{\phi}(p,p^{\prime})=\left(1+\frac{e^{2}}{4\pi\mu} \right)\gamma_{\phi}-\frac{ie^{2}}{4\pi \mu}\epsilon_{\phi\mu\nu}\frac{q^{\mu}}{m}\frac{p^{\nu}}{m}.
\eeq
To renormalize the vertex we use the condition $\Gamma_{\phi}(p,p)=\gamma_{\phi}$, hence the renormalized vertex $\Gamma_{\phi}(p,p^{\prime})$ and the renormalization constant $Z_{1}$ are up to one-loop order:
\beq \label{CSS12} 
\Gamma_{\phi}(p,p^{\prime})=\gamma_{\phi}-\frac{ie^{2}}{4\pi \mu}\epsilon_{\phi\mu\nu}\frac{q^{\mu}}{m}\frac{p^{\nu}}{m} \qquad Z_{1}=1-\frac{e^{2}}{4\pi\mu}.
\eeq

In (3+1) dimensions the one-loop vertex correction leads to the celebrated Schwinger anomalous magnetic moment of the electron. Having (\ref{CSS12}) at our disposal we are ready to calculate the magnetic moment in the Chern-Simons theory. Consider the interaction of the quantized electron field with the classical background electromagnetic field. The invariant amplitude $M$ of the scattering of the electron from the classical static background electromagnetic field $A^{\mu}(x)=(0,\mathbf{A}(\mathbf{x}))$ is given by:
\beq \label{CSS13}
iM(q)=-ie\bar{u}(p^{\prime})\Gamma_{i}(p^{\prime},p)u(p)\underbrace{\int d^{2}x e^{-i\mathbf{q}\cdot\mathbf{x}}A^{i}(\mathbf{x})}_{\widetilde{A}^{i}(\mathbf{q})},
\eeq
where $q=p^{\prime}-p$. We are interested only in the anomalous contribution, which we shall denote $M^{a}$, which comes from the second term in (\ref{CSS12}):
\beq \label{CSS14}
M^{a}(\mathbf{q})=\frac{ie^{3}}{4\pi\mu}\epsilon_{ij}\frac{q^{j}}{m}\frac{p^{0}}{m}\bar{u}(p^{\prime})u(p)\widetilde{A}^{i}(\mathbf{q}),
\eeq
where we used that $q^{0}=(p^{\prime}-p)^{0}\approx 0$. Using $p^{0}\approx m$, (\ref{BE5}) and our definition of the magnetic field (\ref{eq:CLE0}) we obtain the following expression for the anomalous interaction:
\beq \label{CSS15}
\widetilde{V^{a}}(\mathbf{q})=-\frac{M^{a}}{2m}=-\frac{e^{3}}{4\pi m \mu}\underbrace{\int d^{2}x e^{-i\mathbf{q}\cdot \mathbf{x}}B(\mathbf{x})}_{\widetilde{B}(\mathbf{q})},
\eeq
where the first equality comes from the relativistic normalization (\ref{BE5}). In the coordinate space we arrive at:
\beq \label{CSS16}
V^{a}(\mathbf{x})=\frac{eg}{2m}\left(\underbrace{-\frac{e^{2}}{4\pi\mu}}_{S_{in}} \right)B(x),
\eeq
where we introduced the $g$-factor equal 2. Recalling the Pauli equation (\ref{RQM6}) we readily identify the expression in the brackets $S_{in}$ with the induced spin. Why do we claim that the anomalous magnetic moment arises from the induced spin and not from the correction to the $g$-factor (analogously to the (3+1)case)? The answer resides in the spin statistics relation. Recall that due to (\ref{CS16}) the adiabatic interchange of two identical anyons in the Chern-Simons interaction generates the Aharonov-B\"ohm phase\footnote{The difference in the signs between (\ref{CSS17}) and (\ref{CS16}) comes form the difference in the signs of the CS terms in the Lagrangians (\ref{CSS1}) and (\ref{CS3})}:
\beq \label{CSS17}
\Delta \phi= \frac{e^{2}}{2\mu}.
\eeq
Assuming the standard spin-statistics relation the Aharanov-B\"ohm phase:
\beq \label{CSS18}
\exp(i\Delta \phi)=\exp(-2i\pi S_{in} ),
\eeq
where $S_{in}$ is an induced spin of the anyon, we recover the identification from (\ref{CSS16}). Hence the axial part of the vertex correction (\ref{CSS12}) comes from the Pauli interaction of the induced spin of the anyon with the external magnetic field. The total spin $S$ and the spin magnetic moment $\mu_{S}$ of the anyon are, respectively:
\beq \label{CSS19}
S=\frac{1}{2}+S_{in}=\frac{1}{2}-\frac{e^{2}}{4\pi\mu} \qquad \mu_{S}=\frac{e}{m}\left(\frac{1}{2}-\frac{e^{2}}{4\pi\mu}\right).
\eeq

The induced spin $S_{in}$ can be understood as an orbital angular momentum of the classical flux-charge composite particle. Imagine a point magnetic flux situated in the origin, which we are turning on adiabatically from $\Phi_{0}=0$ to some final value $\Phi$. Consider also a charged particle, confined to the circle of some radius $R$ with its center at the origin, which can move without friction on the circle. The motion of the particle is governed by the Newton and the Maxwell equations:
\beq \label{CSS20}
\dot{\mathbf{p}}=e\mathbf{E} \qquad \frac{\partial B}{\partial t}=\epsilon_{jk}\nabla^{j}E^{k}\Longrightarrow \frac{d\Phi}{dt}=\oint d\mathbf{r}\cdot \mathbf{E},
\eeq
where $e$ denotes the charge of the orbiting particle and we used the Stokes theorem in the second expression. The change of the magnetic flux generates the angular electric field, which accelerates the charged particle and gives it a relative angular momentum $L$:
\beq \label{CSS21}
L=\mathbf{r}\times\mathbf{p}=\frac{e\Phi}{2\pi}.
\eeq
In the Chern-Simons theory (\ref{CSS1}) the magnetic flux $\Phi=-\frac{e}{\mu}$ is attached to every particle, thus for the spin of the composite we obtain:
\beq \label{CSS22}
L=-\frac{e^{2}}{2\pi\mu},
\eeq
This result differs by a factor of 2 from (\ref{CSS16}). This subtlety was resolved in \cite{Wilczek:1988}.  

In general the inverse of the corrected fermion propagator is given by (\ref{FP1}) with $\Sigma(p)$:
\begin{eqnarray}
\Sigma_{2}(p)&=&-ie^{2}\int\frac{d^{3}k}{(2\pi)^{3}}\gamma_{\mu}\mathscr{S}(p+k)\gamma_{\nu}\mathscr{D}^{\mu\nu}(k) \nonumber \\
&=&\frac{-e^{2}}{\mu}\int\frac{d^{3}k}{(2\pi)^{3}}\epsilon^{\mu\nu\rho}k_{\rho}\frac{\gamma_{\mu}(\slashed{p}+\slashed{k}+m)\gamma_{\nu}}{[(p+k)^{2}-m^{2}+i\epsilon][k^{2}+i\epsilon]} \label{CSS23}
\end{eqnarray}
at the one-loop level. The integral is UV divergent, so we use the dimensional regularization to calculate it. The result is:
\beq \label{CSS24}
\Sigma_{2}(p)=\frac{e^{2}}{4\pi\mu}\left(2m-(\slashed{p}-m)\frac{\slashed{p}}{m}\left[\frac{m^{2}}{p^{2}}+\frac{2m}{p}\left(1-\frac{m^{2}}{p^{2}} \right)\ln\left(\frac{1+\frac{p}{m}}{1-\frac{p}{m}} \right) \right] \right),
\eeq
where $p=\sqrt{p^{2}}$. Due to (\ref{FP17}) the unrenormalized corrected fermion propagator $S^{\prime}(p)$ near the pole is:
\beq \label{CSS25}
S^{\prime}(p)=\frac{iZ_{2}}{\slashed{p}-m_{ph}} \qquad Z_{2}=1+\frac{\Sigma_{2}(p)}{dp}|_{p=m}+O(e^{4}).
\eeq
Straightforward calculations lead us to:
\beq \label{CSS26}
Z_{2}=1+\frac{e^{2}}{4\pi\mu}+O(e^{4}),
\eeq
hence we have verified that up to one loop $Z_{1}=Z_{2}$, which is the celebrated \emph{Ward identity}.

Finally, the inverse of the unrenormalized corrected Chern-Simons propagator is given by (\ref{VP3}). The polarization tensor $\Pi_{\mu\nu}$ in the MCS theory was calculated to one-loop order in Chapter \ref{ch:perturbative} (\ref{VP3a}), (\ref{VP9}) and it equals the polarization tensor in the pure Chern-Simons theory. The corrected unrenormalized CS propagator $D^{\prime}_{\mu\nu}(q)$ to one-loop order is:
\beq \label{CSS27}
D^{\prime}_{\mu\nu}(q)=\frac{-i\Pi^{(1)}(q)}{\mu^{2}q^{2}}\left(g_{\mu\nu}-\frac{q_{\mu}q_{\nu}}{q^{2}}  \right)-\left(1-\frac{m}{\mu}\Pi^{(2)}(q) \right)\frac{\epsilon_{\mu\nu\rho}q^{\rho}}{\mu q^{2}},
\eeq
with $\Pi^{(1)}(q)$ and $\Pi^{(2)}(q)$ given by (\ref{VP9}). The one-loop corrections generate the transverse part of the Chern-Simons propagator. The renormalization is performed by the substitution (\ref{RC4}), where the counterterms $\delta_{A}$ and $\delta_{\mu}$ are fixed by some renormalization conditions. For example the renormalization condition might be:
\beq \label{CSS28}
D_{\mu\nu}(q)|_{q^{2}\to 0}=-\frac{1}{\mu}\frac{\epsilon_{\mu\nu\rho}q^{\rho}}{q^{2}},
\eeq
where $D_{\mu\nu}(q)$ is the renormalized Chern-Simons propagator. In our problem, we are interested in the behavior of the renormalized propagator for the small momentum transfer $q^{2}<0, \mathbf{q}^{2}<<m^{2}$. The scalar functions $\Pi^{(1)}(q)$ and $\Pi^{(2)}(q)$ from (\ref{VP9}) can be easily expanded up to the order $O(q^{4})$:
\beq \label{CSS29}
\Pi^{(1)}(q)=\frac{e^{2}q^{2}}{12\pi m}+O(q^{4}) \qquad \Pi^{(2)}(q)=\frac{e^{2}}{4\pi m}\left(1+\frac{q^{2}}{12m^{2}} \right)+O(q^{4}). 
\eeq
The first terms in the expansion, of course, coincide with (\ref{VP10a}) and are compensated by the counterterms. Thus for the renormalized Chern-Simons propagator for $q^{2}<<m^{2}$ we get:
\beq \label{CSS30}
D_{\mu\nu}(q)=-\left(1-\frac{e^{2}}{48\pi\mu}\frac{q^{2}}{m^{2}} \right)\frac{\epsilon_{\mu\nu\rho}q^{\rho}}{\mu q^{2}}+O(q^{2}).
\eeq

Let us remind at this point that our main aim in this section is to calculate the non-relativistic limit (neglecting $\frac{\mathbf{P}^{2}}{m^{2}}$ terms, where $\mathbf{P}=\mathbf{p},\mathbf{p^{\prime},\mathbf{q}}$) of the scattering amplitude of two distinguishable anyons in their center of mass frame. The diagrams are given in Fig \ref{graph14}. As we have already demonstrated, the renormalized version of the full CS propagator (\ref{CSS30}) receives quantum corrections from the fermion loop, which are of order $\frac{\mathbf{q}^{2}}{m^{2}}$. These are negligible in the classical limit, hence the diagram F (Fig. \ref{graph14}) does not contribute. We shall calculate the diagrams A-C (Fig. \ref{graph14}) all together by substituting the bare vertex $-ie\gamma_{\mu}$ with the renormalized corrected vertex $-ie\Gamma_{\mu}$. The contribution of the diagrams A-C to the scattering amplitude $M_{A-C}$ can be written in the following form:
\beq \label{CSS31}
M_{A-C}=-\frac{ie^{2}}{\mu}J_{\mu}(p,p^{\prime})\frac{\epsilon^{\mu\nu\rho}q_{\rho}}{q^{2}}J_{\nu}(k,k^{\prime}),
\eeq 
where $q=p^{\prime}-p$ and the currents $J_{\mu}(p,p^{\prime})\equiv\bar{u}(p^{\prime})\Gamma_{\mu}(p,p^{\prime})u(p)$ with $\Gamma_{\mu}(p,p^{\prime})$ given by (\ref{CSS12}). It is easy to compute the currents $J_{\mu}(p,p^{\prime})$ using (\ref{BE5}) and substituting it into (\ref{CSS31}) we get for the non-relativistic scattering amplitude $f_{A-C}$:
\beq \label{CSS32}
f_{A-C}\equiv -\frac{M_{A-C}}{4m^{2}}=-\frac{e^{2}}{\mu m}\left[ \left(1+\frac{e^{2}}{2\pi \mu} \right)+\frac{2i\mathbf{p}\times \mathbf{q}}{\mathbf{q}^{2}} \right].
\eeq

Last but not least are the ladder diagrams D and E from Fig. \ref{graph14}. The computations of these diagrams are tedious, hence here we only present the result obtained in \cite{Szabo:1993}. In \cite{Szabo:1993} the authors investigated the radiative corrections to the non-relativistic anyon-anyon scattering. It was demonstrated in \cite{Szabo:1993} that the ladder diagrams contribution is $M_{D,E}=O(\frac{\mathbf{P}^{2}}{m^{2}})$, where $\mathbf{P}=\mathbf{p},\mathbf{p^{\prime}},\mathbf{q}$. This is negligible in the non-relativistic limit.

To summarize, the scattering of two anyons up to one loop is described by the Feynman diagrams in Fig. \ref{graph14}. The non-relativistic scattering amplitude is given by:
\beq \label{CSS33}
f\equiv -\frac{M}{4m^{2}}=-\frac{e^{2}}{\mu m}\left[ \left(1+\frac{e^{2}}{2\pi \mu} \right)+\frac{2i\mathbf{p}\times \mathbf{q}}{\mathbf{q}^{2}} \right],
\eeq
which entirely comes from the diagrams A, B and C. The diagrams D, E and F do not contribute in the non-relativistic limit. The last expression almost coincides with the scattering amplitude (\ref{CS27}), obtained in the Pauli theory. The only difference is in its real part. Recall that the real part of (\ref{CS27}) had its origin in the Pauli term in the Lagrangian (\ref{CS19}). As we have just shown, the one-loop vertex corrections modify the spin and the magnetic moment of the fermion (\ref{CSS19}). The real part of (\ref{CSS33}) corresponds \emph{exactly} to this modifications, i.e., it comes from the Pauli interaction term $\frac{eg}{2m}B\mu_{S}$ with $\mu_{S}$ given by (\ref{CSS19}).

In conclusion, in this chapter we computed the one-loop corrections to the non-relativistic two-anyon scattering. In the Pauli theory these corrections \emph{vanish}, while in the low energy limit of the Chern-Simons theory there is non-trivial correction coming from the \emph{modified magnetic moment}. Hence, we see that taking the non-relativistic limit (Pauli) of the classical relativistic-invariant theory (CS) and then quantizing it yield a different result to quantizing the theory first and then taking the low energy limit.   
\chapter{Quantum Hall Effect} \label{ch:QHE}
The final chapter of this thesis is devoted to a review of the most successful application of the planar quantum physics- the quantum Hall effect\footnote{Note that at first sight this Chapter is only vaguely connected to the rest of the work. It describes the phenomenon governed by the non-relativistic many-body Hamiltonian. The main reason we have included it in the thesis is that it provides clear and experimentally confirmed evidence of \emph{anyons}, which popped up here and there throughout the previous chapters.}. First we briefly describe this interesting quantum phenomenon. Subsequently the dynamics of a two-dimensional electron gas is examined in some detail. We present the heuristic understanding of both the integer (IQHE) and $\nu=\frac{1}{m}$  fractional (FQHE) quantum Hall effects\footnote{Let us stress that in this work we do not consider FQHE with more general fractions $\nu=\frac{n}{m}$, which were found experimentally. More refined theory is needed to understand these more general fractions.}. The special attention is paid to the Laughlin quasiparticles in the FQHE. Using the Berry's phase technique it is demonstrated that these particles are anyons.  The review is based on the excellent sources \cite{Wilczek:1990, Girvin:1998, Ezawa:2000, Yennie:1987}. 
\section{Basics}
Consider a two-dimensional electron gas (2DEG), i.e., electrons confined to the $xy$ plane, in a homogeneous external magnetic field\footnote{Although QHE is described effectively by the planar physics, it belongs to the realm of the real world with three spatial dimensions. For this reason we use the three-component spatial vectors in this Chapter.} $\mathbf{B}=(0,0,-B_{\bot})$ and a homogeneous external electric field $\mathbf{E}$. Let us assume that the stationary current $\mathbf{j}=\mathsf{const}$ flows in the system and neglect the internal interaction between the electrons. The equation of motion of an electron is:
\beq \label{QHE1}
0=m\mathbf{a}=-e\mathbf{E}-e\mathbf{v}\times\mathbf{B},
\eeq 
where the charge of electron was denoted by $-e$. The stationary current $\mathbf{j}$ can be expressed in terms of the density $\rho$ of the 2DEG and its velocity $\mathbf{v}$ as:
\beq \label{QHE2}
\mathbf{j}=-e\rho \mathbf{v}.
\eeq  
Substituting it into (\ref{QHE1}) we obtain the Ohm's law:
\beq \label{QHE3}
E^{k}=\underbrace{-\frac{B_{\bot}}{e\rho}\epsilon^{kl}}_{r^{kl}}j^{l},
\eeq
where $\epsilon^{ij}$ is the two-dimensional antisymmetric Levi-Civita symbol and $r^{kl}$ is the resistivity tensor, which has the following matrix form:
\beq \label{QHE4}
r=\frac{B_{\bot}}{e\rho}\left(\begin{array}{cc} 0 & -1 \\
																								 1 & 0
																\end{array}\right).								 
\eeq
The current flows perpendicular to the direction of the external electric field, i.e., $r^{xx}=r^{yy}=0$. The absolute value of the non-diagonal element $r^{xy}=-r^{yx}$ called the Hall resistivity $r_{H}$ grows linearly with $B_{\bot}$. This effect was discovered by E. Hall in 1879.

The conductivity tensor $\sigma^{kl}$ is defined by:
\beq \label{QHE5}
j^{k}=\sigma^{kl}E^{l} \qquad \sigma=r^{-1}
\eeq
and in our case looks as:
\beq \label{QHE6}
\sigma^{kl}=\frac{e\rho}{B_{\bot}}\left(\begin{array}{cc} 0 & 1 \\
																								 -1 & 0
																\end{array}\right).
\eeq
Due to the non-vanishing Hall resistivity the system seems to be the perfect conductor ($r^{xx}=r^{yy}=0$) and the perfect insulator ($\sigma^{xx}=\sigma^{yy}=0$) at the same time.

\begin{figure}[ht]
\begin{center}
\includegraphics[width=0.9\textwidth, height=0.6\textwidth]{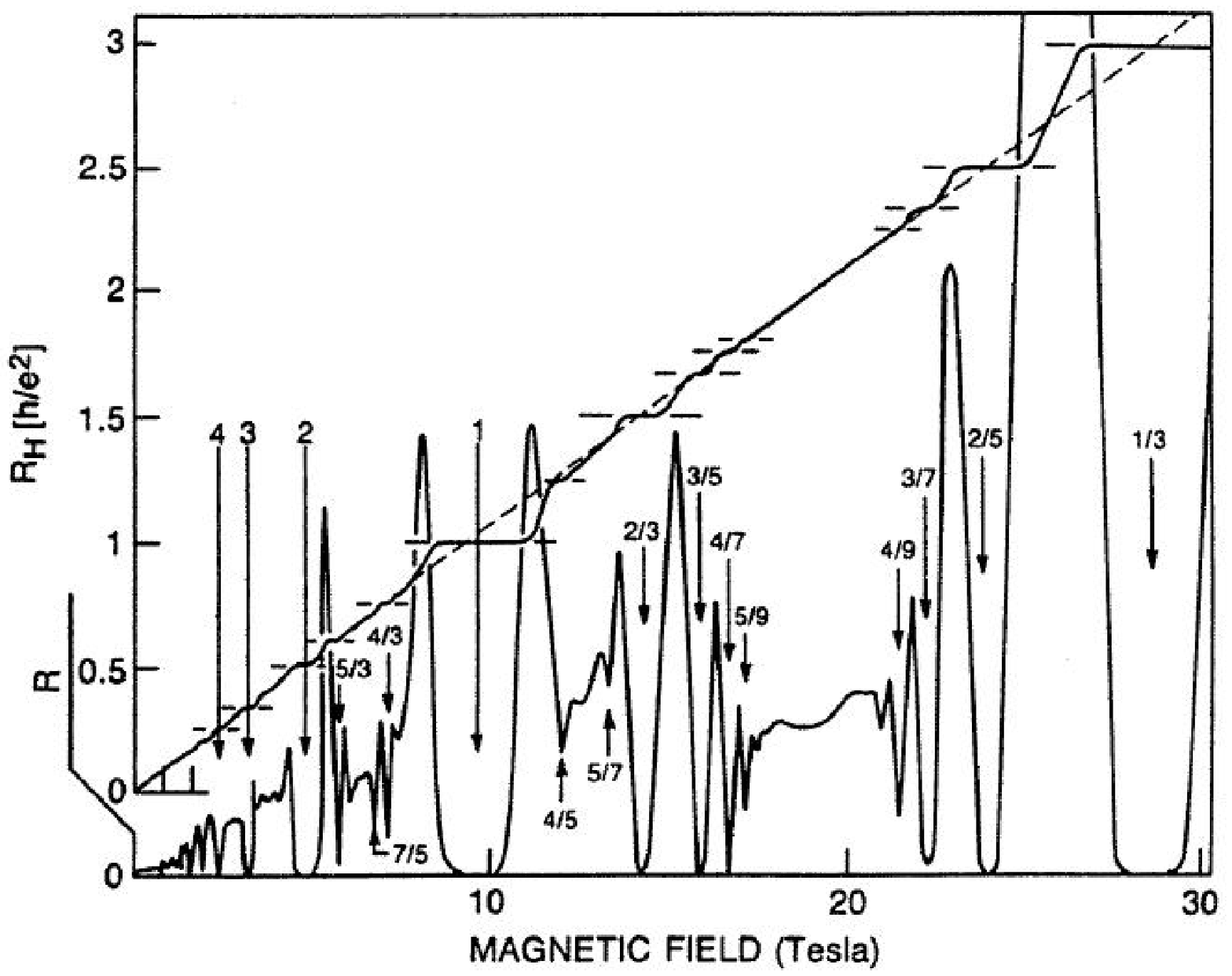}
\caption{The Hall resistivity $R_{H}$ and dissipative resistivity $R$ as functions of the external magnetic field \cite{Girvin:1998}}\label{graph17}
\end{center}
\end{figure}

The experiments performed with the 2DEG in the strong magnetic field at low temperatures show remarkably \emph{different} behavior of the Hall resistivity. The experimental data are shown in Fig.\ref{graph17}.
The Hall resistivity $R_{H}$:
\beq \label{QHE7}
R_{H}=\frac{B_{\bot}}{e\rho}=\frac{1}{\nu}\frac{2\pi \hbar}{e^{2}} \qquad \nu=\frac{2\pi \hbar \rho}{eB_{\bot}}
\eeq
exhibits a series of \emph{plateaus} at the special values of $\nu$. The Hall plateaus are associated with the dips in the dissipative resistivity. It is worth mentioning that the dissipative resistivity $R$ decreases by $13$ orders of magnitude in the plateau regions (see Fig. \ref{graph17}). Quite remarkably the special "magic" values of $\nu$ are the \emph{integers} (with precession $10^{-10}$). This is known as the integer quantum Hall effect (IQHE). The IQHE was experimentally discovered by K. von Klitzing in 1980, who was awarded the Nobel prize in 1985 for this finding. The plateaus are found also for \emph{fractional} values of $\nu=\frac{n}{m}$ with $n$ integer and $m$ odd integer. This is the so called fractional quantum Hall effect (FQHE). It was first observed in 1982 by D. Tsui, H. St\"ormer and A.C. Gossard and B. Laughlin proposed the theoretical description of the phenomenon (for special fractions) in 1983. D. Tsui, H. St\"ormer and B. Laughlin received  the Nobel prize in 1998.

Two remarks are in order at this point. First, the QHE is a very universal phenomenon- it does not depend on the microscopic structure of the system. Due to this fact and the high precision of the quantum Hall quantization it is now used for exact determination of the fine structure constant $\alpha$. Second, although it might seem at the first sight that the IQHE and FQHE have similar theoretical explanation this is not the case. In the following sections we demonstrate it describing the theory of the IQHE and FQHE. 

\section{Quantum dynamics of 2DEG}
Let us try to find a quantum description of the N-electron 2DEG in the external magnetic field $\mathbf{B}=(0, 0, -B_{\bot} )$. The total Hamiltonian of the system $H$ consists of the kinetic part $H_{K}$, the Coulomb part $H_{C}$ and the Zeeman part $H_{Z}$:
\begin{eqnarray} 
H&=&H_{K}+H_{C}+H_{Z}, \label{QHE8} \\ 
H_{K}&=&\frac{1}{2m}\sum_{i=1}^{N}(\mathbf{p}_{i}+e\mathbf{A}(\mathbf{x}_{i}))^{2}, \label{QHE9} \\
H_{C}&=&\frac{e^{2}}{2}\int d^{2}x d^{2}x^{\prime}\frac{\rho(\mathbf{x})\rho(\mathbf{x^{\prime}})}{4\pi \epsilon |\mathbf{x}-\mathbf{x^{\prime}}|}, \label{QHE10} \\
H_{Z}&=&-\frac{1}{2}g\mu_{B}B_{\bot}\int d^{2}x(\rho^{+}(\mathbf{x})-\rho^{-}(\mathbf{x})), \label{QHE11}
\end{eqnarray}
where $\rho(\mathbf{x})$, $\rho^{+}(\mathbf{x})$ and $\rho^{-}(\mathbf{x})$ are the total, the up-spin and the down-spin electron densities respectively; $\mu_{B}$ is the Bohr magneton and $g$ is a g-factor of the electron. To simplify the analysis we assume that the Zeeman gap $\Delta_{Z}=g\mu_{B}B_{\bot}$ is much larger than all other energy scales involved and thus the electrons are frozen in the up-spin states. The Hamiltonian $H$ (without $H_{Z}$) is a very complicated many-body Hamiltonian. It is sufficient to describe both IQHE and FQHE.

To warm up we neglect the Coulomb interaction and look at the \emph{one-particle} Hamiltonian:
\beq \label{QHE12}
\mathscr{H}=\frac{1}{2m}(\mathbf{p}+e\mathbf{A}(\mathbf{x}))^{2}=\frac{1}{2m}\left\{(p^{x}+eA^{x}(\mathbf{x}))^{2}+(p^{y}+eA^{y}(\mathbf{x}))^{2}\right\}.
\eeq
To find the eigenstates it is necessary to fix the gauge, but as we shall show now a large amount of work can be done before gauge fixing. Let us introduce the mechanical momenta $\mathscr{P}^{x}$, $\mathscr{P}^{y}$ and the new coordinates $\mathscr{X}$, $\mathscr{Y}$:
\begin{eqnarray}
\mathscr{P}^{x}=p^{x}+eA^{x}(\mathbf{x}) & \quad & \mathscr{P}^{y}=p^{y}+eA^{y}(\mathbf{x}) \nonumber \\
\mathscr{X}=x+\frac{1}{eB_{\bot}}\mathscr{P}^{y} & \quad & \mathscr{Y}=y-\frac{1}{eB_{\bot}}\mathscr{P}^{x} \label{QHE13} 
\end{eqnarray}
These operators satisfy the following commutation relations:
\beq \label{QHE14}
[\mathscr{P}^{x},\mathscr{P}^{y}]=i\frac{\hbar^{2}}{l_{B}^{2}} \qquad [\mathscr{X},\mathscr{Y}]=-il_{B}^{2}
\eeq 
\beq \label{QHE15}
[\mathscr{X},\mathscr{P}^{x}]=[\mathscr{X},\mathscr{P}^{y}]=[\mathscr{Y},\mathscr{P}^{x}]=[\mathscr{Y},\mathscr{P}^{y}]=0,
\eeq
where $l_{B}=\sqrt{\frac{\hbar}{eB_{\bot}}}$ is the magnetic length- the natural length scale in the system. Two pairs of the hermitian conjugate operators can now be defined:
\begin{eqnarray}
a=\frac{l_{B}}{\sqrt{2}\hbar}\left(\mathscr{P}^{x}+i\mathscr{P}^{y}\right) & \quad & a^{\dag}=\frac{l_{B}}{\sqrt{2}\hbar}\left(\mathscr{P}^{x}-i\mathscr{P}^{y}\right) \nonumber \\
b=\frac{1}{\sqrt{2}l_{B}}\left(\mathscr{X}-i\mathscr{Y}\right) & \quad & b^{\dag}=\frac{1}{\sqrt{2}l_{B}}\left(\mathscr{X}+i\mathscr{Y}\right). \label{QHE16} 
\end{eqnarray} 
These operators satisfy the commutation relations of two independent pairs of creation and annihilation operators:
\beq 
[a,a^{\dag}]=1 \quad [b,b^{\dag}]=1 \quad
[a,b]=0 \quad [a,b^{\dag}]=0. \label{QHE17}
\eeq
The Hilbert space of the system is obtained by multiple application of the creation operators $a^{\dag}$ and $b^{\dag}$ on the vacuum state $\ket{0}$. The Hamiltonian $\mathscr{H}$ can be expressed in terms of the creation and annihilation operators:
\beq \label{QHE18}
\mathscr{H}=\hbar \omega_{C}(a^{\dag}a+\frac{1}{2}),
\eeq
where $\omega_{C}=\frac{\hbar}{ml_{B}}$ is the so called cyclotron frequency. We obtained the celebrated equidistant Landau levels. These are infinitely degenerate because application of $b^{\dag}$ on an arbitrary state does not change the energy.

It is instructive to give the physical interpretation to the operators $\mathscr{X}$ and $\mathscr{Y}$. The position operators are decomposed as in (\ref{QHE13}):
\beq \label{QHE19}
x=\mathscr{X}\underbrace{-\frac{1}{eB_{\bot}}\mathscr{P}^{y}}_{\mathscr{R}^{x}} \quad y=\mathscr{Y}\underbrace{+\frac{1}{eB_{\bot}}\mathscr{P}^{x}}_{\mathscr{R}^{y}}.
\eeq
This decomposition has a physical meaning because ($\mathscr{X}$, $\mathscr{Y}$) part is the constant of motion:
\beq \label{QHE20}
i\hbar\frac{d \mathscr{X}}{dt}=[\mathscr{X},\mathscr{H}]=0 \qquad i\hbar\frac{d \mathscr{Y}}{dt}=[\mathscr{Y},\mathscr{H}]=0
\eeq
and the expectation values $<\mathscr{X}>$ and $<\mathscr{Y}>$ do not evolve in time. On the other hand, the ($\mathscr{R}^{x}$,$\mathscr{R}^{y}$) is a dynamical part of the coordinates. Their equations of motions are:
\beq \label{QHE21}
\frac{d\mathscr{R}^{x}}{dt}=\frac{\hbar}{ml_{B}^{2}}\mathscr{R}^{y} \quad \frac{d\mathscr{R}^{y}}{dt}=-\frac{\hbar}{ml_{B}^{2}}\mathscr{R}^{x}
\eeq
\beq
\mathscr{H}=\frac{\hbar\omega_{C}}{2l_{B}}\mathscr{R}^{2}.
\eeq
Hence the expectation value of ($\mathscr{R}^{x}$, $\mathscr{R}^{y}$) perform the circular motion with the cyclotron frequency with the center at $(<\mathscr{X}>,<\mathscr{Y}>)$:
\beq \label{QHE22}
<\mathscr{R}^{x}>=\sqrt{2n+1}l_{B}\cos(\omega_{c}t) \quad <\mathscr{R}^{y}>=\sqrt{2n+1}l_{B}\sin(\omega_{c}t)
\eeq
for the states from the $n^{th}$ Landau level.

If we switch on the planar homogeneous electric field, the one-particle Hamiltonian $\mathscr{H}$ is modified:
\beq \label{QHE23}
\mathscr{H}\to \mathscr{H}+e\mathbf{x}\cdot \mathbf{E}.
\eeq
The Heisenberg equations of motion for the operators $\mathscr{X}$ and $\mathscr{Y}$ now read:
\beq \label{QHE24}
\frac{d\mathscr{X}}{dt}=-\frac{E^{y}}{B_{\bot}} \qquad \frac{d\mathscr{Y}}{dt}=-\frac{E^{x}}{B_{\bot}}
\eeq
and the non-vanishing drift velocity appears in the system. The current density flowing in the electron gas is:
\beq \label{QHE25}
j^{x}\equiv -e\rho\dot{\mathscr{X}}=\frac{e\rho}{B_{\bot}}E^{y} \qquad j^{y}\equiv -e\rho\dot{\mathscr{Y}}=-\frac{e\rho}{B_{\bot}}E^{x}, 
\eeq
which looks exactly the same as the classical results (\ref{QHE5}),(\ref{QHE6}).

As has been already mentioned above, to find the eigenstates of the Hamiltonian (\ref{QHE12}) it is necessary to fix the gauge. Two choices of gauge are especially convenient for the computations- the symmetric gauge and the Landau gauge. The Landau quantization is well-known \cite{Landau:QM}. One-particle quantization in the symmetric gauge, which is particularly suitable for the QHE analysis, is reviewed in Appendix \ref{appendB}.

Having succeeded in analysis of the one-particle problem with the Hamiltonian (\ref{QHE12}) let us move one step forward and take a look at the kinetic part $H_{K}$ (\ref{QHE9}) of the total many-particle Hamiltonian. We limit our considerations to the lowest Landau level (LLL) states. What is the density of states of the LLL? Due to symmetry the symmetric gauge is useful for analysis of the QHE of the disk geometry, hence imagine that we are putting electrons on the disk of some macroscopic radius $R$. In the Appendix {\ref{appendB}} we show that in the symmetric gauge the one-particle states with definite angular momentum $n$ have non-vanishing probability density located on the ring-bands with the radius $r_{n}=\sqrt{2n}l_{B}$ and the width of the order $l_{B}$. Thus the mean density of states for the fully filled Landau level is:
\beq \label{QHE26}
\rho_{DS}=\frac{N}{S}=\frac{R^{2}}{2l_{B}^{2}\pi R^{2}}=\frac{1}{2\pi l_{B}^{2}},
\eeq 
where $N=\frac{R^{2}}{2l_{B}^{2}}$ is a maximum number of electrons in the LLL, which can be put on the disk of radius $R$ and $S=\pi R^{2}$. It is worth noting here that the "magic" numbers $\nu$ were defined in (\ref{QHE7}) as:
\beq \label{QHE27}
\nu=\frac{2\pi \hbar \rho}{e B_{\bot}}=\frac{\rho}{\rho_{DS}}.
\eeq
Hence the "magic" number $\nu$ has a physical meaning. It is a Landau level \emph{filling factor} of the many-body state.

\section{IQHE}
The tower of plateaus in the IQHE starts at $\nu=1$, i.e., it corresponds to the filled LLL \\ state\footnote{More precisely it corresponds to the fully-filled up-spin frozen LLL.}. In order to understand the phenomenon, it is desirable to construct the many-body wave function of this state. The construction is straightforward in the symmetric gauge, if we neglect the Coulomb interaction. The unnormalized wave function $\Psi(z_{1},z_{2},..z_{N})$ is given by the Slater determinant, constructed from the one-particle wave functions $\psi^{n}_{L=0}(z)$ (\ref{sg7}):
\beq \label{QHE28}
\Psi(z_{1},z_{2},..z_{N})=\underbrace{\left| \begin{array}{ccccc} 
																	1&z_{1}&.&.&z_{1}^{N-1} \\
																	1&z_{2}& & &. \\
																	.& & .& & \\
																	.& & &. & \\
																	1&z_{N}&&.&z_{N}^{N-1}
                                 \end{array}\right|}_{V}\exp(-\sum_{i=1}^{N}|z_{i}|^{2}), 
\eeq 
where $N$ is a maximum number of electrons in the LLL. The determinant $V$ is the well-known Vandermonde determinant, which equals $V=\prod_{i>j}(z_{i}-z_{j})$. Hence the wave function of the filled LLL $\Psi$ can be written as:
\beq \label{QHE29}
\Psi(z_{1},z_{2}...z_{N})=\mathscr{N}\prod_{i>j}(z_{i}-z_{j})\exp(-\sum_{k=1}^{N}|z_{k}|^{2}),
\eeq
where $\mathscr{N}$ is a normalization factor. The wave function (\ref{QHE29}) plays the \emph{central} role in understanding of the IQHE. Although we have constructed (\ref{QHE29}) neglecting the Coulomb interaction among the electrons, it, quite amazingly, takes into account the interaction: the polynomial part of (\ref{QHE29}) holds electrons apart, while the exponential factor is small, if the electrons are spread too much. It is possible to show (using plasma analogy \cite{Wilczek:1990, Girvin:1998}) that the wave function (\ref{QHE29}) favors (in the sense of the probability density) the configurations with the \emph{homogeneous} density distribution $\rho_{DS}=\frac{1}{2\pi l_{B}^{2}}$. Now we are in a position to ask the main questions of the IQHE and to comment on them. The questions are:
\begin{itemize}
\item Why is the Hall resistance at $\nu=1$ quantized to very high accuracy?
\item Why there is a dip in the dissipative resistivity at $\nu=1$?
\item Why does a plateau appear around $\nu=1$?
\end{itemize}   
Here we present short answers and refer the reader to the relevant literature:
\begin{itemize}
\item There are various quite different reasonings, which are used to answer this question. We shall describe neither of them, but refer the reader to the relevant literature. It is believed that the exactness is due to the "edge states" \cite{Girvin:1998}. Alternatively, the topological reasons are used to explain the accuracy of the Hall resistance \cite{Fradkin:1991}.    
\item The wave function (\ref{QHE29}) describes the \emph{filled} Landau level. Thus for electrons it is not possible to scatter \emph{inelastically} (all states are filled)- this corresponds to the vanishing dissipative resistivity \cite{Yennie:1987}.
\item Two previous answers worked only for the \emph{exact} filling $\nu=1$. The experiments show, however, the existence of a plateau: the Hall resistivity does not change by small variation of the filling factor around $\nu=1$ (by changing either magnetic field $B_{\bot}$ or the electron density $\rho$). It is believed that the Hall plateaus are explained by \emph{impurities} (disorder) in the sample. In the presence of impurities the one-particle states of electrons can be divided into two groups: extended states and spatially localized states, which sit at the impurities. The localized states, of course, do not carry the current. If the Fermi energy level of the sample is among the localized states, the variation of the Fermi energy level do not change the Hall current. As the result a plateau appears around $\nu=1$\cite{Wilczek:1990, Ezawa:2000, Girvin:1998}.   
\end{itemize}
It should be mentioned here that the IQHE with $\nu>1$ is described in a similar fashion using as a ground state the state with $\nu$ filled Landau levels.

\section{FQHE}
In the previous section the ground state (\ref{QHE29}) for $\nu=1$ IQHE was constructed by combining antisymmetrically all possible one-particle wave-functions $\psi_{L=0}^{n}$ from the lowest Landau level. It has eventually turned out that (\ref{QHE29}) takes into account the Coulomb interaction. We argue that neglecting the states from the higher Landau levels\footnote{which can be done if the characteristic potential energy $V\thicksim \frac{e^{2}}{l_{B}}$ is much less than the gap between the Landau levels $\hbar \omega_{C}$}, the many-body state (\ref{QHE29}) is a ground state for \emph{every} reasonable potential energy. The reason is simple: the $N$-particle fermionic state is given \emph{uniquely} in the $N$-dimensional LLL Hilbert space.

Experiments, however, show that the Hall plateaus also appear for some \emph{fractional} filling factors $\nu$. The Coulomb interaction plays a decisive role in understanding of the FQHE. It is obvious that the ground state of the kinetic part of the Hamiltonian $H_{K}$ of the partially filled LLL is degenerate. Therefore the Coulomb interaction determines the ground state of the total Hamiltonian $H$.

R. Laughlin \cite{Laughlin:1983} proposed a trial function for the ground state of the FQHE for $\nu=\frac{1}{m}$:
\beq \label{QHE30}
\Psi^{m}(z_{1},z_{2}...z_{N})=\mathscr{N}^{(m)}\prod_{i>j}(z_{i}-z_{j})^{m}\exp(-\sum_{k=1}^{N}|z_{k}|^{2}),
\eeq 
where $\mathscr{N}^{(m)}$ is a normalization factor. Antisymmetry forces $m$ to be an odd integer, i.e., $m=1,3,5,...$. Using plasma analogy Laughlin demonstrated \cite{Laughlin:1983} that the state (\ref{QHE30}) prefers highly homogeneous probability distribution density $\rho=\frac{1}{2\pi l^{2}_{B} m}$, which really corresponds to the filling factor $\nu=\frac{1}{m}$. Moreover, the numerical simulations of systems with small number of electrons displayed excellent overlap of the Laughlin trial function (\ref{QHE30}) with the real ground state. 

The angular momentum operator for a many-body state with all electrons in the lowest Landau level is given by:
\beq \label{QHE31}
L=\sum_{i}\hbar b^{\dag}_{i}b_{i},
\eeq
where $b^{\dag}_{i}$ and $b_{i}$ act on the $i^{th}$ electron. The Laughlin trial wave-function (\ref{QHE30}) is the eigenstate of the operator (\ref{QHE31}) with the eigenvalue:
\beq \label{QHE32}
L^{(m)}=\hbar m \frac{N(N-1)}{2}.
\eeq
The maximal angular momentum of individual electrons $l_{emax}$ is:
\beq \label{QHE33}
l_{emax}=\hbar m (N-1),
\eeq
which defines the spatial size and the filling factor of the system. According to (\ref{sg11}) the electrons occupy the disk of radius $R=\sqrt{2m(N-1)}l_{B}$ with the mean density:
\beq \label{QHE34}
\rho=\frac{N}{2\pi m (N-1)l_{B}^{2}}\rightarrow\frac{1}{m}\rho_{DS} \qquad N>>1
\eeq 
with $\rho_{DS}$ denoting the mean density of the filled LLL. Thus, we have convinced ourselves that (\ref{QHE30}) really corresponds to the filling factor $\nu=\frac{1}{m}$.

Experiments and numerical studies indicate the uniqueness of the FQHE ground state and the finite-energy excitation gap. We shall assume this is true and follow the ingenious Gedankenexperiment due to R. Laughlin to derive the charged excitation\footnote{There appear to be also neutral collective gapful excitations \cite{Girvin:1998}, but we shall not discuss them in this work.} wave function from (\ref{QHE30}). Imagine an infinitely thin solenoid piercing the quantum Hall system at the origin of the coordinate frame with adiabatically\footnote{Adiabatically means that the flux changes at the time scale $\tau>>\frac{\hbar}{\Delta}$, where $\Delta$ is the excitation energy gap of the spectrum} increasing magnetic flux $\Phi$ from $\Phi_{i}=0$ to $\Phi_{f}=\pm\frac{2\pi\hbar}{e}$. Due to its uniqueness, the ground state (\ref{QHE30}) evolves as the eigenstate of the changing Hamiltonian. The solenoid does not produce any magnetic field in the plane, but its flux produces the Aharanov-B\"ohm (AB) phase change. However, if the flux of the solenoid is equal to $\pm\frac{2\pi\hbar}{e}$ (the Dirac flux quantum), the existence of the solenoid is totally unobservable. The AB phase shift in this case is $\pm 2\pi$. Hence, the final Hamiltonian is equivalent to the initial quantum Hall Hamiltonian (\ref{QHE8}) and the ground state (\ref{QHE30}) has evolved into the excited state. 

Putting the Dirac quantum flux by hand (without the adiabatic evolution) is equivalent to the singular gauge transformation of the one-particle wave-function $\psi(r,\phi)$:
\beq \label{QHE35}
\psi(r,\phi)\to \exp[iq\phi]\psi(r,\phi),
\eeq
where $q=+1$ for $\Phi_{f}=+\frac{2\pi \hbar}{e}$ and $q=-1$ for $\Phi_{f}=-\frac{2\pi \hbar}{e}$. For the polynomial part of the LLL function $\psi_{L=0}^{n}(z)$ (\ref{sg7}) the transformation yields:
\beq \label{QHE36}
z^{n}\to \exp[iq\phi]z^{n},
\eeq
which is not analytic and hence does not belong to the LLL. The LLL analyticity (\ref{sg5}) requires that for the adiabatic evolution of the flux:
\begin{eqnarray}
z^{n}\to z^{n+1} &\quad& q=1 \nonumber \\
z^{n}\to \frac{\partial}{\partial z}z^{n} &\quad& q=-1 \label{QHE37a}
\end{eqnarray}
The last transformation acts on the Laughlin trial state (\ref{QHE30}) as:  
\begin{eqnarray}
\ket{\Psi^{m}}\to \prod_{r}b^{\dag}_{r}\ket{\Psi^{m}} &\quad& q=1 \label{QHE37} \\
\ket{\Psi^{m}}\to \prod_{r}b_{r}\ket{\Psi^{m}} &\quad& q=-1 \label{QHE38}.
\end{eqnarray}
The transformation increases (resp. decreases) the angular momenta of all electrons in (\ref{QHE37}) (resp. (\ref{QHE38})) by one unit. The RHS of (\ref{QHE37}) (resp. (\ref{QHE38})) are called quasiholes (resp. quasielectrons) centered at the origin. We shall explain these names shortly by calculating the charge of the excitations. Using (\ref{sg2}) the last expressions can be rewritten as:
\begin{eqnarray}
\Psi^{m}(z_{1},z_{2},...,z_{N})\to \exp(-\sum_{k=1}^{N}|z_{k}|^{2})\prod_{r}z_{r}\prod_{i>j}(z_{i}-z_{j})^{m} &\quad& q=1 \nonumber \\
\Psi^{m}(z_{1},z_{2},...,z_{N})\to \exp(-\sum_{k=1}^{N}|z_{k}|^{2})\prod_{r}\frac{\partial}{\partial z_{r}} \prod_{i>j}(z_{i}-z_{j})^{m} &\quad& q=-1. \label{QHE39}
\end{eqnarray}

The excitations in (\ref{QHE39}) are localized around the origin of the coordinate system frame, but in fact their center can be shifted. The quasihole $\Psi^{m+}_{Z_{0}}$ (resp. quasielectron $\Psi^{m-}_{Z_{0}}$) with the center at $Z_{0}=x_{0}+iy_{0}$ is given by: 
\begin{eqnarray}
\Psi^{m+}_{Z_{0}}(Z_{1},Z_{2},...,Z_{N})&=& \exp(-\frac{\sum_{k=1}^{N}|Z_{k}|^{2}}{4l^{2}_{B}})\prod_{r}(Z_{r}-Z_{0})\prod_{i>j}(Z_{i}-Z_{j})^{m} \label{QHE40} \\
\Psi^{m-}_{Z_{0}}(Z_{1},Z_{2},...,Z_{N}) &=& \exp(-\frac{\sum_{k=1}^{N}|Z_{k}|^{2}}{4l^{2}_{B}})\prod_{r}(2\frac{\partial}{\partial Z_{r}}-\frac{Z^{*}_{0}}{l_{B}^{2}}) \prod_{i>j}(Z_{i}-Z_{j})^{m}, \label{QHE41}
\end{eqnarray}
where the dimensionful variables are used $Z_{i}=x_{i}+iy_{i}$ and $\frac{\partial}{\partial Z_{i}}=\frac{1}{2}(\frac{\partial}{\partial x_{i}}-i\frac{\partial}{\partial y_{i}})$ instead of our definition (\ref{sg3}). Numerical studies of systems with few electrons showed that the states (\ref{QHE40}) and (\ref{QHE41}) have a good overlap with the real localized excited states.

\begin{figure}[ht]
\begin{center}
\includegraphics[angle=0, width=0.5\textwidth, height=0.7\textwidth]{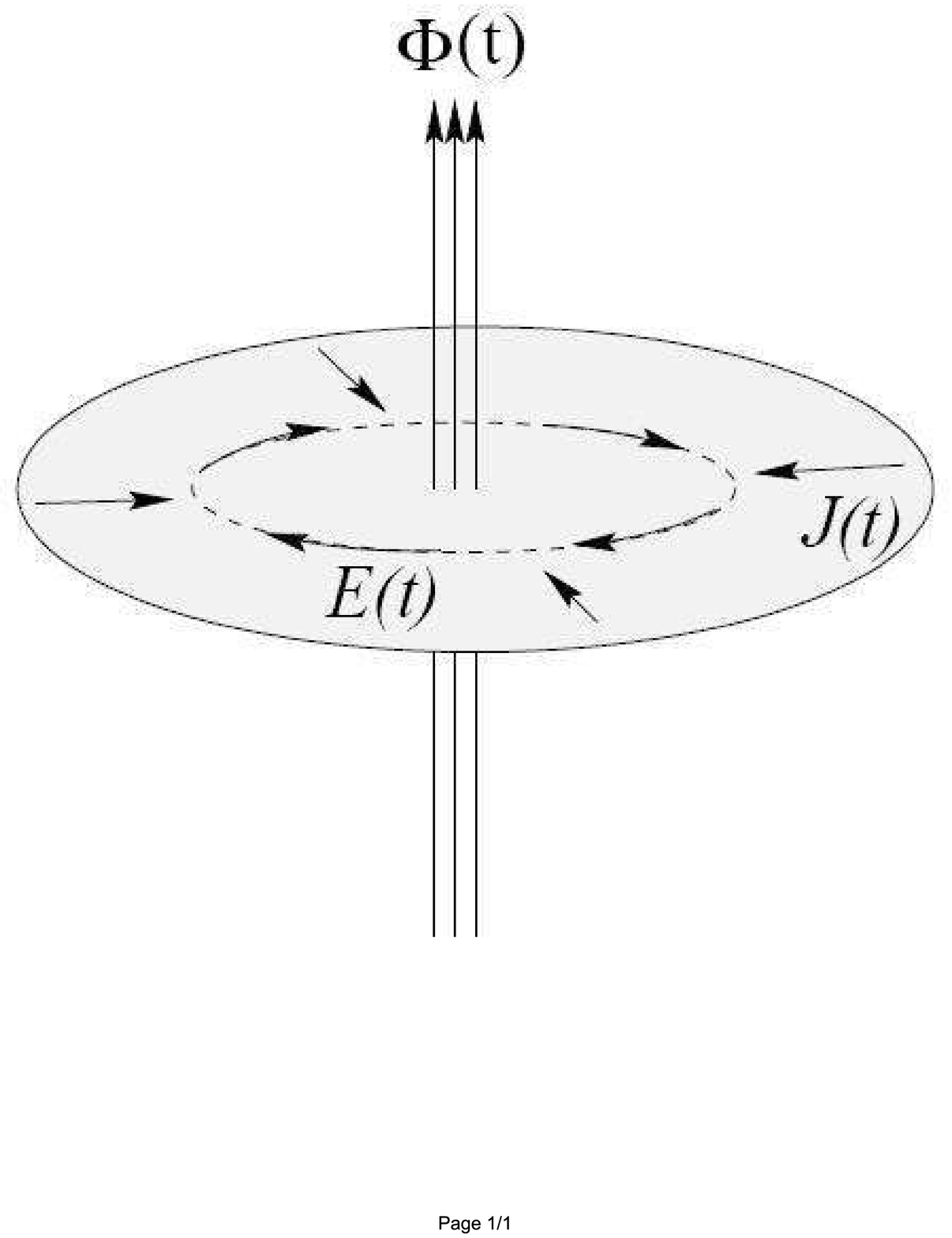}
\caption{Electric field and the Hall current due to the gedanken flux change \cite{Girvin:1998}}\label{graph18}
\end{center}
\end{figure}

What is the charge of the quasihole (\ref{QHE39})? To answer this question let us examine the gauge fields at large distance $L$ in our Gedankenexperiment. The changing flux $\Phi$ generates the circulation of the electric field $E^{\phi}$ in accordance with the Faraday's law (see Fig. \ref{graph18}):
\beq \label{QHE42}
2\pi L E^{\phi}=-\frac{\partial}{\partial t}\Phi,
\eeq
where we take $L>>l_{B}$.
The circulating electric field generates the radial Hall current (see Fig. {\ref{graph18}}):
\beq \label{QHE43}
J^{r}=\frac{e^{2}\nu}{2\pi \hbar}E^{\phi}.
\eeq
Substituting (\ref{QHE43}) into (\ref{QHE42}) and integrating in time (from $\Phi_{i}$ to $\Phi_{f}$) we readily get:
\beq \label{QHE44}
\frac{2\pi \hbar}{e^{2}\nu}\Delta q=\Phi_{D} \quad \Rightarrow \quad \Delta q=\nu e,
\eeq
where $\Delta q$ is a change of charge inside the circle of radius $L$ during the adiabatic evolution, i.e., $\Delta q$ is the charge of the excitation. The surprising fact is that the charge of the quasihole makes only a \emph{fractional} part of the absolute value of the electron charge! Similar consideration leads to the value $\Delta q=-\nu e$ for the quasielectron. It is encouraging that the Laughlin quasiparticles with the fractional electric charge have been recently discovered experimentally \cite{Goldman:2005, Goldman:2007}. 

\section{Anyons in FQHE}
In this final section we examine the main question of this chapter: What is the \emph{statistics} of the Laughlin quasiparticles (\ref{QHE40},\ref{QHE41})? The answer to this question can be found by calculating the Berry's phase \cite{Berry:1984} during the adiabatic transport of the quasiparticle.

Imagine that a quantum system is governed by the Hamiltonian $H_{\mathbf{R}(t)}$, which depends on some time-dependent parameter $\mathbf{R}(t)$. In the case of interest $\mathbf{R}(t)$ will be identified with the position of the quasihole $Z_{0}(t)$ (\ref{QHE40}). Assume that the solution of the time-independent Schr\"odinger equation can be found:
\beq \label{QHE45}
H_{\mathbf{R}(t)}\psi^{i}_{\mathbf{R}(t)}(x)=\epsilon^{i}_{\mathbf{R}(t)}\psi^{i}_{\mathbf{R}(t)}(x),
\eeq
where $\epsilon^{i}_{\mathbf{R}(t)}$ is an element of the \emph{discrete} spectrum and by $x$ we denote a full set of the coordinates. Both $\epsilon^{i}_{\mathbf{R}(t)}$ and $\psi^{i}_{\mathbf{R}(t)}(x)$ implicitly depend on time $t$. Imagine also that we are interested in the evolution of some state $\psi(t,x)$, which is the $i^{th}$ eigenstate of the Hamiltonian $H_{\mathbf{R}(t)}$ at initial time $t=0$:
\beq \label{QHE46}
\psi(t=0,x)=\psi^{i}_{\mathbf{R}(t)}(x).
\eeq
If the Hamiltonian $H_{\mathbf{R}(t)}$ evolves adiabatically\footnote{Adiabatically means that the flux changes at the time scale $\tau>>\frac{\hbar}{\Delta}$, where $\Delta$ is the excitation energy gap of the spectrum} then the state $\psi(t,x)$ is given by:
\beq \label{QHE47}
\psi(t,x)=\psi^{i}_{\mathbf{R}(t)}(x)e^{i\gamma(t)}e^{-\frac{i}{\hbar}\int^{t}_{0}\epsilon^{i}_{\mathbf{R}(t^{\prime})}dt^{\prime}},
\eeq
i.e., is the $i^{th}$ eigenstate of the Hamiltonian $H_{\mathbf{R}(t)}$ at any time. The phase $-\frac{1}{\hbar}\int\epsilon^{i}_{\mathbf{R}(t^{\prime})}dt^{\prime}$ is an ordinary dynamical phase, while the phase $\gamma(t)$ is an additional correcting phase. The phase $\gamma(t)$ is \emph{not} physical, because it is gauge-variant: we can choose the phases of the eigenstates $\psi^{i}_{\mathbf{R}(t)}(x)$ (\ref{QHE45}) differently as time grows and this non-physical choice definitely affects the phase $\gamma$. However, in the special case of the closed path in the parameter space:
\beq \label{QHE48}
\mathbf{R}(t=0)=\mathbf{R}(t=T)
\eeq
the Berry's phase $\gamma_{B}=\gamma(T)$ is \emph{physical} (gauge-invariant) and is given by the integral:
\beq \label{QHE49}
\gamma_{B}=i\int_{0}^{T}\langle\psi(t)\ket{\frac{d}{dt}\psi(t)} dt=i\oint \langle\psi(t)\ket{\frac{d}{d\mathbf{R}}\psi(t)} d\mathbf{R}, 
\eeq 
where the last relation expresses the Berry's phase as a contour integral in the parameter space.

Let us calculate the Berry's phase $\gamma_{B}$ for a quasihole (\ref{QHE40}), the center of which traverses the circular loop of radius $R>>l_{B}$ in the clockwise direction. Denoting by $N^{m+}$ the normalization factor\footnote{The wave functions (\ref{QHE40}), \ref{QHE41}) are unnormalized.} of the quasihole state (\ref{QHE40}):
\begin{eqnarray}
\gamma_{B}&=& i|N^{m+}|^{2}\int_{0}^{T} dt\langle \Psi^{m+}_{Z_{0}(t)} \ket{\frac{d}{dt}\Psi^{m+}_{Z_{0}(t)}}= \nonumber \\
&=& i|N^{m+}|^{2}\int_{0}^{T} dt\bra{\Psi^{m+}_{Z_{0}(t)}}\frac{d}{dt}\sum_{r} \ln(Z_{r}-Z_{0}(t))\ket{\Psi^{m+}_{Z_{0}(t)}}= \nonumber \\
&=& -i|N^{m+}|^{2}\oint dZ_{0}\bra{\Psi^{m+}_{Z_{0}(t)}}\sum_{r} \frac{1}{Z_{r}-Z_{0}(t)}\ket{\Psi^{m+}_{Z_{0}(t)}}= \nonumber \\
&=& -i|N^{m+}|^{2}\oint dZ_{0} \int dx dy\bra{\Psi^{m+}_{Z_{0}(t)}}\sum_{r} \delta(Z-Z_{r})\frac{1}{Z-Z_{0}(t)}\ket{\Psi^{m+}_{Z_{0}(t)}},  \label{QHE50}
\end{eqnarray}
where $Z=x+iy$ and in the last relation we introduced the integration of the delta function. Note that the mean electron density $\rho^{m+}_{Z_{0}(t)}$ in the state $\Psi^{m+}_{Z_{0}(t)}$ is given by:
\beq \label{QHE51}
\rho^{m+}_{Z_{0}(t)}=|N^{m+}|^{2} \int dx dy\bra{\Psi^{m+}_{Z_{0}(t)}}\sum_{r} \delta(Z-Z_{r})\ket{\Psi^{m+}_{Z_{0}(t)}}.
\eeq
Substituting this expression into (\ref{QHE50}) we readily get:
\beq \label{QHE52}
\gamma_{B}=-i\oint dZ_{0}\int dx dy \rho^{m+}_{Z_{0}}(Z)\frac{1}{Z-Z_{0}}.
\eeq
At this point we divide the integral into two parts writing the mean electron density $\rho^{m+}_{Z_{0}}(Z)$ as $\rho^{m}+\delta\rho^{m+}_{Z_{0}}(Z)$, where $\rho^{m}=\nu\frac{eB}{2\pi\hbar}$ is a homogeneous mean electron density of the Laughlin ground state function (\ref{QHE30}) and $\delta\rho^{m+}_{Z_{0}}(Z)$ is a correction due to the presence of the quasihole. The first part of the integral (\ref{QHE52}) with $\rho^{m}$ can be easily calculated using the residue theorem with the result:
\beq \label{QHE53}
\gamma^{(1)}_{B}=-\nu |e| B S,
\eeq
where $S=\pi R^{2}$. The second part of the integral (\ref{QHE50}) $\gamma_{B}^{(2)}$ vanishes due to symmetry. Due to Aharanov and B\"ohm the particle with charge $q$ acquires the magnetic phase $\Delta \phi_{AB}=-q B S$ (\ref{CS14}) encircling the area $S$ in the magnetic field. Hence we recover our old result that the quasihole has the charge $\nu |e|$.

We examine the statistics of the quasiholes in the similar fashion: we take the state $\Psi^{m++}_{Z_{0}, \widetilde{Z}_{0}}$ with two quasiholes with centers at $Z_{0}$ and $\widetilde{Z}_{0}$:
\beq \label{QHE54}
\Psi^{m++}_{Z_{0}, \widetilde{Z}_{0}}=\prod_{r}(Z_{r}-Z_{0})(Z_{r}-\widetilde{Z}_{0}) \Psi^{m}
\eeq
and perform the adiabatic transport of one quasihole (located at $Z_{0}$) around another static quasihole (located at $\widetilde{Z}_{0}$). The calculation of the Berry's phase goes through the same lines here with the result similar to (\ref{QHE52}):
\beq \label{QHE55}
\gamma_{B}=-i\oint dZ_{0}\int dx dy \rho^{m++}_{Z_{0},\widetilde{Z}_{0}}(Z)\frac{1}{Z-Z_{0}},
\eeq
where the mean electron density can be decomposed now into three parts $\rho^{m++}_{Z_{0},\widetilde{Z}_{0}}(Z)=\rho^{m}+\delta \rho^{m+}_{Z_{0}}+\rho^{m+}_{\widetilde{Z}_{0}}$. The first part gives us the Aharanov-B\"ohm phase (\ref{QHE53}), the second part vanishes due to symmetry as before. Finally, the third part, which corresponds to the presence of the static quasihole produces an additional phase shift:
\beq \label{QHE56}
\gamma_{B}^{(3)}=2\pi \nu.
\eeq
The rotation can be interpreted as a double exchange of the two quasiholes (see \\Chapter \ref{ch:CS}). The statistical exchange phase $\Delta \phi$ is thus given by:
\beq \label{QHE57}
\Delta \phi= \pi \nu.
\eeq
Finally we come to the conclusion: the Laughlin quasiholes with $\nu=\frac{1}{m}, \ m=3,5,...$ are the physical realization of \emph{anyons}. It is important to remark that the fractional statistics of the quasiparticles is deeply connected with their fractional charge.  
\chapter{Conclusions}
\label{ch:conc}
In this diploma thesis we investigated the electrodynamics in the plane. The theories of our main interest were the pure $QED_{3}$, defined by the Maxwell Lagrangian, and its extended version called the Maxwell-Chern-Simons theory. The celebrated Chern-Simons term is characteristic only for odd-dimensional spacetimes and gives rise to non-zero photon mass. We tried to make a general review of the basic findings made in this part of the planar physics and paid special attention to the differences between four-dimensional and three-dimensional physics. At this point we should mention that the literature about the three-dimensional quantum electrodynamics and the Chern-Simons term in particular is very vast and the subject is active. The review is by no means comprehensive. Reason for presenting the (slightly extended) diploma thesis publicly is simple: It gave us much pleasure to collect the knowledge on this intriguing subject, and we optimistically believe that this little work might convey it also to others.

In Chapter \ref{ch:classical} the classical version of the electrodynamics was briefly considered. We showed that the Huygens' principle is violated by both the pure Maxwell and the MCS electrodynamics. This indicates the difficulties in calculations of classical radiation of a point charge. The electric potential of the static point charge in the pure Maxwell electrodynamics is logarithmic, i.e., confining. In contrast in the Maxwell-Chern-Simons theory the point charge generates the electric and the magnetic fields in its vicinity.

In Chapter \ref{ch:Lorentz} we developed the Dirac formalism in the plane. First we constructed the minimal $2\times 2$ realization of the Dirac $\gamma$ matrices in three-dimensional spacetime. The Dirac algebra is, in fact, realized by the Pauli matrices. There is no $\gamma_{5}$ matrix in this $2\times2$ realization; i.e., there is no place for continuous chiral symmetry in the theory. Consequently we explicitly found the plane wave solutions of the Dirac equation and derived the Pauli equation for a charged particle in the electromagnetic field.

In the following Chapter \ref{ch:Poincare} we investigated the unitary irreducible representations of the Poincar\'e group in three-dimensional spacetime. We found out that the stability group for massive and massless particles are $SO(2)\sim R$ and $Z_{2}\otimes R$ respectively. These groups are Abelian and their unitary irreducible representations, which are one-dimensional, are labeled by an arbitrary real number. The striking consequence of this fact is that the spin Hilbert space of relativistic particles is one-dimensional and that spin may take arbitrary real values. It opens up the possibility of the existence of anyons in the plane, i.e., particles with arbitrary spin and statistics. Subsequently we also connected various free covariant fields with the unitary irreducible representations of the Poincar\'e group, which provides a particle content to various field theories.

Chapter \ref{ch:quantization} was devoted to the quantization of free fields with the special emphasis on quantization of the gauge fields in the Coulomb gauge. We demonstrated that the pure Maxwell electrodynamics is the theory of massless photons with zero spin. As far as the Maxwell-Chern-Simons theory is concerned, exploiting the well-known Dirac procedure, we showed that it describes massive photons with spin one.

In Chapter \ref{ch:discrete} we were dealing with the subtleties of discrete transformations for massless and massive particles in three-dimensional spacetime. First we illustrated the problems, which arise after employing the naive definitions. Subsequently the general approach was developed, which relies on basic commutation relations. We showed that parity and time-reversal are well-defined for massless particles. For massive particles, however, we were not able to find meaningful parity and time-reversal operators. After that combined symmetry was introduced and it was shown that this is a good symmetry for massive particles. We also demonstrated that there are no difficulties in defining the charge-conjugation transformation for both massive and massless particles.

In Chapter \ref{ch:perturbative} some perturbative calculations were performed in the MCS theory. The theory is superrenormalizable ($e^{2}$ has dimension of mass) and it has only a few diagrams with non-negative superficial degree of divergence $D$. We calculated two of them (one-loop vacuum polarization diagram and one-loop electron self-energy diagram) using dimensional regularization. The vacuum polarization renormalizes the mass and the wave function, but preserves the generic structure of the photon propagator. As for the electron self-energy is concerned we demonstrated that the renormalized fermion mass is a gauge-dependent quantity. We also showed that the infrared divergences may be encountered in higher-order calculations. These problems are cured by using the Landau gauge. We discussed the reasons why this gauge is exceptional. Finally, the running coupling constant was also calculated in the MCS theory in the framework of renormalized perturbative theory with the light photon $\mu\to 0$. It was shown that to one-loop order it has the regular behavior for all energy scales.

In the following Chapter \ref{ch:effect} we examined more thoroughly the CS term induced by massive fermions. First, the notion of the one-loop effective action was introduced. We developed the perturbative expansion of the effective action in the coupling constant $e$ and demonstrated the emergence of the CS term for the slowly varying background electromagnetic field. Subsequently, the powerful non-perturbative "proper-time" technique developed by J. Schwinger was employed for calculation of the effective action in the constant background gauge field.

Chapter \ref{ch:non-perturbative} was devoted to non-relativistic electron-electron interaction in two spatial dimensions. With classical heuristic consideration in mind we showed that it is plausible to assume the existence of bound states. Using first the Pauli equation and then the non-relativistic reduction of the tree-level Feynman diagrams we showed that the presence of the Chern-Simons term in the Lagrangian opens the possibility of an attraction of two electrons.  We also briefly discussed the possibility of the existence of non-relativistic bound states in the theory. 

The pure Chern-Simons theory (without the Maxwell term) also attracts a lot of attention because it describes anyons. In Chapter \ref{ch:CS} we introduced and investigated the pure Chern-Simons theory. The one-loop calculations of the non-relativistic scattering of two fermions was performed first in the Pauli theory and then in the fully relativistic Chern-Simons theory. The amplitudes, obtained by these two different approaches, differ by a term, which comes from the modification of the spin of electron. It is a manifestation of the spin transmutation, which arises in the pure Chern-Simons theory due to the one-loop vertex correction digram.

The final Chapter \ref{ch:QHE} was devoted to arguably the most prominent application of the planar physics- the quantum Hall effect. First we briefly introduced this quantum phenomenon. We presented the theoretical explanation of the integer quantum Hall effect and $\frac{1}{m}$ fractional quantum Hall effect. Subsequently, the charge and the statistics of the Laughlin quasiparticles were examined using the Berry's phase technique. As the main outcome of this Chapter, it turned out that the quasiparticles have fractional charge and are the anyons. Although QHE definitely belongs to the realm of the non-relativistic quantum physics, which was not the main topic of this thesis; it provides clear and experimentally confirmed realization of the anyons, particles which we met few times in this work.

\appendix
\chapter{Aspects of classical and quantum mechanics} \label{appendA}
\section*{Classical mechanics}
Let us assume that the classical gravitation potential $\phi(x)$ in the plane solves the Poisson equation:
\beq \label{eq:ClM0} \Delta\phi(x)=\rho(x),  \eeq
where $\Delta$ is the two-dimensional Laplace operator and $\rho(x)$ is a matter density field. Solving this equation for a point particle with mass $M$ in two spatial dimensions we readily obtain the logarithmic gravitational potential $\phi(r)$: \beq \label{eq:ClM0a}\phi(r)=\frac{M}{2\pi}\ln{r}. \eeq
The logarithmic potential does not have an asymptote for $r\to \infty$, so every two massive particles are bound by gravity. It is a well-known fact that there are only two types of central potentials ($\phi\sim \frac{1}{r}$ and $\phi\sim r^{2}$) in classical mechanics, which create closed orbits for general finite motions. In the case of the logarithmic potential general orbits are not closed. That is why it is not possible to find an explicit solution of the Kepler problem- motion of a point mass in the central logarithmic field. We can try to derive Rutherford's formula for the scattering experiment, but some difficulties arise. First we shall try to calculate a scattering angle $\delta \alpha$ as a function of initial(=final) distance $R$ between a scattering test particle and a source of the mass $M$. Total energy is conserved so we obtain:
\beq \label{eq:ClM0b}
v^{2}_{0}+\frac{M}{\pi}\ln(R)=v^{2}+\frac{M}{\pi}\ln(r), 
\eeq
where $v^{2}_{0}$ resp. $R$ are the initial velocity of the scattering particle resp. the initial distance between the particle and the source. Angular momentum is a pseudoscalar quantity in the planar world:  
\beq \label{eq:ClM2}  
L=xp_{y}-yp_{x}
\eeq     
and is also conserved in the central field:
\beq \label{eq:ClM0c}
v_{0}b=\frac{d\alpha}{dt}r^{2},
\eeq
where $b$ is an impact parameter and $\alpha$ is a polar angle of the particle. Combining (\ref{eq:ClM0b}) and (\ref{eq:ClM0c}) we obtain:
\beq \label{eq:ClM0d}
 \frac{d\alpha}{d r}=\frac{v_{0}b}{r^{2}\sqrt{v_{0}^{2}+\frac{M}{\pi}\ln{\frac{R}{r}}-(\frac{v_{0}b}{r})^{2}}}.
\eeq 
This equation can be formally integrated for the initial distance $r_{in}=R$ and the final distance $r_{f}=R$:
\beq \label{eq:ClM1}
\delta \alpha=2 \int\limits_{r_{min}}^{R}\frac{v_{0}b}{r^{2}\sqrt{v_{0}^{2}+\frac{M}{\pi}\ln{\frac{R}{r}}-(\frac{v_{0}b}{r})^{2}}} dr,
\eeq
where $r_{min}$ is a minimal distance between the particle and the source. It can be easily calculated from (\ref{eq:ClM0b}).  We were not able to calculate this integral. As was mentioned before the logarithmic potential does not have an asymptote for $r\to \infty$, so it is meaningless to take $R\to \infty$ limit in (\ref{eq:ClM1}) because for sufficiently large $R$ we enter relativistic domain, so (\ref{eq:ClM1}) is not applicable anymore.

\section*{Some aspects of non-relativistic quantum mechanics}
Here we shall review some interesting aspects of quantum mechanics in the plane. In analogy with (\ref{eq:ClM2}) the angular momentum $\hat{L}$ operator is a pseudoscalar quantity:
\beq \label{eq:QM0a} \hat{L}=\hat{x_{1}}\hat{p_{2}}-\hat{x_{2}}\hat{p_{1}}, \eeq
which acquires the compact form in polar coordinates: 
\beq \label{eq:QM0b}  \hat{L}=-i\frac{\partial}{\partial\phi}. \eeq
Eigenfunctions of $\hat{L}$ are:
\beq \label{eq:QM0c} \psi^{(m)}(\phi)=\frac{1}{2\pi}e^{i m\phi}, \eeq
where the eigenvalue $m$ is an integer for single-valued polar functions. We know that $\hat{L}$ is a generator of plane rotations ($SO(2)$) in the Hilbert space of the particle. $SO(2)$ is infinitely connected space with universal covering group $R$, the additive group of real numbers. So $SO(2)$ has multivalued representations. If we also want to take multivalued representations into our consideration, $m$ may be any real number\footnote{This corresponds to quantum-mechanical realization anyons}.   

The group $SO(2)$ is Abelian, so the coupling of two angular momenta is trivial:
\beq \label{eq:QM0d} \ket{M=m_{1}+m_{2}}=\ket{m_{1}}\ket{m_{2}}, \eeq
where $\ket{M}$ is an eigenfunction of the total angular momentum and $\ket{m_{1}}$ resp. $\ket{m_{2}}$ are the eigenfunctions of the angular momentum of the first resp. second particle.
Let us now find stationary states of a free particle; i.e., we would like to solve the stationary Schr\"odinger equation:
\beq \label{eq:QM0d0}
\hat{H}\ket{\psi_{E}}=E\ket{\psi_{E}}.
\eeq
The Hamiltonian $H$ in polar coordinates is:
\beq \label{eq:QM0e}
\hat{H}=\frac{\hat{\mathbf{P}}^{2}}{2M}=-\frac{\hbar^{2}}{2M}\left(\frac{\partial^{2}}{\partial r^{2}}+\frac{1}{r}\frac{\partial}{\partial r}+\frac{1}{r^{2}}\frac{\partial^{2}}{\partial \phi^{2}}\right).
\eeq
This Hamiltonian $\hat{H}$ commutes with $\hat{L}$, so we can seek the stationary state with definite angular momentum $m$:
\beq \label{eq:QM0f}
\psi^{(E,m)}(r,\phi)=f^{(E,m)}(r)\frac{1}{2\pi}e^{im\phi},
\eeq 
where $f^{(E,m)}(r)$ is a radial function, that solves:
\beq \label{eq:QM0g}
z^{2}f^{\prime \prime}(z)+zf^{\prime}+(z^{2}-m^{2})f(z)=0,
\eeq
where $z=\kappa r$ and $\kappa^{2}=\frac{2ME}{\hbar}$. In the last equation we recognize the well-known Bessel equation, which has the general solution for integer $m$:
\beq \label{eq:QM0h}
f(\kappa r)=C_{1}J_{m}(\kappa r)+C_{2}N_{m}(\kappa r),
\eeq
where $J_{m}(\kappa r)$ resp. $N_{m}(\kappa r)$ are the Bessel (of the first kind) resp. the Neumann function (Bessel function of the second kind) and $C_{i}$ are some constants. Neumann functions, however, are singular in the origin:
\beq \label{eq:QM0k}  
\begin{array}{cc}
N_{m}(z)\sim \ln(z) & m=0 \\
N_{m}(z)\sim z^{-m} & m\ne0,
\end{array}
\eeq
so they do not solve the original equation (\ref{eq:QM0d0}). Our final solution is: 
\beq \label{eq:QM1}
\psi_{m,\kappa}(r,\phi)=CJ_{m}(\kappa r)\frac{1}{2\pi}e^{im\phi}.
\eeq
These solutions are normalized to delta function $\delta(E-E^{\prime})$. This normalization condition fixes the unknown constant $C$. The spectrum is continuous and non-degenerate.

If we add a central potential $\hat{V}(r)$ to the Hamiltonian $\hat{H}$ we obtain the new Hamiltonian:
\beq \label{eq:QM0l}
\hat{H}=\frac{\hat{\mathbf{P}}^{2}}{2M}+\hat{V}(r),
\eeq
which also commutes with $\hat{L}$. The stationary states may be written as:
\beq \label{eq:QM0m}
\psi^{(E,m)}(r,\phi)=\frac{1}{\sqrt{r}}f^{(E,m)}(r)\frac{1}{2\pi}e^{im\phi},
\eeq
where the radial function $f^{(E,m)}(r)$ solves the 1-dimensional Schr\"odinger equation:
\beq \label{eq:QM0n}
\left[\frac{d^{2}}{dr^{2}}+\frac{2M}{\hbar^{2}}(E-V_{ef}(r,m))\right]f^{(E,m)}(r)=0
\eeq 
in the interval $r\in(0,\infty)$. The function $\frac{1}{\sqrt{r}}f^{(E,m)}(r)$ must be regular in the origin in order to solve the Schr\"odinger equation (\ref{eq:QM0d0}). The effective potential is:
\beq \label{eq:QM0o}
V_{ef}(r,m)=V(r)+\frac{\hbar^{2}}{2M}\frac{m^{2}-\frac{1}{4}}{r^{2}},
\eeq
where the second term is a $m$-dependent centrifugal barrier. The bound states must satisfy the normalization condition:
\beq \label{eq:QM0p}
\int^{\infty}_{0}dr f^{(E,m)}(r)f^({E^{\prime}m})(r)=1
\eeq
and the states from the continuous part of the energy spectrum must be normalized to $\delta(E-E^{\prime})$:
\beq \label{eq:QM0r}
\int^{\infty}_{0}dr f^{*(E,m)}(r)f^({E^{\prime}m})(r)=\delta(E-E^{\prime}).
\eeq

The central potential $V(r)\sim \ln(r)$ is of particular interest because it is a Coulomb potential in the plane. It's not possible, however, to find an explicit analytical solution for the spectrum and the eigenfunctions for this kind of potential\cite{Eveker:1990}. 

It is shown in \cite{Landau:QM}, that the scattering amplitude in the plane can be expressed as follows:
$$ \psi(\rho, \phi)=e^{ikx}+f(\phi)\frac{e^{ik\rho}}{\sqrt{-i\rho}} \quad f(\phi)=\frac{1}{i\sqrt{2\pi k}}\sum^{+\infty}_{m=-\infty}(e^{2i\delta_{m}}-1)e^{im\phi}, $$
where $\delta_{m}$ are the phase factors. 

\chapter{Quantum electron in external magnetic field} \label{appendB}
The quantum evolution of a charged particle in an external magnetic field is governed by the Hamiltonian:
\beq
\mathscr{H}=\frac{1}{2m}(\mathbf{p}+e\mathbf{A}(\mathbf{x}))^{2}=\frac{1}{2m}\left\{(p^{x}+eA^{x}(\mathbf{x}))^{2}+(p^{y}+eA^{y}(\mathbf{x}))^{2}\right\}.
\eeq
In Chapter \ref{ch:QHE} the spectrum of this Hamiltonian was found exploiting the general algebraic relations. To find the eigenstates it is necessary, however, to fix the gauge. In this Appendix we use one particular gauge, which plays  a central role in the QHE calculations- the symmetric gauge. In this gauge both components of the vector potential are treated symmetrically:
\beq \label{sg1}
A^{x}=\frac{1}{2}B_{\bot}y \qquad A^{y}=-\frac{1}{2}B_{\bot}x,
\eeq
which describes a homogeneous magnetic field $B=(0,0,-B_{\bot})$. The two sets of the creation and annihilation operators introduced in (\ref{QHE16}) take the following form in the symmetric gauge:
\begin{eqnarray} \label{sg2}
a=-\frac{i}{\sqrt{2}}\left[z+\frac{\partial}{\partial z^{*}}\right] &\quad & a^{\dag}=\frac{i}{\sqrt{2}}\left[z^{*}-\frac{\partial}{\partial z}\right] \nonumber \\
b=\frac{1}{\sqrt{2}}\left[z^{*}+\frac{\partial}{\partial z}\right] &\quad & b^{\dag}=\frac{1}{\sqrt{2}}\left[z-\frac{\partial}{\partial z^{*}}\right],
\end{eqnarray}
where we introduced the dimensionless complex variables, which are very useful in this treatment:
\begin{eqnarray} \label{sg3}
z=\frac{1}{2l_{B}}(x+iy) &\quad& z^{*}=\frac{1}{2l_{B}}(x-iy) \nonumber \\
\frac{\partial}{\partial z}=l_{B}(\frac{\partial}{\partial x}-i\frac{\partial}{\partial y}) &\quad&
\frac{\partial}{\partial z^{*}}=l_{B}(\frac{\partial}{\partial x}+i\frac{\partial}{\partial y}),
\end{eqnarray}
$l_{B}=\sqrt{\frac{\hbar}{eB_{\bot}}}$ is the magnetic length.

Recalling that $a$ is the annihilation operator of the Landau levels, the lowest Landau level (LLL) condition $a\ket{\psi_{L=0}}=0$ looks in the symmetric gauge as:
\beq \label{sg4}
\left(z+\frac{\partial}{\partial z^{*}}\right)\psi_{L=0}(z,z^{*})=0,
\eeq
where by $\psi_{L=0}$ we denote the general state from the lowest Landau level. This condition can be solved:
\beq \label{sg5}
\psi_{L=0}(z,z^{*})=f(z)e^{-|z|^{2}}=\sum_{n=0}^{\infty}f_{n}\underbrace{z^{n}e^{-|z|^{2}}}_{\bar{\psi}^{n}_{L=0}},
\eeq
where $f(z)$ is some general function of $z$ (only), i.e., $f(z)$ is the analytic function of $z$. The analytic function $f(z)$ can be expanded in the Taylor series around $z=0$ and the general LLL wave function can be written as a linear combination of unnormalized wave functions $\bar{\psi}^{n}_{L=0}$. What is a physical meaning of the number $n$? To answer this question, we express $\bar{\psi}^{n}_{L=0}$ in the polar coordinates:
\beq \label{sg6}
\bar{\psi}^{n}_{L=0}(r,\phi)=e^{in\phi}\left(\frac{r}{2l_{B}} \right)^{n}\exp(-\frac{r^{2}}{4l^{2}_{B}}),
\eeq
where $\phi$ is the polar angle. The physical meaning of $n$ is now clear: $\bar{\psi}^{n}_{L=0}$ are the unnormalized eigenstates of the angular momentum operator and $n$ is the angular momentum quantum number, which differentiates the states in the LLL. Taking care of the normalization, we obtain the LLL eigenstates of angular momentum $\psi^{n}_{L=0}(z,z^{*})$:
\beq \label{sg7}
\psi^{n}_{L=0}(z,z^{*})=\underbrace{\sqrt{\frac{2^{n}}{2\pi l^{2}_{B}n!}}}_{N_{n}}z^{n}e^{-|z|^{2}}.
\eeq
Using (\ref{sg2}) it is easy to demonstrate that:
\beq \label{sg8}
b^{\dag}\psi^{n}_{L=0}=\sqrt{n+1}\psi_{L=0}^{n+1} \qquad b\psi^{n}_{L=0}=\sqrt{n}\psi_{L=0}^{n-1}
\eeq
and hence $b^{\dag}$ increases the angular momentum of the state by one unit and $b$ decreases it by the same amount. 

The whole LLL Hilbert space can now be easily constructed: just take the vacuum state $\ket{0}=\psi^{0}_{L=0}$ and act gradually on it with the creation operators $b^{\dag}$. The result is:
\beq \label{sg9}
\psi^{n}_{L=0}=\frac{1}{\sqrt{n!}}(b^{\dag})^{n}\ket{0}=\frac{1}{\sqrt{2^{n+1}\pi l_{B}^{2}n!}}\left(z-\frac{\partial}{\partial z^{*}} \right)^{n}\exp(-|z|^{2}).
\eeq

The whole one-particle Hilbert space can be constructed applying both the angular momentum creation operators $b^{\dag}$ and the energy creation operators $a^{\dag}$ on the same vacuum as before:
\begin{eqnarray} \label{sg10}
\psi^{n}_{L=m}&=&\frac{1}{\sqrt{(n+m)!m!}}(a^{\dag})^{m}(b^{\dag})^{n+m}\ket{0} \nonumber \\
              &=&\frac{i^{m}}{\sqrt{2^{n+2m+1}\pi l_{B}^{2}m!(n+m)!}}\left(z^{*}-\frac{\partial}{\partial z} \right)^{m}\left(z-\frac{\partial}{\partial z^{*}} \right)^{n+m}\exp(-|z|^{2}).
\end{eqnarray}
This state lies in the $m^{th}$ Landau level and has the angular momentum $n$. Note that the creation operator $a^{\dag}$ not only increases the energy level by one unit, but also simultaneously decreases the angular momentum by one unit. It can be easily understood, expressing the angular momentum operator $L$ in terms of the creation and the annihilation operators:
\beq \label{sg10a}
L=xp^{x}-yp^{y}=\hbar(b^{\dag}b-a^{\dag}a).
\eeq

To gain the physical intuition we go back to the LLL and calculate the probability density of the state $\psi_{L=0}^{n}$:
\beq \label{sg11}
|\psi^{n}_{L=0}|^{2}(r)=N_{n}^{2}\left(\frac{r}{2l_{B}}\right)^{2n}\exp(-\frac{r^{2}}{2l_{B}^{2}})
\eeq
The probability density is angle-independent and as a function of $r$ has a sharp peak at $r_{n}=\sqrt{2n}l_{B}$. The radius $r_{n}$ defines the center of the ring-band, where the probability is non-vanishing. The width of the ring-band is of order $l_{B}$.

We also calculate the probability density current, given by the formula:
\beq \label{sg12}
j^{i}=-\frac{ie\hbar}{2m}\left[\psi^{*}(D^{i}\psi)-(D^{i}\psi)^{*}\psi \right] \qquad D^{i}=\nabla^{i}+\frac{ie}{\hbar}A^{i}.
\eeq
Substituting $\psi^{n}_{L=0}$ into the last relation we obtain that only the angular part of the current is non-vanishing:
\beq \label{sg13}
j^{\phi}(r)=N^{2}\frac{e\hbar}{2m}\left[\frac{2n}{r}-\frac{r}{l_{B}^{2}} \right]\left(\frac{r}{2l}\right)^{2n}\exp\left(-\frac{r^{2}}{2l_{B}^{2}}\right).
\eeq
The current circulates in the clockwise direction (for $r>r_{n}$) and in counterclockwise direction (for $r<r_{n}$) in the ring-band of radius $r_{n}$.    
\bibliographystyle{plain}
\bibliography{dip}
\end{document}